%% file: thesis.tex
\documentclass[12pt,twoside,titlepage]{book}
\input{definitions} 

\newcommand{\mytitle}{Accelerating Discovery: Model-Agnostic Likelihoods for the Reinterpretation of Particle Physics Results and their Application to the Belle~II \texorpdfstring{$\boldsymbol{\BKnn}$}{B->Knunubar} Measurement}

\newcommand{\myname}{Lorenz G\"artner}
\newcommand{\handindate}{4.11.2025}

\newcommand{\changefont}{
    \fontsize{8}{11}\selectfont
}

\fancypagestyle{noheader}{
    \fancyhf{}
    \fancyfoot[LE,RO]{\thepage}
    
}

\fancypagestyle{plain}{
    \fancyhf{}
    \fancyfoot[LE,RO]{\thepage}             
    
}

\graphicspath{{figs/}}  

\begin{document}


\pagestyle{empty}

\begin{titlepage}
\input{titlepages/cover}
\end{titlepage}
\clearpage

\begin{titlepage}
\input{titlepages/title}
\end{titlepage}
\clearpage




\pagestyle{fancy}

\pagenumbering{Roman}

\chapter*{Zusammenfassung}
\thispagestyle{fancy}
\addcontentsline{toc}{chapter}{Zusammenfassung}
\input{chapters/abstract-german}

\chapter*{Abstract}
\thispagestyle{fancy}
\addcontentsline{toc}{chapter}{Abstract}
\input{chapters/abstract}

\tableofcontents


\cleardoublepage
\pagenumbering{arabic}
\setcounter{page}{1}

\chapter{Introduction}
\label{sec:intro}

\input{chapters/introduction}

\chapter{Theoretical background}
\label{sec:theory}
\input{chapters/theory}

\chapter{The Belle II experiment}
\label{sec:experiment}
\input{chapters/experiment}

\chapter{Binned statistical models in particle physics}
\label{sec:stat-models}
\input{chapters/stat-models}

\chapter{Statistical inference in particle physics}
\label{sec:stat-inference}
\input{chapters/stat-inference}

\chapter{Ongoing reinterpretation efforts in particle physics}
\label{sec:reinterpretation-intro}

\input{chapters/reinterpretation-review}

\chapter{Model-agnostic likelihoods for the reinterpretation of particle physics results}
\label{sec:method}
\input{chapters/method}

\chapter{A model-agnostic likelihood for the Belle II \texorpdfstring{$\boldsymbol{\BKnn}$}{B+->K+nunubar} measurement}
\label{sec:reinterpretation knunu mall}
\input{chapters/knunu-mall}

\chapter{A Weak Effective Theory reinterpretation of the Belle II \texorpdfstring{$\boldsymbol{\BKnn}$}{B+->K+nunubar} measurement}
\label{sec:reinterpretation knunu wet}
\input{chapters/knunu-wet}

\chapter{A light new physics reinterpretation of the Belle II \texorpdfstring{$\boldsymbol{\BKnn}$}{B+->K+nunubar} measurement}
\label{sec:reinterpretation knunu bkx}
\input{chapters/knunu-bkx}

\chapter{A combined analysis of \texorpdfstring{$\boldsymbol{B \to K^{(*)} \nu\bar{\nu}}$}{B->K(*)nunubar} decays at Belle~II}
\label{sec:knunu update}
\input{chapters/errors_syst}
\input{chapters/errors_syst_pyhf}
\input{chapters/knunu-update}

\chapter{Conclusion}
\label{sec:conclusion}
\input{chapters/conclusion}

\chapter*{List of Publications}
\addcontentsline{toc}{chapter}{List of Publications}
\input{chapters/publications}


\appendix

\chapter{Theoretical background}
\input{chapters/appendix-theory}

\chapter{Statistical inference in particle physics}
\input{chapters/appendix-stat}

\chapter{Model-agnostic likelihoods for the reinterpretation of particle physics results}
\input{chapters/appendix-method}

\chapter{A model-agnostic likelihood for the Belle II \texorpdfstring{$\boldsymbol{\BKnn}$}{B+->K+nunubar} measurement}
\input{chapters/appendix-knunu-mall}

\chapter{A Weak Effective Theory reinterpretation of the Belle II \texorpdfstring{$\boldsymbol{\BKnn}$}{B+->K+nunubar} measurement}
\input{chapters/appendix-knunu-wet}

\chapter{A light new physics reinterpretation of the Belle II \texorpdfstring{$\boldsymbol{\BKnn}$}{B+->K+nunubar} measurement}
\input{chapters/appendix-knunu-bkx}

\chapter{A combined analysis of \texorpdfstring{$\boldsymbol{B \to K^{(*)} \nu\bar{\nu}}$}{B->K^(*)nunubar} decays at Belle~II}
\input{chapters/appendix-knunu-update}

\chapter*{List of acronyms}
\addcontentsline{toc}{chapter}{List of Acronyms}
\input{chapters/acronym}




\printbibliography

\chapter*{Disclaimer on the use of language tools}
\pagestyle{noheader}
\addcontentsline{toc}{chapter}{Disclaimer on the use of language tools}
\input{chapters/ai-disclaimer}
\clearpage

\chapter*{Acknowledgements}
\addcontentsline{toc}{chapter}{Acknowledgements}
\input{chapters/acknowledgements}
\clearpage

\end{document}

%% file: definitions.tex
\usepackage{amsmath, amssymb}  
\usepackage{bbold}             

\usepackage[utf8]{inputenc}    
\usepackage[english]{babel}    
\usepackage{ragged2e}          
\usepackage{graphicx}          
\usepackage{epstopdf}          
\usepackage{xcolor}            
\usepackage{color}             
\usepackage{fancyhdr}          
\usepackage{hyperref}          
\hypersetup{
  colorlinks = true,
  linkcolor = purple,
  urlcolor = purple,
  citecolor = purple,
  pdfpagemode = {UseNone},
  pdftitle = {Accelerating Discovery: Model-Agnostic Likelihoods for the Reinterpretation of Particle Physics Results and their Application to the Belle~II \texorpdfstring{$\boldsymbol{\BKnn}$}{B->Knunubar} Measurement},
  pdfauthor = {Lorenz G\"artner},
  pdflang = {en-GB}
}
\usepackage[noabbrev, nameinlink, capitalise]{cleveref}          
\usepackage{acronym}           
\makeatletter
\renewcommand*\AC@verridelabel[1]{%
  \ifhmode\else\leavevmode\fi
  \@bsphack
  \protected@write\@auxout{}{\string\AC@undonewlabel{#1}}%
  \ifdefined\ltx@label
    \ltx@label{#1}%
  \else
    \label{#1}%
  \fi
  \AC@overriddenmessage rs{#1}%
  \@esphack
}
\makeatother

  \begingroup
  \newcommand{\ThesisAcronym}[3]{\newacro{##1}[##2]{##3}}%
  \newcommand{\ThesisAcronymPlural}[2]{\newacroplural{##1}{##2}}%
  \input{chapters/acronym-entries}  \endgroup

\usepackage{blindtext}         
\usepackage{subfiles}          
\usepackage{ifthen}            
\usepackage{orcidlink}         
\usepackage{makecell}
\usepackage{pdfpages}
\usepackage[toc,page]{appendix}
\usepackage{parskip}
\usepackage{booktabs}       
\usepackage{tabularx}       
\usepackage{slashed}
\usepackage{rotating}
\usepackage{tikz}
\usepackage[compat=1.1.0]{tikz-feynman} 
\newcolumntype{Y}{>{\centering\arraybackslash}X}

\usepackage[backend=biber,style=numeric-comp,sorting=none,sortcites=true]{biblatex}
\defbibenvironment{notnumbered}
  {\list{}
     {\setlength{\leftmargin}{\bibhang}%
      \setlength{\itemindent}{-\leftmargin}%
      \setlength{\itemsep}{\bibitemsep}%
      \setlength{\parsep}{\bibparsep}}}
  {\endlist}
  {\item}

\usepackage[
  a4paper,
  inner=30mm,   
  outer=20mm,   
  top=25mm,
  bottom=25mm,
  includehead,
  includefoot
]{geometry}

\usepackage[displaymath, mathlines]{lineno}  

\usepackage{fvextra}
\newenvironment{minted}[2][]{%
  \VerbatimEnvironment
  \begin{Verbatim}[breaklines,breakanywhere,#1]%
}{%
  \end{Verbatim}%
}
\newcommand{\usemintedstyle}[1]{}
\graphicspath{{figures/}} 

\newboolean{articletitles}
\setboolean{articletitles}{true} 

\input{belle2-symbols}

\newcommand{\BKnunu}{\ensuremath{{B\to K\nu\bar{\nu}}}\xspace}
\newcommand{\BKsnunu}{\ensuremath{{B\to K^{*}\nu\bar{\nu}}}\xspace}

\newcommand{\BKnn}{\ensuremath{{B^{+}\to K^{+}\nu\bar{\nu}}}\xspace}
\newcommand{\BKX}{\ensuremath{{B^{+}\to K^{+}X}}\xspace}

\newcommand{\BpKpnn}{\ensuremath{{\Bp \to \Kp\nu\bar{\nu}}}\xspace}
\newcommand{\BzKshnn}{\ensuremath{{\Bz \to \KS \nu\bar{\nu}}}\xspace}
\newcommand{\BKstnn}{\ensuremath{{B \to \Kstar\nu\bar{\nu}}}\xspace}
\newcommand{\BpKstnn}{\ensuremath{{\Bp \to \Kstarp\nu\bar{\nu}}}\xspace}
\newcommand{\BzKstnn}{\ensuremath{{\Bz \to \Kstarz\nu\bar{\nu}}}\xspace}

\def\BDT#1{\ensuremath{\mathrm{BDT}_#1}\xspace}

\def\BDT#1{\ensuremath{\mathrm{BDT}_#1}\xspace}

\def\combinationBFdetailed{\ensuremath{\left(2.3 \pm 0.5(\mathrm{stat})^{+0.5}_{-0.4}(\mathrm{syst})\right)\cdot 10^{-5}}\xspace}

\def\combinationsigSM{\ensuremath{2.7}\xspace}

\def\combinationsigO{\ensuremath{3.5}\xspace}

\newcommand{\data}{\ensuremath{\boldsymbol{x}}\xspace}
\newcommand{\yields}{\ensuremath{\boldsymbol{n}}\xspace}
\newcommand{\aux}{\ensuremath{\boldsymbol{a}}\xspace}
\newcommand{\expectation}{\ensuremath{\boldsymbol{\nu}}\xspace}
\newcommand{\estimator}{\ensuremath{\boldsymbol{\hat \theta}}\xspace}
\newcommand{\mapestimator}{\ensuremath{\boldsymbol{\hat \theta}_{\mathrm{MAP}}}\xspace}
\newcommand{\params}{\ensuremath{\boldsymbol{\theta}}\xspace}
\newcommand{\unconstr}{\ensuremath{\boldsymbol{\eta}}\xspace}
\newcommand{\constr}{\ensuremath{\boldsymbol{\chi}}\xspace}

\newcommand{\cdf}{\ensuremath{\text{CDF}}\xspace}

\newcommand{\pois}{\ensuremath{\text{Pois}}\xspace}
\newcommand{\norm}{\ensuremath{\mathcal{N}}\xspace}

\newcommand{\histfactory}{\texttt{HistFactory}\xspace}
\newcommand{\pyhf}{\texttt{pyhf}\xspace}
\newcommand{\EOS}{\texttt{EOS}\xspace}

\newcommand\identity{1\kern-0.25em\text{l}}

%% file: chapters/acronym-entries.tex
\ThesisAcronym{ARICH}{ARICH}{Aerogel Ring-Imaging Cherenkov detector}
\ThesisAcronym{BCL}{BCL}{Bourrely-Caprini-Lellouch}
\ThesisAcronym{BDT}{BDT}{Boosted Decision Tree}
\ThesisAcronym{BSZ}{BSZ}{Bharucha-Straub-Zwicky}
\ThesisAcronym{BSM}{BSM}{Beyond the Standard Model}
\ThesisAcronym{CDC}{CDC}{Central Drift Chamber}
\ThesisAcronym{CDF}{CDF}{Cumulative Distribution Function}
\ThesisAcronym{CKM}{CKM}{Cabibbo-Kobayashi-Maskawa}
\ThesisAcronym{CP}{$CP$}{Charge-Parity}
\ThesisAcronym{d.o.f.}{d.o.f.}{degrees of freedom}
\ThesisAcronym{ECL}{ECL}{Electromagnetic Calorimeter}
\ThesisAcronym{EFT}{EFT}{Effective Field Theory}
\ThesisAcronymPlural{EFT}{Effective Field Theories}
\ThesisAcronym{FAIR}{FAIR}{Findable, Accessible, Interoperable, Reusable}
\ThesisAcronym{FCNC}{FCNC}{Flavour Changing Neutral Current}
\ThesisAcronym{GIM}{GIM}{Glashow-Iliopoulos-Maiani}
\ThesisAcronym{HDI}{HDI}{Highest Density Interval}
\ThesisAcronym{HEP}{HEP}{High-Energy Physics}
\ThesisAcronym{HSR}{HSR}{Highest-Sensitivity Region}
\ThesisAcronym{HS3}{HS3}{HEP Statistics Serialization Standard}
\ThesisAcronym{HTA}{HTA}{Hadronic Tagging Analysis}
\ThesisAcronym{ITA}{ITA}{Inclusive Tagging Analysis}
\ThesisAcronym{KLM}{KLM}{$K_L^0$ and muon detector}
\ThesisAcronym{LHC}{LHC}{Large Hadron Collider}
\ThesisAcronym{MAP}{MAP}{Maximum A Posteriori estimator}
\ThesisAcronym{MC}{MC}{Monte Carlo}
\ThesisAcronym{MCMC}{MCMC}{Markov Chain Monte Carlo}
\ThesisAcronym{mDST}{mDST}{mini Data Summary Table}
\ThesisAcronym{MLE}{MLE}{Maximum Likelihood Estimator}
\ThesisAcronym{OPE}{OPE}{Operator Product Expansion}
\ThesisAcronym{PDF}{PDF}{Probability Density Function}
\ThesisAcronym{PID}{PID}{Particle Identification}
\ThesisAcronym{PMNS}{PMNS}{Pontecorvo-Maki-Nakagawa-Sakata}
\ThesisAcronym{POI}{POI}{Parameter Of Interest}
\ThesisAcronymPlural{POI}{Parameters Of Interest}
\ThesisAcronym{PXD}{PXD}{Pixel Vertex Detector}
\ThesisAcronym{QCD}{QCD}{Quantum Chromodynamics}
\ThesisAcronym{QED}{QED}{Quantum Electrodynamics}
\ThesisAcronym{ROE}{ROE}{Rest-Of-Event}
\ThesisAcronym{ROI}{ROI}{Region Of Interest}
\ThesisAcronymPlural{ROI}{Regions Of Interest}
\ThesisAcronym{SM}{SM}{Standard Model}
\ThesisAcronym{SR}{SR}{Signal Region}
\ThesisAcronym{SVD}{SVD}{Silicon Vertex Detector}
\ThesisAcronym{TOP}{TOP}{Time-Of-Propagation counter}
\ThesisAcronym{UV}{UV}{ultraviolet}
\ThesisAcronym{vev}{vev}{vacuum expectation value}
\ThesisAcronym{VXD}{VXD}{Vertex Detector}
\ThesisAcronym{WET}{WET}{Weak Effective Theory}

%% file: titlepages/cover.tex
\newgeometry{left=25mm, right=25mm, top=25mm, bottom=25mm}
\begin{center}
    
\vspace{0.5cm}
      
\rule{\textwidth}{1.5pt}
\huge
\textbf{\mytitle}
\rule{\textwidth}{1.5pt}

\vfill

\LARGE
\textbf{\myname}
      
\vfill

\includegraphics[width = 0.4\textwidth]{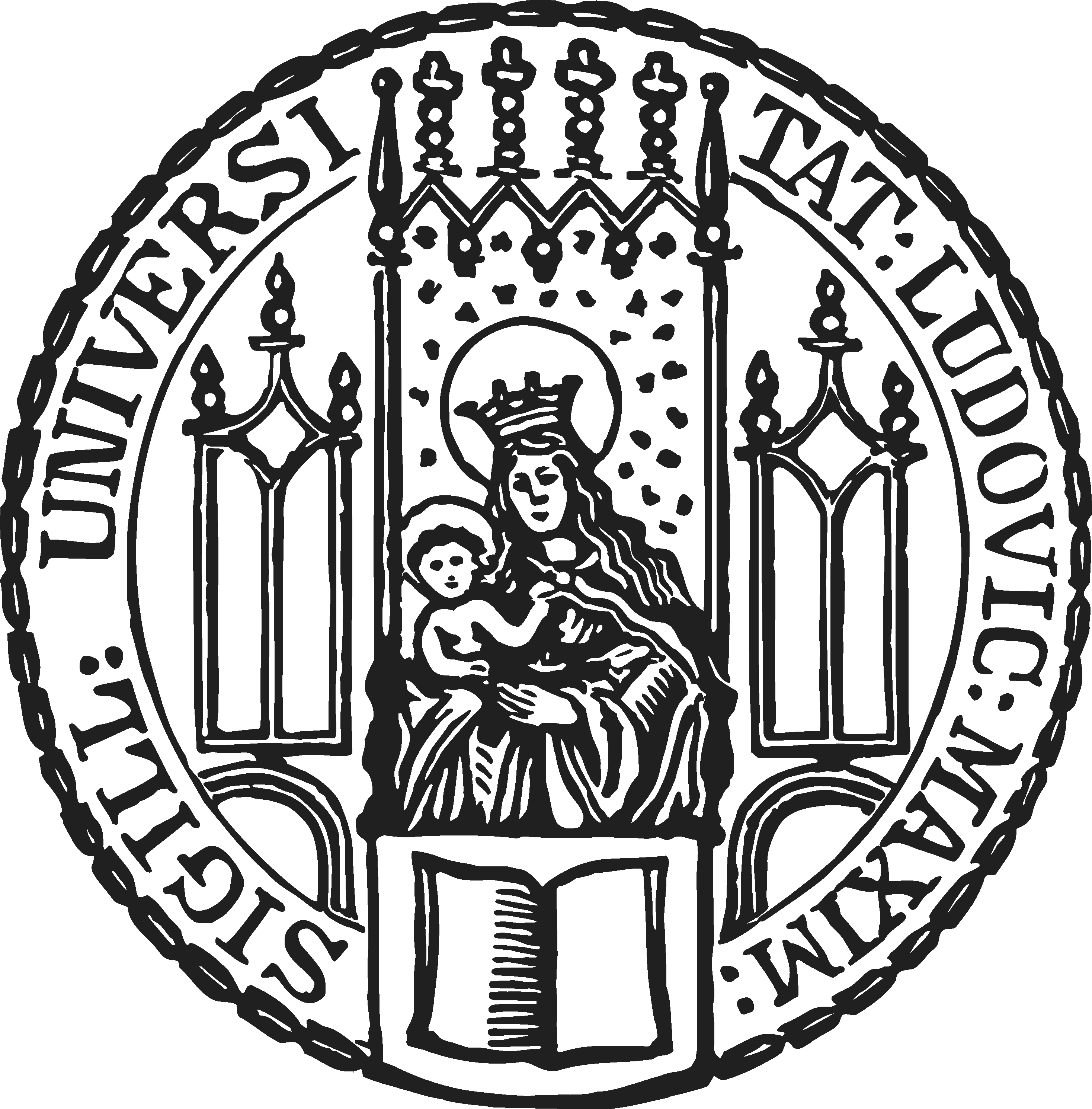}

\vfill

\large
Munich, \handindate

\end{center}
\restoregeometry

%% file: titlepages/title.tex
\begin{center}
    
\vspace{0.5cm}
      
\rule{\textwidth}{1.5pt}
\huge
\textbf{\mytitle}
\vspace{0.5cm}
\begin{flushright}
    \LARGE
    \textbf{\myname}
\end{flushright}
\rule{\textwidth}{1.5pt}

\vfill

\Large
Dissertation \\
an der Fakultät für Physik \\
der Ludwig-Maximilians-Universität \\
München \\

\vspace{0.5cm}

vorgelegt von \\
\myname \\
aus Wien, \"Osterreich \\
      
\vfill

\large
München, den \handindate

\vfill

\end{center}

\newpage

\vspace*{\fill}

\large
Erstgutachter: Prof. Dr. Thomas Kuhr\\
Zweitgutachter: Prof. Dr. Otmar Biebel\\
Tag der mündlichen Prüfung: 17.12.2025

%% file: chapters/abstract-german.tex
Experimentelle Ergebnisse in der Hochenergiephysik sind naturgemäß hypothesenabhängig, da Kollisionsanalysen darauf ausgelegt sind, bestimmte theoretische Rahmenbedingungen zu überprüfen. Die Vielfalt theoretischer Modelle führt zu einer Diskrepanz zwischen dem Tempo der Theorieentwicklung und der experimentellen Validierung. Dies macht eine Neuinterpretation bestehender Ergebnisse im Hinblick auf alternative Hypothesen erforderlich. Bestehende Strategien stoßen an ihre Grenzen: Eine vollständige Neuanalyse erfordert enorme Rechenressourcen und experimentelles Fachwissen, während vereinfachte Methoden durch die Vernachlässigung kinematischer Variationen und systematischer Unsicherheiten an Genauigkeit einbüßen.

Diese Arbeit entwickelt, validiert und wendet eine neuartige Methodik zur Neuinterpretation an, die statistische Genauigkeit bewahrt und gleichzeitig den Rechenaufwand drastisch reduziert, wodurch schnelle Rückschlüsse auf alternative theoretische Parameter durch leicht verteilbare Likelihoods ermöglicht werden.

Die Methode gewichtet die gemeinsame Verteilung von Rekonstruktions- und kinematischen Variablen entsprechend dem Verhältnis der theoretischen Vorhersagen von Alternativ- zu Nullhypothesen neu. Ausgehend von einer histogrammbasierten Wahrscheinlichkeit werden so aktualisierte Vorlagen für jedes theoretische Modell erstellt, wobei Korrelationen und systematische Unsicherheiten erhalten bleiben.
Die Methodik wird anhand von Beispielen validiert und auf die Belle~II $B^+ \to K^+ \nu\bar{\nu}$ Messung ($2.7\sigma$-Abweichung von dem Standardmodell) angewendet. Es werden zwei Analysen vorgestellt: eine Neuinterpretation der Weak Effective Theory, die Wilson-Koeffizienten einschränkt, und eine Suche nach leichter neuer Physik über $B^+ \to K^+ X$ Zweiteilchenzerfälle.

Die Neuinterpretation der Weak Effective Theory liefert die ersten direkten experimentellen Einschränkungen für die Wilson-Koeffizienten $b \to s \nu \bar \nu$. Der dominante Vektorkoeffizient $|C_{\mathrm{VL}}+C_{\mathrm{VR}}|$ erreicht seinen best-fit Wert bei 11.3 (Standard Model: 6.6) mit 95\%-Konfidenzintervallen von $[1.9, 16.2]$, $[0.0, 15.4]$ und $[0.0, 11.2]$ für den Vektor-, Skalar- und Tensorsektor. Die Analyse der leichten neuen Physik ergibt eine best-fit Resonanzmasse von $m_X = 2.1~\text{GeV}$ und einem Verzweigungsverhältnis von ${\mathcal{B}(\BKX) = 9.2 \cdot 10^{-6}}$, was zu einer Signifikanz von $3.0\sigma$ gegenüber der Standard Model-Vorhersage führt. Die 95\%-Konfidenzintervalle betragen $[1.9, 2.7]~\text{GeV}$ bzw. $[3.4, 14.0] \cdot 10^{-6}$. Die Arbeit präsentiert auch den aktuellen Stand der kombinierten Vierkanal-Analyse von Belle~II für $B \to K^{(*)} \nu\bar{\nu}$.

Diese Arbeit schafft einen neuartigen Rahmen für Neuinterpretationen. Die Veröffentlichung der daraus resultierenden modellunabhängigen Likelihoods schafft einen wichtigen Präzedenzfall für offene Wissenschaft in der experimentellen Teilchenphysik. Die Verwendung der Methodik von LHCb zeigt ihre weitreichende Bedeutung. Durch die Ermöglichung einer schnellen theoretischen Validierung demokratisiert dieser Ansatz den Zugang zu experimentellen Daten, erhöht die wissenschaftliche Rentabilität von Investitionen und fördert die Findable, Accessible, Interoperable and Reusable (FAIR) Datenprinzipien in der Hochenergiephysik.

%% file: chapters/abstract.tex
Experimental results in high-energy physics are inherently hypothesis-dependent, as collider analyses are designed to test specific theoretical frameworks. The proliferation of theoretical models creates a mismatch between the pace of theory development and experimental validation, motivating the reinterpretation of existing results in terms of alternative hypotheses. Existing strategies face limitations: full reanalysis requires enormous computational resources and experimental expertise, while simplified methods sacrifice accuracy by neglecting kinematic variations and systematic uncertainties.

This thesis develops, validates, and applies a novel reinterpretation method that preserves statistical rigour while dramatically reducing computational requirements, enabling rapid inference on alternative theoretical parameters through distributable likelihoods.

The method reweights the joint distribution of reconstruction and kinematic variables according to the ratio of theoretical predictions of alternative to null hypotheses. Starting from a histogram-based likelihood, this produces updated templates for any theoretical model while preserving correlations and systematic uncertainties. 
The method is validated through examples and applied to the Belle~II $B^+ \to K^+ \nu\bar{\nu}$ measurement ($2.7\sigma$ tension with the Standard Model). Two analyses are presented: a Weak Effective Theory reinterpretation constraining Wilson coefficients, and a light new physics search via ${B^+ \to K^+ X}$ two-body decays.

The Weak Effective Theory reinterpretation provides the first direct experimental constraints on $b \to s \nu \bar \nu$ Wilson coefficients. The dominant vector coefficient $|C_{\mathrm{VL}}+C_{\mathrm{VR}}|$ peaks at 11.3 (Standard Model: 6.6) with 95\% credible intervals of $[1.9, 16.2]$, $[0.0, 15.4]$, and $[0.0, 11.2]$ for vector, scalar, and tensor sectors. The light new physics analysis yields a best-fit resonance mass of $m_X = 2.1~\text{GeV}$ and branching ratio of $\mathcal{B}(\BKX) = 9.2 \cdot 10^{-6}$, leading to a $3.0\sigma$ significance over the Standard Model prediction. The 95\% credible intervals are $[1.9, 2.7]~\text{GeV}$ and $[3.4, 14.0] \cdot 10^{-6}$, respectively. The thesis also presents the current status on Belle~II's combined four-channel $B \to K^{(*)} \nu\bar{\nu}$ analysis.

This work establishes a novel reinterpretation framework. The public release of the resulting model-agnostic likelihoods establishes an important precedent for open science in experimental particle physics. The method's adoption by LHCb demonstrates its broader impact. By facilitating rapid theoretical validation, this approach democratizes experimental data access, enhances scientific return on investments, and advances Findable, Accessible, Interoperable and Reusable (FAIR) Data principles in high-energy physics.

%% file: chapters/introduction.tex
How do we make progress in science? The foundation of scientific inquiry rests on a systematic approach that has guided discovery for centuries. According to the Oxford Dictionary, the scientific method is defined as~\cite{OED2023scientific}

\begin{quote}
``The scientific method is now commonly represented as ideally comprising some or all of (a) systematic observation, measurement, and experimentation, (b) induction and the formulation of hypotheses, (c) the making of deductions from the hypotheses, (d) the experimental testing of the deductions, and (if necessary) (e) the modification of the hypotheses [...]''
\end{quote}

A crucial aspect of this definition is that hypotheses and their consequent deductions (predictions) must precede experimental testing. The implications of this ordering are profound: experimental results are inherently hypothesis-dependent, since experiments are deliberately designed to test specific theoretical frameworks.

\ac{HEP} exemplifies this principle. Enormous investments in particle accelerators, sophisticated detectors, and complex data analysis pipelines are all guided by theoretical predictions we seek to validate. The patterns we search for in high-dimensional datasets, the statistical methods we employ, and even the very definition of a ``signal'' are all determined by our chosen theoretical hypothesis. Consequently, experimental results are meaningful only within the context of the specific theoretical framework under investigation.

This hypothesis-dependence, however, is often overlooked when interpreting experimental results, creating a critical methodological problem. When new theoretical models emerge with different predictions, their validity may be assessed against experiments that were never designed to test them. Such assessments risk corrupting the scientific method by modifying or excluding hypotheses on inadequate empirical grounds.

Addressing this methodological challenge is the central goal of this work.

The \ac{SM} of particle physics represents the currently accepted paradigm in \ac{HEP} --- a gauge quantum field theory that has withstood empirical tests with extraordinary precision. Despite its remarkable success, extracting concrete predictions from this complex theoretical framework remains computationally challenging. \acp{EFT} address this challenge by approximating the full theory up to a specific energy scale, systematically incorporating effects from both \ac{SM} and potential \ac{BSM} physics while maintaining computational tractability. The theoretical foundations underlying these approaches are discussed in \cref{sec:theory}.

The search for \ac{BSM} physics often focuses on rare particle decays, where \ac{BSM} contributions become relatively more prominent compared to \ac{SM} processes, making deviations easier to detect. A particularly compelling example from flavour physics, which forms the central focus of this work, involves $B \to K^{(*)} \nu \bar{\nu}$ decays studied at Belle~II. These decay channels have gained significant attention following the Belle~II collaboration's recent \BKnn measurement, which exhibits a $2.7\sigma$ tension with \ac{SM} predictions~\cite{Belle-II:2023esi}. The Belle~II experiment is described in \cref{sec:experiment}.

The complexity of particle collider data analysis necessitates sophisticated statistical models and inference techniques to ensure reliable results with high statistical power. However, this very complexity means that not all theoretical hypotheses can be directly tested against experimental data within practical time and computational constraints. Statistical frameworks underlying these challenges are discussed in \cref{sec:stat-models,sec:stat-inference}.

These practical limitations motivate a powerful alternative approach: \textit{reinterpretation} --- the process of interpreting experimental results obtained under one theoretical hypothesis in terms of alternative hypotheses with different predictions. Rather than conducting entirely new measurements for each theoretical model, reinterpretation leverages existing experimental results to constrain or validate new physics scenarios. While the concept itself is not novel, existing methodological approaches have various limitations, which are reviewed in \cref{sec:reinterpretation-intro}.

The novel reinterpretation methodology presented in this work addresses key requirements: accuracy, generality, computational efficiency, and accessibility. The approach enables fast, precise reinterpretations while enabling the preservation and distribution of full statistical models. Both the method and its validation through examples are presented in \cref{sec:method}.

The practical application of this methodology is demonstrated using the aforementioned Belle~II \BKnn measurement~\cite{Belle-II:2023esi}, originally interpreted within the \ac{SM} framework. The analysis is presented in \cref{sec:reinterpretation knunu mall}, followed by reinterpretations in terms of the \ac{WET}~\cite{Aebischer:2017gaw,Jenkins:2017jig,Jenkins:2017dyc} in \cref{sec:reinterpretation knunu wet}. The \ac{WET} encompasses new physics models such as leptoquarks~\cite{PhysRevD.98.055003, Dorsner:2016wpm, Angelescu:2021lln} and heavy $Z'$ bosons~\cite{PhysRevD.104.053007, ParticleDataGroup:2024cfk, Leike:1998wr, Buras:2012jb}.
A reinterpretation in terms of a two-body light new physics model is presented in \cref{sec:reinterpretation knunu bkx}. This encompasses new physics models such as axions~\cite{MartinCamalich:2020dfe,ParticleDataGroup:2024cfk, Peccei:1977hh, Peccei:1977ur}, axion-like particles~\cite{ParticleDataGroup:2024cfk, PhysRevD.102.015023,Guerrera:2022ykl,Bruggisser:2023npd} or other dark-sector mediators~\cite{PhysRevD.101.095006,Datta:2022zng,Abdughani:2023dlr,Berezhnoy:2023rxx}.

The reproducibility and distributability of statistical models represent more than scientific best practices --- they enable powerful combined analyses that can dramatically reduce uncertainties. Building on this capability, the Belle~II collaboration is developing a combined analysis of four different $B \to K^{(*)} \nu \bar{\nu}$ decay channels, whose current status is described in \cref{sec:knunu update}. This combined analysis will significantly enhance sensitivity to \ac{BSM} effects and \ac{WET} parameters, establishing an excellent foundation for future reinterpretation studies at Belle~II and beyond.

%% file: chapters/theory.tex
\footnote{Natural units with $\hbar = c = 1$ are used throughout this work. The Minkowski metric tensor is defined as $g^{\mu\nu} = \text{diag}(1, -1, -1, -1)$. Repeated indices are summed over unless otherwise specified. The Minkowski inner product of two four-vectors $p^\mu$ and $q^\mu$ is defined as $p \cdot q = p^\mu q_\mu = p^0 q^0 - \vec{p} \cdot \vec{q}$.}
Theoretical frameworks provide the mathematical foundation for interpreting experimental observations and testing our understanding of fundamental physics. The \ac{SM} of particle physics represents our most successful description of the fundamental particles and their interactions, excluding gravity. Despite its remarkable experimental validation across energy scales from the eV to the TeV regime~\cite{ParticleDataGroup:2024cfk}, the \ac{SM} cannot account for several observed phenomena, including dark matter, dark energy, neutrino masses, and the baryon asymmetry of the universe. These limitations provide compelling evidence for \ac{BSM} physics.

\ac{FCNC} processes offer a particularly sensitive probe of \ac{BSM} physics. In the \ac{SM}, such transitions are forbidden at tree level and can only proceed through strongly suppressed loop-induced effects, resulting in high sensitivity to potential \ac{BSM} contributions~\cite{Schwartz_2013,Thomson_2013,Peskin:1995ev,Weinberg_1996}. Among these, the $b \to s \nu\bar{\nu}$ transition stands out as theoretically clean due to the absence of photon exchanges, which break factorization in charged lepton modes like $b \to s \ell^+\ell^-$. The rare decays $B \to K^{(*)} \nu\bar{\nu}$, mediated by $b \to s \nu\bar{\nu}$, thus provide an ideal laboratory for precision tests of the \ac{SM} and searches for \ac{BSM} signatures.

\acp{EFT} provide the optimal theoretical framework for analysing such processes. By systematically separating high-energy \ac{BSM} physics from low-energy \ac{SM} dynamics, \acp{EFT} enable model-independent parametrization of new physics effects through Wilson coefficients of higher-dimensional operators. This approach proves particularly powerful when the \ac{BSM} scale lies well above current experimental reach, as it captures the effects of entire classes of \ac{BSM} models through a finite set of effective operators.

For $b \to s \nu\bar{\nu}$ transitions, the \ac{WET} provides the appropriate low-energy description below the electroweak scale. The theoretical prediction for $B \to K^{(*)} \nu\bar{\nu}$ decays requires factorizing short-distance physics (encoded in Wilson coefficients) from long-distance, non-perturbative \ac{QCD} dynamics (captured by hadronic form factors). This factorization enables precision tests of the \ac{SM} and systematic searches for \ac{BSM} signatures through deviations in experimentally measured Wilson coefficients.

This chapter establishes the complete theoretical framework for $B \to K^{(*)} \nu\bar{\nu}$ phenomenology. \Cref{sec:sm} reviews the \ac{SM} foundations, emphasizing the electroweak symmetry breaking mechanism (\cref{sec:ssb}) that generates fermion and gauge boson masses, and the origin of \acp{FCNC} through the \ac{CKM} matrix structure (\cref{sec:fcnc}). \Cref{sec:eft} develops the \ac{EFT} formalism, with detailed treatment of the \ac{WET} Lagrangian (\cref{sec:wet}) and its dimension-six operators relevant for $b \to s \nu\bar{\nu}$ transitions. The non-perturbative evaluation of hadronic matrix elements through form factor parametrizations is discussed in \cref{sec:hadronic-matrix-elements}. Finally, \cref{sec:theory-predictions} combines these elements to derive differential decay rates for $B \to K \nu\bar{\nu}$ (\cref{sec:knunu-theory}) and $B \to K^* \nu\bar{\nu}$ (\cref{sec:ksnunu-theory}), demonstrating the complementary Wilson coefficient sensitivities that enable comprehensive \ac{BSM} searches when both decay modes are analysed simultaneously.

\section{The Standard Model of particle physics}
\label{sec:sm}

\footnote{This chapter follows References~\cite{Schwartz_2013,Thomson_2013,Peskin:1995ev,Weinberg_1996}, unless otherwise specified.}
The \ac{SM} of particle physics is a locally gauged quantum field theory that respects Poincaré invariance. It describes three of the four fundamental forces of nature: the electromagnetic, weak, and strong interactions. Gravity is not included in the \ac{SM}.
The basis of this model is its gauge group 
\begin{equation}
    \mathrm{G}_{SM} = \mathrm{SU}(3)_C \times \mathrm{SU}(2)_W \times \mathrm{U}(1)_Y,
\end{equation}
where $\mathrm{SU}(n)$ is the special unitary group
\begin{equation}
    \mathrm{SU}(n) := \{U \in \mathbb{C}^n | U U^\dagger = U^\dagger U = \mathbb{1}, \det U = 1\}
\end{equation}
and $U(n)$ is the unitary group 
\begin{equation}
    \mathrm{U}(n) := \{U \in \mathbb{C}^n | U U^\dagger = U^\dagger U = \mathbb{1}\}.
\end{equation}
$\mathrm{SU}(3)_C$ is the gauge group of strong interactions, or \ac{QCD}. $\mathrm{SU}(2)_W$ is the gauge group of weak isospin, and $\mathrm{U}(1)_Y$ is the gauge group of hypercharge.
The number of generators of the underlying Lie algebras defines the number of gauge bosons in the \ac{SM}. The number of generators of $\mathrm{SU}(n)$ is $n^2-1$, resulting in a total of 8 \textit{gluons} from $\mathrm{SU}(3)_C$ and 3 $W$ gauge bosons from $\mathrm{SU}(2)_W$. The unitary Lie algebra has $n^2$ generators. Consequently, one gets 1 gauge boson $B$ from $\mathrm{U}(1)_Y$. Gauge bosons carry spin-$1$.

The matter field content of the \ac{SM} comprises fermions, which carry spin-$1/2$. 
They can be characterized according to their interaction properties with the gauge bosons or, in other words, their transformation properties under the \ac{SM} gauge group. 
Under the \ac{SM} gauge group they transform as shown in \cref{tab:fermion transformations}.
\begin{table}
    \renewcommand{\arraystretch}{1.5}
    \caption{Transformation properties of \ac{SM} fermions under the gauge symmetry groups $\mathrm{SU}(3)_C$, $\mathrm{SU}(2)_W$, and $\mathrm{U}(1)_Y$. The numbers indicate the representation dimension under each gauge group (singlet, doublet, or triplet), while the hypercharge $Y$ values are given for $\mathrm{U}(1)_Y$. Left-handed ($L$) fermions transform as doublets under $\mathrm{SU}(2)_W$, while right-handed ($R$) fermions are singlets. The index $i$ labels the three generations of fermions~\cite{Weinberg_1996}.}
    \centering
    \begin{tabularx}{\linewidth}{@{\extracolsep{\fill}}lYYY}
        \toprule \midrule
        Fermions & $\mathrm{SU}(3)_C$ & $\mathrm{SU}(2)_W$ & $\mathrm{U}(1)_Y$\\
        \midrule
        $Q_{iL}$ & $3$ & $2$ & $+1/6$\\
        $u_{iR}$ & $3$ & $1$ & $+2/3$\\
        $d_{iR}$ & $3$ & $1$ & $-1/3$\\
        $L_{iL}$ & $1$ & $2$ & $-1/2$\\
        $l_{iR}$ & $1$ & $1$ & $-1$\\
        $\nu_{iR}$ & $1$ & $1$ & $0$\\
        \midrule \bottomrule
    \end{tabularx}
    \label{tab:fermion transformations}
\end{table}
Since right- and left-handed fermions carry different weak isospin ($\pm1/2$ for doublets and 0 for singlets), the \ac{SM} is a chiral gauge theory~\cite{Buchmuller:2006zu}. 

Left-handed quarks are represented by $Q_{iL}$, right-handed up- and down-type \textit{quarks} by $u_{iR}$ and $d_{iR}$, left-handed \textit{leptons} by $L_{iL}$, and right-handed \textit{leptons} by $l_{iR}$ and $\nu_{iR}$. Each of these fields comes in three generations, as indicated by the subscript $i$,
\begin{equation}
    \begin{aligned}
        Q_{iL} &\in [(u,d)_L, (c,s)_L, (t,b)_L]\\
        u_{iR} &\in [u_R, c_R, t_R]\\
        d_{iR} &\in [d_R, s_R, b_R]\\
        L_{iL} &\in [(\nu_e, e)_L, (\nu_\mu, \mu)_L, (\nu_\tau, \tau)_L]\\
        l_{iR} &\in [e_R, \mu_R, \tau_R]\\
    \end{aligned}
    \label{eq:SM-fermions}
\end{equation}
The \ac{SM} can be readily augmented by including right-handed neutrinos ${\nu_{iR} \in [\nu_{e R}, \nu_{\mu R}, \nu_{\tau R}]}$, which have never been observed.

The \ac{SM} has been constructed out of left-handed fermion doublets and right-handed fermion singlets, because $W^\pm$ bosons couple only to left-handed fermions, and because the $\mathrm{SU}(2)$ group admits a chargeless 1-dimensional singlet representation and a charged multidimensional representation. 

It is an open question why there are exactly three generations of quarks and leptons. However, this number of generations leads to the cancellation of anomalies, i.e. the loss of symmetries of the underlying classical theory after quantization. Anomaly cancellation is a necessity for the consistency of a quantum field theoretical description of nature~\cite{Buchmuller:2006zu}.

\subsection{The Standard Model Lagrangian}
\label{sec:sm-Lagrangian}
The \ac{SM} Lagrangian contains all renormalizable, gauge invariant operators with the specified matter fields~\cite{Grossman:2017thq}. In the most general form, one can write it as 
\begin{equation}
    \mathcal{L}_{SM} = \mathcal{L}_{kin} + \mathcal{L}_{\text{Higgs}} + \mathcal{L}_{Yukawa}.
\end{equation}
The Lagrangian $\mathcal{L}_{kin}$ includes field dynamics through kinetic terms as well as gauge interactions through covariant derivatives and non-Abelian field strengths. 
The Lagrangian $\mathcal{L}_{\text{Higgs}}$ contains the Higgs kinematics and potential terms, creating mass terms for the gauge bosons. 
The Lagrangian $\mathcal{L}_{Yukawa}$ contains all interactions of the Higgs field with fermions, and is hence responsible for generating fermion masses. 
In this section I will focus on the kinetic Lagrangian $\mathcal{L}_{kin}$. I will discuss the other two parts in \cref{sec:ssb}. 

The \ac{SM} symmetry group $\mathrm{G}_{SM}$ can be split into a \ac{QCD} part, $\mathrm{SU}(3)_C$ and an electroweak part $\mathrm{SU}(2)_W \times \mathrm{U}(1)_Y$.
Similarly, one can separate the kinetic Lagrangian,
\begin{equation}
    \mathcal{L}_{kin} = \mathcal{L}_{QCD} + \mathcal{L}_{EW} \, ,
\end{equation}
into a \ac{QCD} part, $\mathcal{L}_{QCD}$, and an electroweak part, $\mathcal{L}_{EW}$.

\ac{QCD} is described by a non-abelian gauge theory, invariant under $\mathrm{SU}(3)$ transformations,
\begin{equation}
    q \to q' = Uq \qquad \text{with} \qquad U = \exp(i \alpha_a T^a),
\end{equation}
where $q$ is a quark multiplet of $\mathrm{SU}(3)$, $\alpha_a$ are real parameters, and $T^a$ are the group generators. For $\mathrm{SU}(3)$ there are 8 generators, which are usually chosen as $T^a = \lambda^a/2$, where $\lambda^a$ are the Gell-Mann matrices (see \cref{sec:generators}). The generators obey the commutation relation
\begin{equation}
    [T^a, T^b] = i f^{abc} T^c,
\end{equation}
where $f^{abc}$ are antisymmetric structure constants. Abelian gauge theories, such as \ac{QED}, are described by $\mathrm{U}(1)$ transformations, which are characterized by a single generator.

The \ac{QCD} contribution to the kinetic Lagrangian is
\begin{equation}
    \mathcal{L}_{QCD} = \bar{q}^i \slashed{D}_{ij} q^j - \frac{1}{4} G^a_{\mu\nu}G_a^{\mu\nu}.
\end{equation}
Here $q^i$ are the quark fields, 
\begin{equation}
    q^i = \left( Q_{1L}^i, Q_{2L}^i, Q_{3L}^i, u_{1R}^i, u_{2R}^i, u_{3R}^i, d_{1R}^i, d_{2R}^i, d_{3R}^i \right)
\end{equation}
with color index $i$ running from 1 to 3, $\bar{q}^i \equiv q^\dagger \gamma^0$ is the adjoint spinor and $\gamma^\mu$ are the Dirac matrices (see \cref{sec:gamma-matrices}). The covariant derivative $\slashed{D}_{ij} = \gamma_\mu D^\mu_{ij}$ contains the gauge fields $A_a^\mu$ and is given by
\begin{equation}
    D^\mu_{ij} = \delta_{ij} \partial^\mu -i g_C T^a_{ij} A_a^\mu,
\end{equation}
where $g_C$ is the \ac{QCD} coupling.
The gauge field $A^\mu = T^a A_a^\mu$ is a vector field, which transforms under $\mathrm{SU}(3)_C$ as
\begin{equation}
    A^\mu \to A^{\prime \mu} = U A^\mu U^\dagger - \frac{i}{g_C} (\partial^\mu U) U^\dagger.
    \label{eq:SU3-gauge-transformation}
\end{equation}
The kinetic terms of the gauge fields enter through the field strength tensor, given by the commutator
\begin{equation}
    G_{\mu\nu} = G_{\mu\nu}^aT_a = \frac{i}{g_C} [D_\mu, D_\nu] = \partial_\mu A_\nu - \partial_\nu A_\mu - i g_C [A_\mu, A_\nu].
\end{equation}
From the last term in the field strength tensor one gets cubic and quartic terms in the gauge fields. This leads to the self-interaction of gluons and is the fundamental reason for confinement. 

The electroweak group is the product $\mathrm{SU}(2)_W \times \mathrm{U}(1)_Y$. Furthermore, the electroweak theory is special because it is a chiral theory. Given the \ac{SM} fermion content in \cref{eq:SM-fermions}, one can write down the covariant derivatives for the chiral fields
\begin{align}
    D_\mu \psi_L &= (\partial_\mu - ig W_\mu - ig' Y B_\mu)\psi_L \, , \\
    D_\mu \psi_R &= (\partial_\mu - ig' Y B_\mu)\psi_R,
\end{align}
where the fermion fields are separated into the vectors
\begin{align}
    \psi_L &= (Q_{1L}, Q_{2L}, Q_{3L}, L_{1L}, L_{2L}, L_{3L}) \, , \\
    \psi_R &= (u_{1R}, u_{2R}, u_{3R}, d_{1R}, d_{2R}, d_{3R}, l_{1R}, l_{2R}, l_{3R})
\end{align}

The $W_\mu = \tau_i W^i_\mu$ are the three $\mathrm{SU}(2)_W$ gauge fields, $\tau_i= \sigma_i/2$ are the generators, where $\sigma_i$ are the Pauli matrices (see \cref{sec:generators}), and $g$ is a coupling constant. The $\mathrm{U}(1)_Y$ gauge field is $B_\mu$, $Y$ is the hypercharge and $g'$ is a coupling constant.

From these one can build the kinetic contribution to the electroweak Lagrangian,
\begin{equation}
    \mathcal{L}_{EW} = i \bar{\psi}_L \slashed{D} \psi_L + i \bar{\psi}_R \slashed{D} \psi_R - \frac{1}{4}W^i_{\mu\nu}W_i^{\mu\nu} - \frac{1}{4} B_{\mu\nu}B^{\mu\nu}.
\end{equation}
The kinetic terms of the gauge fields enter through the field strength tensors
\begin{align}
    W_{\mu\nu} & = W_{\mu\nu}^i T_i = \frac{i}{g} [D_\mu, D_\nu] = \partial_\mu W_\nu - \partial_\nu W_\mu - ig [W_\mu, W_\nu] \, , \\
    B_{\mu\nu} & = \frac{i}{g'Y} [D_\mu, D_\nu] = \partial_\mu B_\nu - \partial_\nu B_\mu.
\end{align}

This completes the kinetic Lagrangian of the \ac{SM} gauge group $\mathrm{G}_{SM}$. One can summarize it as 
\begin{equation}
    \mathcal{L}_{kin} = i \bar{q}^i \slashed{D}_{ij} q^j + i \bar{\psi}_L \slashed{D} \psi_L + i \bar{\psi}_R       \slashed{D} \psi_R - \frac{1}{4} G^a_{\mu\nu}G_a^{\mu\nu} - \frac{1}{4}W^i_{\mu\nu}W_i^{\mu\nu} - \frac{1}{4} B_{\mu\nu}B^{\mu\nu}.
\end{equation}
To make the theory actually describe the physical world, there is one crucial ingredient missing: masses. So far, neither the gauge bosons nor the fermions are considered to be massive. Massless gauge bosons would imply long-range forces, not observed in nature. The weak interaction has short range, requiring massive gauge bosons. How mass terms are generated while preserving gauge invariance is discussed in the next section.

\subsection{Spontaneous symmetry breaking}
\label{sec:ssb}
Mass terms such as $m^2 A_\mu A^\mu$ break gauge invariance, since they are not invariant under the gauge transformations such as in \cref{eq:SU3-gauge-transformation}. Furthermore, fermion mass terms also break gauge invariance, since these mix left- and right-handed fermions, 
\begin{equation}
    m\left( \bar{\psi}_L \psi_R + \bar{\psi}_R \psi_L \right) \, ,
\end{equation}
which transform differently under the gauge group.

Introducing gauge invariant terms, which give rise to masses for the gauge bosons and fermions, is the main purpose of the Higgs mechanism. It is built on the idea of spontaneous symmetry breaking~\cite{Buchmuller:2006zu,Schwartz_2013}. The Higgs mechanism is the simplest way to generate gauge invariant mass terms for the gauge bosons and fermions, by introducing a single complex scalar field $\Phi$.

The newly introduced scalar field $\Phi$ is a doublet under the electroweak gauge group $\mathrm{SU}(2)_W$, with hypercharge $+1/2$, with four real \ac{d.o.f.}.
The Higgs Lagrangian is given by
\begin{equation}
    \mathcal{L}_{\text{Higgs}} = (D_\mu \Phi)^\dagger D^\mu \Phi - V(\Phi^\dagger \Phi).
    \label{eq:Higgs-Lagrangian}
\end{equation}
The covariant derivative $D_\mu$ contains the gauge fields,
\begin{equation}
    D_\mu = \partial_\mu - i g W_\mu - \frac{i}{2} g' B_\mu.
\end{equation}
The beauty of this field lies in its potential, which is given by
\begin{equation}
    V(\Phi^\dagger \Phi) = - \mu^2 \Phi^\dagger \Phi + \lambda (\Phi^\dagger \Phi)^2 \, ,
\end{equation}
which has a non-zero \ac{vev} at $\Phi^\dagger \Phi = v^2/2 \equiv \mu^2 / (2\lambda)$ for $\mu^2 > 0$ and $\lambda > 0$. 
One can take the \ac{vev} to be real and positive, $v > 0$, without loss of generality,
\begin{equation}
    \langle \Phi \rangle = \frac{1}{\sqrt{2}} \begin{pmatrix} 0 \\ v \end{pmatrix}.
\end{equation}
In the vicinity of the \ac{vev}, the theory is not invariant under the gauge transformations of $\mathrm{SU}(2)_W \times \mathrm{U}(1)_Y$, but only under the electromagnetic $\mathrm{U}(1)_{\text{em}}$ transformations. 
This is colloquially known as breaking of the gauge symmetry from $\mathrm{SU}(2)_W \times \mathrm{U}(1)_Y$ to $\mathrm{U}(1)_{\text{em}}$. The generator of the electromagnetic transformation is the $Q = T_3 + Y$ charge, where $T_3$ is the third component of the weak isospin and $Y$ is the weak hypercharge. This is known as spontaneous symmetry breaking~\cite{Schwartz_2013}.

In the unitarity gauge, the Higgs field can be written as
\begin{equation}
    \Phi = \frac{1}{\sqrt{2}} \begin{pmatrix} 0 \\ v + h(x) \end{pmatrix},
\end{equation}
where $h(x)$ is a real scalar field. The Lagrangian \cref{eq:Higgs-Lagrangian} then becomes
\begin{align}
\mathcal{L}_{\text{Higgs}} 
    &= \frac{1}{2} (\partial_\mu h)^2 
    + \frac{(v+h)^2}{8} \left[ g^2 \left( W_\mu^1 W^{1\mu} + W_\mu^2 W^{2\mu} \right) 
    + \left( g W_\mu^3 - g' B_\mu \right)^2 \right] \notag \\
    &\quad - \left[ \lambda v^2 h^2 + \lambda v h^3 + \frac{\lambda}{4} h^4 \right].
    \label{eq:Higgs-Lagrangian-expanded}
\end{align}
One can identify the mass terms for the gauge bosons and the Higgs field. The Higgs mass is easily read off as $m_h^2 = 2 \lambda v^2$. 
To diagonalize the gauge boson mass terms, one performs a change of basis
\begin{align}
    W_\mu^\pm &= \frac{1}{\sqrt{2}} \left( W_\mu^1 \mp i W_\mu^2 \right), \\
    Z_\mu &= \frac{1}{\sqrt{g^2 + g'^2}} \left( g W_\mu^3 - g' B_\mu \right), \\
    A_\mu &= \frac{1}{\sqrt{g^2 + g'^2}} \left( g' W_\mu^3 + g B_\mu \right),
\end{align}
which are the three massive weak fields $W^\pm$, $Z$, and the massless photon field $A$. Often this is written in terms of the Weinberg angle $\theta_W$, defined as
\begin{equation}
    \tan \theta_W = \frac{g'}{g} \qquad \text{or} \qquad \sin^2 \theta_W = \frac{g'^2}{g^2 + g'^2}.
\end{equation}

The mass terms for the gauge bosons are then
\begin{equation}
    \mathcal{L}_{\text{Higgs}} \supset m_W^2 W_\mu^+ W^{-\mu} + \frac{1}{2} m_Z^2 Z_\mu Z^\mu.
\end{equation}
The masses of the $W^\pm$ and $Z$ bosons are
\begin{equation}
    m_W = \frac{1}{2} g v, \qquad m_Z = \frac{1}{2} \sqrt{g^2 + g'^2} \, v = \frac{m_W}{\cos \theta_W}.
\end{equation}

The interaction terms between the Higgs field and the gauge bosons are given by
\begin{equation}
    \mathcal{L}_{\text{Higgs}} \supset 
    2 \frac{m_W^2}{v} \, h \, W_\mu^+ W^{-\mu} 
    + \frac{m_W^2}{v^2} \, h^2 W_\mu^+ W^{-\mu}
    + \frac{m_Z^2}{v} \, h \, Z_\mu Z^\mu 
    + \frac{m_Z^2}{2 v^2} \, h^2 Z_\mu Z^\mu.
\end{equation}
This shows that the Higgs field couples to the gauge bosons, which is crucial for the mass generation of the gauge bosons. Furthermore, from \cref{eq:Higgs-Lagrangian-expanded} one can see that the Higgs field has cubic and quartic self-interaction terms. In the mass-diagonal basis, the kinetic Lagrangian $\mathcal{L}_{kin}$ contains interaction terms between the $W^\pm$, $Z$ and the photon fields of the form
\begin{equation}
    W^\pm W^\mp A, \qquad W^\pm W^\mp Z, \qquad W^\pm W^\mp A A, \qquad W^\pm W^\mp Z Z, \qquad W^\pm W^\mp A Z.
\end{equation}
The coupling constant $e$ of the photon is related to the weak coupling constant $g$ and the Weinberg angle $\theta_W$ by
\begin{equation}
    e = g \sin \theta_W = g' \cos \theta_W.
\end{equation}
Neutral-only interactions of the form $ZZA$, $ZZZ$, or similar are not present at tree level. These interactions will be discussed in more detail in \cref{sec:fcnc}.

Note that the $\mathrm{SU}(2)_W \times \mathrm{U}(1)_Y$ gauge symmetry is never really broken during spontaneous symmetry breaking. The Higgs mechanism can also be described in a gauge invariant way, without applying the unitary gauge~\cite{Buchmuller:2006zu}.
This can also be seen by looking at the available \ac{d.o.f.}. Before symmetry breaking there is a complex Higgs doublet (4 real \ac{d.o.f.}) plus 4 massless vector fields (8 \ac{d.o.f.}), so 12 \ac{d.o.f.} in total.
After symmetry breaking, there is one real Higgs field (1 \ac{d.o.f.}) plus three massive vector fields (9 \ac{d.o.f.}) plus one massless vector field (2 \ac{d.o.f.}), so again 12 \ac{d.o.f.} in total.

The Higgs mechanism is not only responsible for the mass generation of the gauge bosons, but also for the mass generation of the fermions. The fermion mass terms are generated through Yukawa interactions with the Higgs field,
\begin{equation}
    \mathcal{L}_{Yukawa} = - \sum_{i,j} \left( y_{ij}^u \bar{Q}_{iL} \tilde{\Phi} u_{jR} + y_{ij}^d \bar{Q}_{iL} \Phi d_{jR} + y_{ij}^l \bar{L}_{iL} \Phi l_{jR} + \text{h.c.} \right) \, ,
\end{equation}
where $\tilde{\Phi} = i \sigma_2 \Phi^*$ is the conjugate Higgs doublet and $y_{ij}^u$, $y_{ij}^d$, and $y_{ij}^l$ are the Yukawa couplings. I exclude right-handed neutrinos here. The Higgs field couples to the left-handed quark doublets $Q_{iL}$ and lepton doublets $L_{iL}$, as well as to the right-handed quark singlets $u_{iR}$, $d_{iR}$ and lepton singlets $l_{iR}$.
After spontaneous symmetry breaking, the Yukawa interactions lead to mass terms for the fermions,
\begin{equation}
    \mathcal{L}_{Yukawa} \supset - \sum_{i,j} \left( \frac{ v}{\sqrt{2}} \bar{u}_{iL} y_{ij}^u u_{jR} + \frac{v}{\sqrt{2}} \bar{d}_{iL} y_{ij}^d d_{jR} + \frac{v}{\sqrt{2}} \bar{l}_{iL} y_{ij}^l l_{jR} + \text{h.c.} \right).
\end{equation}
The Yukawa couplings are arbitrary complex matrices, which can be diagonalized by a bi-unitary transformation,
\begin{equation}
    \begin{aligned}
        y_{ij}^u &= U_{ik}^u M_{kl}^u (K^u)^\dagger_{lj}, \\
        y_{ij}^d &= U_{ik}^d M_{kl}^d (K^d)^\dagger_{lj}, \\
        y_{ij}^l &= U_{ik}^l M_{kl}^l (K^l)^\dagger_{lj},
    \end{aligned}
\end{equation}
where $U^f$ and $K^f$ are unitary matrices, and $M^f$ are diagonal matrices with the fermion masses on the diagonal, and $f=u,d,l$.
The basis of the fermion fields can be freely changed, such that the Yukawa couplings are diagonal,
\begin{equation}
    \begin{aligned}
        u_{iL} &\to U_{ij}^u u_{jL}, \\
        d_{iL} &\to U_{ij}^d d_{jL}, \\
        l_{iL} &\to U_{ij}^l l_{jL}, \\
        u_{iR} &\to K_{ij}^u u_{jR}, \\
        d_{iR} &\to K_{ij}^d d_{jR}, \\
        l_{iR} &\to K_{ij}^l l_{jR}.
    \end{aligned}
    \label{eq:fermion-basis-change}
\end{equation}
This is known as the mass basis. In the mass basis, the Yukawa interactions become
\begin{equation}
    \mathcal{L}_{Yukawa} \supset - \sum_{i} \left(m_i^u \bar{u}_{iL} u_{jR} + m_i^d \bar{d}_{iL} d_{jR} + m_i^l \bar{l}_{iL} l_{jR} + \text{h.c.} \right) \, ,
\end{equation}
where the masses of the fermions are proportional to the Yukawa couplings and the Higgs \ac{vev},  
\begin{equation}
    m_i^u = \frac{v}{\sqrt{2}} M^u_{ii}, \qquad m_i^d = \frac{v}{\sqrt{2}} M^d_{ii}, \qquad m_i^l = \frac{v}{\sqrt{2}} M^l_{ii}.
\end{equation} 

The interaction terms between the physical gauge bosons and the fermions can be separated into neutral currents involving $A_\mu$ and $Z_\mu$ and charged currents involving $W^\pm$. The neutral current interactions are given by
\begin{equation}
    \mathcal{L}_{kin,NC} = e A_\mu J^\mu_{em} + \frac{e}{\sin\theta_W} Z_\mu J_Z^\mu.
\end{equation}
The electromagnetic current $J^\mu_{em}$ is given by
\begin{equation}
    J^\mu_{em} = \sum_f ( \bar{f}_L \gamma^\mu Q f_L + \bar{f}_R \gamma^\mu Q f_R ) \, ,
\end{equation}
where the sum runs over all fermions. The electromagnetic current does not mix left- and right-handed fermions, since it is a vector current.
The weak neutral current $J_Z^\mu$ is given by
\begin{equation}
    J_Z^\mu = \sum_f \frac{1}{\cos\theta_W} \left(\bar{f}_L \gamma^\mu T_3 f_L - \sin^2\theta_W J^\mu_{em} \right) \, .
\end{equation}
The transformations of \cref{eq:fermion-basis-change} leave the neutral current interactions invariant, making them flavour diagonal.

The charged current does not mix left- and right-handed fermions either, but it only couples to left-handed fermions. The interactions are given by
\begin{equation}
    \mathcal{L}_{kin,CC} = 
    \frac{e}{\sqrt{2}\sin\theta_W} \left[
    \bar{u}_{iL} \gamma^\mu V^{ij} d_{jL} W_\mu^+ +
    \bar{\nu}_{iL} \gamma^\mu e_{iL} W_\mu^+ + \text{h.c.}
\right] \, ,
\end{equation}
where $V \equiv U_u^\dagger U_d$ is the \ac{CKM} matrix. 
The \ac{CKM} matrix $V^{ij}$ is a unitary matrix, which describes the flavour-changing transitions between the different quark flavours. It can be written as 
\begin{equation}
    V = \begin{pmatrix}
        V_{ud} & V_{us} & V_{ub} \\
        V_{cd} & V_{cs} & V_{cb} \\
        V_{td} & V_{ts} & V_{tb}
    \end{pmatrix}.
\end{equation}
The \ac{CKM} matrix is a complex unitary matrix, which can be parametrized in terms of three mixing angles and one phase. The phase is responsible for \ac{CP} violation in the \ac{SM}~\cite{Schwartz_2013}.

For leptons, there is no equivalent of the \ac{CKM} matrix, if neutrinos are massless. This is because one is free to rotate the neutrinos in the same way as the leptons in \cref{eq:fermion-basis-change}. If neutrinos are massive, charged leptons and neutrinos need to be diagonalized separately. In this case, one can define a \ac{PMNS} matrix, which is analogous to the \ac{CKM} matrix for leptons~\cite{Schwartz_2013}.
The charged current interactions are only present for left-handed fermions, since the $W^\pm$ bosons only couple to left-handed fermions. The right-handed fermions do not couple to the $W^\pm$ bosons, but only to the neutral currents.

\subsection{Flavour changing neutral currents}
\label{sec:fcnc}

The \ac{CKM} matrix only appears in the charged current interactions. As a result, it is not possible to have \acp{FCNC} at tree level in the \ac{SM}. This is a consequence of the gauge structure of the theory: neutral currents (mediated by the $Z$ boson, photon, and gluons) are diagonal in the mass eigenstate basis, while only charged currents (mediated by $W^\pm$ bosons) involve the \ac{CKM} mixing matrix~\cite{Schwartz_2013,Thomson_2013}.

\ac{FCNC} transitions can occur at loop level, but the loop contributions themselves are highly suppressed by the \ac{GIM} mechanism~\cite{Glashow:1970gm}. The \ac{GIM} mechanism describes how loop amplitudes from different intermediate quarks tend to cancel each other, resulting in additional suppression proportional to mass-squared differences between the intermediate quarks~\cite{Schwartz_2013,Peskin:1995ev}.

\begin{figure}[h]
    \centering
        \begin{tikzpicture}[scale=1.4, transform shape]
            \begin{feynman}
                \vertex (b1) {$b$};
                \vertex [right=of b1] (b2);
                \vertex [right=of b2] (b3);
                \vertex [right=of b3] (b4) {$s$};
            
                \vertex at ($(b2)!0.5!(b3)!0.8cm!90:(b3)$) (g1);
                \vertex [above=5em of b3] (g2);
                \vertex [above=4em of b4] (l1) {\(\nu_l\)};
                \vertex [above=2em of l1] (l2) {\(\overline{\nu}_l\)};
            
                \diagram* {
                  (b1) -- [fermion] (b2) -- [fermion, edge label={$u,c,t$}] (b3) -- [fermion] (b4),
                  (b2) -- [boson, quarter left, edge label={$W$}] (g1) -- [boson, quarter left] (b3),
                  (g1) -- [boson, edge label={$Z$}] (g2),
                  (l2) -- [fermion, bend right] (g2) -- [fermion, bend right] (l1),
                };
            \end{feynman}
        \end{tikzpicture}
        \begin{tikzpicture}[scale=1.4, transform shape]
            \begin{feynman}
                \vertex (b1) {$b$};
                \vertex [right=of b1] (b2);
                \vertex [right=of b2] (b3);
                \vertex [right=of b3] (b4) {$s$};
            
                \vertex [above=3em of b2] (g1);
                \vertex [above=3em of b3] (g2);
                \vertex [above=4em of b4] (l1) {\(\nu_l\)};
                \vertex [above=2em of l1] (l2) {\(\overline{\nu}_l\)};
            
                \diagram* {
                  (b1) -- [fermion] (b2) -- [fermion,edge label={$u,c,t$}] (b3) -- [fermion] (b4),
                  (b2) -- [boson, edge label={$W$}] (g1),
                  (g2) -- [boson, edge label={$W$}] (b3),
                  (g1) -- [fermion, edge label={$l$}] (g2),
                  (g2) -- [fermion] (l1),
                  (l2) -- [fermion,bend right] (g1),
                };
            \end{feynman}
        \end{tikzpicture}
        \caption{Feynman diagrams for the $b \to s \nu \bar \nu$ transition. The left diagram is the penguin diagram, while the right diagram is the box diagram. The $W$ bosons are exchanged in both diagrams, while the $Z^0$ boson is only exchanged in the penguin diagram. The neutrino $\nu_l$ can be any of the three active neutrinos, $\nu_e$, $\nu_\mu$, or $\nu_\tau$.}
        \label{fig:fcnc-diagrams}
\end{figure}
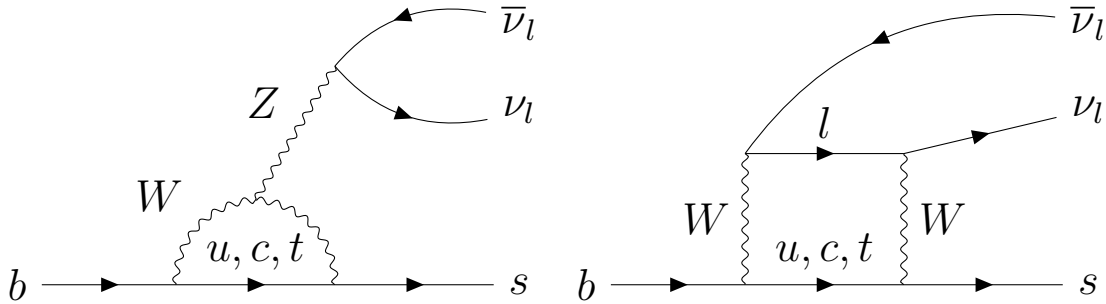

In this work, the focus lies on one \ac{FCNC} transition in particular, the $b \to s \nu \bar \nu$ transition. There are two main contributions to this transition: the box diagram and the penguin diagram, shown in \cref{fig:fcnc-diagrams}. Both diagrams involve two flavour-changing currents and hence the interaction is additionally suppressed by a factor of $|V_{ts} V_{tb}^*|^2$ or $|V_{cs} V_{cb}^*|^2$. The transition via the $u$ quark is even more strongly suppressed by the smallness of the \ac{CKM} matrix elements $|V_{us} V_{ub}^*|^2$~\cite{ParticleDataGroup:2024cfk}.

\section{Effective field theories}
\label{sec:eft}

The central idea of \acp{EFT} relevant to this work is the separation of scales: high-energy modes are removed from the spectrum, leaving an effective description in terms of light fields and a tower of higher-dimensional local operators suppressed by powers of the heavy scale $\Lambda$. This procedure preserves the symmetries of the underlying theory, most notably gauge invariance, and provides a controlled expansion in $E/\Lambda$, where $E$ is the typical energy of the process under consideration.

These types of theories are only applicable to energies far below the scale of new physics $E \ll \Lambda$. In the high-energy, \ac{UV} regime, the expansion parameter $E/\Lambda$ becomes large and the theory loses its predictive power.

The construction of an \ac{EFT} follows a systematic procedure~\cite{Buchalla:1995vs,Buras:1998raa}. For a separation scale $\mu$, one identifies the relevant light \ac{d.o.f.} that are active (i.e.~on-shell) at the energies $E< \mu$. Heavy \ac{d.o.f.}, above the separation scale, are integrated out, generating higher-dimensional operators that encode their effects. The resulting effective Lagrangian takes the form
\begin{equation}
    \mathcal{L}_{\text{EFT}} = \mathcal{L}_{d \leq 4} + \sum_{d>4} \sum_i \mathcal{C}_i^{(d)}(\mu) \mathcal{O}_i^{(d)} + \text{h.c.}\,,
\end{equation}
where $\mathcal{L}_{d \leq 4}$ contains the renormalizable terms (kinetic terms, gauge interactions, and mass terms for the light \ac{d.o.f.}), $\mathcal{O}_i^{(d)}$ are gauge-invariant operators of mass dimension $d$, and $\mathcal{C}_i^{(d)}(\mu)$ are the corresponding Wilson coefficients that encode the \ac{HEP} and depend on the separation scale. The expansion is organized by powers of $(E/\Lambda)^d$, making it a controlled approximation when $E \ll \Lambda$.

This \ac{OPE} separates the short-distance (perturbative) physics at scales higher than $\mu$ (encoded in the Wilson coefficients) from the long-distance (usually non-perturbative) physics lower than $\mu$ (encoded in the matrix elements of the operators). The renormalization scale is typically chosen to be of the order of the mass of the decaying hadron, to suppress large logarithms in perturbative calculations. For $B$ decays this is around the $b$-quark mass $\mu_b \sim 4.2~\text{GeV}$.

Physical amplitudes however do not depend on the arbitrary scale $\mu$. The $\mu$ dependence of the Wilson coefficients $\mathcal{C}_i^{(d)}(\mu)$, has to be cancelled by the matrix elements of the operators $\langle \mathcal{O}_i \rangle(\mu)$~\cite{Buchalla:1995vs}. Hence, what belongs to the Wilson coefficients and what to the matrix elements is a matter of the concrete choice of $\mu$. 
This means that a change in $\mu$ implies renormalization of both Wilson coefficients and the matrix elements of the operators~\cite{Buchalla:1995vs}. Since this renormalization is typically approached as a perturbative expansion in the electromagnetic and strong coupling constants, $\alpha$ and $\alpha_s$, it typically depends on the renormalization scheme.
This scheme dependence is also cancelled between the Wilson coefficients and the matrix elements.

The Wilson coefficients $\mathcal{C}_i^{(d)}(\mu)$ are connected to the full theory through a matching procedure at the scale $\mu \sim \Lambda$, where the \ac{EFT} is required to reproduce the amplitudes of the full theory order by order in perturbation theory. These coefficients encode all effects of the heavy \ac{d.o.f.} and are process-independent, making the \ac{EFT} framework particularly powerful for phenomenological applications.

A crucial aspect of \ac{EFT} calculations is the treatment of renormalization group evolution~\cite{Buchalla:1995vs}. The Wilson coefficients satisfy renormalization group equations that govern their scale dependence:
\begin{equation}
    \mu \frac{d}{d\mu} \vec{\mathcal{C}}(\mu) = \gamma^T(\alpha_s(\mu)) \vec{\mathcal{C}}(\mu),
\end{equation}
where $\gamma(\alpha_s)$ is the anomalous dimension matrix and $\vec{\mathcal{C}}(\mu)$ represents the vector of Wilson coefficients. The solution of these equations allows the evolution from the matching scale $\Lambda$ down to the low-energy scale relevant for the physical process of interest.

\subsection{Weak Effective Theory}
\label{sec:wet}

The \ac{WET} is a specific \ac{EFT} that describes physics below the electroweak scale, typically around $\Lambda = M_W \sim 100$ GeV~\cite{Aebischer:2017gaw,Jenkins:2017jig,Jenkins:2017dyc}. 
This is a large scale compared to the $B$-physics scale $\mu_b \sim 4.2~\text{GeV} \ll \Lambda$, which is the relevant scale for $B$ meson decays. Thus, the weak interaction can be thought of as local at this scale, as propagators of the $W$ and $Z$ bosons can be approximated by a point-like interaction. This factorizes the weak interaction (encoded in the Wilson coefficients) from the electromagnetic and strong interaction (encoded in the hadronic matrix elements of the operators).
In this framework, the $W$, $Z$, and Higgs bosons, along with the top quark, are integrated out as heavy \ac{d.o.f.}. The remaining light fields are the photon, gluons, and the five light quarks ($u$, $d$, $s$, $c$, $b$) and three charged leptons ($e$, $\mu$, $\tau$) along with their associated neutrinos.
If \ac{BSM} physics exists at a scale of $\Lambda$ or above, these interactions will also be approximately local, and hence encoded in the Wilson coefficients.

The \ac{WET} Lagrangian is organized as an expansion in operators of increasing mass dimension:
\begin{equation}
    \mathcal{L}_{\text{WET}} = \mathcal{L}_{\text{QED} + \text{QCD}} + \sum_{d>4} \sum_i \mathcal{C}_i^{(d)}(\mu_b) \mathcal{O}_i^{(d)} + \text{h.c.}\, ,
\end{equation}
where $\mathcal{L}_{\text{QED} + \text{QCD}}$ contains the renormalizable interactions of \ac{QED} and \ac{QCD} (with five active quark flavours). 
The dimension-five operators are responsible for neutrino masses via the Weinberg operator. For $b \to s \nu \bar{\nu}$ transitions, these operators effectively contribute at dimension-six, are generated only at loop level, and are strongly experimentally constrained~\cite{Giunti:2014ixa,Felkl:2021uxi}, leading to further suppression. These are therefore typically neglected.
Quark flavour-changing interactions first appear at dimension-six, leading to corrections to \ac{SM} processes, and are the primary focus in most phenomenological studies. 
Since the mass of the initial on-shell $B$ meson limits the maximum momentum transfer in this process, the matrix elements of
operators with mass-dimension seven or above are suppressed by at least a factor of $M_B / M_W \simeq 0.06$, which
are hence commonly neglected in these types of analyses.

For flavour-changing neutral current processes like $b \to s \nu \bar{\nu}$, it suffices to discuss the $sb\nu\nu$ \emph{sector} of the \ac{WET}~\cite{Gartner:2024muk}.
It is spanned by a subset of local operators of mass-dimension six, which is closed under the renormalization group to leading order in the Fermi constant $G_F$~\cite{Aebischer:2017ugx}.
The corresponding Lagrangian density for the $sb\nu\nu$ sector reads~\cite{Felkl:2021uxi} \footnote{The superscript $(6)$ on the operators and Wilson coefficients is dropped for brevity.}
\begin{equation}
    \mathcal{L}_\text{WET} \supset - \frac{4 G_\text{F}}{\sqrt{2}} \frac{\alpha}{2 \pi} V_{t s}^* V_{t b}
    \sum_i C_i(\mu_b) O_i + \text{h.c.}\,,
    \label{eq:wet-lagrangian}
\end{equation}
with $\alpha$ the fine structure constant.
Due to the normalization in terms of $G_F$ the newly defined Wilson coefficients
\begin{equation}
    C_i(\mu_b) = \frac{\sqrt{2}}{4 G_F} \frac{2 \pi}{\alpha} \frac{1}{V_{t s}^* V_{t b}} \mathcal{C}_i(\mu_b)
\end{equation}
are dimensionless.
They are generally complex-valued and defined in the modified minimal subtraction ($\overline{\text{MS}}$) scheme~\cite{Felkl:2021uxi}.

Assuming massless neutrinos, the full set of dimension-six operators for the $sb\nu\nu$ sector
is given by~\cite{Felkl:2021uxi},
\begin{equation}
\begin{aligned}
\mathcal{O}_{\mathrm{VL}} &=\left(\overline{\nu_L} \gamma_\mu \nu_L\right)\left(\overline{s_L} \gamma^\mu b_L\right) \\
\mathcal{O}_{\mathrm{VR}} &=\left(\overline{\nu_L} \gamma_\mu \nu_L\right)\left(\overline{s_R} \gamma^\mu b_R\right) \\
\mathcal{O}_{\mathrm{SL}} &=\left(\overline{\nu_L^c} \nu_L\right)\left(\overline{s_R} b_L\right) \\
\mathcal{O}_{\mathrm{SR}} &=\left(\overline{\nu_L^c} \nu_L\right)\left(\overline{s_L} b_R\right) \\
\mathcal{O}_{\mathrm{TL}} &=\left(\overline{\nu_L^c} \sigma_{\mu \nu} \nu_L\right)\left(\overline{s_R} \sigma^{\mu \nu} b_L\right),
\end{aligned}
\label{eq:operators}
\end{equation}
with
$
    \nu_L^c
         \equiv C \overline{\nu_L}^T
$ 
and where $C=i\gamma^2\gamma^0$ is the charge conjugation operator and ${\sigma^{\mu\nu} \equiv \frac{i}{2} [\gamma^\mu, \gamma^\nu]}$. 
In the above, the subscripts $\text{V},\text{S},\text{T}$ represent vector, scalar, and tensor operators, respectively.
The subscripts $\text{L},\text{R}$ denote the chirality of the quark current.
The operators are defined as sums over the neutrino flavours, since this is a property that one currently cannot determine experimentally.
If one assumes the existence of only left-handed massless neutrinos, all operators
except $\text{VL}$ and $\text{VR}$ vanish.

In the \ac{SM}, only the vector operator $\mathcal{O}_{\mathrm{VL}}$ receives a non-zero Wilson coefficient at the matching scale~\cite{Parrott:2022zte,EOS:v1.0.16},
\begin{equation}
    C_{\mathrm{VL}}^{\mathrm{SM}} =6.6 \pm 0.1, \quad C_i^{\mathrm{SM}} = 0 \quad \text{for} \quad i \neq \mathrm{VL}.
    \label{eq:WET-SM-point}
\end{equation}
This is due to the fact that the charged current interactions in the \ac{SM} only couple to left-handed fermions, and hence only generate left-handed vector operators at low energies.
New physics beyond the \ac{SM} can modify these Wilson coefficients, potentially generating non-zero contributions to the scalar and tensor operators or modifying the vector coefficients.

\section{Hadronic matrix elements}
\label{sec:hadronic-matrix-elements}

Having established the \ac{WET} framework and the relevant dimension-six operators for $b \to s \nu \bar{\nu}$ transitions in \cref{sec:wet}, I now turn to the evaluation of the corresponding matrix elements. These matrix elements represent the second crucial component of the factorized amplitude structure discussed in \cref{sec:eft}, where the short-distance physics (encoded in the Wilson coefficients) is separated from the long-distance, non-perturbative \ac{QCD} dynamics.

These are matrix elements of the effective operators between initial and final quark states. Since quarks are bound inside hadrons, one needs to evaluate the matrix elements between hadronic states, e.g.~the initial $B$ meson and final $K^{(*)}$ meson. This gives rise to the so-called hadronic matrix elements, which, due to their hadronic nature, contain the non-perturbative \ac{QCD} dynamics.

The evaluation of these matrix elements is inherently non-perturbative since it involves the binding of quarks into hadrons at energy scales where $\alpha_s \sim 1$. This necessitates the use of non-perturbative techniques such as lattice \ac{QCD} calculations. The discussion of lattice \ac{QCD} calculations is beyond this work, for further details see for example Reference~\cite{Lellouch:2011qw}.

Since the hadronic matrix elements are Lorentz tensors, they can be decomposed into Lorentz-invariant form factors multiplied by Lorentz structures that depend on the momenta and spin configurations of the initial and final states. 
The form factor approach provides a systematic way to parametrize these matrix elements in terms of Lorentz-invariant functions that depend only on the momentum transfer $q^2$.

For $B \to K$ transitions, the vector operators in \cref{eq:operators} $\mathcal{O}_{\mathrm{VL}}$ and $\mathcal{O}_{\mathrm{VR}}$ give rise to matrix elements of the form~\cite{Felkl:2021uxi,Gubernari:2018wyi}
\begin{equation}
    \langle K(k) | \bar{s} \gamma^\mu b | B(p) \rangle = f_+(q^2) \left[ (p + k)^\mu - \frac{M_B^2 - M_K^2}{q^2} q^\mu \right] + f_0(q^2) \frac{M_B^2 - M_K^2}{q^2} q^\mu,
    \label{eq:vector-matrix-element}
\end{equation}
where $q = p - k$ is the momentum transfer, $q^2 = {(p - k)}^2$ is the invariant mass squared of the dineutrino system, and $f_+(q^2)$ and $f_0(q^2)$ are the vector form factors, satisfying the constraint $f_+(0) = f_0(0)$. $M_B$ and $M_K$ are the masses of the $B$ meson and kaon, respectively.

The scalar operators $\mathcal{O}_{\mathrm{SL}}$ and $\mathcal{O}_{\mathrm{SR}}$ involve matrix elements of scalar currents~\cite{Felkl:2021uxi,Gubernari:2018wyi}:
\begin{equation}
    \langle K(k) | \bar{s} b | B(p) \rangle = f_0(q^2) \frac{M_B^2 - M_K^2}{m_b - m_s},
    \label{eq:scalar-matrix-element}
\end{equation}
where $m_b$ and $m_s$ are the $b$ and $s$ quark masses, respectively. Interestingly, this matrix element is also proportional to $f_0(q^2)$, which connects the scalar and vector sectors.

The tensor operator $\mathcal{O}_{\mathrm{TL}}$ requires the evaluation of~\cite{Felkl:2021uxi,Gubernari:2018wyi}
\begin{equation}
    \langle K(k) | \bar{s} \sigma^{\mu\nu} b | B(p) \rangle = f_T(q^2) \frac{i}{M_B + M_K} \left[ {(p + k)}^\mu q^\nu - q^\mu {(p + k)}^\nu \right],
    \label{eq:tensor-matrix-element}
\end{equation}
which defines the tensor form factor $f_T(q^2)$.

Pseudoscalar, axial vector, and pseudotensor currents do not contribute to $B \to K$ transitions, by virtue of the symmetries of \ac{QCD}. In particular, the invariance of the strong interaction under $C$, $P$, and \ac{CP} transformations implies that these $B \to K$ matrix elements vanish.

For $B \to K^*$ transitions involving a vector meson in the final state, the parametrization is more complex due to the additional polarization \ac{d.o.f.} of the $K^*$ meson~\cite{Felkl:2021uxi,Gubernari:2018wyi}. 
The vector operators give rise to four independent form factors: $V(q^2)$, $A_0(q^2)$, $A_1(q^2)$, $A_2(q^2)$. The scalar operators contribute through the pseudoscalar current, proportional to $A_0(q^2)$, while the scalar current vanishes by Lorentz structure constraints. The tensor operators introduce three additional form factors $T_1(q^2)$, $T_2(q^2)$, and $T_3(q^2)$, with the constraint $T_1(0) = T_2(0)$. It is common to replace the form factors $A_2(q^2)$ and $T_3(q^2)$ with the linear combinations $A_{12}(q^2)$ and $T_{23}(q^2)$, defined in \cref{eq:a12,eq:t23}. The complete parametrization for all operator types is given in \cref{sec:form-factors}.

The increased complexity of $B \to K^*$ transitions (seven independent form factors compared to three for $B \to K$) reflects the richer kinematic structure available when the final state carries spin, providing enhanced sensitivity to new physics through different angular distributions and $q^2$ dependencies.

The form factors $f_+(q^2)$, $f_0(q^2)$, $f_T(q^2)$ for $B \to K$ transitions and $V(q^2)$, $A_0(q^2)$, $A_1(q^2)$, $A_2(q^2)$, $T_1(q^2)$, $T_2(q^2)$, $T_3(q^2)$ for $B \to K^*$ transitions are non-perturbative objects that must be determined using lattice \ac{QCD} calculations.
In the past, lattice \ac{QCD} calculations were limited to high $q^2 \gtrsim 10~\text{GeV}^2$ due to computational limitations, such as finite volume effects, but recently for the $B \to K^{(*)}$ form factors these limitations were overcome and lattice \ac{QCD} calculations down to $q^2=0$ have become available.
The results from these two methods can be combined using a suitable series expansion. In this work, form factors are parametrized using the \ac{BCL}~\cite{Bourrely:2008} and the \ac{BSZ}~\cite{Bharucha:2015bzk} parametrizations, which respect analyticity and unitarity constraints.
These parametrizations provide a model-independent approach to form factor extrapolation, as they are based on fundamental principles of quantum field theory rather than specific assumptions about the underlying hadronic dynamics.

The series expansion is typically performed in terms of a conformal variable 
\begin{equation}
    z(q^2,t_0) = \frac{\sqrt{t_+ - q^2} - \sqrt{t_+ - t_0}}{\sqrt{t_+ - q^2} + \sqrt{t_+ - t_0}},
\end{equation}
where $t_+ = (M_B + M_{K^{(*)}})^2$ and $t_0$ is a free parameter.
This variable $z$ maps the physical $q^2$ region onto a small interval, improving the convergence of the series expansion.

The \ac{BSZ} parametrization expresses each form factor as a series expansion in the conformal variable $z(q^2, t_0)$~\cite{Bharucha:2015bzk},
\begin{equation}
    f(q^2) = \frac{1}{1 - q^2/m_{\text{res}}^2} \sum_{k=0}^{K} a_k^{(f)} [z(q^2,t_0) - z(0,t_0)]^k,
\end{equation}
where $m_{\text{res}}$ is the mass of the lightest resonance with appropriate quantum numbers, using $t_0 \equiv t_+ (1 - \sqrt{1 - t_-/t_+})$ with $t_- = (M_B - M_{K^{(*)}})^2$, and the coefficients $a_k$ are fit parameters determined from lattice \ac{QCD} data. The series is typically truncated at order $K=2$ or $K=3$, limiting the number of parameters. The \ac{BSZ} parametrization incorporates known pole structures and ensures proper analytical behaviour, making it particularly suitable for extrapolating lattice \ac{QCD} results from high $q^2$ to the full kinematic range. Unitarity constraints can be imposed to further reduce the number of free parameters and improve the stability of extrapolations~\cite{Bharucha:2015bzk}.

The \ac{BCL} parametrization follows a similar approach but emphasizes the implementation of unitarity constraints more systematically~\cite{Bourrely:2008}. In the \ac{BCL} framework, the form factors are expressed as
\begin{equation}
    f(q^2) = \frac{1}{1 - q^2/m_{\text{res}}^2} \sum_{k=0}^{K-1} b_k \left[z(q^2,t_0)^k - (-1)^{(k-K)} \frac{k}{K} z(q^2,t_0)^K \right],
\end{equation}
where the coefficients $b_k$ are constrained by unitarity bounds that limit their allowed values. These constraints reduce the parameter space and improve the stability of extrapolations.

\section{Theory predictions for \texorpdfstring{$B \to K^{(*)} \nu \bar{\nu}$}{B->K(*)nunubar}}
\label{sec:theory-predictions}
The theoretical prediction for $B \to K^{(*)} \nu \bar{\nu}$ decay rates requires combining the \ac{WET} framework with precise determinations of the hadronic form factors. While one cannot achieve a completely model-independent theoretical description, it is nevertheless possible to capture the effects of a large number of \ac{BSM} theories under mild assumptions. Here, one assumes that potential new \ac{BSM} particles and force carriers have masses at or above the scale of electroweak symmetry breaking, making the \ac{WET} description appropriate.
All relevant numerical inputs to predictions in this work are summarized in \cref{tab:sm parameters}.
\begin{table}
  \caption{The input parameters used for calculating predictions. The parameters are taken from References~\cite{EOS:v1.0.16,ParticleDataGroup:2024cfk,Buras:2014fpa,Becirevic:2023aov}. Here $\alpha$ is evaluated at the $B$-physics scale $\mu_b$. The running of CKM elements is neglected, as their scale dependence is very mild.}
  \centering
  \begin{tabularx}{\linewidth}{@{\extracolsep{\fill}}ll}
    \toprule \midrule
    $C_{\mathrm{VL}}^{\mathrm{SM}}$ & $6.6\pm 0.1$ \\
    $1/\alpha(\mu_b)$ & $133.00\pm0.15$ \\
    $|V_{tb}V_{ts}^*|$ & $0.041 \pm 0.001$ \\
    $\tau_{B^0}$ & $1.519 \pm 0.005\, \text{ps}$ \\ 
    $\tau_{B^+}$ & $1.641 \pm 0.004\, \text{ps}$ \\ 
    \midrule \bottomrule
  \end{tabularx}
  \label{tab:sm parameters}
\end{table}

\subsection{Observables for \texorpdfstring{\BKnunu}{B->Knunubar}}
\label{sec:knunu-theory}
Since the $B$ meson is a pseudoscalar, the decay is isotropic in the $B$ meson rest frame, and the only kinematically free variable is the squared dineutrino invariant mass $q^2$. The kinematic range extends from $q^2_{\min} = 0$ (assuming massless neutrinos) to $q^2_{\max} = {(M_B - M_K)}^2$.

The decay $B \to K \nu \bar \nu$ is governed by the \ac{WET} Lagrangian as described by
\cref{eq:wet-lagrangian,eq:operators}.
The differential decay rate for $B \to K \nu \bar{\nu}$ in terms of the \ac{WET} Wilson coefficients is given by~\cite{Gratrex:2015hna,Felkl:2021uxi,Gartner:2024muk}
\begin{equation}
    \begin{aligned}
      \frac{d \Gamma}{d q^{2}}
      & =
      3
      \left(\frac{4 G_\text{F}}{\sqrt{2}} \frac{\alpha}{2 \pi} \right)^2 \left|V_{ts}^* V_{tb}\right|^2
      \frac{\sqrt{\lambda_{BK}} q^{2}}{(4 \pi)^{3} M_{B}^{3}}\\
      &\cdot\left[\frac{\lambda_{BK}}{24 q^{2}}\left|f_{+}(q^2)\right|^{2}\left|C_{\mathrm{VL}}+C_{\mathrm{VR}}\right|^{2}\right.\\
      &\phantom{\cdot}+\frac{\left(M_{B}^{2}-M_{K}^{2}\right)^{2}}{8\left(m_{b}-m_{s}\right)^{2}}\left|f_{0}(q^2)\right|^{2}\left|C_{\mathrm{SL}}+C_{\mathrm{SR}}\right|^{2} \\
      &\phantom{\cdot}\left.+\frac{2 \lambda_{BK}}{3\left(M_{B}+M_{K}\right)^{2}}\left|f_{T}(q^2)\right|^{2}\left|C_{\mathrm{TL}}\right|^{2}\right],
    \end{aligned}
\label{eq:width}
\end{equation}
where $M_B$ and $M_K$ are the masses of the $B$ meson and kaon, respectively, $m_b(\mu_b)$ and $m_s(\mu_b)$ are the $\overline{\text{MS}}$ scheme quark masses evaluated at the $B$-physics scale $\mu_b$, and $\lambda_{BK} \equiv \lambda(M_B^2, M_K^2, q^2)$ is the Källén function
\begin{equation}
    \lambda(a,b,c) = a^2 + b^2 + c^2 - 2ab - 2ac - 2bc.
\end{equation}

As can be inferred from \cref{eq:width}, the decay is only sensitive to the magnitude of three linear combinations
of Wilson coefficients:
\begin{equation}
    |C_{\mathrm{VL}}+C_{\mathrm{VR}}|, \quad |C_{\mathrm{SL}}+C_{\mathrm{SR}}|, \quad |C_{\mathrm{TL}}|.
    \label{eq:BToKnunu-wc-sensitivity}
\end{equation}
This represents a reduction from the 5-dimensional Wilson coefficient space to a 3-dimensional observable space, highlighting the limited sensitivity of current measurements to the full parameter space of the \ac{WET}.

\begin{figure}[t]
    \centering
    \includegraphics[width=0.8\linewidth]{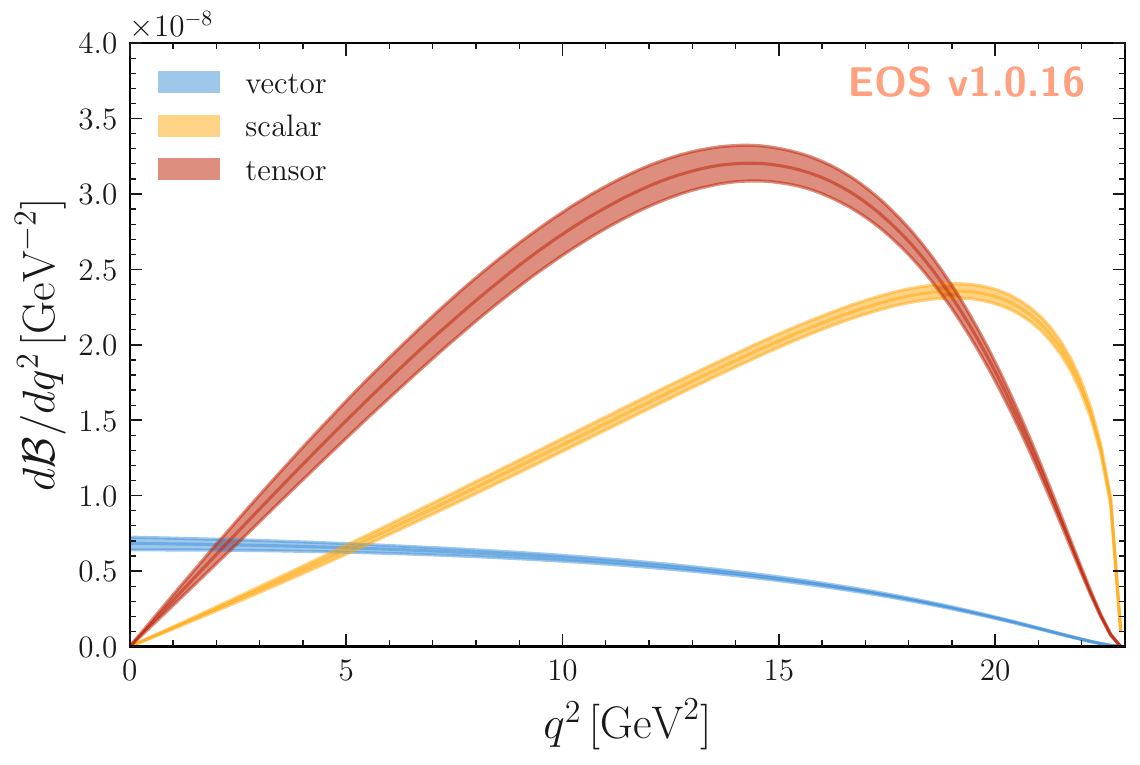}
    \caption{%
        Illustration of the variety of shapes of the $B \to K \nu \bar \nu$ branching ratio due to purely vector, scalar, or tensor interactions.
        Each curve corresponds to setting a single non-zero Wilson Coefficient in \cref{eq:width} to unity while keeping
        all other coefficients at zero. These predictions are obtained using the \EOS software~\cite{EOSAuthors:2021xpv,EOS:v1.0.16}.
    }
    \label{fig:knunu-theory}
\end{figure}

The branching ratio, $\mathcal{B}$, is connected to the decay rate simply through $\mathcal{B} = \tau \Gamma$, where $\tau$ is the decay lifetime~\cite{Schwartz_2013}.
To highlight the individual contributions of vector, scalar, and tensor terms from \cref{eq:width} to the differential branching ratio, an illustration where individual left-handed Wilson coefficients are set to unity is shown in \cref{fig:knunu-theory}. 
The values for the corresponding $8$ hadronic parameters are extracted from a joint theoretical
prior \ac{PDF} comprised of the 2021 lattice world average based on results by the Fermilab/MILC and HPQCD
collaborations~\cite{FlavourLatticeAveragingGroupFLAG:2021npn},
and recent results by the HPQCD collaboration~\cite{Parrott:2022rgu} and are reported in \cref{tab:hadronic parameters full}.

\begin{table}[ht]
    \caption{The hadronic parameters used to parametrize the $f_{+}(q^2)$, $f_{0}(q^2)$ and $f_{T}(q^2)$ form factors in the \ac{BSZ} parametrization~\cite{Bharucha_2016}, based on results by the Fermilab/MILC and HPQCD collaborations~\cite{FlavourLatticeAveragingGroupFLAG:2021npn, Parrott:2022rgu}. The table shows the mean values (first row) and the covariance matrix (upper triangular part).}
    \centering
    \begin{tabularx}{\linewidth}{@{\extracolsep{\fill}}YYYYYYYY}
    \toprule \midrule
    $a_0^{0 /+}$ & $a_1^+$ & $a_2^+$ & $a_1^0$ & $a_2^0$ & $a_0^T$ & $a_1^T$ & $a_2^T$ \\
    0.3380 & -0.8845 & -0.1109 & 0.3742 & 0.0787 & 0.3269 & -0.9625 & -0.2780 \\
    \midrule 0.0001 & 0.0007 & 0.0010 & 0.0008 & 0.0013 & 0.0001 & 0.0006 & 0.0008 \\
    & 0.0093 & 0.0216 & 0.0057 & 0.0102 & 0.0004 & 0.0062 & 0.0147 \\
    & & 0.0672 & 0.0085 & 0.0171 & 0.0001 & 0.0112 & 0.0425 \\
    & & & 0.0062 & 0.0113 & 0.0006 & 0.0047	& 0.0066 \\
    & & & & 0.0220 & 0.0010 & 0.0081 & 0.0128 \\
    & & & & & 0.0005 & 0.0034 & 0.0058 \\
    & & & & & & 0.0319 & 0.0619 \\
    & & & & & & & 0.1441 \\
        \midrule \bottomrule
    \end{tabularx}
    \label{tab:hadronic parameters full}
\end{table}

\subsection{Observables for \texorpdfstring{\BKstnn}{B->K*nunubar}}
\label{sec:ksnunu-theory}
The pseudoscalar $B$ meson decays to the vector meson $K^*$ --- this decay is isotropic in the $B$ meson rest frame. However, the $K^*$ meson is an unstable resonance that subsequently decays to $K \pi$. The decay $K^* \to K \pi$ introduces an angular dependence on the angle between the direction of flight of the $B$ and the $K$ in the $K^*$ rest frame.
This means that the decay rate in \cref{eq:widthKs} exhibits an angular dependence. However, here I integrate out this angular dependence, leaving $q^2$ as the only kinematically free variable. The kinematic range extends from $q^2_{\min} = 0$ (assuming massless neutrinos) to $q^2_{\max} = {(M_B - M_{K^*})}^2$.

The decay $B \to K^* \nu \bar \nu$ is governed by the \ac{WET} Lagrangian as described by
\cref{eq:wet-lagrangian,eq:operators}.
Its differential decay rate reads~\cite{Gratrex:2015hna,Felkl:2021uxi,Gartner:2024muk}
\begin{multline}
    \frac{d \Gamma}{d q^2}
        = 3 \left(\frac{4 G_\text{F}}{\sqrt{2}} \frac{\alpha}{2 \pi} \right)^2 \left|V_{t s}^* V_{t b}\right|^2
        \frac{\sqrt{\lambda_{B K^*}} q^2}{(4 \pi)^3 M_B^3}\\
        \cdot
          \left[|\mathcal{A}_V|^2\left|C_{\mathrm{VL}}+C_{\mathrm{VR}}\right|^2\right.
        + |\mathcal{A}_A|^2 \left|C_{\mathrm{VL}}-C_{\mathrm{VR}}\right|^2
        + |\mathcal{A}_P|^2\left|C_{\mathrm{SR}}-C_{\mathrm{SL}}\right|^2
        + \left. |\mathcal{A}_T|^2 \left|C_{\mathrm{TL}}\right|^2\right] \, ,
\label{eq:widthKs}
\end{multline}
where the reduced amplitudes multiplying the Wilson coefficients read
\begin{equation}
\begin{aligned}
    |\mathcal{A}_V|^2
        & = \frac{\lambda_{B K^*}|V(q^2)|^2}{12\left(M_B+M_{K^*}\right)^2} \, , \\
    |\mathcal{A}_A|^2 
         &= \frac{8 M_B^2 M_{K^*}^2}{3 q^2}\left|A_{12}(q^2)\right|^2
         +\frac{\left(M_B+M_{K^*}\right)^2\left|A_1(q^2)\right|^2}{12} \, ,\\
    |\mathcal{A}_P|^2 
        & = \frac{\lambda_{B K^*}}{8\left(m_b+m_s\right)^2}\left|A_0(q^2)\right|^2 \, , \\
    |\mathcal{A}_T|^2
         &= \frac{32 M_B^2 M_{K^*}^2\left|T_{23}(q^2)\right|^2}{3\left(M_B+M_{K^*}\right)^2} +\frac{4 \lambda_{B K^*}\left|T_1(q^2)\right|^2}{3 q^2} 
     +\frac{4\left(M_B^2-M_{K^*}^2\right)^2\left|T_2(q^2)\right|^2}{3 q^2} \, .
\end{aligned}
\end{equation}

One readily finds that the dependence of the observables on the Wilson coefficients is very different in \cref{eq:width,eq:widthKs}, respectively.
Compared to $B\to K\nu\bar{\nu}$ decays, the differential $B\to K^*\nu\bar{\nu}$ decay rate exhibits additional sensitivity to the quantities
\begin{equation}
    |C_{\mathrm{VL}} - C_{\mathrm{VR}}|, ~ |C_{\mathrm{SL}} - C_{\mathrm{SR}}|.
    \label{eq:BToKstarnunu-wc-sensitivity}
\end{equation}
As a consequence, a simultaneous analysis of both decays allows constraining a total of five real-valued (out of ten total real-valued) parameters in the $sb\nu\nu$ sector. Assuming all \ac{WET} Wilson coefficients to be real-valued, this corresponds to constraining the magnitudes of all Wilson coefficients.

\begin{figure}[t]
    \centering
    \includegraphics[width=\linewidth]{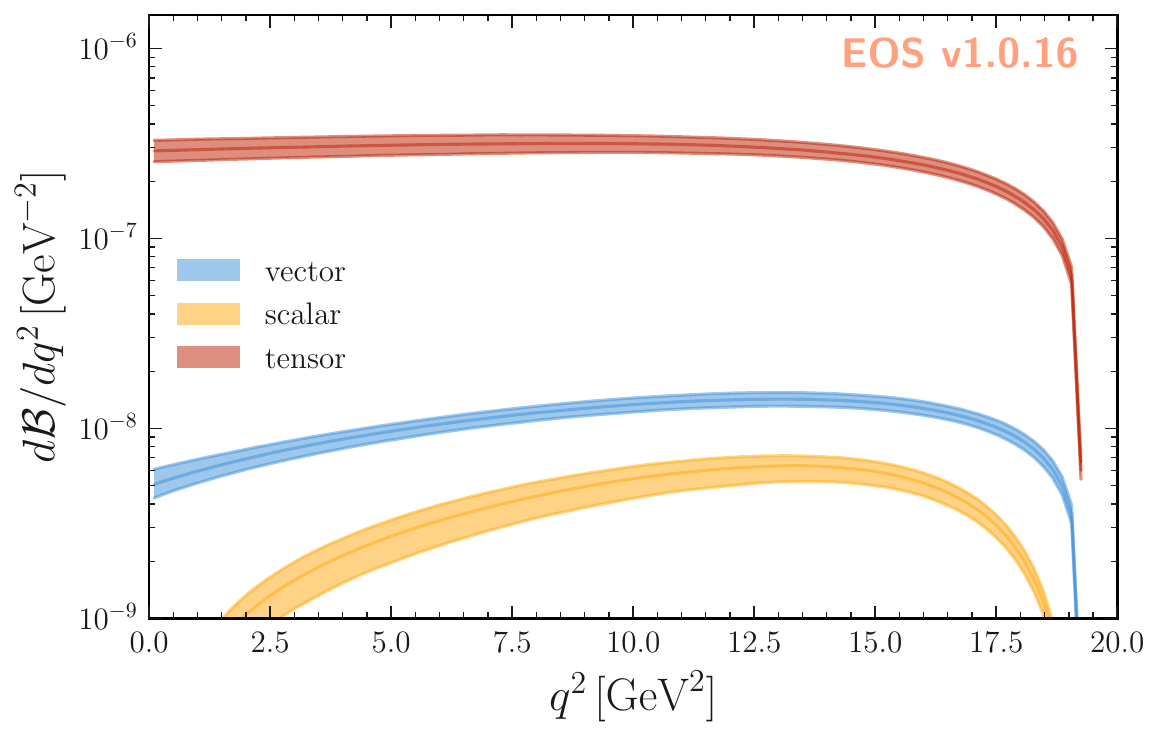}
    \caption{
        Illustration of the variety of shapes of the $B \to K^* \nu \bar \nu$ branching ratio due to purely vector, scalar, or tensor interactions.
        Each curve corresponds to setting a single (left-handed) non-zero Wilson coefficient in \cref{eq:widthKs} to unity while keeping all other coefficients at zero. These predictions are obtained using the \EOS software~\cite{EOSAuthors:2021xpv,EOS:v1.0.16}. The values for the corresponding $19$ hadronic parameters arise from the Gaussian likelihood provided in Reference~\cite{Gubernari:2023puw}.
    }
    \label{fig:ksnunu-theory}
\end{figure}

To understand the individual contributions to the differential decay rate in \cref{eq:widthKs} by vector, scalar, and tensor operators, I provide an illustration of their relative sizes and their shapes in \cref{fig:ksnunu-theory}. This is achieved by setting their respective Wilson coefficients to unity. The values for the corresponding $19$ hadronic parameters arise from the Gaussian likelihood provided in Reference~\cite{Gubernari:2023puw} and are reported in \cref{tab:hadronic parameters kstar full} of \cref{sec:kstar form factors}.

\subsection{The \EOS software package}
\label{sec:eos}
New physics contributions can significantly modify the $B \to K^{(*)} \nu \bar{\nu}$ prediction by introducing non-zero Wilson coefficients for the scalar and tensor operators or by modifying the vector coefficients. The shape of the $q^2$ spectrum also provides additional discriminating power between different new physics scenarios, as vector, scalar, and tensor interactions lead to characteristically different kinematic distributions.

The \EOS software package~\cite{EOSAuthors:2021xpv,EOS:v1.0.16} provides a comprehensive framework for computing theoretical predictions for $B \to K^{(*)} \nu \bar{\nu}$ and many other processes. It incorporates the \ac{WET} framework, allowing users to specify Wilson coefficients at a chosen scale and compute the corresponding decay rates and distributions. 

For the $sb\nu\nu$ sector, \EOS provides implementations of all five dimension-six operators in \cref{eq:operators}, allowing for arbitrary values of the Wilson coefficients $C_{\mathrm{VL}}$, $C_{\mathrm{VR}}$, $C_{\mathrm{SL}}$, $C_{\mathrm{SR}}$, and $C_{\mathrm{TL}}$. This flexibility makes it particularly suitable for exploring \ac{BSM} scenarios that deviate from the \ac{SM} prediction in \cref{eq:WET-SM-point}.

For $B \to K^{(*)} \nu \bar{\nu}$ decays \EOS can compute a wide range of observables:

\textbf{Integrated observables:}
\begin{itemize}
\item Total branching ratios $\mathcal{B}(B \to K^{(*)} \nu \bar{\nu})$
\item Partial branching ratios in specified $q^2$ bins
\item Ratios of branching ratios $\mathcal{R}_{K^*} = \mathcal{B}(B \to K^* \nu \bar{\nu})/\mathcal{B}(B \to K \nu \bar{\nu})$
\item Longitudinal polarization fraction $F_L$ for $B \to K^* \nu \bar{\nu}$
\end{itemize}

\textbf{Differential observables:}
\begin{itemize}
\item Differential branching ratios $d\mathcal{B}/dq^2$
\item Normalized $q^2$ spectra and shape parameters
\item Angular distributions for $B \to K^* \nu \bar{\nu}$
\end{itemize}

The package incorporates many hadronic form factor results from different lattice \ac{QCD} collaborations. These form factors are implemented using multiple different parametrizations, with full covariance matrices to properly account for correlations between different form factors.

Furthermore, it is possible to combine different likelihoods, for example from different lattice \ac{QCD} collaborations, in a statistically consistent manner. This, in turn, leads to more precise determinations of the form factors and hence more accurate predictions for the decay rates.

The package is actively maintained and regularly updated to incorporate new lattice \ac{QCD} results, improved theoretical calculations, and additional \ac{BSM} models. All calculations are documented with complete references to the underlying theoretical framework, ensuring reproducibility and traceability of results. The implementation of \cref{eq:width} and partially \cref{eq:widthKs} was one of my contributions to the \EOS package.

This comprehensive implementation makes \EOS a valuable tool for precision flavour phenomenology, enabling both theoretical predictions and experimental interpretation within a unified, well-validated framework.

%% file: chapters/experiment.tex
Experimental validation of theoretical predictions requires sophisticated experimental facilities capable of producing and studying rare processes with high precision. This chapter presents the Belle~II experiment and its underlying infrastructure, which provides the experimental data foundation for all the reinterpretation studies discussed in subsequent chapters.

The Belle~II experiment, operating at the SuperKEKB electron-positron collider in Japan, is a $B$ factory designed to study flavour physics at unprecedented precision. Building upon the success of its predecessor Belle, Belle~II aims to collect 50 times more data, enabling searches for rare processes with branching ratios as small as $10^{-6}$ or below. This sensitivity is essential for studying flavour-changing neutral current processes like $B \to K^{(*)} \nu\bar{\nu}$.

The Belle~II experiment offers unique capabilities for studying rare $B$ meson decays that are sensitive to \ac{BSM} physics. Operating at the $\Upsilon(4S)$ resonance, which decays dominantly to $B\bar{B}$ pairs, provides a clean experimental environment with well-defined initial conditions and low combinatorial backgrounds compared to hadron colliders. Understanding the experimental setup, data collection methods, and analysis techniques is essential for appreciating both the complexity of extracting rare signals from collision data and the opportunities for model-agnostic reinterpretation that preserve this experimental information.

This chapter establishes the experimental foundation for the measurements and reinterpretations presented in subsequent chapters. \Cref{sec:b-factories} introduces $B$ factories as precision intensity-frontier experiments. \Cref{sec:superkekb} describes the SuperKEKB accelerator. \Cref{sec:detector} presents the Belle~II detector system, covering the tracking detectors (\cref{sec:pxd,sec:svd,sec:cdc}), particle identification systems (\cref{sec:particle-identification}), electromagnetic calorimeter (\cref{sec:ecl}), $K_L^0$ and muon detection (\cref{sec:klm}), and the trigger system (\cref{sec:trigger}). Finally, \cref{sec:computing} discusses the computing and software infrastructure, including event reconstruction (\cref{sec:reconstruction}) and \ac{MC} simulation (\cref{sec:simulation}) that enables both data processing and physics analysis.

\section{$B$ factories}
\label{sec:b-factories}
$B$ factories are high-luminosity particle colliders specifically designed to produce and study $B$ mesons, which are composed of a bottom $b$ quark and a lighter $u$ or $d$ quark. In these factories, $B$ mesons are produced by colliding electrons and positrons at a centre-of-mass energy of around 10.58~GeV, to produce $\Upsilon(4S)$ mesons at the highest possible rate. These decay into $B\bar{B}$ meson pairs with a branching ratio of $\mathcal{B}>0.96$ at $95\%$ confidence level~\cite{ParticleDataGroup:2024cfk}. 

Whereas experiments at the \ac{LHC} probe the outcomes of proton-proton collisions at the energy frontier --- with large centre-of-mass energies of up to 14~TeV --- $B$ factories are optimized for the intensity frontier --- producing large, clean data sets with a well-defined initial state. This enables precision studies of rare and suppressed flavour processes, offering high sensitivity to indirect effects of \ac{BSM} physics.

The original motivation for $B$ factories was to test the mechanism of \ac{CP} violation in the $B$ meson system, as predicted by the \ac{CKM} matrix. Two $B$ factories, Belle and BaBar, independently observed this prediction for the first time~\cite{BaBar:2001pki,Belle:2001zzw}. 

The successor of Belle, Belle~II, now targets an even deeper understanding of flavour physics~\cite{Belle-II:2018jsg}: \textit{Are there new \ac{CP}-violating phases in the quark sector?}; \textit{Does nature have a left-right symmetry, and are there \ac{BSM} \ac{FCNC}?}; \textit{Are there sources of \ac{BSM} lepton flavour violation?}.
The experiment also addresses questions that go beyond the realm of flavour physics, such as: \textit{Is there a dark sector of particle physics at the same mass scale as ordinary matter?}.

A key capability of $B$ factories is their ability to study time-dependent \ac{CP} violation through the analysis of entangled $B^0 \bar{B}^0$ pairs produced in $\Upsilon(4S)$ decays. In the neutral $B$ meson system, the flavour eigenstates and mass eigenstates are not identical, leading to quantum oscillations between $B^0$ and $\bar{B}^0$. This mixing proceeds via flavour-changing neutral current box diagrams (see \cref{sec:fcnc}). Because the $B^0 \bar{B}^0$ pair is produced in a coherent quantum state, the decay of one meson at a given time projects the other into a definite flavour state. By reconstructing the flavour of both mesons and measuring the time difference between their decays, one can extract time-dependent decay rates. These measurements provide access to both the magnitudes and phases of \ac{CKM} matrix elements, including the \ac{CP}-violating phase~\cite{Thomson_2013, Belle-II:2018jsg}. Given the short lifetimes of neutral $B$ mesons, such studies are only possible at asymmetric-energy colliders, where the boost of the $\Upsilon(4S)$ along the beam line enables resolution of the two decay vertices.

\section{Accelerator: SuperKEKB}
\label{sec:superkekb}
SuperKEKB is a high-luminosity, asymmetric-energy electron-positron collider located at the High-Energy Accelerator Research Organization (KEK) in Tsukuba, Japan. It operates at a centre-of-mass energy corresponding to the mass of the $\Upsilon(4S)$ resonance. The design luminosity of SuperKEKB is $6 \cdot 10^{35}~\text{cm}^{-2}~\text{s}^{-1}$ --- about 30 times higher than that achieved by its predecessor, KEKB~\cite{superkekb, 10.1093/ptep/pts083, Belle-II:2024vuc}.

\begin{figure}
    \centering
    \includegraphics[width=0.8\textwidth]{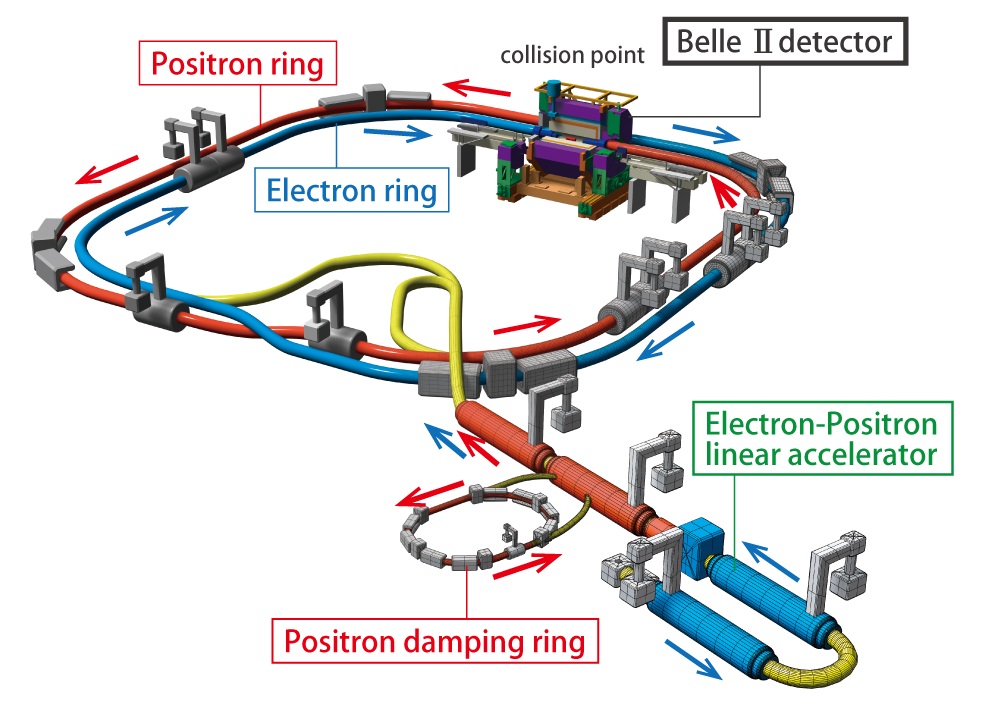}
    \caption{A schematic view of the SuperKEKB collider complex. Electron and positron rings intersect at the Belle~II detector~\cite{superkekb}.}
    \label{fig:superkekb}
\end{figure}

At SuperKEKB, electrons and positrons are generated and accelerated through a multi-stage process before colliding at the interaction point within the Belle~II detector. Electrons are produced by thermionic emission from an electron gun and are directly injected into the High-Energy Ring (HER) at $7~\text{GeV}$. Positrons are created when high-energy electrons strike a tungsten target, producing electron-positron pairs. The resulting positron beam has a large emittance --- the particles are spread out in position and angle. To reduce this, positrons are passed through a damping ring, where synchrotron radiation reduces their emittance, making the beam narrower and more collimated. The cooled positrons are then injected into the Low Energy Ring (LER) at $4~\text{GeV}$. Both beams circulate in opposite directions, steered and focused by magnets and accelerated by radiofrequency (RF) cavities, and eventually collide at the Belle~II detector's interaction point.
An illustration is shown in \cref{fig:superkekb}.

The asymmetric beam energies lead to a boost of the centre-of-mass system, which allows for measuring the time-dependent \ac{CP} symmetry violation by creating a displacement of decay vertices~\cite{Belle-II:2018jsg}.

\section{The Belle~II detector}
\label{sec:detector}
The Belle~II detector is a large-acceptance magnetic spectrometer designed for precision studies of flavour physics in $e^+e^-$ collisions at the SuperKEKB collider in Japan~\cite{ADACHI201846, Abe:2010gxa}. It is optimized for measurements in the $B$, $D$, and $\tau$ sectors and searches for \ac{BSM} physics, designed to operate at high luminosities and background conditions.

\begin{figure}
    \centering
    \includegraphics[width=\textwidth]{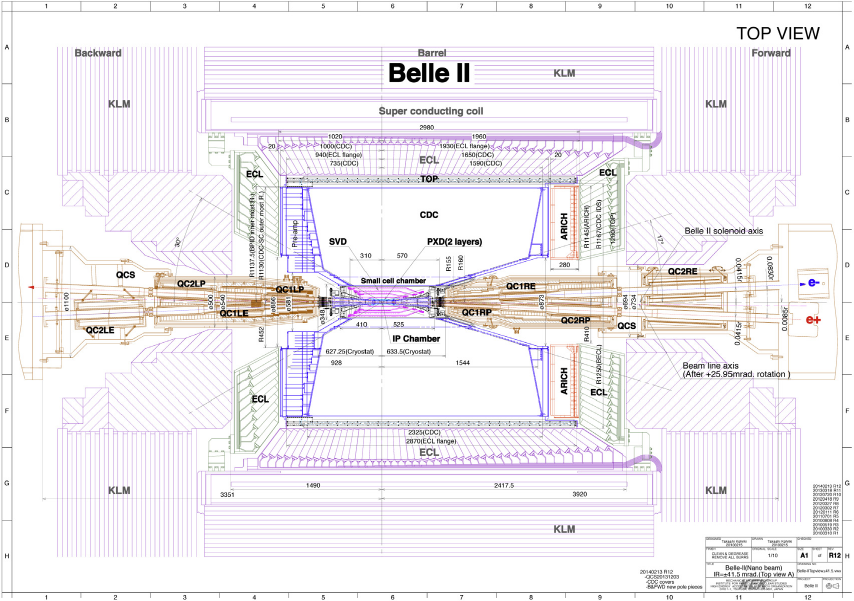}
    \caption{A technical design drawing of the Belle~II detector~\cite{ADACHI201846}.}
    \label{fig:belle2design}
\end{figure}

The detector is composed of several subdetectors arranged in a cylindrical geometry around the interaction point, see \cref{fig:belle2design}. Starting from the innermost layer and moving outward, the tracking system includes a high-resolution \ac{PXD} (\cref{sec:pxd}) and \ac{SVD} (\cref{sec:svd}), which together enable precise vertex reconstruction. These are followed by a large \ac{CDC} (\cref{sec:cdc}) that measures charged particle momenta and provides energy loss information for particle identification.

Surrounding the tracking system are specialized detectors for particle identification (\cref{sec:particle-identification}): the \ac{TOP} in the barrel region and the \ac{ARICH} in the forward endcap. The \ac{ECL} (\cref{sec:ecl}), measures photon and electron energies and contributes to event triggering and particle identification. The entire system is enclosed in a 1.5~T superconducting solenoid magnet, providing the magnetic field necessary for charged particle momentum measurements.
The outermost component, the \ac{KLM} (\cref{sec:klm}), identifies muons and neutral kaons that interact in iron absorber layers, which simultaneously serve as the return yoke for the superconducting magnet.

\subsection{Pixel vertex detector}
\label{sec:pxd}
The Belle~II \ac{PXD} comprises the innermost two layers of the Belle~II tracking system and is essential for precisely reconstructing the decay vertices of short-lived particles, especially $B$ mesons~\cite{ADACHI201846,Abe:2010gxa}. Accurate vertex determination is critical for studies of \ac{CP} violation and rare decays.

The \ac{PXD} uses DEPFET (DEPleted Field Effect Transistor) pixel sensors, which combine charge collection and signal amplification in each pixel. This design enables a high signal-to-noise ratio and a minimal material thickness, reducing multiple scattering and preserving tracking precision.

Structured as two cylindrical layers located approximately 14 mm and 22 mm from the interaction point, the \ac{PXD} is positioned to detect charged particles as early as possible. This proximity maximizes vertex resolution but also exposes the detector to high background rates and radiation levels. These challenges are mitigated through the use of finely segmented pixels, fast readout electronics, and radiation-hardened construction.

As the luminosity of the accelerator increases, the \ac{PXD} must handle higher data rates and occupancy. To manage this in the future, a data reduction scheme will use information from the outer tracking detectors to define \acp{ROI}. Only pixel data within these \acp{ROI} are processed, allowing the system to retain essential tracking information while significantly reducing data throughput.

\subsection{Silicon vertex detector}
\label{sec:svd}
The \ac{SVD} is the second-innermost component of the Belle~II tracking system, surrounding the \ac{PXD} and contributing to precise reconstruction of charged particle trajectories and decay vertices~\cite{ADACHI201846,Abe:2010gxa}. Together with the \ac{PXD} (\cref{sec:pxd}) and \ac{CDC} (\cref{sec:cdc}), it forms the core of the Belle~II tracking system.

The \ac{SVD} consists of four layers of double-sided silicon strip sensors positioned at radii between 38 mm and 135 mm. Each sensor provides 2-dimensional hit information via orthogonal strip readouts, enabling accurate three-dimensional track reconstruction. This complements the high-resolution but localized information from the \ac{PXD}.

With its large acceptance, the \ac{SVD} improves tracking efficiency and vertexing, especially for low-momentum particles and decay chains with displaced vertices. To minimize multiple scattering, the detector is built with a low material budget.
Operating in a high-radiation environment, it employs radiation-tolerant electronics. 
The \ac{SVD} also plays a key role in the \ac{ROI} determination that reduces \ac{PXD} data volume in real time.

\subsection{Central drift chamber}
\label{sec:cdc}
The \ac{CDC} is the primary tracking detector of Belle~II~\cite{ADACHI201846,Abe:2010gxa}. It reconstructs charged tracks and measures their momenta, provides particle identification via measuring the energy loss, enables identification of low-momentum particles that do not reach outer detectors, and supplies reliable trigger signals for charged particles.

To satisfy these requirements, the \ac{CDC} is designed with large geometrical acceptance, fine spatial resolution, and allows for stable operation under high-luminosity conditions. It consists of a cylindrical volume filled with a low-mass helium-ethane gas mixture to minimize multiple scattering, and incorporates sense wires arranged in alternating axial and stereo layers of wires. Axial wires are aligned parallel to the beam axis and provide measurements in the transverse plane, while stereo wires are tilted at a small angle and supply position information along the beam direction.

As charged particles pass through the \ac{CDC}, they ionize gas molecules, leaving behind a trail of ionization electrons along their paths. These electrons drift toward thin sense wires held at high voltage, under the influence of an electric field. When the electrons reach a sense wire, they create a measurable electrical signal. From the arrival time of these signals and the drift velocity of the electrons in the gas, the distance of closest approach between the particle and the wire can be determined. With many such measurements across multiple layers of wires, the full trajectory of a charged particle can be reconstructed. The particle momentum and charge is inferred from the curvature of the track in a 1.5~T solenoid magnetic field, while the amount of ionization per unit length, or energy loss, provides information on the particle identity.

\subsection{Particle identification system}
\label{sec:particle-identification}
The detector employs a sophisticated \ac{PID} system designed to distinguish between different charged particle species, such as pions, kaons, and protons, over a wide momentum range. This capability is essential for accurate reconstruction of decay processes and reducing background contamination.

Two complementary Cherenkov detectors comprise the core of the \ac{PID} system: the \ac{TOP} and the \ac{ARICH} detector~\cite{ADACHI201846,Abe:2010gxa}. 

\subsubsection{Barrel: Time-of-propagation counter}
The \ac{TOP} detector surrounds the barrel region of the tracking system and exploits the Cherenkov effect, where charged particles travelling faster than the speed of light in a radiator material emit photons at a characteristic angle.
Quartz bars serve as radiators and light guides. Cherenkov photons produced by a traversing particle propagate via total internal reflection to photodetectors. By precisely measuring the arrival time and spatial distribution of these photons, the detector reconstructs the Cherenkov angle and thus infers the particle velocity. When combined with momentum measurements from the tracking system, this allows for identification of particle species in the intermediate momentum range.

\subsubsection{End-cap: Aerogel Ring-Imaging Cherenkov detector}
Located in the forward endcap region, the \ac{ARICH} detector complements the barrel-region \ac{TOP} detector by providing \ac{PID} in the forward acceptance. It employs a dual-layer aerogel radiator with varying refractive indices to produce Cherenkov photons, which are detected by arrays of hybrid avalanche photodiodes. The radius of the resulting Cherenkov rings depends on the particle velocity, enabling separation of charged particles. While the \ac{TOP} detector offers excellent \ac{PID} in the barrel region, the \ac{ARICH} detector covers the forward angles.
The dual-layer aerogel structure enhances photon yield and angular resolution, improving \ac{PID} efficiency in the forward direction.

\subsection{Electromagnetic calorimeter}
\label{sec:ecl}
The \ac{ECL} measures the energy and position of electromagnetic showers~\cite{Belle-II:2018jsg, ADACHI201846, Abe:2010gxa, Shwartz_2017}. It is responsible for particle identification, triggering, and the reconstruction of neutral particles, which decay into photons.

The \ac{ECL} consists of scintillating crystals in the barrel and the two endcap sections, covering 90\% of the solid angle in the centre-of-mass frame. When a photon or electron enters the calorimeter, it initiates an electromagnetic shower through cascading processes of bremsstrahlung and pair production. High-energy electrons emit photons via bremsstrahlung when deflected by atomic nuclei, while high-energy photons convert to electron-positron pairs. This cascade multiplies until the particles reach low enough energies to be absorbed through ionization and excitation of the crystal lattice. The scintillation light is produced in proportion to the total deposited energy. The resulting signals provide precise measurements of both the total deposited energy and the shower position~\cite{ParticleDataGroup:2024cfk}.

In addition to energy measurement, the \ac{ECL} contributes to particle identification by helping to distinguish between electrons and hadrons based on shower shape and energy-momentum correlation.

\subsection{\texorpdfstring{$K_0^L$ and $\mu$ detection}{K0L and mu detection}}
\label{sec:klm}
The \ac{KLM} forms the outermost component of the detector~\cite{ADACHI201846, Abe:2010gxa}. Its primary functions are muon identification, detection of $K_L^0$ mesons through hadronic interactions, and providing trigger and background rejection capabilities.

The \ac{KLM} consists of an alternating sandwich of iron plates and active detector elements, located outside the superconducting solenoid. In the barrel region, the active layers are composed of resistive plate chambers.  To withstand higher background rates, the endcaps and innermost barrel layers use plastic scintillator strips with wavelength-shifting fibres, read out by silicon photomultipliers.

Muons are identified based on their ability to penetrate through the detector and absorber material. As charged particles, muons traverse the entire detector, including the \ac{ECL} and solenoid magnet, and continue into the \ac{KLM}. They pass through the detector layers without initiating hadronic or electromagnetic showers. This distinguishes them from hadrons and electrons, which typically stop or shower earlier.

Neutral kaons of the $K_L^0$ type, on the other hand, do not leave a signal in the inner tracking systems or calorimeter due to their long lifetimes and lack of charge. Instead, they are detected in the \ac{KLM} via hadronic interactions with the absorber material, which produce visible secondary particles in the active detector layers.

\subsection{Trigger}
\label{sec:trigger}
The trigger system is responsible for selecting potentially interesting events from the large number of particle collisions, reducing the data rate to a manageable level~\cite{Belle-II:2018jsg, BAHR2025170279}. Its primary goal is to maintain high efficiency for events of physics interest, while rejecting background and uninformative events in real time.

Belle~II employs a two-level trigger system: the hardware-based Level-1 (L1) trigger and the software-based High-Level Trigger (HLT). The L1 trigger operates with low latency and uses information from the \ac{ECL}, \ac{CDC}, \ac{KLM} and \ac{TOP} to identify charged tracks, electromagnetic clusters, and muon-like signatures. It reduces the event rate from the initial collision frequency of about 200~MHz to about 30~kHz.

Events accepted by the L1 trigger are processed by the HLT, which performs more refined reconstruction and selection using full detector information, except for \ac{PXD} information. The HLT leverages real-time software algorithms similar to offline reconstruction. The HLT reduces the data rate further to approximately 10~kHz, making storage and subsequent analysis feasible. Further, it identifies \acp{ROI} for the \ac{PXD} readouts to reduce the data flux.

The selection logic of the trigger system is governed by predefined trigger menus, which define criteria for accepting events based on reconstructed objects such as high-momentum tracks, electromagnetic clusters, and muon candidates. These menus are optimized for various physics goals, including rare decay searches, flavour tagging, and calibration processes.

\section{Computing and software}
\label{sec:computing}

The core software framework used at Belle~II is \texttt{basf2} (Belle Analysis Software Framework 2), a modular C++ framework designed for reconstruction, simulation, and analysis~\cite{basf2}. It supports both real-time data processing and offline analysis. 

Once raw data are recorded by the detector and passed through the data acquisition system (DAQ), calibration constants, dependent on accelerator and detector conditions, are determined.
Subsequently, the raw data are processed, and the result is stored in smaller \ac{mDST} files, subsets of which serve as inputs to most physics analyses. This reconstruction step is discussed in \cref{sec:reconstruction}. For individual analyses, or topically related groups of such, data undergo \textit{skimming}, to include more analysis-specific information~\cite{Abe:2010gxa}.

Observed collision data must be compared to well-understood theoretical expectations to enable meaningful interpretation. This is mostly done with simulated \ac{MC} data, produced to replicate both the underlying physics processes and the detector response as accurately as possible (see \cref{sec:simulation}).

\subsection{Reconstruction}
\label{sec:reconstruction}
Reconstruction is the process of converting raw detector signals (or their simulated equivalents) into high-level physics objects such as tracks, clusters, and \ac{PID} likelihoods. It is conventionally divided into tracking and \ac{PID}, where each step combines data from multiple subdetectors~\cite{Abe:2010gxa, Belle-II:2018jsg}.

\paragraph{Tracking}
The tracking process combines information from the \acp{VXD} (\ac{PXD} and \ac{SVD}), and \ac{CDC} to reconstruct charged particle trajectories. This is crucial for determining the momentum and charge of particles, as well as their decay vertices. 
Track finding collects hits or hit clusters to form track candidates by applying pattern recognition algorithms. 

In the \ac{CDC}, a global algorithm uses all hits simultaneously and applies a Legendre transformation in conformal space to efficiently find circular trajectories from the interaction point. A local algorithm complements this by identifying displaced or short tracks via a cellular automaton that builds track segments iteratively. The outputs are merged into a final track list~\cite{Bertacchi_2021}.

Track finding in the \ac{VXD} uses a cellular automaton to link space points in the detectors into track candidates. Possible combinations are reduced by applying geometric selection criteria, which are defined through a so-called sector map. The sector map is a data-driven lookup table that encodes geometrically allowed hit combinations based on \ac{MC} information~\cite{Belle-II:2018jsg}.
Track candidates from the \ac{VXD} and \ac{CDC} are combined by extrapolating \ac{CDC} tracks to the \ac{VXD} layers and matching them to \ac{VXD} tracks~\cite{ADACHI201846}.

Track fitting estimates the kinematic \acp{POI} of track candidates, such as momentum and position at the vertex. This is achieved using a Kalman filter-based algorithm, which iteratively combines predictions from a model of the particle's motion with measurements from the detector. The Kalman filter accounts for uncertainties in both the model and the measurements.
For the fit, a specific particle hypothesis must be assumed~\cite{Belle-II:2018jsg}.

\paragraph{Particle identification}

After tracks have been identified, \ac{PID} is performed by combining information from multiple subdetectors. For each reconstructed track, subsystem-specific likelihoods are calculated for various particle hypotheses using observables such as ionization energy loss ($dE/dx$) in the \ac{CDC}, time-of-propagation in the \ac{TOP} detector, Cherenkov angle measurements in the \ac{ARICH}, electromagnetic shower profiles in the \ac{ECL}, and hit patterns in the \ac{KLM}. These observables are compared to expected responses derived from calibration and simulation. 
The subsystem likelihoods are then combined into global \ac{PID} likelihoods for six different particle species: $\pi^\pm$, $K^\pm$, $p^\pm$, $d^\pm$, $e^\pm$, $\mu^\pm$. Differences of log likelihoods are used to construct \ac{PID} values, which are used to classify the particle type of each track. Neutral \ac{PID} is also possible using the \ac{ECL} and \ac{KLM} clusters. Thereby, one can identify photons and $K^0_L$ mesons~\cite{Belle-II:2018jsg,ADACHI201846}.


All reconstructed data are stored in the \ac{mDST} format~\cite{Abe:2010gxa}, which includes a compact yet comprehensive set of reconstructed objects. These serve as the primary input to most physics analyses.


\subsection{Simulation}
\label{sec:simulation}
Simulated data, also known as \ac{MC} data, is a crucial part of any \ac{HEP} experiment, replicating the full detector response to known physics processes. This ensures accurate comparison between theory and experiment.
These simulations provide controlled datasets to interpret experimental data, evaluate detector performance, and design analysis strategies. Further, they allow estimation of signal efficiencies, background modelling, and validation of reconstruction algorithms. Large sets of generic \ac{MC} samples, which simulate all known \ac{SM} processes, are produced to represent the real data as faithfully as possible, whereas specialized, smaller sets of \ac{MC} samples are generated for individual analyses~\cite{Abe:2010gxa}.

For Belle~II~\cite{Belle-II:2018jsg}, the simulation chain begins with event generation using physics event generators such as \texttt{EvtGen}~\cite{evtgen} to model decays of $B$ and $D$ mesons into exclusive final states. \texttt{PYTHIA}~\cite{pythia} is used for inclusive decay final states and for the continuum production of light quark pairs. 

The generated events are then passed to a detailed \texttt{GEANT4}-based detector simulation~\cite{GEANT4:2002zbu}, which models the interaction of particles with the full Belle~II detector geometry and materials, including energy deposition, multiple scattering, and detector response.

After the physics interactions are simulated, the digitization step emulates the response of the Belle~II electronics. This converts the simulated signals into the same digital format used by the real detector hardware. The digitized output is then processed using the same reconstruction algorithms as real collision data, resulting in reconstructed tracks, clusters, and subsequently particle candidates.

To accurately reflect real experimental conditions, simulated events are overlaid with beam-induced backgrounds and noise sampled from randomly triggered data events. This ensures that the simulation accounts for realistic detector occupancy and noise effects. 

Large \ac{MC} samples are produced for signal and background processes under varying beam and detector configurations to provide sufficient statistical power for precision measurements.

The output of the \ac{MC} simulation consists of \ac{mDST} files, which are analysed in the same way as the real data. The \ac{MC} data additionally contains the true event information. A unified reconstruction chain for both real and simulated data guarantee direct comparability and accurate modelling of detector effects.

%% file: chapters/stat-models.tex
``Why does one need so much collision data? Wouldn't one new physics event suffice?''~--- This question is asked quite frequently, and it reveals a common misconception about particle physics searches.
My answer usually starts with the probabilistic nature of quantum mechanics, which makes every measurement inherently uncertain. It continues with the challenges posed by theoretical predictions that come with their own uncertainties, and it ends with a discussion about the limitations in modelling our detectors --- instruments that, despite their sophistication, cannot perfectly capture every detail of particle interactions. But the core message is always the same: distinguishing a genuine signal from random fluctuations in a sea of background processes requires not just one or two events, but statistically significant evidence accumulated over millions or billions of collisions. This is the foundation of this and the following chapter: the extraction of physics insights from experimental data relies fundamentally and entirely on statistics.

Modern \ac{HEP} analyses employ sophisticated statistical models capable of handling complex, multi-dimensional data while rigorously accounting for various sources of uncertainty --- both statistical fluctuations inherent to the measurement process and systematic effects arising from theoretical and experimental limitations.
This chapter introduces the binned likelihood models that form the mathematical backbone of contemporary \ac{HEP} statistical analyses. These models provide the language in which experimental results are expressed, constraints are communicated, and limits are set. Understanding this framework is essential for the reinterpretation methods presented in later chapters, as reinterpretation fundamentally involves translating experimental constraints from one theoretical context into another using the same statistical machinery.


\Cref{sec:probability} reviews probability theory, including frequentist and Bayesian interpretations, and introduces likelihood-based inference. \Cref{sec:counting} derives the Poisson distribution and binned likelihood functions that arise naturally in counting experiments. \Cref{sec:histfactory} presents the \histfactory specification --- a standardized framework for binned likelihood models in particle physics --- detailing its modifier types for parametrizing systematic uncertainties and its Bayesian extension. \Cref{sec:pyhf} introduces \pyhf, a modern Python implementation of the \histfactory formalism, providing portable JSON-based models and efficient inference through automatic differentiation. Finally, \cref{sec:histfactory-hacks} presents extensions developed for this work: methods for arbitrary parameter correlations (\cref{sec:histfactory-correlations}) and custom modifiers (\cref{sec:histfactory-custom-modifiers}) that enable the flexible reinterpretation approaches in later chapters.

\section{Probability and random variables}
\label{sec:probability}
The fundamental concept of uncertainty in measurements can have many origins. For example, it could derive from a lack of knowledge of a system, e.g.~the exact details of a particle detector. On the other hand, it could also stem from a more fundamental unpredictability such as the nondeterministic property of quantum mechanics. A measurement which depends on such undetermined factors is said to be \textit{random}.

To which degree this randomness manifests itself can be classified with the concept of \textit{probability}. Probability can be defined with three simple axioms. Consider a sample space $S$, where to each subset $A \subset S$ one assigns a real number $p(A)$ called the probability, satisfying the axioms~\cite{Cowan1998}
\begin{itemize}
    \item $p(A) \geq 0$ for every subset $A\subset S$
    \item For two disjoint subsets $A,B$ ($A\cap B = \emptyset$), $p(A\cup B) = p(A) + p(B)$
    \item The probability assigned to the sample space is unity $p(S)=1$
\end{itemize}
A variable that takes a specific value for every element of the sample space $S$ is called a random variable.

The \textit{conditional probability} is defined as
\begin{equation}
    p(A|B) = \frac{p(A\cap B)}{p(B)},
\end{equation}
provided $p(B) \neq 0$.

Two subsets are said to be \textit{independent} if their joint probability factorizes:
\begin{equation}
    p(A\cap B) = p(A)p(B).
\end{equation}
This independence condition immediately implies that ${p(A|B)=p(A)}$ and ${p(B|A)=p(B)}$, meaning knowledge of one event provides no information about the other.

Using the symmetry property $A\cap B = B\cap A$, the joint probability can be expressed in two equivalent ways:
\begin{equation}
    p(A\cap B) = p(A|B) p(B) = p(B \cap A) =  p(B|A) p(A).
\end{equation}
Rearranging this fundamental relationship yields \textit{Bayes' theorem}:
\begin{equation}
    p(A|B) = \frac{p(B|A)p(A)}{p(B)}.
\end{equation}
In this formulation, $p(A)$ represents the \textit{prior probability}, $p(B|A)$ the \textit{likelihood}, $p(B)$ the \textit{marginal likelihood}, and $p(A|B)$ the \textit{posterior probability}~\cite{Cowan1998}.


\subsection{Interpretations of probability}
Given a function satisfying the above axioms, it is still necessary to specify how to interpret the elements of the sample space and how to assign and interpret the probability values. There are two main interpretations: the \textit{frequentist} and \textit{subjective} interpretations~\cite{Cowan1998}.

One important point to keep in mind is that physics does not care about the interpretation of probability --- conclusions about the physical world from measurements should not depend on it.

\subsubsection{Frequentist probability}
The \textit{frequentist} interpretation assigns probability to the relative number of occurrences of some event $A$, in the limit of infinite trials~\cite{Cowan1998, Young_Smith_2005, James:2006zz},
\begin{equation}
    p(A) = \lim_{N\to\infty}\frac{N_A}{N}
    \label{eq:frequentist-probability}
\end{equation}
where $A$ occurs $N_A$ times in $N$ trials. This is a natural interpretation for particle physics, where processes are counted over large numbers of independent collisions. However, it is not applicable when repeated measurements are infeasible, and prior information can only be incorporated through the data itself, typically via constraints or auxiliary measurements.

\subsubsection{Subjective probability}
\label{sec:bayesian-probability}
In the \textit{subjective}, or \textit{Bayesian}, interpretation, probability is associated with the degree of belief in some hypothesis $A$~\cite{gelman2013bayesian, James:2006zz, Cowan1998},
\begin{equation}
    p(A) = \textrm{degree of belief that hypothesis $A$ is true}.
\end{equation}
This interpretation can handle situations where the frequentist approach fails, such as assigning probabilities to theories defined by unknown constants. 

The subjective interpretation is closely related to Bayes' theorem, formulated as
\begin{equation}
    p(\text{theory}|\text{data}) = \frac{ p(\text{data}|\text{theory})p(\text{theory})}{p(\text{data})}.
\end{equation}
A key advantage is the method of prior updating: posteriors from one analysis can serve as priors for subsequent data, iteratively refining one's degree of belief.

The main criticism is that results depend on the prior, whose assignment is not always clear. Further, computing $p(\text{data})$ is often intractable, so one typically samples from the posterior: $\text{theory} \sim p(\text{theory}|\text{data})$.

\subsubsection{The common ground}
Given either interpretation, the goal of any statistical analysis is to make inferences about the underlying probability distribution of the data. The mathematics behind the two interpretations is the same, and Bayes' theorem is valid under both.

The physics lies in the model of the observed data $p(\text{data}|\text{theory})$, which is common ground for both interpretations. Frequentist inference is based on the likelihood alone, while Bayesian inference additionally incorporates prior beliefs $p(\text{theory})$. In both cases, obtaining the best possible model of the data is paramount.

\section{The nature of statistical inference}
Consider the measurement of a random variable $x$, distributed according to some \ac{PDF} (see \cref{sec:pdfs}) $p( x | \params)$, depending on a set of unknown parameters $\params = (\theta_1, \ldots, \theta_m)$.

For a statistical analysis, most effort should go into the modelling of the assumed data-generating process
\begin{equation}
    x \sim p( x | \params).
\end{equation}
The model should be chosen such that it is as close as possible to the true data-generating process.

When treating $p( x| \params)$ as a function of $\params$ for fixed observed data $x$, it is called the \textit{likelihood} function. If the hypothesized \ac{PDF} and parameter values are correct, one expects a large value for the likelihood. Conversely, parameter values far away from the true values should yield a low probability for the measurements obtained.
It is important to note that the likelihood is not a probability distribution of $\params$, as it is generally not normalized~\cite{Cowan1998},
\begin{equation}
    \int dx ~ p( x | \params) = 1 \quad \forall \params \ , \quad \text{but} \quad \int d\params ~ p( x | \params) \neq 1.
\end{equation}

In particle physics the power lies in repeated experiments. Consider a \textit{sample} of $N$ independent measurements of $x$, $\data = (x_1, \ldots, x_N)$. The likelihood function is then given by
\begin{equation}
    p(\data | \params) = \prod_{i=1}^N p(x_i|\params).
    \label{eq:sample-likelihood}
\end{equation}
This equation highlights an important property of likelihood methods: \textit{the power of combination}. It demonstrates the possibility to combine independent measurements simply by multiplying their likelihoods.
This implies that likelihoods of completely different samples $\data$ and $\boldsymbol{y}$, depending on parameters $\params_x$ and $\params_y$, respectively, can be combined,
\begin{equation}
p(\data, \boldsymbol{y} | \params_x \cup \params_y) = p(\data | \params_x) p(\boldsymbol{y} | \params_y),
\end{equation}
to improve constraints on parameters, which both measurements are sensitive to, ${\boldsymbol{\rho} = \params_x \cap \params_y}$. This has very powerful consequences for parameter estimates and their uncertainties.

Since the likelihood function is commonly a product of small numbers, its actual value may become very small and hence hard to deal with computationally. Due to this reason, it is useful to consider twice the negative log-likelihood, $-2\log p(\data | \params)$, to avoid dealing with very small numbers. The factor of $2$ is a convenience, explicable when considering twice the negative log-likelihood for normally distributed data points, dropping all constant terms,
\begin{equation}
-2\log p(\data | \params) = \sum_{i=1}^N \frac{(x_i - \mu(\params))^2}{\sigma_i^2},
\end{equation}
which is the well known $\chisq(\params)$ variable, which is distributed according to a $\chisq$-distribution. Here $\sigma_i$ corresponds to the uncertainty of $x_i$.

\section{Counting experiments}
\label{sec:counting}
In particle physics one performs counting experiments. For example, given that one particle $X$ decays to particle $Y$ with a probability $P$, the probability to observe $n$ $X\to Y$ events out of $N$ decays of $X$ can be calculated.

If all the decays were \textit{distinguishable} (the order of events matters), the probability is~\cite{Cowan1998},
\begin{equation}
    p(n|N,P) = P^n(1-P)^{N-n},
\end{equation}
which simply is the product of any event with probability $P$ happening $n$ times out of $N$ distinguishable trials.

For \textit{indistinguishable} events (only the overall number of a particular decay matters), there is an additional factor accounting for the combinatorics. Adding this factor results in the \textit{binomial distribution}~\cite{Cowan1998},
\begin{equation}
    p(n|N,P) = \frac{N!}{n!(N-n)!}P^n(1-P)^{N-n}.
    \label{eq:binomial}
\end{equation}

What if there are not only two possible outcomes? One might want to consider $X$ decaying into a full alphabet of particles, $X\to A$, or $X\to B$, \ldots, $X\to Z$ each happening with a probability $P_A, P_B, \ldots, P_Z$. Then the probability to get $n_A$ $X\to A$ events, $n_B$ $X\to B$ events, \ldots, is given by the \textit{multinomial distribution},
\begin{equation}
    p(n_A, n_B, \ldots, n_Z|N,P_A, P_B, \ldots, P_Z) = \frac{N!}{n_A!n_B!\ldots n_Z!}P_A^{n_A}P_B^{n_B} \ldots P_Z^{n_Z},
    \label{eq:multinomial}
\end{equation}
where $N = \sum_{i=A}^Z n_i$ and $\sum_{i=A}^Z P_i=1$.

In principle, one could perform analyses in particle physics with the binomial distribution of \cref{eq:binomial} or its extension, the multinomial distribution of \cref{eq:multinomial}. In general, one wants to compare the \textit{expected number} of $X\to Y$ events to the actually observed ones. The expected number of events is simply $\nu = NP$, where commonly $N$ is very large (experiments take a lot of data) and $P$ is very small (actually observing $X\to Y$ is very rare).


In the limit of large $N$ and small $P$, with some finite $\nu=NP$, the binomial distribution in \cref{eq:binomial} becomes the \textit{Poisson distribution}~\cite{Cowan1998},
\begin{equation}
    p(n|\nu) = \frac{\nu^n}{n!}e^{-\nu}.
\end{equation}

\subsection{Binned likelihoods}
Given a total number of $N$ observations of a random variable $x$ distributed according to a \ac{PDF} $p(x|\params)$, in the case of $N$ becoming very large, it becomes difficult to evaluate the negative log-likelihood for all points in the sample $\data$.
One often simplifies by binning the data in a histogram of $B$ bins. This results in a set of observed data yields $\yields = (n_1, \ldots, n_B)$ and expected yields $\expectation = (\nu_1, \ldots, \nu_B)$ in each bin. The expected yields are given by
\begin{equation}
    \nu_b(\params) = N \int_{\text{bin}~b} dx ~ p(x|\params),
\end{equation}
where the integral boundaries correspond to the bin boundaries of the given bin.

Since the individual observations, and hence also the bins are independent, and given that each event must fall into any of the bins, the full \ac{PDF} is given by a multinomial distribution (\cref{eq:multinomial}) with $P_b = \nu_b/N$~\cite{Cowan1998},
\begin{equation}
    p(\yields | \expectation) = \frac{N!}{n_1! \ldots n_B!}\left(\frac{\nu_1}{N}\right)^{n_1}\ldots\left(\frac{\nu_B}{N}\right)^{n_B}.
\end{equation}
Empty bins, i.e.~$n_i=0$ for some $i$, are of no concern, but expectation values $\nu_i=0$ would result in $p(\yields | \expectation)=0$.

So far, the discussion has considered the situation where the total number of events was fixed to $N$. But the total number of events might be a random variable itself, distributed according to a Poisson distribution,
\begin{equation}
    p(N|\nu_{tot}) = \frac{\nu_{tot}^{N}}{N!}e^{-\nu_{tot}}.
\end{equation}
Here  $N = \sum_{b=1}^{B}n_b$ and $\nu_{tot} = \sum_{b=1}^{B}\nu_b$, where the expected yields are now a function of the expected total yields,
\begin{equation}
    \nu_b(\nu_{tot}, \params) = \nu_{tot} \int_{\text{bin}~b} dx ~ p(x|\params).
\end{equation}

In this case the full \ac{PDF} is given by
\begin{equation}
    p(\yields | \expectation) = \frac{\nu_{tot}^{N}}{N!}e^{-\nu_{tot}} \frac{N!}{n_1! \ldots n_B!}\left(\frac{\nu_1}{\nu_{tot}}\right)^{n_1}\ldots\left(\frac{\nu_B}{\nu_{tot}}\right)^{n_B},
\end{equation}
where the probability of an event landing in bin $i$ is now determined through the total expected yield, $P_b = \nu_b / \nu_{tot}$.

Interestingly, the above equation can be simplified to
\begin{equation}
    p(\yields | \expectation(\params)) = \prod_{b=1}^B \pois (n_b | \nu_b) = \prod_{b=1}^B \frac{\nu_b(\params)^{n_b}}{n_b!}e^{-\nu_b(\params)}.
    \label{eq:binned poisson PDF}
\end{equation}
Hence, it is found that a multinomial \ac{PDF} with an overall Poisson constraint on the total number of observations is equivalent to the product of Poisson \acp{PDF} for individual histogram bins.

\section{\histfactory}
\label{sec:histfactory}
Data analyses in particle physics have many common features: they are based on counting experiments which compare collider data with simulated \ac{MC} data, are subject to statistical and systematic fluctuations (as any analysis of data), and many are histogram-based due to the overall large data yields. Simulated data are easily classifiable into categories, such as \textit{signal} (events that one is interested in) and \textit{background} (events that one is less interested in) in simple cases.

To unify data analyses in particle physics to a certain degree, a mathematical specification of \textit{closed-world}\footnote{
    Closed-world statistical models contain a set of given building blocks for model building, whereas open-world models allow for custom components to be added~\cite{Cranmer:2021urp}.
} statistical models was developed: \histfactory~\cite{histfactory}. It has become a gold standard in the particle physics community, across experiments, since it is flexible enough to fit the purpose of many analyses, while providing a limited set of building blocks, which maintains simplicity.

The \histfactory model applies in cases where,
\begin{itemize}
    \item one conducts an analysis of binned data,
    \item one can separate the events into categories of physics processes (\textit{samples}\footnote{A \histfactory specific wording, not to be confused with a statistical sample.}),
    \item uncertainties can be modelled as either shape or normalization variations to the histogram.
\end{itemize}
The hierarchy of a \histfactory model consists of $C$ \textit{channels} --- each a distinct binned distribution with $B_c$ bins. Each channel consists of $S_c$ \textit{samples} --- distinct physics processes. Furthermore, variations due to statistical or systematic effects are specified as parametrized histogram variations, named \textit{modifiers}.

The basis of the \histfactory model is the Poisson \ac{PDF} derived in \cref{eq:binned poisson PDF} \cite{histfactory,Heinrich:2021gyp},
\begin{equation}
    p_{\text{data}}(\yields | \params) = \prod_{c \in C} \prod_{b \in B_c} \pois (n_{cb} | \nu_{cb}(\params)).
    \label{eq:data-pdf}
\end{equation}
The expected yields are given by nominal expected yields for each channel $c$, sample $s$ and bin $b$, $\nu_{csb}^{\text{nom}}$, subject to parametrized variations, due to given uncertainties in the expectation. Nominal yields are commonly obtained from \ac{MC} simulation and subsequent reconstruction and event selection. These parametrized variations can either be multiplicative, $\boldsymbol{\kappa}(\params)$, or additive, $\boldsymbol{\Delta}(\params)$. The resulting expected yields are given by
\begin{equation}
    \nu_{cb}(\params) = \sum_{s \in S_c} \nu_{csb} = \sum_{s \in S_c} \left(\prod_{\kappa \in \boldsymbol{\kappa}} \kappa_{csb}(\params) \right) \left( \nu_{csb}^{\text{nom}} + \sum_{\Delta \in \boldsymbol{\Delta}} \Delta_{csb}(\params) \right).
    \label{eq:expected-yields}
\end{equation}

The model parameters, controlling the variations, fall into two categories: \textit{unconstrained} and \textit{constrained} parameters,
\begin{equation}  
    \params = (\unconstr, \constr).
\end{equation}

Unconstrained parameters, \unconstr, are free --- their variation only changes the likelihood value through the variation of the expected yields. Constrained parameters, \constr, are subject to given constraint \acp{PDF}, given \textit{auxiliary data} \aux,
\begin{equation}
    p_{constr.}(\aux | \constr) = \prod_{\chi \in \constr} p(a_\chi | \chi).
    \label{eq:constraint-pdf}
\end{equation}
Note that this is a product of individual constraint terms for each parameter $\chi \in \constr$. Hence, each parameter is considered independent by assumption. How one can nevertheless deal with correlated parameters is discussed in \cref{sec:histfactory-hacks}. 

One can also separate the model parameters into two alternative categories: \textit{\acp{POI}} and \textit{nuisance parameters},
\begin{equation}
    \params = (\boldsymbol{\mu}, \boldsymbol{\psi}).
\end{equation}

Parameters of interest, $\boldsymbol{\mu}$, as the name suggests, are parameters from which valuable knowledge is extracted. Nuisance parameters, $\boldsymbol{\psi}$, are commonly not of interest, but are needed to parametrize the model --- these are usually used to describe sources of uncertainties. Although not strictly true for any model, in the context of this work, \acp{POI} are always unconstrained and nuisance parameters are always constrained parameters. 

Within the frequentist interpretation of statistics, constraint terms can only be introduced through auxiliary measurements. This can be for example a control region measurement, constraining the normalization of one or multiple samples in the model.

The full \histfactory statistical model is given as the product of \cref{eq:data-pdf,eq:constraint-pdf},
\begin{equation}
    p(\yields, \aux | \unconstr, \constr) = p_{\text{data}}(\yields | \unconstr, \constr) ~ p_{constr.}(\aux | \constr) = \prod_{c \in C} \prod_{b \in B_c} \pois (n_{cb} | \nu_{cb}(\unconstr, \constr)) ~ \prod_{\chi \in \constr} p(a_\chi | \chi).
    \label{eq:histfactory-pdf}
\end{equation}

\subsection{Modifier types}
\label{sec:histfactory-modifiers}
In the \histfactory statistical model, parametrized variations to binned distributions are controlled through \textit{modifiers}. Since the \histfactory model is closed-world, only a fixed set of modifications are available within the framework. They can be classified into multiplicative, $\boldsymbol{\kappa}(\params)$, or additive, $\boldsymbol{\Delta}(\params)$ modifiers, as in \cref{eq:expected-yields}. Each modifier is parametrized by one or a set of parameters, with a corresponding constraint term. Furthermore, each modifier requires certain input specifications, which are used to compute the auxiliary data, depending on the type. Modifier input specifications commonly include variations of the histogram. Modifier parameters then interpolate between the nominal histogram, the variations, and beyond.

The available modifier types in the \histfactory specification and their respective constraint terms are listed here \cite{histfactory,pyhf}.

\paragraph{Uncorrelated shape --- \texttt{shapesys}}
This is a bin-wise, multiplicative modifier for a fixed channel $\bar c$ and sample $\bar s$, $\kappa_{\bar c \bar s b}$, introducing one parameter per bin. The required input is the absolute uncertainty on the yield of each bin $\sigma_{\bar c \bar s b}$, for the given channel and sample.
The amount of auxiliary data $a_{\bar c \bar s b}$ per bin is determined by asking ``which effective number of entries would give that relative uncertainty''. That is given by
\begin{equation}
    \frac{\sqrt{a_{\bar c \bar s b}}}{a_{\bar c \bar s b}} = \frac{\sigma_{\bar c \bar s b}}{\nu_{\bar c \bar s b}^{\text{nom}}} \rightarrow a_{\bar c \bar s b} = \left(\frac{\nu_{\bar c \bar s b}^{\text{nom}}}{\sigma_{\bar c \bar s b}}\right)^2
\end{equation}
The corresponding constraint term is
\begin{equation}
    p(\aux | \boldsymbol{\kappa_{\bar c \bar s}}) = \prod_{b \in B_{\bar c}} \pois(a_{\bar c \bar s b} | a_{\bar c \bar s b} \kappa_{\bar c \bar s b}).
\end{equation}

\paragraph{Statistical uncertainty --- \texttt{staterror}}
Expected bin yields are often derived from \ac{MC} datasets. Consequently, they necessarily carry an uncertainty due to the finite sample size of the datasets. In principle, one could add an \textit{uncorrelated shape} modifier for each sample, to account for this uncertainty, but this would yield an excessively large number of parameters with limited utility. To circumvent this, one can quadratically combine the overall statistical uncertainty per bin, over the collection of all samples~\cite{BARLOW1993219}. This is exactly what the \texttt{staterror} modifier does, adding a bin-wise, multiplicative modifier, for a given channel $\bar c$, $\kappa_{\bar c b} = \sum_{s\in S_c}\kappa_{\bar c \bar s b}$. The required input is the absolute uncertainty on the yield of each bin $\sigma_{\bar c s b}$, for each of the samples in a given channel to be included.

The constraint term is a product of normal \acp{PDF} over bins,
\begin{equation}
    p(\aux | \boldsymbol{\kappa_{\bar c}}) = \prod_{b \in B_{\bar c}} \mathcal{N}(a_{\bar c b}=1 | \kappa_{\bar c b}, \sigma_{\bar c b}^{rel}),
\end{equation}
where the auxiliary data are $a_{\bar c b}=1$, and
\begin{equation}
    \sigma_{\bar c b}^{rel} = \sqrt{\sum_{s\in S_c} \sigma_{\bar c s b}^2} \Bigg/ \sum_{s\in S_c} \nu_{\bar c s b}^{\text{nom}}.
\end{equation}

\paragraph{Correlated shape uncertainty --- \texttt{histosys}}
Systematic effects mostly affect the whole histogram in a correlated manner, affecting both shape and normalization of the histogram. By varying certain parameters of underlying assumptions on either the theoretical or experimental model by one standard deviation (assuming Gaussianity), one can compute upward, $\boldsymbol{\Delta}^{\text{up}}$, and downward, $\boldsymbol{\Delta}^{\text{dn}}$, variations of the nominal bin yields. These variations serve as the input data for the \texttt{histosys} modifier. 
This is an additive modifier, where a single interpolation parameter $\alpha$ controls the modification to the expected bin yields,
\begin{equation}
    \boldsymbol{\Delta}(\alpha) = 
    \begin{cases}
        \alpha |\boldsymbol{\Delta}^{\text{up}} - \boldsymbol{\nu}^{\text{nom}}| \quad &\alpha \geq 1\\
        \sum_{n=0}^6 k_n \alpha^n \quad &|\alpha| < 1\\
        \alpha |\boldsymbol{\Delta}^{\text{dn}} - \boldsymbol{\nu}^{\text{nom}}| \quad &\alpha \leq -1
    \end{cases} \, ,
\end{equation}
where the parameters $k_n$ are fixed by the boundary conditions at $\alpha=0,\pm 1$, as well as the continuity conditions (matching of first and second derivative) at $\alpha=\pm 1$. 
The corresponding constraint term for $\alpha$ is a normal distribution with unit width and auxiliary data $a=0$,
\begin{equation}
    p(a | \alpha) = \mathcal{N}(a=0 | \alpha, \sigma=1).
\end{equation}

\paragraph{Normalization uncertainty  --- \texttt{normsys}}
As evident from the name, with this modifier one can introduce an uncertainty on the overall normalization of a particular sample in a given channel $\bar c$. This is a multiplicative modifier $\kappa_{\bar c s}(\alpha)$, parametrized by a single parameter $\alpha$, which interpolates between the upward and downward multiplicative factor at one standard deviation, $\kappa_{\bar c s}(+1) = \kappa^{\text{up}}$ and $\kappa_{\bar c s}(-1) = \kappa^{\text{dn}}$, given as input data. The nominal value is $\kappa(0)=1$.
The corresponding constraint term for $\alpha$ is a normal distribution with unit width and auxiliary data $a=0$,
\begin{equation}
    p(a | \alpha) = \mathcal{N}(a=0 | \alpha, \sigma=1).
\end{equation}

\paragraph{Luminosity uncertainty --- \texttt{lumi}}
Histograms created from simulated \ac{MC} samples are commonly scaled to the overall luminosity of the data itself, for a truthful comparison to the overall theoretically predicted rate. This is a global multiplicative modifier, with a luminosity scale $\kappa_L$ and uncertainty $\sigma_L$ as inputs. Global here means that it affects all samples to which the modifier is added in the same way. The constraint term is a normal distribution,
\begin{equation}
    p(\kappa_L | \kappa) = \mathcal{N}(\kappa_L | \kappa, \sigma_L).
\end{equation}

\paragraph{Unconstrained normalization --- \texttt{normfactor}}
This modifier scales the histogram by a free parameter $\mu$. This is commonly used to measure the signal strength of a given process, as the \ac{POI}. There is no input data needed. Furthermore, there is no constraint term, such that a deviation from the nominal value $\mu=1$ only affects the data likelihood.

\paragraph{Data-driven shape --- \texttt{shapefactor}}
This is a free, bin-wise multiplicative modifier for a fixed channel $\bar c$ and sample $\bar s$, $\kappa_{\bar c \bar s b}$. No additional input data are needed, and no constraint term is added to the likelihood. This is for example used to model the shape of a background process which is not well known, but can be constrained by data in a control region. The parameters $\kappa_{\bar c \bar s b}$ are then fitted simultaneously in the signal and control region.

\subsection{Bayesian \histfactory}
\label{sec:histfactory-bayesian}
A well-defined statistical model is the basis for any inference problem. In \cref{sec:bayesian-probability} it was discussed that Bayesian inference is based on the posterior. Starting from \histfactory likelihood $p\left( \yields, \aux | \unconstr, \constr\right)$ in \cref{eq:histfactory-pdf}, one can formulate a posterior model $p\left( \unconstr, \constr \vert \yields, \aux \right)$, given that corresponding priors are defined for all parameters $\unconstr$ and $\constr$. Given that parameters in the \histfactory model are independent, it is possible to separate the prior distributions for constrained and unconstrained parameters, $p(\unconstr)$ and $p(\constr)$, respectively. 
The posterior model is then given by
\begin{equation}
        p\left( \unconstr, \constr \vert \yields, \aux \right) \propto p\left( \yields, \aux | \unconstr, \constr\right) ~ p\left( \unconstr \right) ~ p\left( \constr\right),
\end{equation}
where the normalization is just the total integral over the right-hand side. Separating the \histfactory likelihood into its \textit{data} and \textit{constraint} parts as in \cref{eq:histfactory-pdf}, one obtains
\begin{equation}
    p\left( \unconstr, \constr \vert \yields, \aux \right) \propto p_{\text{data}}\left( \yields \vert \unconstr, \constr \right) ~ p\left( \constr | \aux \right) ~ p\left( \unconstr \right).
\end{equation}
where $p\left( \constr | \aux \right) \propto p(\aux | \constr) p\left( \constr\right)$ has been defined. This last step is a classic example of \textit{Bayesian updating}~\cite{gelman2013bayesian}, where the posterior of a previous measurement is used as a prior for a subsequent one. In this case, the constraint likelihood of \cref{eq:histfactory-pdf}, is used to update the \ac{PDF} of the constraint parameters given in $p(\constr)$\footnote{This is sometimes referred to as an \textit{ur-prior}.}. 

Since the constraint terms are a set of default \acp{PDF}, as discussed in \cref{sec:histfactory-modifiers}, the priors $p(\constr)$ can also be set to default \acp{PDF}~\cite{Feickert:2023hhr}. It is convenient to choose \textit{conjugate priors}, which are priors that yield a posterior of the same type as the prior itself. This is especially useful for the \histfactory model, since it allows for an algebraic expression of the posterior, which can be computed efficiently.
The constraint likelihoods are either normal or Poisson \acp{PDF}. Conjugate priors are discussed in \cref{sec:conjugate-priors}.

\subsection{\histfactory in Python: \pyhf}
\label{sec:pyhf}
The classic implementation of the \histfactory model is within the \texttt{ROOT} framework \cite{root}. Implemented through \texttt{RooFit} and \texttt{RooStats}, the ROOT-based \histfactory provides a flexible configuration via XML files and extensive C++/Python interoperability. However, this approach often requires a full ROOT environment, entails non-trivial software dependencies, and is less convenient to modern analysis workflows.

In contrast to the original \texttt{ROOT}-based implementation, \pyhf (pure Python \histfactory) is a modern, \texttt{ROOT}-independent reimplementation of the \histfactory statistical model~\cite{Heinrich:2021gyp,pyhf}. Written entirely in Python, \pyhf is designed to support binned likelihood fits in a lightweight and modular environment. It enables analyses to be developed, shared, and executed without the need for the \texttt{ROOT} framework, facilitating integration into contemporary scientific software ecosystems. 

\pyhf offers two minimizers to choose from --- \texttt{scipy}-based minimizers~\cite{2020SciPy-NMeth} for fast and robust minimization, without uncertainty quantification, and \texttt{minuit}~\cite{James:1994vla} for a more thorough analysis and detailed results.
A key feature of \pyhf is its support for gradient-based optimisation and automatic differentiation through backends such as \texttt{JAX}~\cite{jax2018github}. These capabilities allow efficient computation of likelihoods and parameter gradients, which is particularly advantageous for large-scale parameter inference (\cref{sec:parameter-estimation}) and enables compatibility with modern machine learning tools and hardware acceleration.

Model definitions in \pyhf are specified using a structured JSON schema, making them portable, human-readable, and version-controllable. This declarative approach promotes reproducibility and transparency, and aligns with \ac{FAIR} principles in data preservation and open science. JSON-based models can be easily shared and reused across platforms.

\pyhf provides a streamlined and efficient implementation of frequentist hypothesis testing using the profile likelihood ratio as the core test statistic (\cref{sec:pointwise-comparison}). 
\pyhf leverages asymptotic formulae derived in Reference~\cite{Cowan:2010js} to approximate the sampling distributions of various test statistics. This allows for fast calculation of $P$-values and discovery significances without resorting to computationally expensive pseudo-experiments (\cref{sec:pointwise-comparison}). This approach provides scalable inference, particularly useful for setting confidence intervals and exclusion limits (\cref{sec:limit-setting}).

As an example, a simple \pyhf model with one bin is shown below, with a single channel \texttt{singlechannel}, containing two samples: \texttt{signal} and \texttt{background}.

\begin{minted}[fontsize=\small, breaklines]{json}
    {
    "channels": [
        {
        "name": "singlechannel",
        "samples": [
            {
            "name": "signal",
            "data": [10],
            "modifiers": [
                { "name": "mu", "type": "normfactor", "data": null }
            ]
            },
            {
            "name": "background",
            "data": [74],
            "modifiers": [
                {
                "name": "bkg_shape_unc",
                "type": "histosys",
                "data": {
                    "lo_data": [70.0],
                    "hi_data": [80.0]
                }
                }
            ]
            }
        ]
        }
    ]
    }
\end{minted}

Nominal bin values are provided for both samples. The \texttt{signal} includes a normalization factor modifier \texttt{mu} representing the \ac{POI}. The \texttt{background} includes a shape systematic uncertainty modelled with a \texttt{histosys} modifier, defined by low and high variation templates (\cref{sec:histfactory-modifiers}). More details on the syntax of adding modifiers, can be found in the \href{https://pyhf.readthedocs.io}{\pyhf documentation} \cite{pyhf}.

As \pyhf is the implementation of the \histfactory model, there is \texttt{bayesian pyhf}, which is the implementation of the Bayesian \histfactory model~\cite{Feickert:2023hhr}. It takes the \pyhf likelihood and translates it into a joint posterior distribution as described in \cref{sec:histfactory-bayesian}. The posterior model is implemented in the \texttt{PyMC}~\cite{AbrilPla2023PyMCAM} framework, which implements a suite of \ac{MCMC} sampling algorithms
\footnote{
    A good overview of sampling algorithms can be found in Reference~\cite{gelman2013bayesian}.
} 
to finally obtain posterior samples.

\subsection{Extended features for \histfactory and \pyhf}
\label{sec:histfactory-hacks}
As with any model or software, \histfactory and \pyhf also come with limitations and drawbacks. In this section I want to discuss two of these limitations and provide solutions to them.

\subsubsection{Arbitrarily correlated uncertainties}
\label{sec:histfactory-correlations}
In \pyhf, model parameters --- particularly nuisance parameters associated with systematic uncertainties --- are, by default, treated as statistically uncorrelated across different measurements or channels. This assumption simplifies the construction of the likelihood function by modelling each nuisance parameter with its own independent constraint term. However, in realistic analyses, some systematic effects are shared across multiple samples or channels and must be correlated to avoid under- or overestimating the total uncertainty. 

To introduce \textit{complete correlations} in \pyhf, the same modifier name is used across all affected samples and channels within the JSON-based model specification. When \pyhf parses the model, it identifies modifiers with matching names and treats them as a single global nuisance parameter, applying their effect consistently wherever referenced. A prerequisite for this is, that the parameters share the same constraint term.
\textit{Complete anti-correlation} can be achieved by swapping the up- and down-variations in the modifier definition of one of the modifiers and then correlating them by the same naming scheme.

While \pyhf supports full correlations between nuisance parameters through shared modifier names, it currently lacks support for defining arbitrary correlations between parameters, such as specifying non-diagonal correlation matrices. This limitation means users cannot directly encode partial correlations or more complex dependence structures among parameters within the model. As a result, all parameter correlations must in principle be either full (via shared modifiers) or absent, which may restrict modelling flexibility in analyses requiring nuanced treatment of correlated uncertainties.

To solve this issue, I have developed \texttt{pyhfcorr}~\cite{pyhfcorr} --- a tool to arbitrarily correlate parameters in a \pyhf JSON format. It is based on eigendecomposing the covariance matrix of the parameters, resulting in new independent parameters as superpositions of the original ones~\cite{Gartner:2024muk}.

Eigendecomposition is a useful method for decorrelating a set of parameters by a unitary transformation. 
One starts with a covariance matrix $C$, which is symmetric. Hence, one can always decompose it as 
\begin{equation}
    C = USU^T
\end{equation}
where $UU^T = \identity$. The columns of the transformation matrix $U$ are the eigenvectors of $C$.
The eigenvalues, $s_i = S_{ii}$, are the variances in the rotated space. The standard deviations are $\sigma_i = \sqrt{s_i}$.

If one wants to incorporate the variances, one can define a new transformation matrix $Z = U \sqrt{S}$. $Z$ is column-wise composed of the eigenvectors of $C$, each of which is scaled by the corresponding standard deviation.

The \pyhf modifier parameters, $\boldsymbol{p}$, describe the contribution of each of these scaled eigenvectors. Hence, to rotate from these parameters to the standard deviation vector for the correlated parameters, $\boldsymbol{\alpha}$, one can apply
\begin{equation}
    \boldsymbol{\sigma}_\alpha = Z \boldsymbol{p}.
\end{equation}
In that way, the modifier parameters are all interpretable in the same way as the usual \pyhf modifier parameters, i.e.~$p_i = \pm 1$ corresponds to a shift of $\pm \sigma_i$ along the $i^{\text{th}}$ eigenvector direction. 

This approach allows the specification of a correlation matrix for a set of modifier parameters in the \pyhf model, which are rotated into a new basis of uncorrelated parameters. Based on this, a new \pyhf model in terms of the rotated parameters can be defined. The number of parameters is always conserved with this method. Furthermore, the output of the \texttt{pyhfcorr} tool is a valid \pyhf model, such that the full inference suite of \pyhf can still be applied without drawbacks.

With \texttt{pyhfcorr} one can extend an existing \pyhf JSON specification by a \texttt{correlations} entry. There one can provide the parameter names and correlation matrix of the to-be correlated parameters.
As an example, to correlate two parameters \texttt{a} and \texttt{b}, present in the \pyhf model, with a correlation coefficient of $0.5$, one simply adds:

\begin{minted}[breaklines]{python}
    spec = {
        "channels" : ...,
        "correlations": [
            {
                "name": "decorrelate_a_b",
                "vars": ["a", "b"],
                "corr": [[1.0, 0.5], [0.5, 1.]],
            }
        ]
    }

    new_spec = pyhfcorr.decorrelate.decorrelate(spec)
    new_model = pyhf.Model(new_spec)
\end{minted}

This results in a new \pyhf model with parameters \texttt{a} and \texttt{b} replaced by parameters \texttt{decorrelate\_a\_b[0]} and \texttt{decorrelate\_a\_b[1]}, corresponding to the parameters in the decorrelated space.

\subsubsection{Custom modifiers}
\label{sec:histfactory-custom-modifiers}
\begin{figure}
    \centering
    \includegraphics[width=0.4\textwidth]{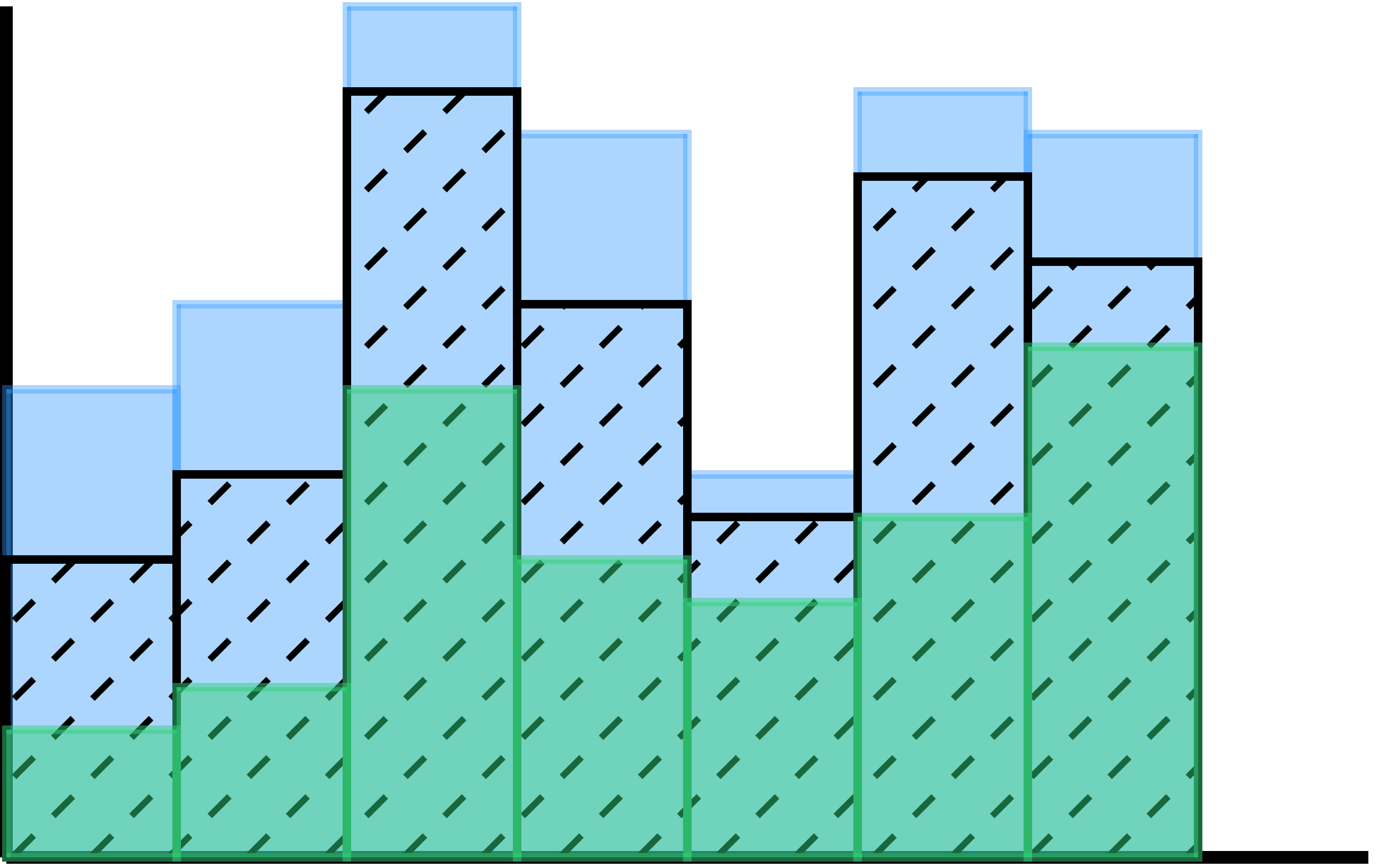}
    \includegraphics[width=0.4\textwidth]{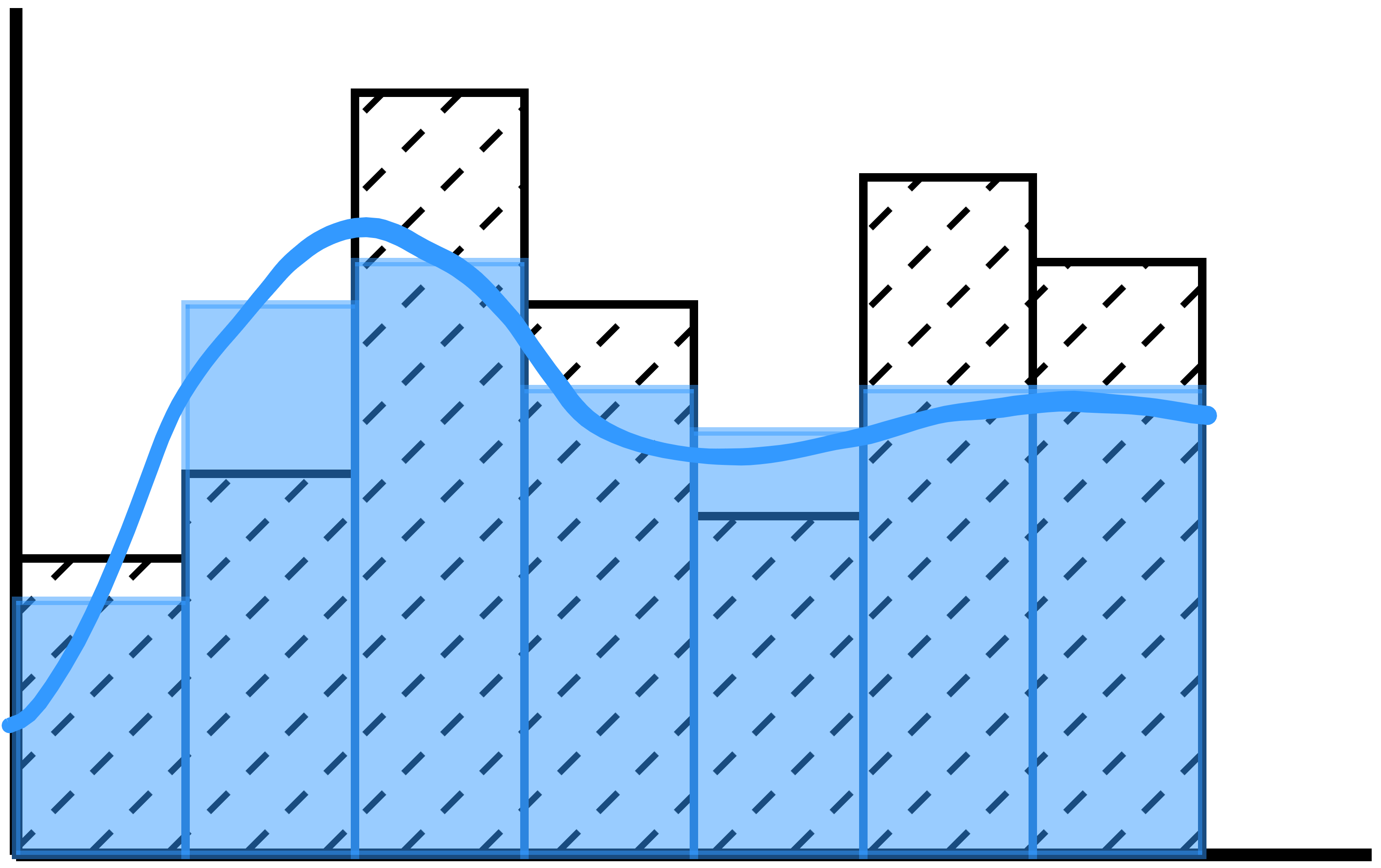}
    \caption{
        This figure illustrates two types of modifiers in \pyhf. In both panels, the nominal expected yields are represented by the black dashed histogram.
        \textit{Left:} An additive modifier that alters the expected yields by interpolating between the nominal yields and their variations, as is typical for the default \pyhf modifiers. These variations are depicted by the blue and green histograms.
        \textit{Right:} A custom modifier that scales the expected yields in a specific channel using a parametrized function. The function is shown as blue line. The resulting variations, shown by the blue histogram, follow the prescribed custom function.
    }
    \label{fig:pyhf-custom-modifiers}
\end{figure}

\pyhf is a closed-world statistical model, with a fixed set of modifiers. In some cases, however, it may be desirable to introduce \textit{custom modifiers} --- that is, parametrized functions not included in the default set, which directly control the expected yields in a given channel. This is illustrated in \cref{fig:pyhf-custom-modifiers}.

There are multiple approaches to map the output of a function onto variations within a histogram. Here, I implement one variant of a custom modifier in \pyhf, where a parametrized function returns bin-wise weights for a given sample. This function can be defined either directly in Python or as a string expression using the \texttt{numexpr} package~\cite{numexpr}.
The custom modifier operates multiplicatively, ensuring compatibility with other modifiers acting on the same sample. Formally, the custom function can be expressed as
\begin{equation}
    \kappa_{\bar c \bar s b}(\params) = \nu_{\bar c \bar s b}(\params) / \nu^{\text{nom}}_{\bar c \bar s b} \, ,
\end{equation}
where $\nu_{\bar c \bar s b}(\params)$ denotes the new bin yields computed as a function of the parameters $\params$. The strength of this approach lies in the fact that $\nu_{\bar c \bar s b}(\params)$ is an \textit{arbitrary} function, enabling users to construct an “open world” \histfactory model~\cite{Cranmer:2021urp}.\footnote{All models remain binned.} This flexibility is particularly valuable, as it extends beyond the standard \histfactory statistical model implemented in \pyhf, allowing users to incorporate any necessary custom modifications. 
A limitation of this approach is that, to distribute the \pyhf model, both the JSON specification and the implementation of the custom function are required --- unless the function is defined using \texttt{numexpr}.

An example of how the custom modifier can be added to a \pyhf model is shown below. The modifier is implemented as part of the \texttt{redist} package~\cite{redist_v1.0.4}, which I will discuss further in \cref{sec:method}.

\begin{minted}[fontsize=\small, breaklines]{python}
    import pyhf
    import numpy as np
    from scipy import stats
    from redist import custom_modifier

    def custom_func(pars):
        x = np.linspace(0, 5, 50)
        return lambda: stats.norm.pdf(x, loc=pars['mu'], scale=pars['sig'])

    spec = {
        "channels": [{
            "name": "singlechannel",
            "samples": [{
                "name": "signal",
                "data": [1.]*50,
                "modifiers": [{
                    "name": "custom_modifier",
                    "type": "custom",
                    "data": {"expr": 'gauss_func'}
                }]
            }]
        }]
    }

    new_params = {
        'mu':  {'inits': (1.,), 'bounds': ((-5, 5),), 'paramset_type': 'unconstrained'},
        'sig': {'inits': (1.,), 'bounds': ((-5, 5),), 'paramset_type': 'unconstrained'}
    }

    modifier_set = custom_modifier.add(
        'custom', ['mu', 'sig'], new_params, namespace={'gauss_func': custom_func}
    )

    model = pyhf.Model(spec, validate=False, modifier_set=modifier_set)
\end{minted} 

The custom function \texttt{custom\_func} returns a Gaussian \ac{PDF}, which is then used to scale the expected yields in the channel \texttt{singlechannel}. The custom modifier is added to the \pyhf JSON specification as a modifier of type ``\texttt{custom}'', an arbitrary identifier chosen by the user.
This identifier is also provided to \texttt{custom\_modifier.add}, which registers the custom modifier within the \pyhf model. 
The custom function accepts a dictionary of parameters, defined in the \texttt{new\_params} dictionary and passed to \texttt{custom\_modifier.add}. This dictionary allows specification of initial values, parameter bounds, constraint types, and can even include correlated parameters. Correlations among parameters are internally handled via decorrelation techniques, as discussed in \cref{sec:histfactory-correlations}.
The custom function is referenced within the namespace as \texttt{gauss\_func}, which is also supplied to \texttt{custom\_modifier.add}. 

\begin{figure}
    \centering
    \includegraphics[width=0.8\textwidth]{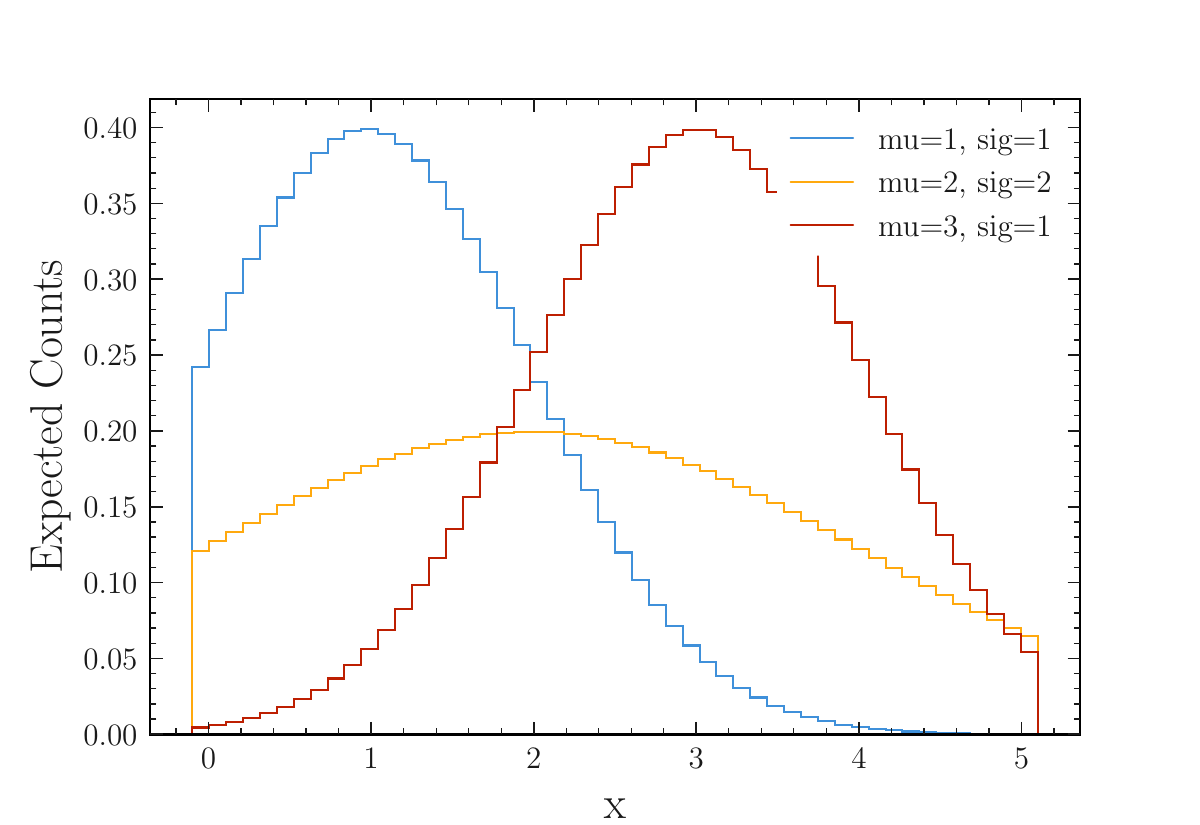}
    \caption{Expected yields of the custom modifier example for three different parameter values of \texttt{mu} and \texttt{sig}.}
    \label{fig:custom-modifier-example}
\end{figure}

The \pyhf model can then be constructed with an extended modifier set (including the custom modifier) and the JSON specification.
The parameters \texttt{mu} and \texttt{sig} are now model parameters.
The custom function is evaluated during model fitting, where the parameters \texttt{mu} and \texttt{sig} are optimized to fit the data. The expected yields of the above example are shown in \cref{fig:custom-modifier-example}. The expected yields are computed for three different parameter values of \texttt{mu} and \texttt{sig}, demonstrating the flexibility of the custom modifier.

%% file: chapters/stat-inference.tex
Building upon the foundational concepts of probability and binned likelihood models established in the previous chapter, this chapter addresses their practical application to extracting physics knowledge from experimental data. Statistical inference provides the mathematical framework for drawing scientifically rigorous conclusions about theoretical parameters from observations while properly accounting for both statistical and systematic uncertainties.

Modern particle physics experiments present interesting challenges: high-dimensional parameter spaces, intricate correlations between observables, and the demanding task of distinguishing genuine signals from background processes. This chapter develops a comprehensive statistical inference framework tailored to address these complexities, with particular emphasis on techniques essential for \ac{BSM} searches and precision measurements.

\Cref{sec:parameter-estimation} covers parameter estimation from both frequentist (maximum likelihood) and Bayesian (maximum a posteriori) perspectives. I discuss fundamental properties of estimators including bias, consistency, and the Cramér-Rao bound, demonstrating their equivalence under uniform priors. \Cref{sec:model-comparison} addresses point-wise comparison through hypothesis testing and global model comparison using Bayes factors, including the Neyman-Pearson lemma, nested hypotheses, and weighted $P$-values. \Cref{sec:goodness-of-fit} formulates the goodness-of-fit testing framework. Finally, \cref{sec:limit-setting} covers interval estimation via Bayesian credible intervals and frequentist confidence intervals using Neyman construction, with emphasis on exclusion limits central to \ac{BSM} searches.

\section{Parameter estimation}
\label{sec:parameter-estimation}

Any empirical scientific endeavour involves extracting information from data. One common objective is to infer the numerical values of parameters believed to govern the data-generating process --- such as the mass of a particle, specific coupling constants, or branching ratios. This section discusses some of the basic ideas that flow into the procedure of parameter estimation~\cite{Cowan1998}.

The goal is to learn about the underlying \ac{PDF} from a sample $\data = (x_1, \ldots, x_N)$ by constructing a likelihood model $p(\data|\params)$ depending on parameters $\params = (\theta_1, \ldots, \theta_M)$. The procedure of determining these parameters is known as \textit{parameter estimation}, yielding \textit{estimators} $ \estimator(\data)$. Since estimators are functions of random data, they are themselves random variables with a sampling distribution $p(\estimator | \params_0)$ depending on the true parameter values $\params_0$.

Key properties of estimators include~\cite{Cowan1998}:
\begin{itemize}
    \item \textbf{Bias}: $\boldsymbol{b} = \boldsymbol{E}[\estimator] - \params_0$. An \textit{unbiased} estimator has $\boldsymbol{b} = 0$.
    \item \textbf{Consistency}: An estimator is \textit{consistent} if $\lim_{N\to\infty} \boldsymbol{E}[\estimator] = \params_0$.
    \item \textbf{Variance}: The estimator variance $\boldsymbol{V}[\estimator]$ quantifies the spread of estimates across repeated experiments.
\end{itemize}

The \textit{Cramér-Rao bound}~\cite{James:2006zz,Young_Smith_2005} establishes a fundamental limit: the variance of any unbiased estimator satisfies $\boldsymbol{V}[\estimator] \geq \boldsymbol{I}(\params)^{-1}$, where $\boldsymbol{I}(\params)$ is the Fisher information~\cite{James:2006zz}. 
An estimator achieving this bound is optimal.
A proof of the Cramér-Rao bound can be found in \cref{sec:cramer-rao-bound}. 
Fisher information is described in more detail in \cref{sec:fisher-information}.

\subsection{Method of maximum likelihood}
\label{sec:method-of-maximum-likelihood}
An intuitive way to find the best estimator is to choose the one that maximizes the probability of observing the obtained data
\begin{equation}
    \estimator(\data) = \text{argmax}_{\params} p(\data | \params).
    \label{eq:mleestimator}
\end{equation}
$\estimator(\data)$ is the \textit{\ac{MLE}} of $\params$.

The \ac{MLE} corresponds to the point in parameter space where the likelihood function is maximized. This can be found by solving the condition
\begin{equation}
    \nabla  p(\data | \params) |_{\params = \estimator} = 0.
    \label{eq:mle-condition-diff}
\end{equation}
The roots of these equations are the parameter values that give the maximal probability for the observed data.

\subsubsection{Asymptotic properties of maximum likelihood estimators}
\label{sec:mle-asymptotics}
The properties of any estimator can be broadly classified into two categories: \textit{asymptotic} properties, which are valid in the limit of large sample sizes ($N \to \infty$), and \textit{finite-sample} properties, which describe behaviour for a fixed and typically small number of observations~\cite{James:2006zz}. This section focuses on the asymptotic properties of the \ac{MLE}, which possesses several attractive features.

In the asymptotic limit, the \ac{MLE} has the following properties~\cite{Cowan1998,Young_Smith_2005}: it is \textit{consistent} and \textit{asymptotically normal}, which means that in the large sample limit $\sqrt{N}(\estimator - \params_0)$ converges to a normal distribution with minimum variance,
\begin{equation}
    \sqrt{N}(\estimator - \params_0) \to^d \mathcal{N}(\boldsymbol{0}, \boldsymbol{I}(\params_0)^{-1}) \ .
    \label{eq:asymptotic-normality}
\end{equation}
These properties are proven in \cref{sec:consistency-mle} and \cref{sec:asymptotic-normality-mle}.

\subsection{Maximum a posteriori estimate}
In the Bayesian interpretation of probability, parameters are treated as fundamentally random variables, rather than fixed but unknown quantities (frequentist interpretation). Accordingly, there is no notion of a single \textit{true parameter} value. All available knowledge about the parameters, after observing the data, is encoded in the \textit{posterior distribution} --- a multidimensional \ac{PDF} over the parameter space that may also capture correlations among parameters.

Nonetheless, often a particular point in a parameter space is desired. The most common choice in this case is the \textit{posterior mode} or \textit{\ac{MAP}},
\begin{equation}
    \mapestimator(\data) = \text{argmax}_{\params} p(\params | \data).
    \label{eq:mapestimator}
\end{equation}
This estimate corresponds to the point in parameter space where the posterior distribution is maximized. Hence, the \ac{MAP} estimator satisfies
\begin{equation}
    \nabla_{\params}  p(\params | \data) |_{\params = \mapestimator} = 0.
\end{equation}

\subsection{Equivalence of maximum likelihood and maximum a posteriori estimators}
The choice between frequentist and Bayesian interpretations of probability should not change the underlying physics of the data-generating process. It is therefore of interest to identify the conditions under which the \ac{MLE} and the \ac{MAP} yield identical results.

Starting from the definition of the \ac{MAP} estimator in \cref{eq:mapestimator}, and substituting the expression for the posterior via Bayes' theorem, the result is
\begin{equation}
    \mapestimator(\data) = \text{argmax}_{\params} p(\params | \data) = \text{argmax}_{\params}\left( \frac{p(\data | \params) p(\params)}{p(\data)} \right) \ .
\end{equation}
Given that the $\text{argmax}$ function is invariant under scaling by a constant factor, the marginal likelihood $p(\data)$ can be dropped,
\begin{equation}
    \mapestimator(\data) = \text{argmax}_{\params}\left( p(\data | \params) p(\params) \right) \ .
\end{equation}
If the prior $p(\params)$ is constant (i.e.~uniform over the parameter space), it does not influence the location of the maximum, and the \ac{MAP} estimator reduces to the \ac{MLE}
\begin{equation}
    \mapestimator(\data) = \text{argmax}_{\params} p(\params | \data) = \text{argmax}_{\params} p(\data | \params) = \estimator(\data) \ .
\end{equation}
 Therefore, in the special case of a uniform prior, the \ac{MAP} and \ac{MLE} estimators are equivalent.

\section{Model comparison}
\label{sec:model-comparison}
Scientific progress is based on iteratively improving theoretical models, given observed data. But given a set of different theories that claim to describe the data-generating process, which one should be chosen, and how?
How can the performance of one model over another be quantified? 

Answering these questions will be the goal of this section. A distinction is made between two complementary approaches to model comparison: the point-wise and the global comparison. The point-wise compares models at their respective best-fit point (\cref{sec:pointwise-comparison}), whereas the global comparison evaluates the average performance over the full parameter space (\cref{sec:global-comparison}).

\subsection{Point-wise model comparison: hypothesis testing}
\label{sec:pointwise-comparison}
The goal of most analyses is to test how well a predicted model or \textit{hypothesis} describes the data. A hypothesis $H$ would, for example, be the \ac{PDF} of a random variable $x$, $p(x|H)$. 

Commonly, a benchmark model --- the null hypothesis, $H_0$ --- is chosen, which predicts a \ac{PDF} $p(x|H_0)$. One then wants to compare the performance to alternative hypotheses $H_1, H_2,\ldots$ which predict different \acp{PDF} $p(x|H_1), \ p(x|H_2),\ldots$

For example, in particle physics, the null hypothesis could be the \ac{SM} expectation of some process, excluding any contribution from a given new physics process. The alternative hypotheses would include the contribution of the given new physics process that one is targeting to observe.

To measure the agreement between a given prediction and the observed data, it would be helpful to have a function of the data that reflects this. This function is called the \textit{test statistic} $t(\data)$.
Since the test statistic is a function of a random variable, it itself is a random variable. Hence, each hypothesis $H_i$ predicts a \ac{PDF} for the test statistic, $p(t|H_i)$. Commonly, smaller values of the test statistic correspond to better agreement between the prediction and the data.

\begin{figure}
    \centering
    \includegraphics[scale=0.6]{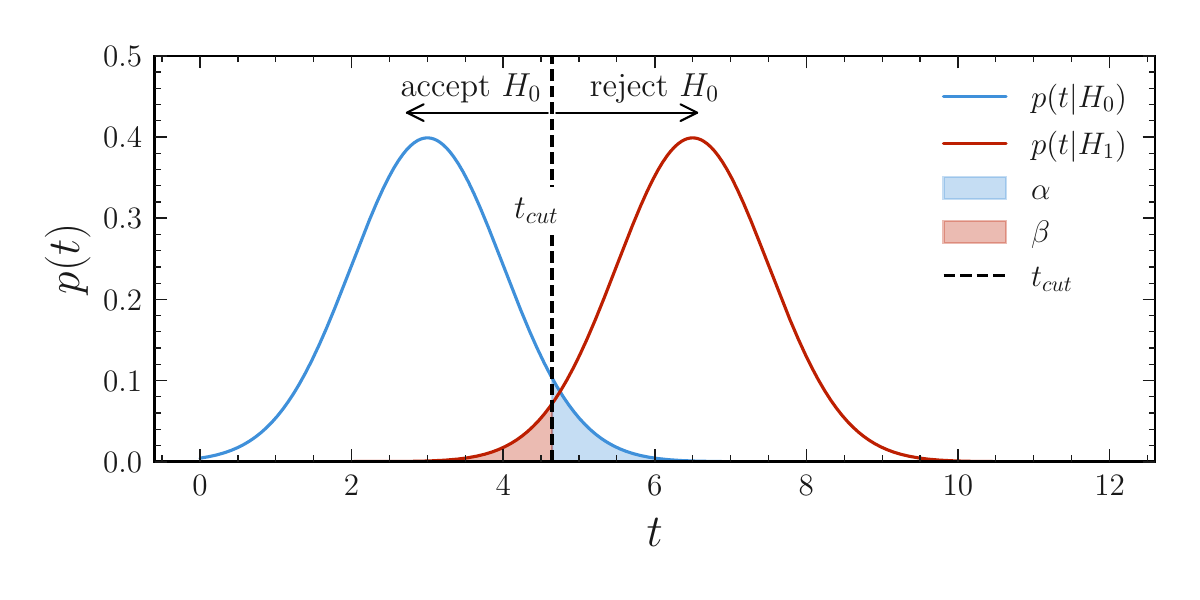}
    \caption{The \acp{PDF} of the test statistic, given two different hypotheses. The chosen value of $\alpha=0.05$ defines $t_{cut}$, which separates the rejection region (right) from the acceptance region (left). The integrals of the shaded regions correspond to $\alpha$ and $\beta$, respectively.}
    \label{fig:teststat}
\end{figure}

The test statistic can be used to define a \textit{cut} or \textit{decision boundary} $t_{cut}$, which separates the \textit{rejection region}, $t > t_{cut}$, from the \textit{acceptance region}, $t < t_{cut}$.
If the observed value of the test statistic falls into the rejection region, one rejects the null hypothesis. If, on the other hand, the measured test statistic falls into the acceptance region, one rejects the alternative hypothesis. For an illustration see \cref{fig:teststat}.
The decision boundary is usually chosen such that the probability for $t$ to be observed there, under the assumption of the null hypothesis, is some chosen value $\alpha$,
\begin{equation}
    \alpha = \int_{t_{cut}}^{\infty} dt \ p(t|H_0) = 1-F(t_{cut}| H_0),
    \label{eq:pvalue}
\end{equation}
which is also known as the \textit{size} of the test. Here, $F(t_{cut}| H_0)$ is the \ac{CDF} (see \cref{sec:cdfs}) up to the cut. Typical values for $\alpha$ are $\alpha=0.05$ or $\alpha=0.1$.

Any hypothesis test involves two types of errors: rejecting $H_0$ when it is true (\textit{Type I error}, probability $\alpha$), and failing to reject $H_0$ when $H_1$ is true (\textit{Type II error}, probability $\beta = \int^{t_{cut}}_{-\infty}dt \, p(t|H_1)$). The quantity $1-\beta$ is the \textit{power} of the test. An ideal test statistic maximizes power by clearly separating the \acp{PDF} of $H_0$ and $H_1$.

\begin{figure}
    \centering
    \includegraphics[scale=0.6]{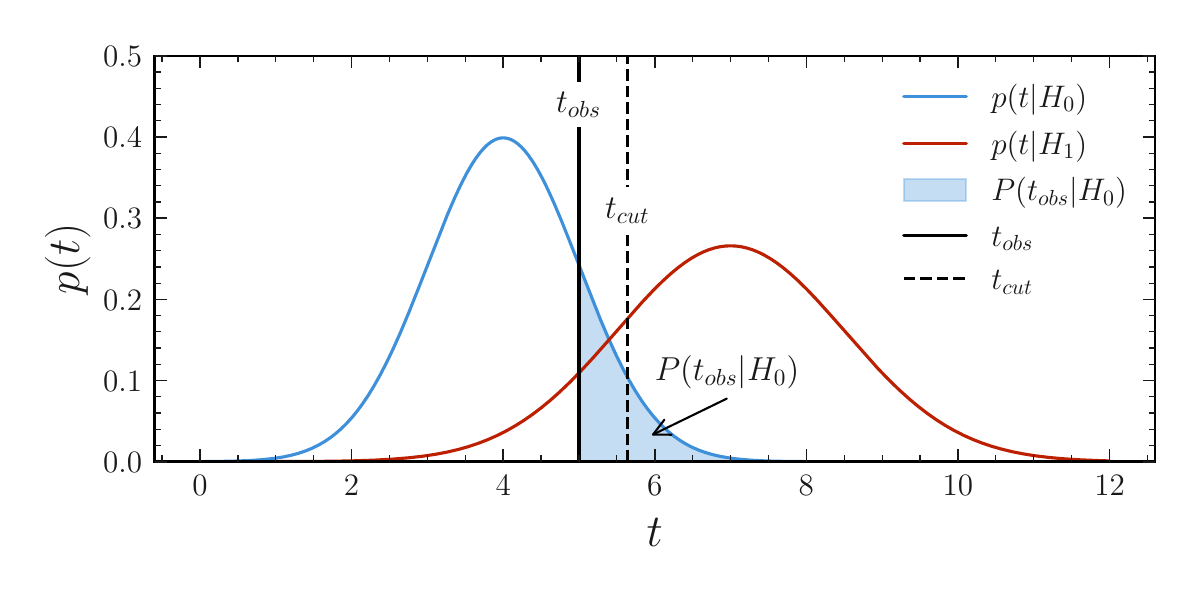}
    \caption{The \acp{PDF} of the test statistic, given two different hypotheses. The chosen value of $\alpha=0.05$ defines $t_{cut}$, which separates the rejection region from the acceptance region. The observed value of the test statistic $t_{obs}$ defines the lower bound of the integration for the $P$-value.}
    \label{fig:pvalue}
\end{figure}

Given a set of observations, an observed value for the test statistic, $t_{obs}$, can be calculated. Subsequently, the probability of obtaining a result as compatible or less with $H_0$ than the one actually \textit{observed} can be determined,
\begin{equation}
    P(t_{obs}| H_0) = \int_{t_{obs}}^{\infty}dt \ p(t|H_0) = 1-F(t_{obs}| H_0),
\end{equation}
which is known as the \textit{$P$-value}.\footnote{Not to be confused with the symbol for probability used in the previous sections} This is illustrated in \cref{fig:pvalue}. If $P(t_{obs}| H_0)< \alpha$, one rejects the null hypothesis. If $P(t_{obs}| H_0) > \alpha$, one does not reject the null hypothesis.

\subsubsection{Weighted $P$-values}
\label{sec:weighted-pvalues}
If there is low statistical power, but one observes $P(t_{obs}| H_0) < \alpha$, should the null hypothesis still be rejected? The power of the test is a measure of how well the test can discriminate between the null and the alternative hypothesis. If the power is low, the test is not sensitive enough to prioritize one hypothesis over the other. In this case, the null hypothesis should not be rejected. To protect against this case, weighted $P$-values can be considered.

Consider the case, where the hypothesis to be excluded is the null hypothesis $H_0$, which describes the case where signal plus background is observed ($H_0=s+b$), according to some model. The alternative hypothesis $H_1$ describes the case where background only is observed ($H_1=b$).

A $P$-value can be defined
\begin{equation}
    P(t_{obs}| s+b) = \int_{t_{obs}}^{\infty}dt \, p(t|s+b),
\end{equation}
which is the probability of obtaining a measurement result as compatible or less with the $s+b$ hypothesis.

Given the choice of a rejection region, or equivalently a value of $\alpha$, the $s+b$ hypothesis is excluded if
\begin{equation}
    P(t_{obs}| s+b) < \alpha.
\end{equation}
The problem with this approach is that hypotheses to which there is little or no sensitivity might be excluded. By sensitivity, it is meant that the power of the test is low, and the \acp{PDF} of the test statistic are almost indistinguishable, or in other words largely overlapping.
This could be the case if, for example, there are very few signal events compared to the number of background events. In this case, the \acp{PDF} will be very similar.

To protect against this problem, $P(t_{obs}| s+b)$ can be compared to
\begin{equation}
    P(t_{obs}| b) = \int^{\infty}_{t_{obs}}dt \, p(t|b),
\end{equation}
which is the probability of obtaining a measurement result as compatible or less with the $b$ hypothesis.
One can define the \textit{weighted} $P$-value
\begin{equation}
    CL_s \equiv \frac{P(t_{obs}| s+b)}{P(t_{obs}| b)} < \alpha
    \label{eq:cls}
\end{equation}
as exclusion condition~\cite{read2002}. This condition takes the $b$ hypothesis into account, which is equivalent to saying ``the $s+b$ hypothesis is excluded only if the result is sufficiently more compatible with the $b$ hypothesis''. If there is similar compatibility of the measurement with both of the hypotheses, i.e.~a lack of sensitivity, $s+b$ should not be excluded.

If the two \acp{PDF} $p(t|s+b)$ and $p(t|b)$ are widely separated, then $P(t_{obs}| b)$ will be close to unity and $CL_s \approx P(t_{obs}| s+b)$. Contrarily, if the two \acp{PDF} are largely overlapping, then $P(t_{obs}| b)\approx P(t_{obs}| s+b)$ and $CL_s\approx 1$.

\Cref{eq:cls} shows that $CL_s \geq P(t_{obs}| s+b)$ in all possible cases. The hypotheses excluded by $CL_s < \alpha$ are a subset of the ones excluded by $P(t_{obs}| s+b)<\alpha$. Hence, the criterion $CL_s<\alpha$ limit is \textit{weaker} than the one defined by $P(t_{obs}| s+b)<\alpha$. This is the reason why the $CL_s$ criterion is classified as \textit{conservative}.

\subsubsection{Functional form of the test statistic}
In general, one wants to find a test statistic with optimal power for a given size $\alpha$. The \textit{Neyman-Pearson lemma} states that the likelihood ratio,
\begin{equation}
    \lambda = \frac{p(\data|H_0)}{p(\data|H_1)} \ ,
    \label{eq:neyman-likelihood-ratio}
\end{equation}
leads to the most powerful rejection region. A proof can be found in \cref{sec:proof-npl}. Based on this a test statistic can be defined as
\begin{equation}
    t = - 2 \log \lambda \ .
\end{equation}
Most commonly used test statistics are based on the likelihood ratio~\cite{Cowan:2010js}.

The null hypothesis is rejected if the data are too compatible with the alternative hypothesis,
\begin{equation}
    t = - 2 \log \lambda > t_{cut} \ ,
\end{equation}
A nice property of the likelihood ratio test is that it is invariant under reparametrization of the data,
\begin{equation}
    \frac{p\left(y \mid H_0\right)}{p\left(y \mid H_1\right)}=\frac{|J_f|^{-1} p\left(x=f^{-1}(y) \mid H_0\right)}{|J_f|^{-1} p\left(x=f^{-1}(y) \mid H_1\right)}=\frac{p\left(x \mid H_0\right)}{p\left(x \mid H_1\right)}.
\end{equation}
The Jacobian factors $|J_f|^{-1}$ cancel. This is a very useful property, as it allows the choice of the most convenient parameterization for the problem, without any loss of power.

\subsubsection{Distribution of the test statistic}
One technical question remains --- what is the \ac{PDF} of the test statistic under a hypothesis $H$, parametrized by $\theta$, $p(t | \theta)$?

There is no unique answer to this question. It depends on the particular analysis case under consideration. A general, but also computing intensive method of finding $p(t | \theta)$ for a general functional form of $t$ is through \ac{MC} methods~\cite{James:2006zz}. For a given parameter point $\theta$, the procedure goes as follows:
\begin{enumerate}
    \item Sample a large set of data points from the likelihood
    \begin{equation}
        \data \sim p(\data | \theta) \, .
    \end{equation}
    \item For each sampled dataset $\data$, compute the test statistic $t(\data)$.
    \item Construct the empirical distribution of the test statistic $p(t | \theta)$ from the computed values, through kernel density estimation or histograms.
\end{enumerate}
This procedure can be computationally intensive, especially for high-dimensional parameter spaces and functional forms of the test statistic, which involve optimisation steps in the evaluation (e.g.~likelihood ratios).

In \cref{sec:mle-asymptotics} the asymptotic behaviour of maximum likelihood estimators was discussed. Similarly, one can utilize asymptotic approximations to derive a functional form for the \ac{PDF} of a test statistic. 
References~\cite{Cowan:2010js,wald,wilks} provide further discussions and collection of asymptotic distributions for a number of important test statistics.

\subsubsection{Nested hypotheses}
\label{sec:profiled-likelihood}
In the case where the null hypothesis is a special case of the alternative hypothesis ${H_0 \subset H_1}$, $H_0$ is said to be \textit{nested} within $H_1$.

For example the \ac{POI} could be a signal strength\footnote{A multiplicative factor of the overall expected signal yields.} $\mu$ and the nuisance parameters could be the background normalization $\psi$. Assume the null hypothesis to be the background only hypothesis, $H_0: \mu_0=0$, and the alternative hypothesis to be the signal plus background hypothesis, $H_1: \mu>0$. The value of $\psi$ is not of interest.

To compare the hypotheses, it is necessary to compare how well the data are described by the \ac{POI} of the null hypothesis compared to any other \ac{POI} value $\mu_0$, irrespective of the nuisance parameters. In this case the likelihood ratio test statistic is~\cite{Cowan:2010js}
\begin{equation}
    t(\mu_0) = -2 \log \lambda(\mu_0)= -2 \log \frac{p(\data | \mu_0, \hat{\hat{\psi}})}{p(\data | \hat \mu,\hat \psi)}.
    \label{eq:nested-likelihood-ratio-test-statistic}
\end{equation}
The $\hat \mu$ and $\hat \psi$ are the \textit{unconditional} ML estimator, which maximize the likelihood function. The $\hat{\hat{\psi}}$ is the \textit{conditional} ML estimators, which maximize the likelihood for the specified $\mu_0$. The procedure of finding the conditional ML estimators is known as \textit{profiling}~\cite{Cowan:2010js}.

From the definition of $\lambda(\mu)$, it follows that $0<\lambda(\mu)<1$, with $\lambda(\mu) \to 1$ implying good agreement between data and hypothesized $\mu$. Hence, increasing compatibility between data and $\mu$ occurs as $t(\mu) \to 0$.

This test statistic does not distinguish between $\mu_0 > \hat \mu$ or $\mu_0 < \hat \mu$. Hence, one can achieve a small $P$-value through either too large or too small values of $\mu_0$. This results in a two-sided confidence interval (\cref{sec:confidence-intervals}). 

Wald~\cite{wald} demonstrated that for a single \ac{POI}, the test statistic can be approximated as
\begin{equation}
    t(\mu) = \frac{(\hat{\mu} - \mu)^2}{\sigma^2} + \mathcal{O}(1/\sqrt{N}) \ ,
    \label{eq:wald-test-statistic}
\end{equation}
where $\sigma^2$ is the variance of the estimator $\hat{\mu}$. In the large-sample limit, this variance is given by the inverse of the Fisher information, $\sigma^2 = \boldsymbol{I}(\mu)^{-1}$ (see \cref{sec:mle-asymptotics}).

If the estimator $\hat{\mu}$ is normally distributed and the $\mathcal{O}(1/\sqrt{N})$ correction term is neglected, the test statistic $t(\mu)$ asymptotically follows a $\chisq$-distribution with one \ac{d.o.f.}. This result is known as \textit{Wilks' theorem}~\cite{wilks}. A proof is provided in \cref{sec:likelihood-ratio-test-statistic-normal}.

\subsubsection{Significance}
\label{sec:significance}
Often one converts $P$-values into \textit{significance levels} $Z$, which describe the number of standard deviations from a Gaussian distributed variable \cite{Cowan:2010js},
\begin{equation}
    P(t_{obs}| H_0) = 1-\Phi(Z) \ ,
\end{equation}
where $\Phi$ is the \ac{CDF} of a standard normal distribution, $\mathcal{N}(0,1)$.
Inverting the above equation, the significance level is given by
\begin{equation}
    Z = \Phi^{-1}\left(1-P(t_{obs}| H_0)\right) \ ,
\end{equation}
where $\Phi^{-1}$ is the inverse of $\Phi$.
For a normally distributed test statistic,
\begin{equation}
    t_{obs} - E[t|H_0] = Z \sigma \ ,
\end{equation}
inverting the definition of $\Phi$.

The significance level hence corresponds to the number of standard deviations from the mean of the \ac{PDF} predicted by $H_0$. Commonly used benchmarks for the significance level are $Z=3$ ($P(Z=3)=1.35\cdot10^{-3}$) for \textit{evidence} and $Z=5$ ($P(Z=5)=2.87\cdot10^{-7}$) for an \textit{observation}. In the latter case, this translates to: ``for repeated experiments one expects one out of $1/2.87\cdot10^{-7} = 3.48\cdot10^{6}$ measurement outcomes to show a deviation as large as the one observed''.

\subsection{Global model comparison: Bayes factors}
\label{sec:global-comparison}
A global model comparison focuses on the model, independently of a given parameter point. 
Here, the objective is to compare probabilities of a theoretical model, or hypothesis\footnote{In this context, the term \textit{model} is more commonly used. Here, models are referred to as hypotheses, to draw the parallels with the pointwise comparison in \cref{sec:pointwise-comparison}.} $H$, given the observed data $\data$, $p(H | \data)$. Using Bayes theorem, one can write 
\begin{equation}
    p(H | \data) = \frac{p(\data | H) p(H)}{p(\data)}.
\end{equation}
If one wants to compare two hypotheses $H_0$ and $H_1$, one common option is to compare the ratio
\begin{equation}
    \frac{p(H_1 | \data)}{p(H_0 | \data)} =  \frac{p(\data | H_1 )}{p(\data | H_0)} \frac{p(H_1)}{p(H_0)}.
\end{equation}
Commonly, one assigns the same probability to either hypothesis, such that $p(H_1) = p(H_0)$.
The \textit{marginal likelihood} or \textit{evidence} $p(\data | H)$ is obtained by marginalizing the likelihood over the full, or a subset of the parameter space, weighted by the prior,
\begin{equation}
    p(\data | H) = \int d^M \theta ~ p(\data | \theta, H) p(\theta | H).
    \label{eq:marginal-likelihood}
\end{equation}
This is the probability of the observed data assuming $H$ is the true underlying hypothesis.
The Bayes factor is defined as the ratio of marginal likelihoods \cite{gelman2013bayesian},
\begin{equation}
    B = \frac{p(\data | H_1 )}{p(\data | H_0)}.
\end{equation}

A Bayes factor of $B>1$ favours hypothesis $H_1$, whereas $B<1$ favours $H_0$. More specifically, Jeffreys \cite{jeffreys1961theory} has introduced a scale for interpreting the strength of evidence provided by the Bayes factor, shown in \cref{tab:jeffreys-scale}.

\begin{table}[ht]
    \renewcommand{\arraystretch}{1.5}
    \caption{Jeffreys's scale for interpreting Bayes factors. The ranges of $\log_{10} B$ indicate the strength of evidence in favour of hypothesis $H_1$ over $H_0$~\cite{jeffreys1961theory}.}
    \centering
    \begin{tabularx}{\linewidth}{@{\extracolsep{\fill}}YY}
        \toprule \midrule
        $\log_{10} B$ & Evidence for $H_1$ over $H_0$ \\
        \midrule
        $0.0 - 0.5$ & Barely worth mentioning \\
        $0.5 - 1.0$ & Substantial \\
        $1.0 - 1.5$ & Strong \\
        $1.5 - 2.0$ & Very strong \\
        ${}>2.0$ & Decisive \\
        \midrule \bottomrule
    \end{tabularx}
    \label{tab:jeffreys-scale}
\end{table}

When interpreting the Bayes factor, there is one important aspect to consider: the prior sensitivity. Whereas the support of priors for unconstrained parameters is less critical for the posterior result, it matters largely for the Bayes factor \cite{kass1995}. This is because the \textit{average} likelihood in the \textit{full} parameter space is being computed. Hence, if the parameter space is increased to regions where the likelihood is very small, a smaller marginal likelihood will be obtained. This is in general an advantage, as it penalizes overly complex models. On the other hand, it ties the prior choice even tighter with the model.

To illustrate this, consider the case of a model hypothesis $H_1$, parametrized by one parameter $\theta$, and another unparametrized model hypothesis $H_0$ \cite{Karamanis2023}. In addition, assume that the model hypothesis $H_0$ is nested within $H_1$, in other words there exists a parameter point $\theta_0$ where $p(\data | \theta_0, H_1) = p(\data | H_0)$.
Take the prior of $\theta$ to be uniform with a width of $\Delta \theta$ and the likelihood $p(\data | \theta, H_1)$ to be sharply peaked around the value $\bar \theta$, with a width of $\delta \theta$, which lies in the region of support of the prior. Under these assumptions, the marginal likelihood can then be approximated as
\begin{equation}
    p(\data | H_1) = \int d\theta  \ p(\data | \theta, H_1) p(\theta | H_1) \approx p(\data | \bar \theta, H_1) \frac{\delta \theta}{\Delta \theta}.
\end{equation}
The Bayes factor is
\begin{equation}
    B = \frac{p(\data | H_1)}{p(\data | H_0)} \approx \frac{p(\data | \bar \theta, H_1)}{p(\data | \theta_0, H_1)} \frac{\delta \theta}{\Delta \theta} \ .
\label{eq:bayes-factor-occam}
\end{equation}
The ratio 
\begin{equation}
    \frac{p(\data | \bar \theta, H_1)}{p(\data | \theta_0, H_1)} \geq 1 \ ,
\end{equation}
since the data can always be fit equally well or slightly better with an additional parameter. 

The second term in \cref{eq:bayes-factor-occam}, $\delta \theta / \Delta \theta$, sometimes referred to as the \textit{Occam's razor} term, penalizes the more complex model $H_1$ for including parameter space which is unfavoured by the data. Hence, the Bayes factor only favours more complex models, when the likelihood ratio compensates for the penalty of the Occam's razor term.

\section{Goodness-of-fit}
\label{sec:goodness-of-fit}
Goodness-of-fit testing can be interpreted as a hypothesis test against the optimal model --- namely, the model that reproduces the data exactly.
A well motivated measure for the goodness-of-fit is the $P$-value, calculated from the \ac{PDF} of the test statistic $p(t_{\text{gof}})$, where
\begin{equation}
    P_{gof} = \int_{t_{obs}}^\infty dt_{\text{gof}} \ p(t_{\text{gof}}) = 1-\cdf_p(t_{obs}) \ ,
    \qquad
    t_{\text{gof}} = -2 \ln \frac{p(\data | \hat{\params})}{p_{\rm sat}}.
\end{equation}
Here $p(\data | \hat{\params})$ is the best-fit likelihood. The saturated likelihood, $p_{\rm sat}$, is the best possibly achievable likelihood \cite{Cousins2013GeneralizationOC}. 

The value of $t_{\text{gof}}$ indicates how close one comes to the saturated (``perfect'') value of the likelihood (smaller is better, $t_{\text{gof}}=0$ would correspond to a ``perfect'' fit). The saturated likelihood has zero free parameters.

For a Gaussian likelihood with $N$ independent data points with means $\mu_i$ and variances $\sigma_i^2$, the saturated model sets $x_i=\mu_i$, and the goodness-of-fit statistic reduces to the familiar $\chisq$ form $t_{\text{gof}}=\sum_i (x_i-\mu_i)^2/\sigma_i^2$. Each standardized residual contributes a $\chisq$ with 1 \ac{d.o.f.}. By independence $t_{\text{gof}}\sim\chisq_N$, and fitting $m$ parameters reduces the \ac{d.o.f.} to $N-m$. A derivation is given in~\cref{sec:likelihood-ratio-test-statistic-normal}.

For the \histfactory likelihood $p(\yields, \aux | \unconstr, \constr)$ (\cref{sec:histfactory}) the saturated likelihood is
\begin{equation}
    p_{\rm sat}(\tilde{\constr}) = \mathrm{Pois}(\boldsymbol{n} \mid \boldsymbol{n}) ~ C(\boldsymbol{a} \mid \tilde{\constr}),
\end{equation}
which is the product of the Poisson likelihood with expectation values equal to the observed data counts and the constraint likelihood with parameters $\tilde{\constr}$ maximizing the constraint term. In the asymptotic limit, $p(t_{\text{gof}})$ will equal a $\chisq$-distribution with $N-m$ \ac{d.o.f.}, where in this case $N$ is the number of bins in the model and $m$ the number of free parameters $\unconstr$.

\section{Limit setting}
\label{sec:limit-setting}
In interval estimation, the goal is to find a parameter range
\begin{equation}
    \theta_a \leq \theta \leq \theta_b \ ,
\end{equation}
which contains the true value of the parameter $\theta_0$ with a certain probability $1-\alpha$,
\begin{equation}
    P(\theta_a \leq \theta_0 \leq \theta_b) = 1-\alpha \ .
    \label{eq:interval-probability}
\end{equation}
In an idealized setting, one assumes access to the sampling distribution of the estimator $\theta$, given the true parameter value $\theta_0$, denoted by $p(\theta | \theta_0)$.
Under this assumption, the interval can be determined by solving
\begin{equation}
    P(\theta_a \leq \theta_0 \leq \theta_b) = \int_{\theta_a}^{\theta_b} d\theta \, p(\theta | \theta_0) = 1-\alpha \ .
    \label{eq:confidence-interval}
\end{equation}
Hence, one could directly determine the interval boundaries $\theta_a$ and $\theta_b$ from the parameter \ac{PDF}. Note that such an interval is not necessarily unique. However, since the true distribution $p(\theta | \theta_0)$ is typically unknown in practice, this represents an idealized scenario.

The Bayesian approach to interval estimation comes closest to this scenario. Here one does not have access to the \ac{PDF} $p(\theta | \theta_0)$, but to the parameter \ac{PDF} $p(\theta | \data)$. In this case the obtained interval is known as a \textit{credible interval}. This is discussed in \cref{sec:credible-intervals}.

In the frequentist case, one constructs intervals for a large range of parameter values, utilizing the \ac{PDF} of a test statistic $t(\data)$. This is known as \textit{Neyman construction} and results in \textit{confidence intervals} of the parameter. This is discussed in \cref{sec:confidence-intervals}.

\subsection{Credible intervals}
\label{sec:credible-intervals}
Bayesian credible intervals are defined as the range of parameter values that contain the true parameter value with a certain probability $1-\alpha$, given the observed data. Given the posterior distribution $p(\theta | \data)$, the credible interval can be found by determining the interval boundaries $\theta_a$ and $\theta_b$ such that
\begin{equation}
   \int_{\theta_a}^{\theta_b} d\theta \, p(\theta | \data) = 1 - \alpha \ .
   \label{eq:credible-interval}
\end{equation}
The credible interval is then given by $[\theta_a, \theta_b]$. Hence, Bayesian intervals provide a direct probabilistic interpretation, namely, the probability that the true parameter value lies within the interval is equal to $1-\alpha$.

Generally, this interval is not unique. There are multiple options for defining the boundaries, leading to different credible intervals:
\begin{itemize}
    \item \textbf{\ac{HDI}}: This is the smallest possible interval, containing $1-\alpha$ probability. It has the property that the probability density within the interval is never lower than that outside the interval~\cite{gelman2013bayesian}. This includes all parameter values $\theta: p(\theta | \data) > W$, for some $W$, such that~\cite{kruschke2015doing}
    \begin{equation}
        \int_{\theta: p(\theta | \data) > W} d\theta \, p(\theta | \data) = 1-\alpha \ .
    \end{equation}

    \item \textbf{Central interval}: This is an interval $[\theta_a,\theta_b]$, centred around a specific value, usually the posterior mode or mean. 

    \item \textbf{Equal-tailed interval}: This interval ensures the same probability mass in the upper and lower tails of the posterior distribution. One calculates the boundaries such that
    \begin{equation}
        P(\theta < \theta_a | \data) = P(\theta > \theta_b | \data) = \frac{\alpha}{2} \ .
    \end{equation}

        \item \textbf{Upper/lower interval}: These one-sided intervals specify either an upper or lower bound on the parameter, containing a given posterior probability mass. For an upper interval,
        \begin{equation}
        P(\theta < \theta_b | \data) = 1-\alpha,
        \end{equation}  
        and for a lower interval,
        \begin{equation}
        P(\theta > \theta_a | \data) = 1-\alpha.
        \end{equation}
        These are particularly useful in setting exclusion bounds.
\end{itemize}

In the case of a multidimensional parameter space $\params \in \mathcal{R}^M$, the above results are easily generalizable. In such cases the credible interval spans over a region $R \subset \mathcal{R}^M$ in parameter space, such that
\begin{equation}
    \int_R d^M \theta \, p(\params | \data) = 1- \alpha \ .
\end{equation}
Further, one can obtain credible intervals for a subset of parameters $\params' \subseteq \params$ by marginalizing the posterior over all but the \acp{POI}.

\subsection{Confidence intervals}
\label{sec:confidence-intervals}
Frequentist confidence intervals are constructed by inverting the probability statement in \cref{eq:interval-probability} --- that is, instead of asking what is the probability that a fixed interval contains the true parameter, one constructs an interval that, in repeated experiments, would contain the true value with the specified coverage probability. 
This method is known as \textit{Neyman construction}~\cite{James:2006zz}. 

This inversion can be achieved by summarizing the data with a test statistic $t(\data)$, with a \ac{PDF} $p(t | \theta)$, and then solving for the interval boundaries $t_a(\theta)$ and $t_b(\theta)$ for a given point $\theta$,
\begin{equation}
    P(t_a < t < t_b | \theta) = \int_{t_a}^{t_b}dt \,  p(t | \theta) = 1- \alpha.
    \label{eq:neyman-construction}
\end{equation}
Repeating this for a range of $\theta$ values, one constructs the confidence intervals in $t$-space $[t_a(\theta), t_b(\theta)]$. 

The uniqueness of these intervals can be given by further specifying interval properties such as:
\begin{itemize}
    \item \textbf{Central interval}: Here the interval is constructed to have equal probability in the upper and lower tails,
    \begin{equation}
        \int_{-\infty}^{t_a}dt \,  p(t | \theta) = \int_{t_b}^{\infty} dt \,  p(t | \theta) = \frac{\alpha}{2}.
    \end{equation}
    \item \textbf{Upper/lower interval}: These intervals specify either an upper or lower bound on the parameter. For an upper bound,
        \begin{equation}
        \int_{-\infty}^{t_b}dt \,  p(t | \theta) = 1-\alpha,
        \end{equation}
        and for a lower interval,
        \begin{equation}
        \int_{t_a}^{\infty} dt \,  p(t | \theta) = 1-\alpha.
        \end{equation}
        These are particularly useful in setting exclusion bounds.
\end{itemize}

\begin{figure}
    \centering
    \includegraphics[width=0.8\textwidth]{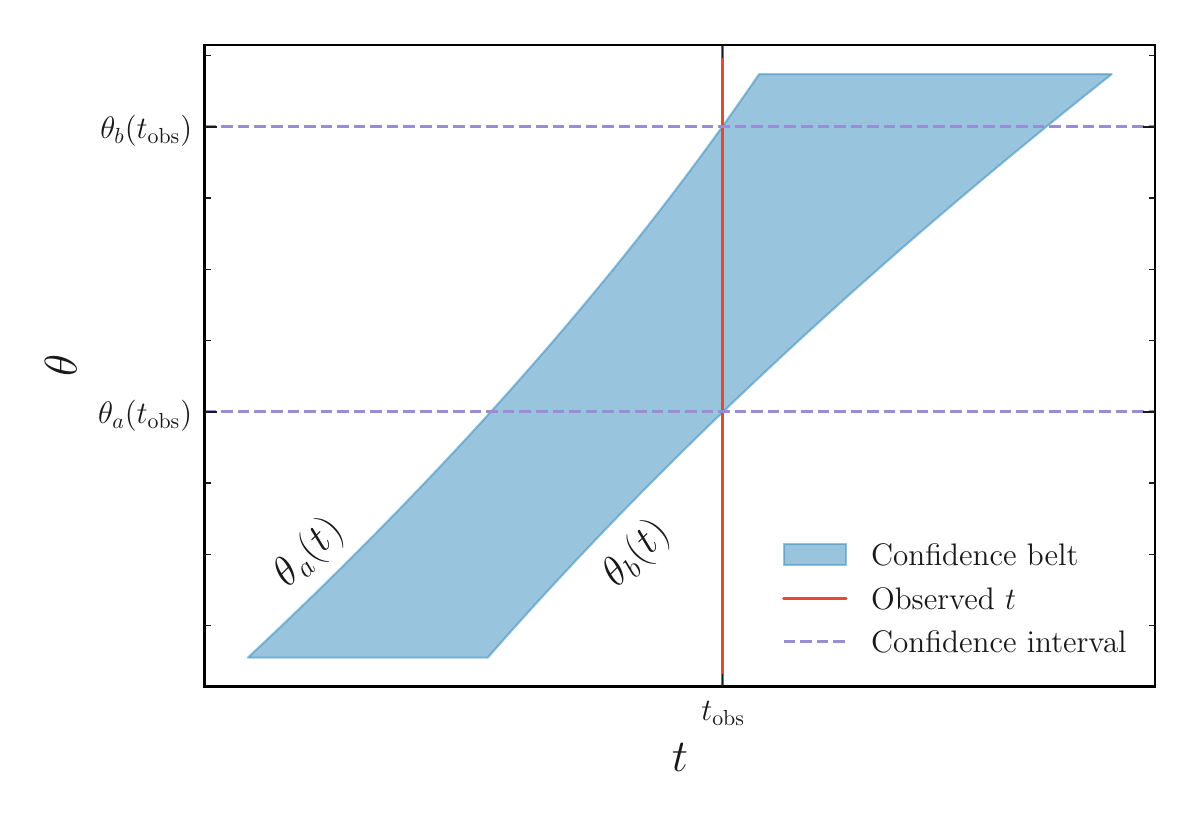}
    \caption{Neyman construction of confidence intervals. The confidence interval is constructed horizontally by scanning through values of $\theta$, determining $t_a(\theta)$ and $t_b(\theta)$ such that $P(t_a(\theta) < t < t_b(\theta)) = 1-\alpha$.}
    \label{fig:neyman}
\end{figure}

Suppose the confidence intervals $[t_a(\theta), t_b(\theta)]$ have been constructed for a range of $\theta$ values. These can be inverted to obtain the confidence intervals in $\theta$-space, $[\theta_a(t), \theta_b(t)]$. The required confidence interval for $\theta$ is then given by the intersection of an observed value $t_{\rm obs}$ with $t_a(\theta)$ and $t_b(\theta)$, or equivalently by $[\theta_a(t_{\rm obs}), \theta_b(t_{\rm obs})]$. This is illustrated in \cref{fig:neyman}.

To verify that the interval $[\theta_a, \theta_b]$ is truly the desired confidence interval, suppose a large number of experiments are performed. The proportion of values $t(\data)$ falling into a range $[t_a(\theta_0), t_b(\theta_0)]$, for the true value $\theta_0$, is $1-\alpha$ by definition. 
Therefore, the proportion of intervals $[\theta_a, \theta_b]$, corresponding to the $t$ values in $[t_a(\theta_0), t_b(\theta_0)]$, covering the true value $\theta_0$ is $1-\alpha$, that is
\begin{equation}
    P(\theta_a < \theta < \theta_b) = 1-\alpha.
\end{equation}

A method which gives an interval $[\theta_a, \theta_b]$, satisfying \cref{eq:confidence-interval}, possesses the property of \textit{coverage}.
As long as \cref{eq:neyman-construction} is satisfied, the Neyman construction method has exact coverage~\cite{James:2006zz}.

The extension to a multidimensional parameter space is more complicated for the frequentist case. This is mainly due to the fact that one must find the distribution of the test statistic in multiple dimensions. In a limited number of cases, this can be approximated asymptotically~\cite{James:2006zz}. If asymptotics are not applicable, the \ac{MC} approach is only feasible for a small number of parameters.

%% file: chapters/reinterpretation-review.tex
The reinterpretation of published analysis results has emerged as a transformative methodology in \ac{HEP}, allowing theorists and experimentalists alike to extract maximal value from existing datasets. By reconstructing and modifying statistical models, this practice enables comprehensive exploration of \ac{SM} and \ac{BSM} scenarios without necessitating new data collection or analysis designs. Reinterpretation also facilitates precision measurements, model comparisons, and global fits~\cite{Cranmer:2021urp}.

Collider experiments are unique, resource-intensive endeavours. Once an analysis is published, the original authors often move on, making further direct access to their insight and methods difficult. Thus, a structured strategy for the preservation and reuse of analysis data products and methods is essential to long-term scientific productivity~\cite{Butterworth:2929207,Bailey:2022tdz,LHCReinterpretationForum:2020xtr}.

This chapter reviews the essential components of reinterpretation infrastructure: data preservation (\cref{sec:data-preservation}), analysis preservation (\cref{sec:analysis-preservation}), statistical model serialization (\cref{sec:reinterpretation-statistical-models}), classification of reinterpretation approaches (\cref{sec:reinterpretation-types}), and available public tools (\cref{sec:reinterpretation-tools}). Finally, \cref{sec:reinterpretation-challenges} discusses remaining challenges and future goals for the community. 

\section{Data preservation}
\label{sec:data-preservation}
Data are the currency of the modern \ac{HEP} community. Experimental collaborations highly value their collected collision data and rarely release it to the public. This is understandable, given the amount of work and dedication needed to run colliders, detectors, and finally collect the data.

Experimental collaborations, such as ATLAS and CMS, have begun releasing data through the CERN Open Data portal~\cite{CERNOpenData}, often adhering to the \ac{FAIR} data principles. Due to the complexity of \ac{HEP} analyses, however, \ac{FAIR} compliance alone does not ensure reproducibility of measurement results~\cite{Bailey:2022tdz}.

Collaborations not only benefit from large computing resources, allowing them to process the data in the first place, but also from internal expertise, which allows for correct analyses and interpretations. These computation- as well as knowledge-bottlenecks limit the usability of large open datasets for the wider \ac{HEP} community.

Smaller, analysis-specific data products are more restricted in their versatility, but benefit greatly from their usability and accessibility. In this case, it is recommended to preserve~\cite{Butterworth:2929207}:
\begin{itemize}
    \item \textbf{Event-level data}, including both reconstructed data and \ac{MC} simulations.
    \item \textbf{Generator-level data}, necessary for fast reinterpretation pipelines and Rivet-based workflows~\cite{rivet}.
    \item \textbf{Analysis code and logic}, which includes the scripts and frameworks used to perform the selection and statistical interpretation.
    \item \textbf{Detector performance and response}: Either through unfolding or via response matrices.
    \item \textbf{Full statistical models}, which are the full probability models $p(\data | \params)$, with transparent treatment of nuisance parameters and their correlations.
    \item \textbf{Analysis data products}, which encode derived data, such as selection efficiencies and limits.
\end{itemize}
Having these data products in addition to a full analysis preservation, discussed in \cref{sec:analysis-preservation}, ensures the long-term reproducibility of experimental results and increases their credibility. 

\section{Analysis preservation}
\label{sec:analysis-preservation}
Analysis preservation allows for the processing of the preserved data products (\cref{sec:data-preservation}). This enables, on the one hand, the reproduction of the analysis data products or results, and on the other hand, modifications to the analysis and/or input data, enabling reinterpretation. 

Either the full analysis workflow --- from the initial collider data to the final result --- or a lightweight approximation of the analysis can be preserved, each coming with its own set of advantages and disadvantages.

\paragraph{Full analysis preservation}
This involves storing the complete analysis software stack, including detector simulation, reconstruction, and physics analysis code, ideally within containerized or virtualized environments. While this is the most accurate form of preservation, it is also the most computationally expensive and hardest to maintain due to the diversity of constantly evolving software ecosystems (even within a single experiment).

\paragraph{Lightweight preservation}
More suitable for rapid reinterpretation over large model spaces, lightweight formats omit full reconstruction details in favour of simplified implementations. These are typically based on fast simulations, cut-and-count approximations, or generator-level analysis, and are easier to share publicly.

Both methods serve complementary roles. Lightweight approaches enable rapid exploration of high-dimensional parameter spaces and identification of regions compatible with data, with combinations of multiple analyses further boosting sensitivity at minimal computational cost. Full analysis preservation becomes valuable once interesting regions are identified and precision is prioritized.

This work presents a reinterpretation method that bridges the gap between generator-level and reconstruction-level observables, combining the computational efficiency of lightweight approaches with accurate results. Reinterpretation is particularly valuable for exploring large theory parameter spaces rather than for precision measurements at individual points. Since experimental analyses are optimized for specific scenarios (typically the \ac{SM} or particular \ac{BSM} models), reinterpretation studies sacrifice some sensitivity compared to dedicated analyses. However, combining multiple lightweight analyses can achieve competitive or even superior sensitivity across broader parameter regions --- a capability unattainable with resource-intensive full-stack approaches. The identification of interesting parameter space regions through reinterpretation should motivate dedicated experimental analyses, thereby strengthening the feedback loop between theoretical and experimental communities in the search for new physics.

\section{Statistical models and their preservation}
\label{sec:reinterpretation-statistical-models}
Reinterpretation efforts are mostly hindered by the lack of access to full statistical models. This lack of availability can be attributed to the technical challenges that come with the serialization of statistical models. These challenges lie mainly in the complexity of the models and the need for high customization to the underlying analysis.
The earliest reusable models were often implemented in \texttt{RooFit}~\cite{root} and stored in \texttt{RooWorkspaces}, which offered a structured way to encode complex likelihoods within the \texttt{ROOT} framework. However, these formats were tightly coupled to C++/\texttt{ROOT} environments, limiting portability and long-term preservation.
The introduction of the \pyhf library~\cite{Heinrich:2021gyp,pyhf}, a pure Python implementation of the \histfactory framework~\cite{histfactory}, has enabled the preservation and reuse of full statistical models in a platform-independent, JSON-based format.

To support long-term preservation and interoperability of statistical models, the \ac{HS3}~\cite{HS3github} has been proposed. 
\ac{HS3} defines a standard JSON-based schema for encoding statistical models, ensuring compatibility with existing tools such as \pyhf~\cite{Heinrich:2021gyp,pyhf}, \texttt{RooFit}~\cite{root}, and \texttt{zfit}~\cite{Eschle2020}. 
It also supports the inclusion of metadata and documentation that describes the analysis context. The specification is maintained in a public repository by the community~\cite{HS3github}, promoting transparency and ease of access.

\section{Types of reinterpretation}
\label{sec:reinterpretation-types}
One can classify reinterpretation efforts as follows~\cite{Bailey:2022tdz,Gartner:2024muk}\footnote{This section is based on the review article~\cite{Bailey:2022tdz}, reformulated in Reference~\cite{Gartner:2024muk}.}:
\begin{itemize}
    \item \textit{Kinematic reinterpretation} or \textit{recasting}, which includes testing an alternative physics process with different kinematic distributions. This typically requires propagating the new signal hypothesis through the detector simulation or applying response matrices to account for changes in efficiencies and acceptance regions. It is the most demanding form of reinterpretation and often relies on external tools.
    \item \textit{Model updating}, which refines either theoretical predictions or experimental inputs. The main goal is an overall reduction of uncertainties. Technically, this can be viewed as a subclass of kinematic reinterpretation where the analysis logic remains unchanged.
    \item \textit{Combinations} of datasets and measurements across experiments or channels. This is crucial for reducing uncertainties on shared parameters or for deriving global parameter constraints, where different channels may have different sensitivities to some parameters. For such combinations, it is necessary that the underlying model assumptions are mutually consistent.
\end{itemize}

\section{Public tools for reinterpretation}
\label{sec:reinterpretation-tools}

Several public frameworks have been developed to facilitate the reinterpretation of \ac{HEP} data. A representative overview is provided in Reference~\cite{LHCReinterpretationForum:2020xtr}.

\begin{description}
    \item[\textbf{RECAST}] (full\footnote{See \cref{sec:analysis-preservation}.})
    Allows theorists to submit signal models, which are recast by experimental workflows within experimental collaborations, using full simulation and statistical frameworks. Provides high accuracy at the cost of internal computational resources~\cite{Cranmer:2010hk}.
    
    \item[\textbf{CheckMATE}] (full)
    Performs full \ac{MC} simulations including detector effects, and applies implemented experimental analyses with their original cuts. It yields ratios of predicted to excluded cross-sections~\cite{DREES2015227, DERCKS2017383}.
    
    \item[\textbf{MadAnalysis 5}] (full)
    Allows both detector-level recasting and generator-level analyses with user-defined cuts or integrated public analyses. Input includes \ac{MC} events and analysis code. Computational cost varies from moderate to high depending on simulation depth. Inference outputs include $CL_S$ exclusion limits~\cite{Conte:2012fm, doi:10.1142/S0217751X18300272}.
    
    \item[\textbf{GAMBIT / ColliderBit}] (full/lightweight)
    Embedded within the global-fitting framework GAMBIT, ColliderBit handles event generation and fast detector simulation, integrating the results into global fits. Provides simplified likelihoods for calculating exclusion limits~\cite{KVELLESTAD2020103769}.

    \item[\textbf{Rivet}] (lightweight)  
    Designed for validating \ac{MC} generators. Applies truth-level particle-level analyses to generator-level events, relying on published measurements or unfolded distributions. It provides truth and detector level distributions~\cite{rivet}.
    
    \item[\textbf{Contur}] (lightweight)  
    Builds on Rivet and utilizes unfolded (particle-level) differential cross-sections to derive exclusion contours in \ac{BSM} model space, using existing measurements~\cite{Butterworth_2017}.
    
    \item[\textbf{pyhf}] (full)
    Performs reinterpretation via JSON-serialized full statistical models (\histfactory), allowing complete likelihood recalculations without \ac{MC} simulation. Requires a \pyhf JSON model, signal templates, and background data. Computational cost is low to moderate, depending on the complexity of likelihood evaluations. Requires full simulation and analysis for additional signal models~\cite{Heinrich:2021gyp,histfactory}.
    
    \item[\textbf{mapyde}] (full/lightweight)
    Strings together event generation, fast detector simulation, analysis selection, and likelihood interpretation in an automated pipeline. Input includes model parameters, detector acceptances and selection efficiencies, and \pyhf models. It requires high computational cost but offers end-to-end reproducibility~\cite{Stark:2023ont,10.21468/SciPostPhysCodeb.27-r0.5}.

    \item[\textbf{SModelS}] (lightweight)
    Decomposes \ac{BSM} models into simplified model spectra~\cite{Chatrchyan_2013} and confronts them with a database of experimental limits from \ac{LHC} searches, using efficiency maps and published cross-section constraints. Provides a simplified likelihood for inference. Operates with low computational cost as no \ac{MC} generation is involved, and is ideal for rapid screening of large model spaces~\cite{Kraml_2014, AMBROGI2020106848}.
\end{description}

Each tool occupies a distinct niche in the reinterpretation landscape. Lightweight methods excel at speed and broad survey capability, while full-stack or hybrid tools offer fidelity and statistical rigour. Tools such as \pyhf bridge between these paradigms by facilitating direct likelihood reinterpretation without simulating events.

The reinterpretation framework presented in this work occupies a unique position in this landscape, combining the computational efficiency of lightweight approaches with the statistical rigour of full likelihood-based methods. Unlike purely generator-level tools such as \texttt{Rivet} and \texttt{Contur}, this method incorporates detector-level effects through parametrized response functions, bridging the gap between truth-level and reconstruction-level observables without requiring full detector simulation. This approach maintains compatibility with \pyhf statistical models while enabling rapid exploration of high-dimensional parameter spaces --- a capability that distinguishes it from resource-intensive full-stack tools like \texttt{CheckMATE} or \texttt{RECAST}.

\section{Challenges and future goals}
\label{sec:reinterpretation-challenges}
Despite major progress in reinterpretation infrastructure, significant challenges remain. A key difficulty is the diversity of analysis implementations, statistical frameworks, and data formats. Even within a single experiment, variations in software environments and detector simulation chains make consistent preservation and reuse difficult~\cite{Bailey:2022tdz}.

Another challenge is the resource-intensive nature of full reinterpretations. Detector-level reinterpretation requires access to full simulation and reconstruction chains, which is often impractical due to computing demands. This motivates the development of lightweight reinterpretation approaches, such as fast simulation or parameterized efficiencies.
However, such methods may sacrifice precision or require validation against results based on fully-preserved analyses.

Machine-learning-based analyses introduce new complexities for reinterpretation. Sharing a trained model requires not only the architecture and weights, but also a detailed specification of input features, preprocessing, and software dependencies~\cite{Bailey:2022tdz}. Standardization and tooling for machine learning preservation are still in early stages.

Future goals include~\cite{Bailey:2022tdz}:
\begin{itemize}
    \item Establishing community-wide standards for analysis preservation and statistical model sharing.
    \item Facilitating reinterpretation-aware design of new analyses, including public documentation of selection logic and systematics.
    \item Creating a centralized, searchable portal of recastable analyses with accompanying statistical models and metadata.
    \item Enabling reinterpretation-driven model testing in global fits to identify gaps in experimental coverage.
\end{itemize}

%% file: chapters/method.tex
\footnote{
The material presented in this chapter has been published in the European Physical Journal C
~\cite{Gartner:2024muk}. While the content largely follows that of the referenced paper, editorial modifications have been made, including restructuring of the presentation, rephrasing of portions of the text, and expansion of selected explanations. As this project is the result of collaborative work, the specific contributions of each co-author are documented in \cref{app:method-contributions}.
}
Building upon the review of existing reinterpretation approaches, this chapter presents a novel method that addresses many of the limitations identified in current reinterpretation strategies. The approach bridges the gap between computationally expensive full reanalysis and potentially biased simplified methods by introducing a technique based on so-called \textit{model-agnostic likelihoods}.

This reinterpretation method enables direct inference on theoretical parameters while correctly accounting for changes in the kinematic distributions. The method offers several key advantages: inference is typically very fast; it requires only minimal additional information beyond the likelihood itself; it is general in terms of applicability across different models and analyses; and the resulting likelihoods are easily distributable, making them a powerful tool for enhancing the long-term impact and reusability of \ac{HEP} measurements.

The development of this method represents a significant step forward in making experimental results more accessible to the broader theoretical physics community, thereby increasing the scientific return on the substantial investments in experimental particle physics.

This chapter is structured as follows: I first introduce the method in \cref{sec:method paper reweighting method}, followed by a discussion of its implementation in \cref{sec:method paper implementation}. In \cref{sec:method paper examples}, I present applications of the method to two toy examples from flavour physics, the \BKnunu and $B \to K^* \nu \bar \nu$ decays. Finally, I discuss the significance and implications of the method in \cref{sec:method paper discussion}.

\section{Reinterpretation through reweighting}
\label{sec:method paper reweighting method}
The reinterpretation method described here is based on updating the distributions of the observable variables, given changes in the underlying kinematic distribution.

The \ac{PDF} of reconstructed events $p(x)$ results from folding the \ac{PDF} of a theoretical kinematic prediction $p(z)$ with the conditional distribution $p(x|z)$ and the indicator function $\identity_\varepsilon(x)$,
\begin{equation}
    p(x) = \frac{1}{\varepsilon}\int dz ~ \identity_\varepsilon(x) ~ p(x|z) ~ p(z).
\end{equation}
Here, the reconstruction variable $x$ represents one or multiple observable variables, and the kinematic variable $z$ represents one or multiple kinematic \ac{d.o.f.}.
The function $\identity_\varepsilon(x)$ models the selection criteria for a reconstructed event, and $p(x|z)$ is the conditional probability of measuring a reconstructed configuration $x$ given an underlying particle configuration $z$. 
The overall reconstruction efficiency $\varepsilon$ acts as a normalization factor for the \ac{PDF} $p(x)$. 
The \ac{PDF} $p(z)$ corresponds to the normalized kinematic distribution of a theoretical prediction $\sigma(z)$,
\begin{equation}
    p(z) = \frac{\sigma(z)}{\sigma}.
\end{equation}

The number density of expected events, given a total integrated luminosity $L$, is ${\nu(x) = L ~ \sigma ~\varepsilon ~p(x)}$ and reads
\begin{equation}
    \nu(x) = L \int dz ~ \varepsilon(x|z) ~ \sigma(z) = \int dz ~ \nu(x,z),
\end{equation}
where both reconstruction and selection are combined into $\varepsilon(x|z) = \identity_\varepsilon(x) ~ p(x|z)$, and $\nu(x,z) = L ~ \varepsilon(x|z) ~ \sigma(z)$ can be thought of as a \textit{joint number density}, analogous to the joint \ac{PDF} $p(x,z) = p(x|z) ~ p(z)$. This represents a \textit{folding procedure} in which the theoretical prediction is convolved with $\varepsilon(x|z)$ (and scaled by $L$) to obtain the expected number of events~\cite{Cowan1998}. In this way, the kinematic distribution can be mapped to the reconstructed space, taking both reconstruction and selection efficiencies into account.

The reinterpretation task involves determining the number density $\nu_1(x)$ of an \textit{alternative} theoretical prediction $\sigma_1(z)$. This can be obtained by reweighting the joint number density $\nu_0(x,z)$ according to the kinematic \textit{null} distribution $\sigma_0(z)$ via
\begin{equation}
    \begin{aligned}
    \nu_1(x)  &= L \int dz ~ \varepsilon(x|z) ~ \sigma_1(z) \\
            &= L \int dz ~ \varepsilon(x|z) ~ \sigma_0(z) ~ \frac{\sigma_1(z)}{\sigma_0(z)}\\
            &= \int dz ~ \nu_0(x,z) ~ w(z).
    \end{aligned}
    \label{eq:reweight}
\end{equation}

The weight factor $w(z)$ is simply the ratio of the theoretically predicted alternative kinematic distribution to the null distribution.

This reweighting process solely requires knowledge of the joint null number density $\nu_0(x,z)$. Together with the weight factor, this is sufficient to predict the number density according to an alternative theory.

\subsection{Discrete reweighting}
\label{sec:method paper reweighting-method-discrete}
In practical applications, the continuous joint number density is typically not analytically obtainable and requires estimation through \ac{MC} simulations.
To address this, one can discretize the reweighting method by representing the joint number density as a multidimensional matrix $\nu_{xz}$ in bins of $x \times z$. This is done alongside the binning of the theoretically predicted distribution $\sigma_{z}$ and weight factor $w_{z}$ in the kinematic \ac{d.o.f.} $z$. Consequently, the discrete joint number density has dimension $\dim(x) \cdot \dim(z)$. 

The discrete joint number density can be obtained by integrating the continuous joint number density over each $x \times z$ bin,
\begin{equation}
    \nu_{xz} = \int_{\text{bin } x} \int_{\text{bin }  z} dx' ~ dz' ~ \nu(x',z'),
\end{equation}
where the integral boundaries are the bin boundaries of each $x$ and $z$ bin, respectively. The binned weights are given by 
\begin{equation}
    w_z = \frac{\sigma_{1,z}}{\sigma_{0,z}} = \frac{\int_{\text{bin }  z} dz' ~ \sigma_1(z')}{\int_{\text{bin }  z} dz' ~ \sigma_0(z')} ~ .
    \label{eq:binned-weights}
\end{equation}

The reweighting step of \cref{eq:reweight} becomes
\begin{equation}
    \nu_{1,x} = \sum_{z} ~ \nu_{0,xz} ~ w_{z}.
    \label{eq:reweight_discrete}
\end{equation}

This discrete approach offers an advantage due to its lower computational cost compared to event-by-event reweighting.
The cost of this simplification is a loss of accuracy due to the binning in the kinematic \ac{d.o.f.} $z$. However, this loss is controllable by increasing the number of bins. 
The binning should be chosen such that the joint number density and weight function factorize approximately in each bin.
In principle, an arbitrarily fine binning can be chosen such that the uncertainty due to this loss is negligible compared to other sources of uncertainty (provided enough \ac{MC} samples are available; see \cref{app:kinematic-binning}).

Crucially, only a fixed set of samples from the joint \ac{PDF} $p_0(x,z)$ based on the null prediction is required, and no new samples need to be produced for the reinterpretation.
Therefore, publishing the (binned) joint null number density and knowledge of the underlying kinematic null distribution is sufficient to perform the reinterpretation of a given measurement.

\subsection{Limitation}
\label{sec:method paper reweighting-method-limitations}
The proposed reweighting approach is a lightweight and accurate way of obtaining new signal templates given a joint number density and the kinematic null distribution. Still, it does have one main limitation:
if the phase space contains regions with $w(z) \gg 1$, the effective \ac{MC} sample size decreases.
Put differently, large weights are assigned to regions that are sparsely populated with \ac{MC} samples obtained from the null distribution.
Furthermore, if the null distribution lacks support in a region of phase space, $\sigma_0(z) \to 0$, it can happen that $w(z) \to \infty$ when reweighting to an alternative distribution. In this case, no \ac{MC} samples exist in the given region. The only solution is to re-analyse new samples produced according to the alternative distribution.

\section{Implementation of the reweighting method}
\label{sec:method paper implementation}
Using the reweighting method, one can construct likelihood functions for particle physics analyses directly parametrized in terms of theory parameters. Although the reweighting method is independent of any likelihood formalism, it is showcased here in terms of the \histfactory formalism~\cite{histfactory} as a baseline statistical model. For a review of the \histfactory formalism, I refer to \cref{sec:histfactory}. In terms of implementation, I use the \pyhf~\cite{pyhf,Heinrich:2021gyp} package, the Python implementation of the \histfactory model, introduced in \cref{sec:pyhf}.

To obtain the data likelihood $p_\text{data}$ (see \cref{eq:data-pdf}) for any theoretical model, one needs to calculate the event rates of the corresponding signal template. This requires an implementation of \cref{eq:reweight_discrete}, and hence an extension of the \pyhf codebase.
Since the method prescribes changes to event rates, these can be applied via a multiplicative modifier. Therefore, \pyhf is extended with a \textit{custom modifier} (see \cref{sec:histfactory-custom-modifiers}) that computes the modifications to the nominal event rates according to the procedure in \cref{eq:reweight_discrete}. This modifier is expressed as a function of the underlying theory parameters of the alternative kinematic distribution.

The constraint likelihood $p_{constr.}$ (see \cref{eq:constraint-pdf}) contains all constraint terms of these underlying theory parameters, which can be correlated in general. By definition, modifier parameters in \pyhf are treated as uncorrelated. To correctly include correlated parameter constraints in the statistical model, the theory parameters are decorrelated using matrix eigendecomposition (discussed in \cref{sec:histfactory-correlations}). One normally-constrained \pyhf modifier parameter is then assigned to each of these decorrelated parameters.

\section{Application of the reweighting method to toy examples}
\label{sec:method paper examples}
A central aim of this work is to motivate the experimental \ac{HEP} community to make use of the proposed method.
This will, in turn, enable subsequent model-agnostic phenomenological analyses of experimental results.
Here, I detail the full reinterpretation of two toy examples from low-energy flavour physics.


In these toy examples, no experimental data are used, but rather two datasets of simulated samples, where one dataset acts as \textit{real} data. The \textit{\ac{MC} data} is produced according to the \ac{SM} prediction. The \textit{real data} is produced by assuming that \ac{BSM} physics affects the example process. 
To simulate detector and other analysis selection effects,
observables in both datasets are smeared according to estimated detector resolutions, and event yields are scaled by an efficiency map.

By comparing the produced datasets using either Bayesian or frequentist methods, the aim is to recover the chosen \ac{BSM} parameters starting from the \ac{SM}. This is possible only because the likelihood is made a function of the theory parameters, allowing shape changes in the kinematic distribution due to \ac{BSM} physics to be directly taken into account.

As a general result, the goal is to compute the posterior distribution in the space of theory parameters, given the two simulated datasets and prior parameter constraints. To obtain a posterior distribution from a \histfactory likelihood, I use the Bayesian \histfactory formalism, which I discuss in \cref{sec:histfactory-bayesian}, and is implemented in \texttt{bayesian pyhf}~\cite{Feickert:2023hhr}.

\subsection{\texorpdfstring{\BKnunu}{B->Knunubar}}
\label{sec:method paper knunu}
The recent measurements of the total rate of \BKnunu decays by the Belle~II collaboration~\cite{Belle-II:2021rof,Belle-II:2023esi} hint at an excess of signal events compared to the \ac{SM} expectation. This has triggered substantial interest in the \ac{HEP} phenomenology community to interpret this excess as a sign of \ac{BSM} physics and to extract the corresponding model parameters~\cite{Fridell:2023ssf,PhysRevD.109.075008, Gabrielli:2024wys}.
In this subsection, I study the performance of the proposed approach using simulated \BKnunu data.

\subsubsection{Weak Effective Theory parametrization}
\label{sec:method paper knunu-theory}
To capture the effects of \ac{BSM} physics in \BKnunu decays, I work within the \ac{WET}, which is a low-energy effective field theory that describes both \ac{SM} and potential \ac{BSM} effects using a common set of parameters. Details about the \ac{WET} are given in \cref{sec:wet}.
Here, it is assumed that potential new \ac{BSM} particles and force carriers have masses at or above the scale of electroweak symmetry breaking.

The $sb\nu\nu$ \ac{WET} Lagrangian is constructed from the five mass-dimension 6 operators given in \cref{eq:wet-lagrangian,eq:operators}.
Presently, the only measured observable is the differential decay rate for $B\to K \nu \bar{\nu}$, discussed in \cref{sec:knunu-theory} and simulated in this example. The decay rate in terms of the Wilson coefficients of the operators is given in \cref{eq:width}. From this prediction, it is clear that this decay is sensitive only to linear combinations of vector, scalar, and tensor Wilson coefficients: ${|C_{\mathrm{VL}}+C_{\mathrm{VR}}|, ~ |C_{\mathrm{SL}}+C_{\mathrm{SR}}|, ~ |C_{\mathrm{TL}}|}$ (see \cref{eq:BToKnunu-wc-sensitivity}).
The kinematic shape of the differential branching ratio, due to the vector, scalar and tensor interactions, is illustrated in \cref{fig:knunu-theory}.

The hadronic matrix elements of the contributing operators are described by the three independent hadronic form factors $f_{+}(q^2)$, $f_{0}(q^2)$ and $f_{T}(q^2)$, as discussed in \cref{sec:hadronic-matrix-elements}.
In this work, the form factors are parametrized following the BSZ parametrization~\cite{Bharucha:2015bzk} (see \cref{sec:hadronic-matrix-elements}), which is truncated at order $K=2$.
The values and uncertainties for the corresponding 8 hadronic parameters are extracted from a joint theoretical prior \ac{PDF} comprised of the 2021 lattice world average based on results by the Fermilab/MILC and HPQCD collaborations~\cite{FlavourLatticeAveragingGroupFLAG:2021npn} and recent results by the HPQCD collaboration~\cite{Parrott:2022rgu}; see \cref{tab:hadronic parameters full}.
Correlations between the hadronic parameters are taken into account through their respective covariance matrices and implemented as discussed in \cref{sec:method paper implementation}.

For the \ac{SM} prediction, only the vector left-handed Wilson coefficient $C_{\mathrm{VL}}$ is non-zero (see \cref{eq:WET-SM-point}), leading to a predicted branching ratio of~\cite{EOSAuthors:2021xpv,EOS:v1.0.16}
\begin{equation}
    \mathcal{B}(B^+ \to K^+ \nu \bar{\nu})_{\mathrm{SM}} = (4.8 \pm 0.2) \cdot 10^{-6}.
    \label{eq:knunu-sm-br}
\end{equation}
The theoretical uncertainty is dominated by the form factor uncertainties from lattice \ac{QCD}, with smaller contributions from \ac{CKM} matrix elements and higher-order \ac{QCD} corrections.

The Belle~II experiment, which found the first evidence for this decay, observes more events than expected in the \ac{SM}.
The ratio of observed to expected events is ${4.6 \pm 1.3}$~\cite{Belle-II:2023esi} and the measured branching ratio is
\begin{equation}
    \mathcal{B}(B^+ \to K^+ \nu \bar{\nu})_{\mathrm{exp}} = (2.3 \pm 0.7) \cdot 10^{-5}.
\end{equation}
For later use, I define a benchmark point in the space of Wilson coefficients that roughly reproduces the measured branching ratio after correcting for the efficiency.
Assuming all Wilson coefficients to be real, it reads
\begin{equation}
    \label{eq:WET-BSM-benchmark}
    \begin{aligned}
        |C_{\mathrm{VL}}| & = 10\,, \qquad &|C_{\mathrm{VR}}| =  4\,, \\
        |C_{\mathrm{SL}}| & =  3\,, \qquad &|C_{\mathrm{SR}}| =  1\,, \\
        |C_{\mathrm{TL}}| & =  1\,.
    \end{aligned}
\end{equation}
As shown in \cref{eq:width}, the decay rate of \BKnunu is only sensitive to three linear
combinations of Wilson coefficients shown in \cref{eq:BToKnunu-wc-sensitivity}.
The projection of this benchmark point onto this subspace reads
\begin{equation}
    \label{eq:WET-BSM-benchmark-Knunu}
    \begin{aligned}
        |C_{\mathrm{VL}}+C_{\mathrm{VR}}| & = 14\,, \\
        |C_{\mathrm{SL}}+C_{\mathrm{SR}}| & =  4\,, \\
        |C_{\mathrm{TL}}|        & =  1\,.
    \end{aligned}
\end{equation}

\subsubsection{Datasets}
\label{sec:method paper knunu-data}

\begin{table}[t]
    \renewcommand{\arraystretch}{1.5}
    \caption{The number of \BKnunu samples produced for this study, corresponding to an equivalent of $362~\text{fb}^{-1}$ and $50~\text{ab}^{-1}$ integrated luminosity at the SuperKEKB collider. \textit{Generated} and \textit{reconstructed} samples correspond to the numbers prior and post efficiency correction.}
    \centering
    \begin{tabularx}{\linewidth}{@{\extracolsep{\fill}}YYYY}
        \toprule \midrule
    Luminosity & $B \overline{\kern -0.18em B} $ events & \makecell{\ac{MC} generated \textbackslash \\ reconstructed} &\makecell{Data generated \textbackslash \\ reconstructed} \\
    \midrule
    $362~\text{fb}^{-1}$ & $\sim 3.87 \cdot 10^8$ & \makecell{$1.86\cdot 10^3$ \\ $2.41\cdot 10^2$} & \makecell{ $1.05\cdot 10^4$ \\ $1.14\cdot 10^3$} \\ 
    $50~\text{ab}^{-1}$   & $\sim 5.35 \cdot 10^{10}$ & \makecell{$2.57\cdot 10^5$ \\ $3.21\cdot 10^4$} & \makecell{$1.45 \cdot 10^6$ \\ $1.58 \cdot 10^5$}\\ 
    \midrule \bottomrule
    \end{tabularx}
    \label{tab:samples}
\end{table}


To make this example as realistic as possible, the setup is designed similarly to what has been done in the Belle~II analysis.

The \ac{MC} data are produced according to the \ac{SM} prediction (\textit{null} hypothesis) of the differential branching ratio $d\mathcal{B}(B \to K \nu \bar \nu)/dq^2$, where the Wilson coefficients are set to the values in \cref{eq:WET-SM-point}.
The number of samples produced is equivalent to the number of signal events seen or expected for a given integrated luminosity.
Samples are produced for $362~\text{fb}^{-1}$ integrated luminosity, which corresponds to the total collider data used in Reference~\cite{Belle-II:2023esi}, and for $50~\text{ab}^{-1}$ integrated luminosity corresponding to the total target luminosity of the Belle~II experiment~\cite{Belle-II:2018jsg}. 
The estimated number of $B \overline{\kern -0.18em B}$ events is multiplied by the \ac{SM} branching ratio of \cref{eq:knunu-sm-br}
to obtain a rough estimate for the number of \ac{MC} events to produce, prior to the efficiency correction (see below).

The real data are produced according to a \ac{BSM} prediction (\textit{alternative} hypothesis).
Following the observations of more events than predicted in the \ac{SM}~\cite{Belle-II:2023esi},
the previously defined benchmark point in \cref{eq:WET-BSM-benchmark} is used. The estimated number of $B \overline{\kern -0.18em B}$ events is multiplied by the \ac{BSM} branching ratio, $\mathcal{B}(B \to K \nu \bar \nu) \approx 2.71 \cdot 10^{-5}$~\cite{EOSAuthors:2021xpv,EOS:v1.0.11}, to obtain a rough estimate for the number of data events to produce, prior to the efficiency correction (see below).

The number of produced events is listed in \cref{tab:samples}.
Unless stated otherwise, all numerical values, figures, and tables provided in the following are obtained
from studies that assume the $50~\text{ab}^{-1}$ sample size.
Samples of the decay's probability distribution for both the null and the alternative hypothesis are produced
using the \EOS software in version 1.0.11~\cite{EOS:v1.0.11} (see \cref{sec:eos}).

To simulate the detector resolution, the $q^2$ value of each sample is shifted by a value drawn from a normal distribution of width $1~\text{GeV}^2$. This roughly corresponds to the Belle~II detector resolution. 

Furthermore, an efficiency map is applied to the samples according to the function 
\begin{equation}
    \varepsilon(q^2) = 0.4 ~ \exp\left(- 5 ~ q^2/M_B^2\right),
    \label{eq:eff-knunu}
\end{equation}
which mimics the efficiency obtained in Reference~\cite{Belle-II:2023esi}.

\begin{figure}[ht]
    \centering
    \includegraphics[width=0.8\linewidth]{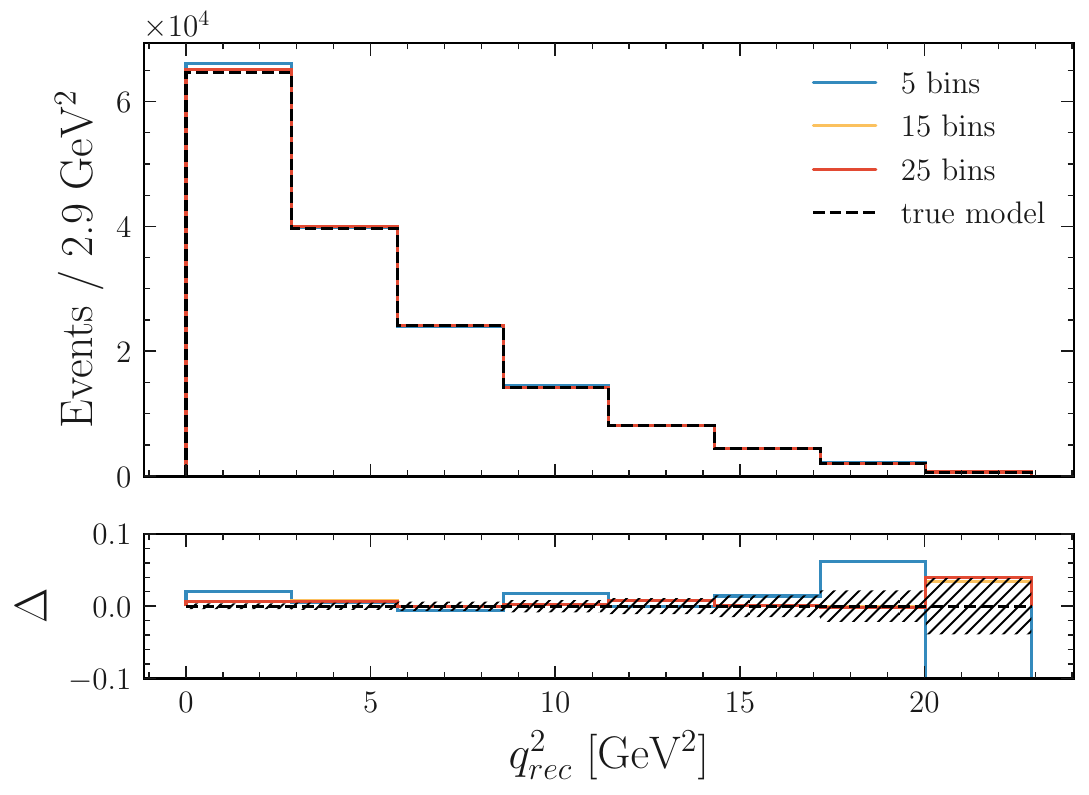}
    \caption{The null histogram yields, reweighted to the benchmark point in \cref{eq:WET-BSM-benchmark}, as a function of kinematic bins (red). The histogram yields of the true model, sampled from the probability density function at the benchmark point (black). The bottom plot shows the relative difference $\Delta$ of the reweighted models to the true model. The statistical uncertainty of the true model is shown as the hatched region.}
    \label{fig:knunu-binning}
\end{figure}

The reconstruction variable is chosen to be the reconstructed momentum transfer $q_{\rm rec}^2$, obtained from the kinematic variable $q^2$ by applying detector and efficiency corrections. 
The binnings of the kinematic and reconstruction variables differ:
\begin{itemize}
\item For the reconstruction variable, a balance needs to be struck between the number of events in each bin (to ensure sufficient statistical power) and the number of bins (to ensure sensitivity to differences in the shape of the $q^2$ distribution).
Eight equally-spaced bins are chosen for the reconstruction variable.
\item For the kinematic variable, the number of bins is determined as follows.
The convergence of the expected yields from the reweighted model is studied as the number of kinematic bins increases.
For this study, the detector resolution smearing is removed.
This is done for 100 randomly chosen theoretical models (see \cref{app:kinematic-binning} for further details).
These models correspond to normally-distributed variations $\sim \mathcal{N}(0, 10)$ of the \ac{WET} parameters with respect to the \ac{SM} parameter point. A width of $\sigma=10$ is chosen to ensure that the models are sufficiently different from the \ac{SM} point while still being within a reasonable range of the \ac{WET} parameters.
The aim is to ensure convergence at the level of $1\%$ accuracy. It is found that using 24 equally-spaced bins for the kinematic variable ensures this aim.
\Cref{fig:knunu-binning} illustrates this procedure for the benchmark point in \cref{eq:WET-BSM-benchmark}, where the changes for 5, 15 and 25 bins in the kinematic variable are shown.
\end{itemize}

\begin{figure}[ht]
    \centering
    \includegraphics[width=0.8\linewidth]{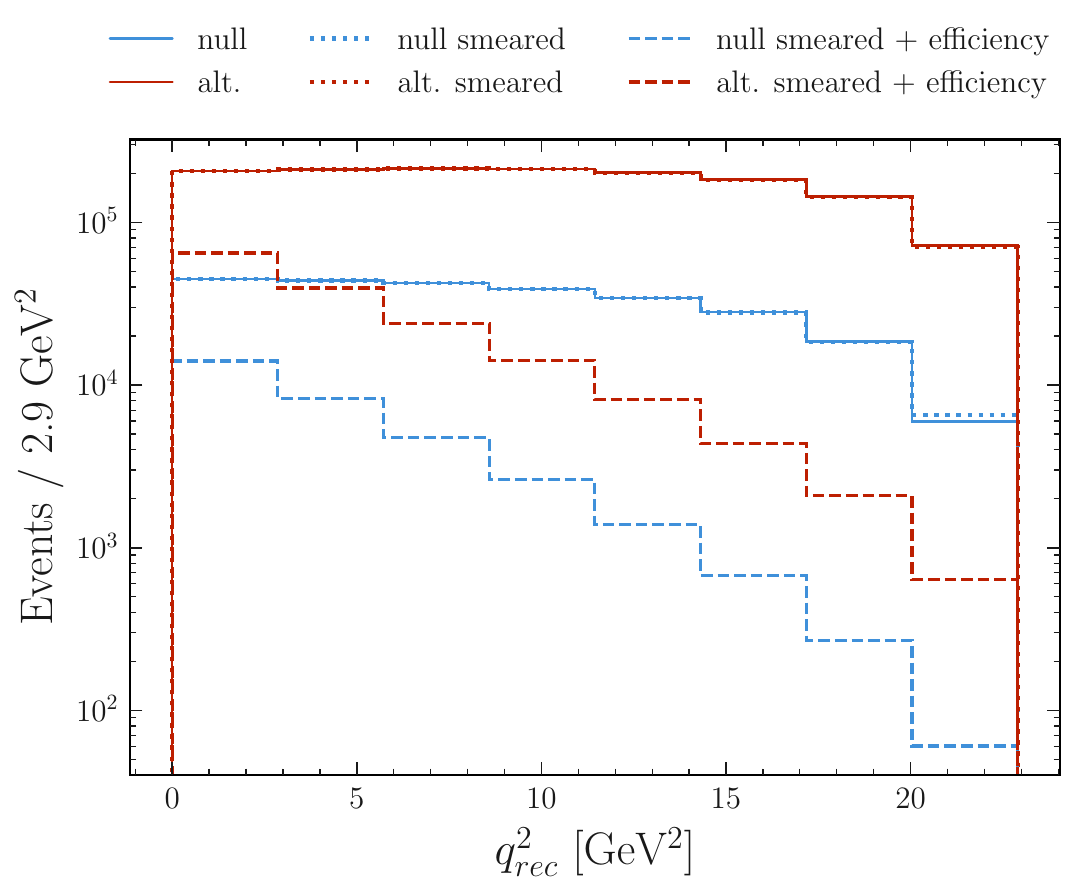}
    \caption{Both the null/\ac{SM} (blue) and alternative/\ac{BSM} (red) \BKnunu datasets, according to the pure theoretical prediction, after detector resolution smearing and efficiency correction.}
    \label{fig:knunu-data-sets}
\end{figure}

Both datasets, according to the null (\ac{SM}) and alternative (\ac{BSM}) hypotheses, and their corresponding changes after detector resolution smearing and efficiency correction, are shown in \cref{fig:knunu-data-sets}.

\begin{figure}[ht]
    \centering
    \includegraphics[width=0.8\linewidth]{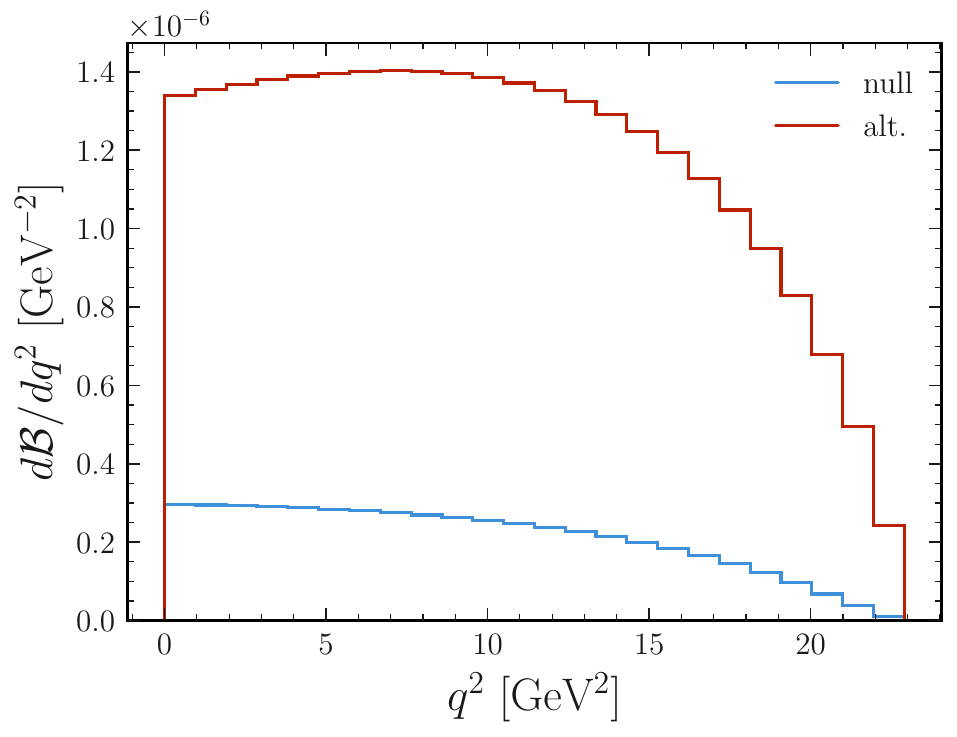}
    \caption{The bin-integrated null/\ac{SM} (blue) and alternative/\ac{BSM} (red) predictions for the differential branching ratio $d\mathcal{B}(B \to K \nu \bar \nu)/dq^2$.}
    \label{fig:knunu-dist}
\end{figure}

\begin{figure}[ht]
    \centering
    \includegraphics[width=0.8\linewidth]{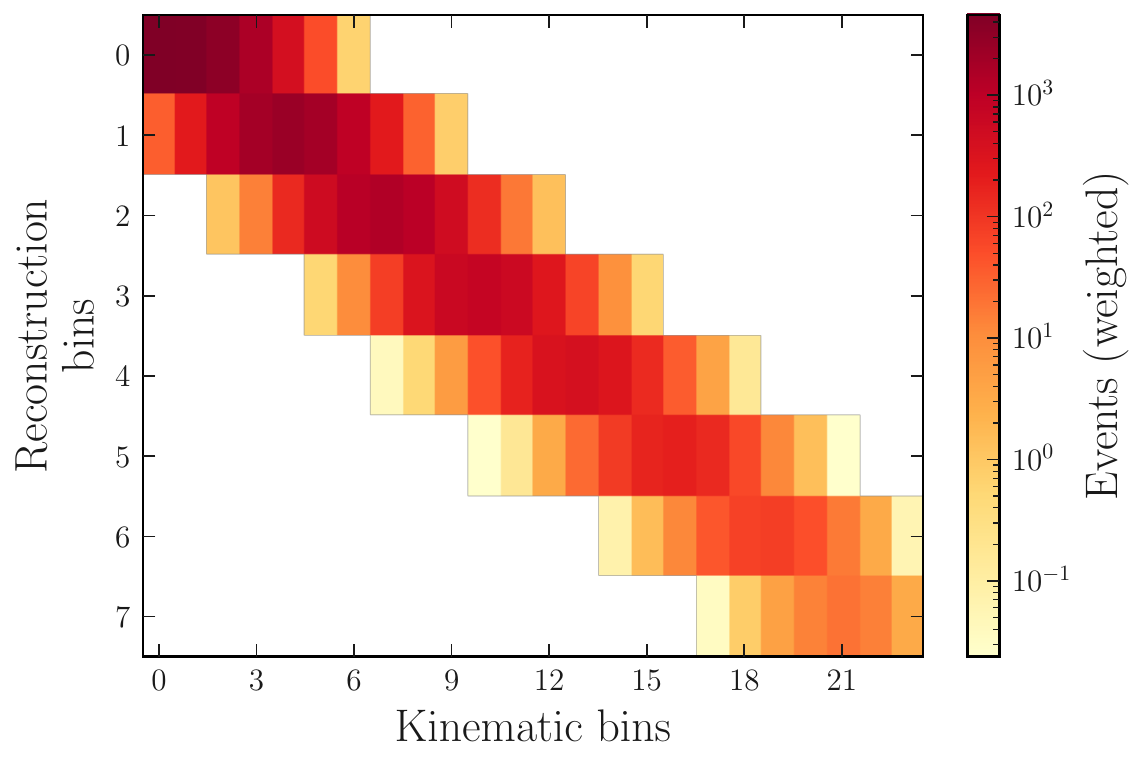}
    \caption{The null joint number density, showing the 8 bins of the reconstruction variable on the vertical axis and the 24 bins of the kinematic variable on the horizontal axis.}
    \label{fig:knunu-map}
\end{figure}

The null (\ac{SM}) and alternative (\ac{BSM}) predictions have also been calculated using the \EOS software~\cite{EOSAuthors:2021xpv,EOS:v1.0.11} and are shown in \cref{fig:knunu-dist} .
The null joint number density is obtained by binning the \ac{MC} data in a 2-dimensional histogram of the reconstruction variable $q_{\rm rec}^2$ against the kinematic variable $q^2$, as shown in \cref{fig:knunu-map}.

\subsubsection{Full statistical model}
\label{sec:method paper stat-model-K}
To build the posterior according to the statistical model described in \cref{sec:method paper implementation}, all parameters of the likelihood and their corresponding priors are collected.

The theoretical parameters include the \ac{WET} parameters and the hadronic parameters.
The \ac{WET} parameters correspond to the three independent linear combinations of Wilson coefficients
that enter the theoretical description of $B\to K \nu \bar \nu$ decays; see \cref{eq:width}.
These are $|C_{\mathrm{VL}} + C_{\mathrm{VR}}|$, $|C_{\mathrm{SL}} + C_{\mathrm{SR}}|$, and $|C_{\mathrm{TL}}|$.
While the Wilson coefficients are --- in general --- complex-valued parameters,
the overall phase of the \ac{WET} Lagrangian \cref{eq:wet-lagrangian} is not observable.
Moreover, an inspection of the differential decay rate in \cref{eq:width} shows only sensitivity to the absolute values of the three linear combinations of Wilson coefficients.
Hence, each linear combination is represented as a positive real-valued number. 
Their prior is chosen as the uncorrelated product of uniform distributions with support 
\begin{equation}
    \begin{aligned}
        & 5 < |C_{\mathrm{VL}}+C_{\mathrm{VR}}| < 20,\\
        & 0 < |C_{\mathrm{SL}}+C_{\mathrm{SR}}| < 15,\\
        & 0 < |C_{\mathrm{TL}}| < 15.
    \end{aligned}
\end{equation}
These ranges are motivated with the benchmark point, \cref{eq:WET-BSM-benchmark-Knunu}, in mind and wide enough to ensure that the posterior can explore the parameter space sufficiently.
The prior of the hadronic parameters, discussed in \cref{sec:method paper knunu-theory}, is a multivariate normal distribution, which is implemented as a sequence of independent univariate normal distributions, as discussed at the end of \cref{sec:method paper implementation}.

The experimental constraint includes one parameter per bin of the reconstruction variable, representing the statistical uncertainty of the \ac{MC} yields.
The priors for these parameters are normal distributions $\mathcal{N}(1,1/\sqrt{N_b})$, where $N_b$ is the total yield in reconstruction bin $b$.
For the purpose of this proof-of-concept study, further (systematic) sources of uncertainty are not accounted for.

\subsubsection{Reinterpretation results}
\label{sec:method paper knunu-results}
\begin{figure}
    \centering
    \includegraphics[width=\textwidth]{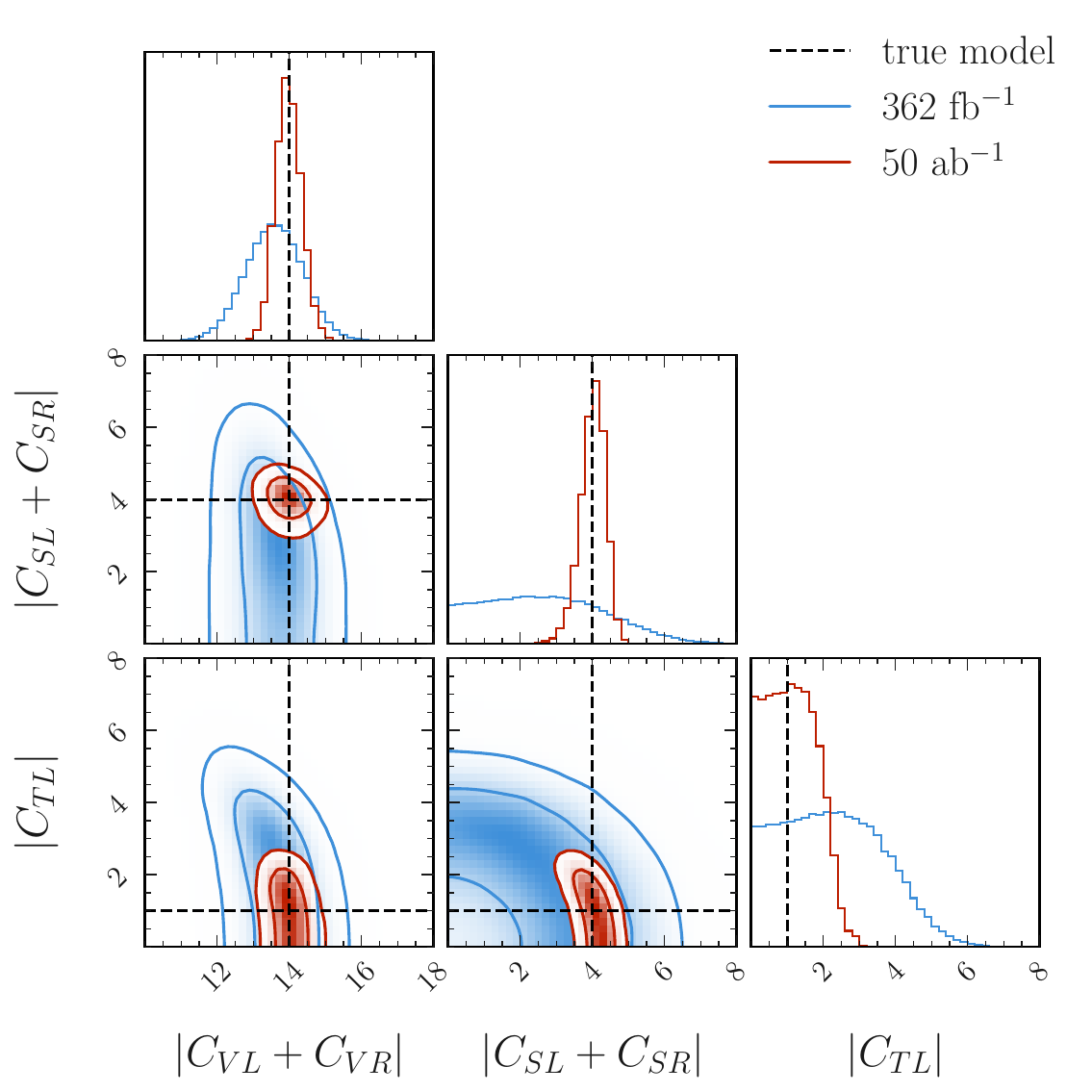}
    \caption{The marginalized posterior distributions, obtained by \ac{MCMC} sampling from the \BKnunu likelihood. On the diagonal, the 1-dimensional marginal distributions of the Wilson coefficients in \cref{eq:BToKnunu-wc-sensitivity} are shown. The contours on the 2-dimensional plots correspond to $68\%$ (inner) and $95\%$ (outer) probability. The dashed lines indicate the true underlying model (\cref{eq:WET-BSM-benchmark-Knunu}).}
    \label{fig:knunu-posterior}
\end{figure}

Having built a model-agnostic likelihood function from the toy data, I investigate the potential of the approach to constrain the Wilson coefficients.
Using \ac{MCMC} sampling, the 3-dimensional marginal posterior distribution of the Wilson coefficients is obtained.
The posterior is sampled using 8 independent chains\footnote{
    Chains are initialized randomly in the parameter space and run independently. This allows for better exploration of the parameter space and provides a way to assess convergence of the sampling~\cite{gelman2013bayesian}.
}, each with 100\,000 samples.
The values at the mode of the full posterior agree with those of the benchmark point outlined in \cref{eq:WET-BSM-benchmark}. 
The full set of 2-dimensional marginalizations of this posterior and the resulting intervals at $68\%$ and $95\%$ probability are shown in \cref{fig:knunu-posterior}. The posterior mode and the $95\%$ \ac{HDI} (see \cref{sec:credible-intervals}) for the magnitudes of the Wilson coefficients are shown in \cref{tab:method paper wet results}.
\begin{table}[ht]
    \renewcommand{\arraystretch}{1.5}
    \caption{The mode of the posterior, and \ac{HDI} at 95\% for the (sums of the) \ac{WET} Wilson coefficients in \cref{eq:width}, derived from the posterior in \cref{fig:knunu-posterior}.}
    \centering
    \begin{tabularx}{\linewidth}{@{\extracolsep{\fill}}lYYYY}
        \toprule
        \midrule
        \textbf{Parameters} & \multicolumn{2}{c}{\textbf{362$~\text{fb}^{-1}$}} & \multicolumn{2}{c}{\textbf{50$~\text{ab}^{-1}$}} \\
        \cmidrule(lr){2-3} \cmidrule(lr){4-5}
        & \textbf{Mode} & \textbf{95\% HDI} & \textbf{Mode} & \textbf{95\% HDI} \\
        \midrule
        $
        \begin{aligned}
            &|C_\mathrm{VL}+C_\mathrm{VR}|\\
            &|C_\mathrm{SL}+C_\mathrm{SR}|\\
            &|C_\mathrm{TL}|
        \end{aligned}
        $ & $
        \begin{aligned}
            13.9&\\
            4.2& \\
            0.5&
        \end{aligned}
        $ & $
        \begin{aligned}
            [12.0, ~15.1]&\\
            [0.0, ~5.5]& \\
            [0.0, ~4.6]&
        \end{aligned}
        $ & $
        \begin{aligned}
            14.0&\\
            4.0& \\
            1.0&
        \end{aligned}
        $ & $
        \begin{aligned}
            [13.3, ~14.7]&\\
            [3.3, ~4.7]& \\
            [0.0, ~2.2]&
        \end{aligned}
        $ \\
        \midrule
        \bottomrule
    \end{tabularx}
    \label{tab:method paper wet results}
\end{table}

The marginal posterior is found to peak at the benchmark point given in \cref{eq:WET-BSM-benchmark-Knunu}.
The posterior mode agrees well with the benchmark point, especially for the $50~\text{ab}^{-1}$ sample, where the statistical precision is higher.
The $95\%$ \ac{HDI} contains the benchmark point for all three parameters in both datasets.
The deviations from the benchmark point observed in the $362~\text{fb}^{-1}$ sample are due to statistical fluctuations, arising from a combination of variance and potential bias. These deviations are substantially reduced in the $50~\text{ab}^{-1}$ sample, where the likelihood provides much stronger constraints on the parameters.
An anti-correlation between the scalar and tensorial Wilson coefficients is visible in their marginalized 2-dimensional distribution. This behaviour is expected, as both the tensor and scalar terms in \cref{eq:width} contribute most significantly at larger values of $q^2$, where the efficiency (\cref{eq:eff-knunu}) is low.
Moreover, this anti-correlation weakens as the statistical power of the data increases, demonstrating that it is driven primarily by limited statistics rather than fundamental parameter degeneracies.

Overall, the observed agreement with the benchmark Wilson coefficients provides a successful closure test for the reinterpretation method.

\subsection{Combination of \texorpdfstring{\BKnunu}{B->Knunubar} and \texorpdfstring{$B \to K^* \nu \bar \nu$}{B->K*nunubar}}
\label{sec:method paper ksnunu}

A limitation of studying solely the \BKnunu process is that its sensitivity to the \ac{WET} Wilson coefficients is limited to the three linear combinations shown in \cref{eq:BToKnunu-wc-sensitivity}.
In this example, I showcase the power of combining data on \BKnunu and $B \to K^* \nu \bar \nu$ decays. These decays exhibit \emph{complementary sensitivity} to the Wilson coefficients due to their different hadronic spin and orbital angular momentum configurations. This example is inspired by the ongoing Belle~II analysis of the $B \to K^* \nu \bar \nu$ decay discussed in \cref{sec:knunu update}.
For the sake of simplicity in this example, effects of additional kinematic variables in the decay chain
$B\to K^*(\to K \pi)\nu\bar\nu$, such as the helicity angle $\theta_K$ of the kaon and the $K \pi$ invariant mass, are neglected. For the application of the proposed method to real-world examples, all kinematic variables should be included in the joint number density for full sensitivity to variations in these variables.

Moreover, this example demonstrates from a technical perspective that the method and its implementation also work for combined \pyhf models, providing full access to the complementarity in sensitivity.

\subsubsection{\texorpdfstring{$B \to K^* \nu \bar \nu$}{B->K*nunubar} Weak Effective Theory parametrization}
\label{sec:method paper ksnunu-theory}

The decay $B \to K^* \nu \bar \nu$ is governed by the same \ac{WET} Lagrangian as described by
\cref{eq:wet-lagrangian,eq:operators}.
Its differential decay rate in terms of the Wilson coefficients of the operators is given in \cref{eq:widthKs}, discussed in \cref{sec:ksnunu-theory}. Compared to \BKnunu decays, the differential $B\to K^*\nu\bar\nu$ decay rate
exhibits additional sensitivity to the quantities $|C_{\mathrm{VL}} - C_{\mathrm{VR}}|, ~ |C_{\mathrm{SL}} - C_{\mathrm{SR}}|$ (see \cref{eq:BToKstarnunu-wc-sensitivity}). As a consequence, a simultaneous analysis of both decays allows constraining the magnitudes of all Wilson coefficients, assuming all Wilson coefficients are real-valued. For an illustrative example, this assumption is applied here.
The kinematic shape of the differential branching ratio, due to the vector, scalar and tensor interactions, is illustrated in \cref{fig:ksnunu-theory}.

The hadronic matrix elements of the \ac{WET} operators in this decay are expressed in terms of seven independent form factors $V(q^2)$, $A_0(q^2)$, $A_1(q^2)$, $A_{12}(q^2)$, $T_1(q^2)$, $T_2(q^2)$ and $T_{23}(q^2)$,  as discussed in \cref{sec:hadronic-matrix-elements,sec:form-factors}.
In this work, these form factors are parametrized following the BSZ parametrization~\cite{Bharucha:2015bzk}, which is truncated at order ${K=2}$.
The values for the corresponding $19$ hadronic parameters arise from the Gaussian likelihood provided in Reference~\cite{Gubernari:2023puw}; see \cref{tab:hadronic parameters kstar full}.
Correlations between the hadronic parameters are taken into account through their covariance matrix and implemented as discussed in \cref{sec:method paper implementation}.

For the \ac{SM} prediction, only the vector left-handed Wilson coefficient $C_{\mathrm{VL}}$ is non-zero (see \cref{eq:WET-SM-point}), leading to a predicted branching ratio of~\cite{EOSAuthors:2021xpv,EOS:v1.0.16}
\begin{equation}
    \mathcal{B}(B^+ \to K^{*+} \nu \bar{\nu})_{\mathrm{SM}} = (9.3 \pm 0.9) \cdot 10^{-6}.
    \label{eq:ksnunu-sm-br}
\end{equation}
The theoretical uncertainty is dominated by the form factor uncertainties from lattice \ac{QCD}, with smaller contributions from \ac{CKM} matrix elements and higher-order \ac{QCD} corrections.

\subsubsection{Datasets}
\label{sec:method paper ksnunu-data}
\begin{table}[t]
    \renewcommand{\arraystretch}{1.5}
    \caption{The number of $B \to K^* \nu \bar \nu$ samples produced for this study, corresponding to an equivalent of $50~\text{ab}^{-1}$ integrated luminosity at the SuperKEKB collider. \textit{Generated} and \textit{reconstructed} samples correspond to the numbers prior and post efficiency correction.}
    \centering
    \begin{tabularx}{\linewidth}{@{\extracolsep{\fill}}YYYY} 
    \toprule \midrule
    Luminosity & $B \overline{\kern -0.18em B}$ events & \makecell{\ac{MC} generated \textbackslash \\ reconstructed} &\makecell{Data generated \textbackslash \\ reconstructed} \\
    \midrule 
    $362~\text{fb}^{-1}$ & $\sim 3.87 \cdot 10^8$ & \makecell{$3.61\cdot 10^3$ \\ $8.46\cdot 10^2$} & \makecell{ $1.05\cdot 10^4$ \\ $2.49\cdot 10^3$} \\ 
    $50~\text{ab}^{-1}$   & $\sim 5.35 \cdot 10^{10}$ & \makecell{$4.99\cdot 10^5$ \\ $1.16\cdot 10^5$} & \makecell{$1.45 \cdot 10^6$ \\ $3.43 \cdot 10^5$}\\ 
    \midrule \bottomrule
    \end{tabularx}
    \label{tab:samplesKs}
\end{table}

To produce the $B \to K^* \nu \bar \nu$ datasets, the same procedure as in \cref{sec:method paper knunu-data} is adapted.

The \ac{MC} data are produced according to the \ac{SM} prediction (\textit{null} hypothesis).
The number of samples is calculated by multiplying the estimated number of $B \overline{\kern -0.18em B}$ events in a $50~\text{ab}^{-1}$ Belle~II dataset with the predicted \ac{SM} branching ratio of \cref{eq:ksnunu-sm-br}.

The real data are produced according to the \ac{BSM} prediction at the benchmark point in \cref{eq:WET-BSM-benchmark} (\textit{alternative} hypothesis). The number of data samples is calculated by multiplying the estimated number of $B \overline{\kern -0.18em B}$ events in a $50~\text{ab}^{-1}$ Belle~II dataset by the predicted \ac{BSM} branching ratio, $\mathcal{B}(B \to K^* \nu \bar \nu) \approx 2.72 \cdot 10^{-5}$~\cite{EOSAuthors:2021xpv,EOS:v1.0.11}.

The number of samples is listed in \cref{tab:samplesKs}.
\ac{MC} samples of the decay's probability distribution for both the null and the alternative hypothesis are produced using the \EOS software in version 1.0.11~\cite{EOS:v1.0.11}.

The efficiency map in this case is chosen to be 
\begin{equation}
    \varepsilon(q^2) = 0.3\left(1-0.08\exp\left(- 2.5 ~ q^2/M_B^2\right) \right),
\end{equation}
which is an approximate expectation for an inclusive $B \to K^* \nu \bar \nu$ analysis. 
For the sake of simplicity, it is assumed that the efficiency is independent of the helicity angle $\theta_K$ and the $K\pi$ invariant mass.

Ten bins in the reconstruction variable ($q_{\rm rec}^2$) are chosen, and it is found that 25 bins in the kinematic variable ($q^2$) provide sufficient accuracy using the procedure described in \cref{app:kinematic-binning,sec:method paper knunu-data}.

\begin{figure}[ht]
    \centering
    \includegraphics[width=0.8\linewidth]{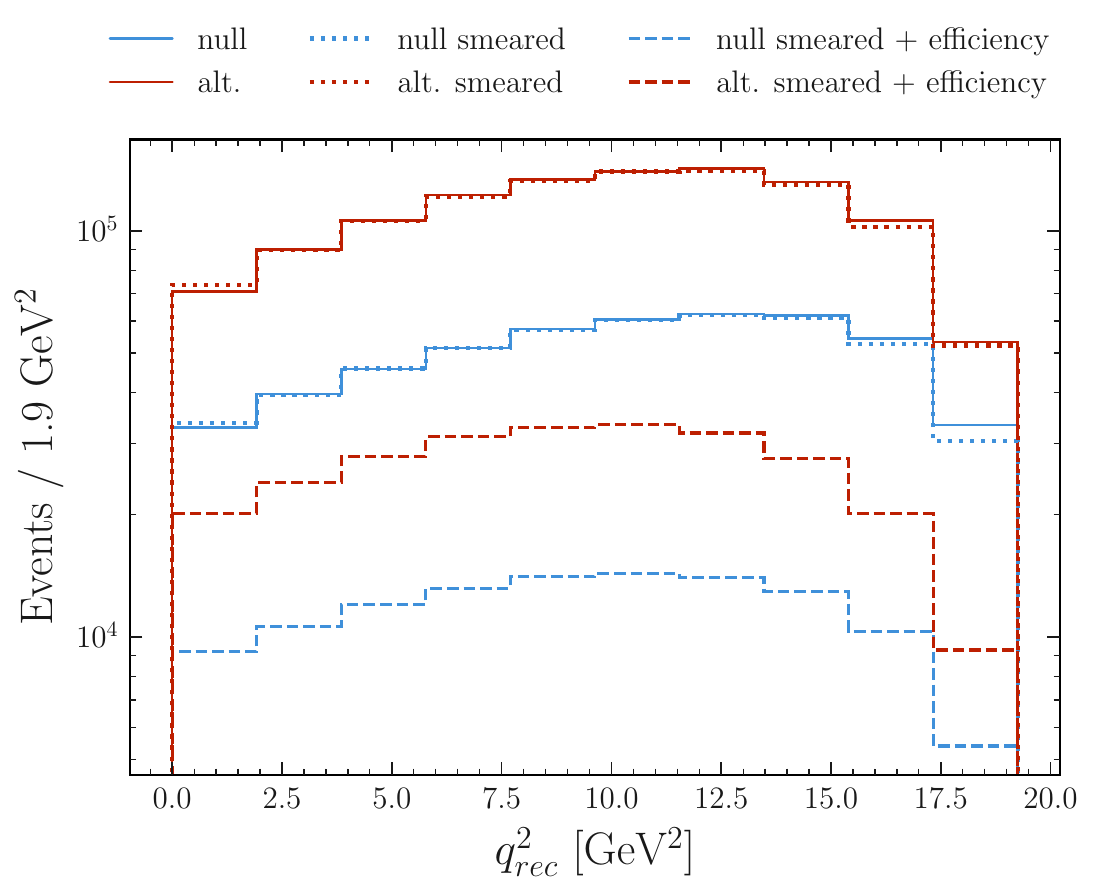}
    \caption{Both the null/\ac{SM} (blue lines) and alternative/\ac{BSM} (red lines) $B \to K^* \nu \bar \nu$ datasets, according to the pure theoretical prediction, after detector resolution smearing and efficiency correction.}
    \label{fig:kstarnunu-data-sets}
\end{figure}

Both datasets, according to the null (\ac{SM}) and alternative (\ac{BSM}) hypotheses, and their corresponding changes after detector resolution smearing and efficiency correction, are shown in \cref{fig:kstarnunu-data-sets}.

\subsubsection{Full statistical model}
\label{sec:method paper ksnunu-stat-model}
To derive the statistical model encompassing \BKnunu and $B \to K^* \nu \bar \nu$, individual posteriors for each channel are constructed following the methodology outlined in \cref{sec:method paper stat-model-K}. The combined posterior arises from their product, with the \ac{WET} theory parameters being the only parameters shared by the individual posteriors.

These parameters correspond to the five magnitudes of Wilson coefficients that enter the theoretical description of $B\to K \nu \bar \nu$ and $B\to K^* \nu \bar \nu$ decays (see \cref{eq:width,eq:widthKs}). They are $C_{\mathrm{VL}}$, $C_{\mathrm{VR}}$, $C_{\mathrm{SL}}$, $C_{\mathrm{SR}}$, and $C_{\mathrm{TL}}$. Their prior is chosen as the uncorrelated product of uniform priors with support 
\begin{equation}
    \begin{aligned}
        &5 < |C_{\mathrm{VL}}| < 15,\\
        &0 < |C_{\mathrm{VR}}| < 10,\\
        &0 < |C_{\mathrm{SL}}| < 10,\\
        &0 < |C_{\mathrm{SR}}| < 10,\\
        &0 < |C_{\mathrm{TL}}| < 10.
    \end{aligned}
\end{equation}
These ranges are motivated with the benchmark point, \cref{eq:WET-BSM-benchmark}, in mind and wide enough to ensure that the posterior can explore the parameter space sufficiently.

The hadronic parameters, describing the $B\to K$ and $B\to K^*$ form factors are discussed in \cref{sec:method paper knunu-theory,sec:method paper ksnunu-theory}. Their prior is a multivariate normal distribution, which is implemented as a sequence of independent univariate normal distributions, as discussed at the end of \cref{sec:method paper implementation}.

In the context of \histfactory models, the combined likelihood of \BKnunu and ${B \to K^* \nu \bar \nu}$ is a combination at the \textit{channel} level, as discussed in \cref{sec:method paper implementation}. One custom modifier is added to each channel, which are functions of the same \ac{WET} parameters but different hadronic parameters.

\subsubsection{Reinterpretation results}
\label{sec:method paper ksnunu-results}
\begin{figure}
    \centering
    \includegraphics[width=\textwidth]{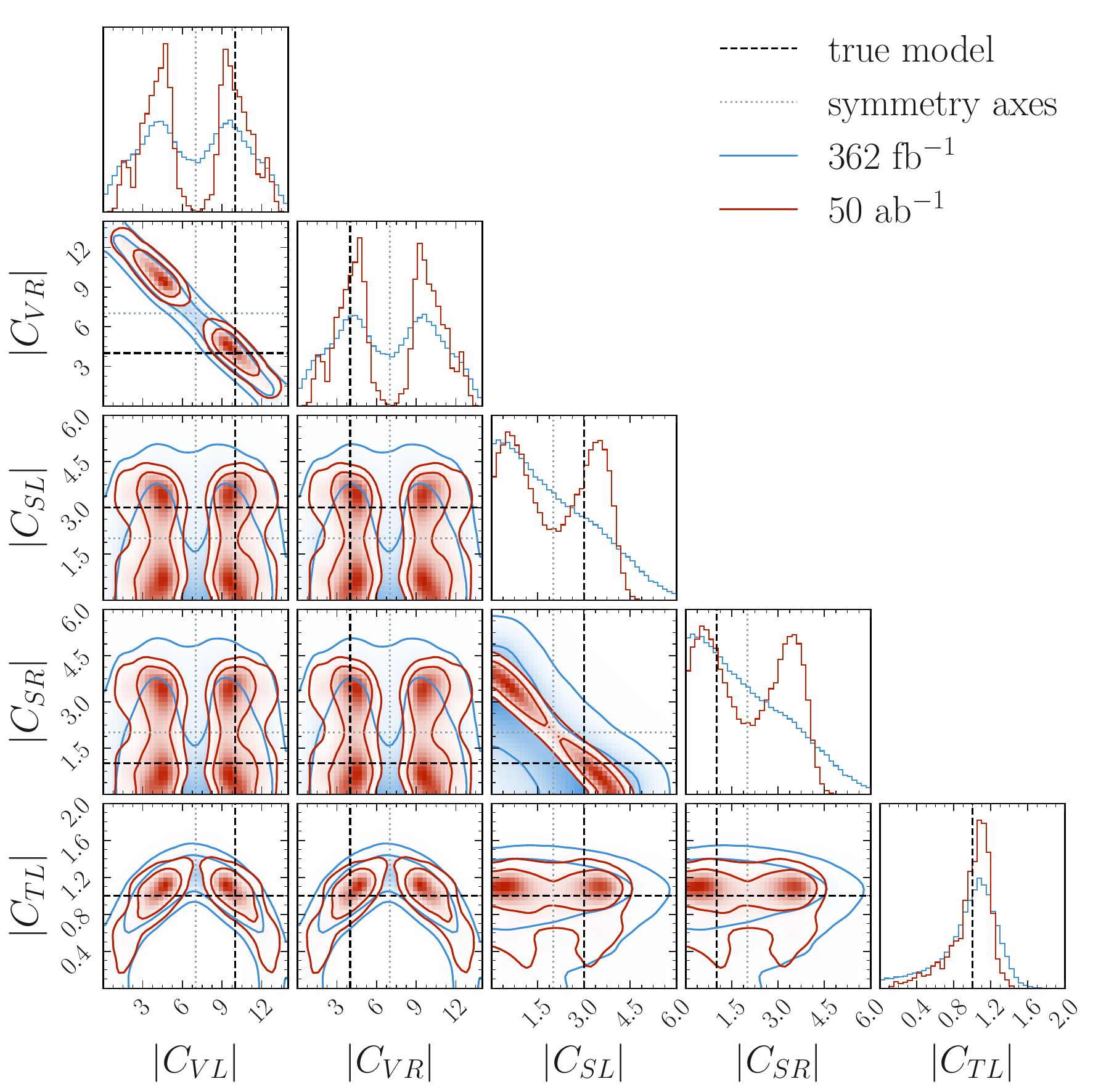}
    \caption{
    The marginal posterior distributions, obtained by \ac{MCMC} sampling from the combined \BKnunu and $B \to K^* \nu \bar \nu$ likelihood. On the diagonal, the 1-dimensional marginal distributions of the Wilson coefficients appearing in \cref{eq:width,eq:widthKs} are shown. The contours on the 2-dimensional plots correspond to $68\%$ (inner) and $95\%$ (outer) probability. The dashed lines indicate the true underlying model. The dotted lines indicate the symmetry axes of the global likelihood.}
    \label{fig:combination-posterior}
\end{figure}

From the constructed model-agnostic likelihood function, I investigate the power of constraining the full set of Wilson coefficients appearing in \cref{eq:width,eq:widthKs}, under the assumption that they are real-valued.
The decay rates in \cref{eq:width,eq:widthKs} exhibit two discrete symmetries: one under the exchange ${C_{\mathrm{VL}} \leftrightarrow C_{\mathrm{VR}}}$ and another under the exchange ${C_{\mathrm{SL}} \leftrightarrow C_{\mathrm{SR}}}$. The combination of both symmetries leads to a four-fold ambiguity in the extraction of the Wilson coefficients from data and therefore a multimodal posterior density.
To avoid computational issues when sampling from the posterior, one of the four fully-equivalent modes is selected for sampling. This is achieved by imposing the additional constraints $C_{\mathrm{VL}} > C_{\mathrm{VR}}$ and $C_{\mathrm{SL}} > C_{\mathrm{SR}}$.
\ac{MCMC} sampling is used, and the chains are initialized with the mode of the full posterior. The posterior is sampled using 8 independent chains, each with 100\,000 samples.
The values at the mode of the full posterior align with those of the benchmark point outlined in \cref{eq:WET-BSM-benchmark}.
To obtain the full multimodal posterior, the original symmetry is restored manually.
From the symmetrized samples, the 5-dimensional marginal posterior distribution of the Wilson coefficients is obtained.
This is illustrated in \cref{fig:combination-posterior} by showing the full set of 1- and 2-dimensional marginalizations and the resulting regions at $68\%$ and $95\%$ probability.\footnote{The binning of the posterior samples is not aligned with the symmetry axis, leading to small visual asymmetries.} The posterior mode and the $95\%$ \ac{HDI} (see \cref{sec:credible-intervals}) for the magnitudes of the Wilson coefficients are shown in \cref{tab:method paper wet results combination}.
\begin{table}[ht]
    \renewcommand{\arraystretch}{1.5}
    \caption{The mode of the posterior, and \ac{HDI} at 95\% for the (sums of the) \ac{WET} Wilson coefficients in \cref{eq:width,eq:widthKs}, derived from the posterior in \cref{fig:combination-posterior}.}
    \centering
    \begin{tabularx}{\linewidth}{@{\extracolsep{\fill}}lYYYY}
        \toprule
        \midrule
        \textbf{Parameters} & \multicolumn{2}{c}{\textbf{362$~\text{fb}^{-1}$}} & \multicolumn{2}{c}{\textbf{50$~\text{ab}^{-1}$}} \\
        \cmidrule(lr){2-3} \cmidrule(lr){4-5}
        & \textbf{Mode} & \textbf{95\% HDI} & \textbf{Mode} & \textbf{95\% HDI} \\
        \midrule
        $
        \begin{aligned}
            &|C_\mathrm{VL}|\\
            &|C_\mathrm{VR}|\\
            &|C_\mathrm{SL}|\\
            &|C_\mathrm{SR}|\\
            &|C_\mathrm{TL}|
        \end{aligned}
        $ & $
        \begin{aligned}
            10.0&\\
            4.0&\\
            3.2&\\
            0.9&\\
            1.0&
        \end{aligned}
        $ & $
        \begin{aligned}
            [6.9, ~13.2]&\\
            [0.6, ~6.8]&\\
            [0.8, ~5.5]&\\
            [0.0, ~2.2]&\\
            [0.2, ~1.5]&
        \end{aligned}
        $ & $
        \begin{aligned}
            10.0&\\
            4.0&\\
            3.0&\\
            1.0&\\
            1.0&
        \end{aligned}
        $ & $
        \begin{aligned}
            [8.4, ~12.7]&\\
            [1.3, ~5.6]&\\
            [2.1, ~4.2]&\\
            [0.0, ~1.8]&\\
            [0.4, ~1.3]&
        \end{aligned}
        $ \\
        \midrule
        \bottomrule
    \end{tabularx}
    \label{tab:method paper wet results combination}
\end{table}

Combining results for \BKnunu and $B \to K^* \nu \bar \nu$ enables probing of all five Wilson coefficient magnitudes. The posterior modes align well with the benchmark point in \cref{eq:WET-BSM-benchmark} for both datasets, with the $50~\text{ab}^{-1}$ dataset showing improved precision. The $95\%$ \acp{HDI} contain the benchmark point in all cases, with substantial narrowing at higher luminosity.

The combination significantly improves constraints on $|C_{\mathrm{TL}}|$ compared to \BKnunu alone (see \cref{fig:knunu-posterior,tab:method paper wet results}), while the scalar sector shows the most dramatic precision gain between datasets. For the smaller $362~\text{fb}^{-1}$ dataset, overlapping peaks and magnitude sampling lead to asymmetric scalar distributions with broader tails. The tail of $|C_{\mathrm{TL}}|$ toward lower values persists, consistent with the previous example in \cref{fig:knunu-posterior}.

This study provides a successful closure test for the reinterpretation method and demonstrates the power of combining measurements within this framework. 

The combined analysis demonstrates the enhanced constraining power achievable through the reinterpretation method when applied to multiple complementary measurements. However, to fully appreciate the value of this approach, it is essential to understand the limitations of conventional reinterpretation strategies and the biases they introduce.

\subsection{The necessity for reinterpretation}
\label{sec:method paper necessity}
The availability of open datasets for most particle physics results is currently very limited, although improving thanks to the popularity of statistical approaches such as \histfactory~\cite{histfactory} and tools such as \pyhf~\cite{pyhf,Heinrich:2021gyp}. These current limitations regularly hinder theorists from fully interpreting existing experimental results in their \ac{BSM} analyses. In particular, \ac{BSM} changes to the distribution of the reconstruction variable are routinely neglected. In fact, the most common approach in a \ac{BSM} analysis is to constrain the ratio of the \ac{BSM} prediction to the \ac{SM} prediction from branching ratio measurements or upper limits. This approach is valid only if \ac{BSM} changes to the shape of the kinematic variable distribution can be accounted for by systematic experimental uncertainties in the reconstruction space. As shown in the following, this does not hold for measurements of the \BKnunu branching ratio.

To illustrate the issue, the results in \cref{sec:method paper knunu} based on simulated data are compared with those obtained from naive rescaling of the branching ratio. In the language of the presented reinterpretation method, the latter corresponds to using only a single bin in the kinematic \ac{d.o.f.}, covering the full kinematic range. This translates to a single weight applied to all bins of the reconstruction variable, corresponding to the ratio of the alternative to the null prediction integrated over the full kinematic range. Therefore, an additional ``naive'' \BKnunu posterior is constructed that deviates from the setup in \cref{sec:method paper knunu} only by using a single bin in the kinematic range.

\begin{figure}
    \centering
    \includegraphics[width=\textwidth]{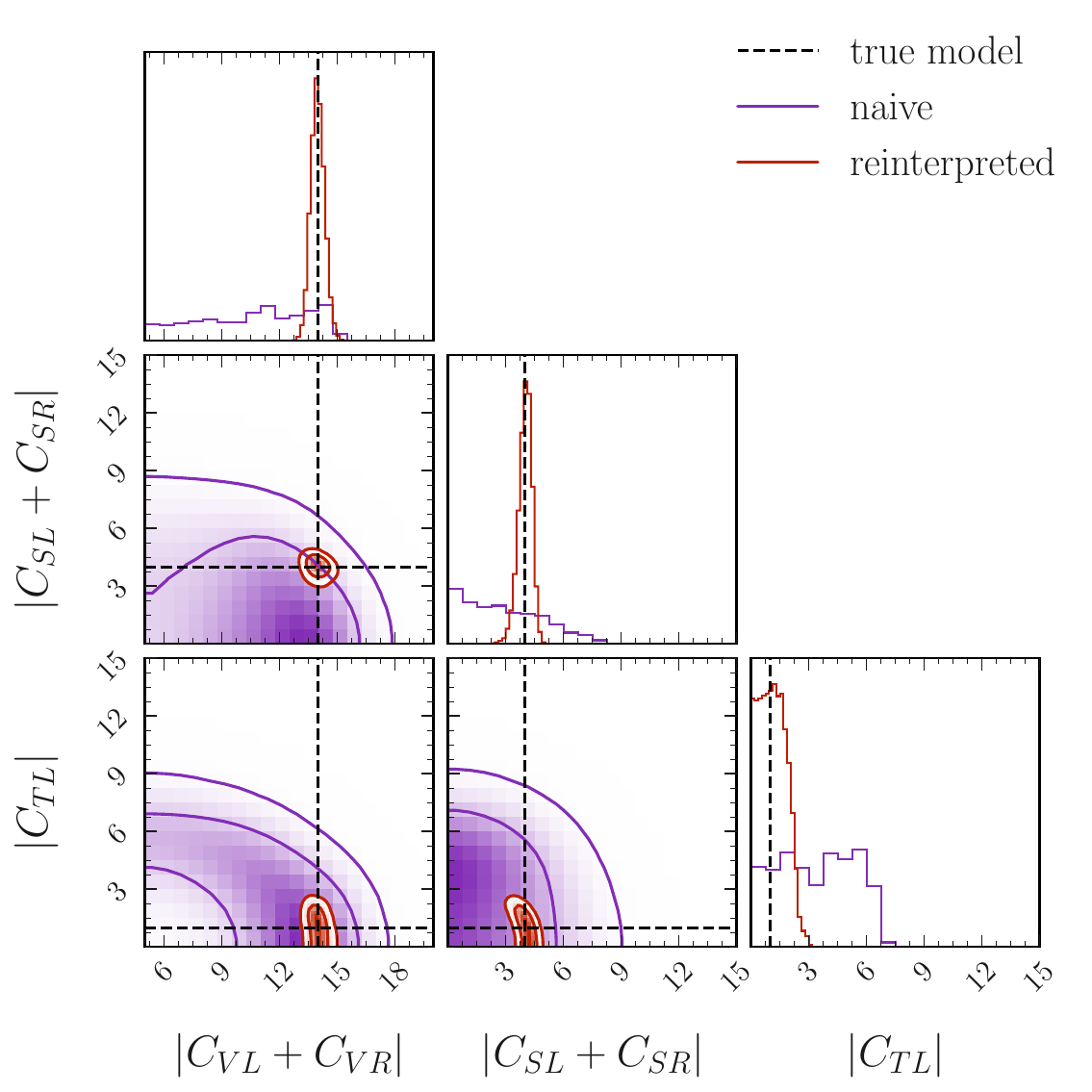}
    \caption{The comparison of the posterior distribution resulting from a model with only one bin in the kinematic \ac{d.o.f.} to the proposed reinterpretation method, respecting shape changes in the kinematic distribution (see \cref{fig:knunu-posterior}).}
    \label{fig:knunu-posterior-compare}
\end{figure}

After sampling from this ``naive'' posterior, the marginal distributions for the Wilson coefficients are compared to those presented in \cref{fig:knunu-posterior}. This comparison is shown in \cref{fig:knunu-posterior-compare}.
A striking difference in the overall shape of the distributions and the central intervals at $68\%$ ($95\%$) probability is found.
Clearly, the ``naive'' procedure fails to validate, yielding large deviations from the benchmark point in \cref{eq:WET-BSM-benchmark-Knunu} in all three sectors.
The results illustrate that the approach is essential for faithful reinterpretation of \BKnunu experimental results. 

It should be emphasized that the method provides a means to ensure \emph{accurate} interpretation of the existing likelihood beyond the assumptions of the underlying signal model. This does not imply, however, that this interpretation is more \emph{precise} than a naive \ac{BSM} interpretation. Put differently, the approach eliminates bias introduced by using an incorrect template for the decay's kinematic distribution but at the expense of potentially larger uncertainties on the theory parameters.

\section{Discussion and significance of the method}
\label{sec:method paper discussion}
I present a novel reinterpretation method for particle physics results that is simple in its application and requires only minimal information in addition to published likelihoods.

The proposed method avoids biases that are introduced in naive reinterpretation of data with a negligible increase in compute time.
As such, it provides most of the benefits of reinterpretation using full analysis preservation.
Therefore, this method provides a good trade-off between accuracy and speed, and has the potential to improve the accuracy of global effective field theory fits to many analysis results.

To showcase the method, it is applied to a simulated dataset of the \BKnunu decay, inspired by the recent Belle~II analysis~\cite{Belle-II:2023esi} but without resorting to using any public or private Belle~II data.

Using the two examples discussed in \cref{sec:method paper examples}, I validate the method by successfully recovering the benchmark theory point from the underlying synthetic data.
This outcome underscores the accuracy and self-consistency of the approach.
The bias introduced by naive rescaling of the \BKnunu branching ratio is further investigated.
For the benchmark point, a sizable bias is found when determining the \ac{WET} Wilson coefficients without applying the reinterpretation method. 

In conclusion, the work presented in this chapter illustrates both the ease of applicability and the urgent necessity of
shape-respecting reinterpretation over the traditional approach.
It is hoped that this work will motivate experimental collaborations and analysts to consider the future reinterpretability of their results, and to publish the necessary material (for further details, see \cref{app:recipe}). This will, in turn, enable the entire community to use these analysis results accurately.

%% file: chapters/knunu-mall.tex
\footnote{
The material presented in this chapter has been submitted to PRD~\cite{belle2pub88}. While the content largely follows that of the referenced paper, editorial modifications have been made, including restructuring of the presentation, rephrasing of portions of the text, and expansion of selected explanations. 
As this project is the result of collaborative work of the Belle~II collaboration, individual contributions of the proponents of this work, and further collaboration members are documented in \cref{sec:wet-contributions}.
}
Having established the theoretical framework for model-agnostic reinterpretation in the previous chapter, I now proceed to its first practical application to real experimental data. This and the following chapters present the first-ever implementation of the reinterpretation method presented in \cref{sec:method} to actual Belle~II data, specifically applied to the \BKnn analysis\footnote{Charge-conjugate channels are implied.}~\cite{Belle-II:2023esi}.

The Belle~II collaboration reported a $2.7\sigma$ excess over the \ac{SM} expectation in the \BKnn decay channel~\cite{Belle-II:2023esi}, which could indicate contributions from \ac{BSM} physics. The original analysis quantified this excess through a signal strength parameter $\mu$ relative to the \ac{SM} prediction. A reinterpretation within the \ac{WET} framework enables a more detailed characterization of the potential new physics by resolving contributions from different operator structures.

This work not only demonstrates the real-world feasibility of the approach but also establishes important precedents for future applications of Belle~II data. The choice of the Belle~II \BKnn analysis lies in its recent experimental prominence and the theoretical interest it has generated in the particle physics community.

The \ac{SM} branching ratio of the \BKnn decay used in the Belle~II analysis~\cite{Belle-II:2023esi} is predicted to be~\cite{Parrott:2022zte}
\begin{equation}
\mathcal{B}(\BKnn) =
\left(5.58 \pm 0.37 \right)\cdot 10^{-6}\, ,
\label{eq:brf}
\end{equation}
which includes a contribution of $\left(0.61 \pm 0.06\right)\cdot 10^{-6}$ from the long-distance double-charged-current ${B^+\to \tau^+(\to K^+\bar{\nu}_\tau)\nu_\tau}$ decay.
In this measurement, only the short-distance contribution was considered as signal, with a corresponding branching ratio of $(4.97\pm 0.37) \cdot 10^{-6}$~\cite{Parrott:2022zte}. These can be distinguished due to a different kinematic signature.

The \BKnn decay has been studied by the CLEO, \textit{BABAR}, Belle, and Belle~II collaborations~\cite{PhysRevLett.86.2950,PhysRevD.87.111103,PhysRevD.87.112005,PhysRevD.82.112002,PhysRevD.96.091101,Belle-II:2021rof,Belle-II:2023esi}, with the latest measurement by the Belle~II collaboration finding the first evidence for this decay at \combinationsigO\ standard deviations.
This result, based on the \ac{SM} prediction from Ref.~\cite{Parrott:2022zte} and hadronic parameters from the HPQCD collaboration~\cite{Parrott:2022rgu} as a model ansatz, exceeds the \ac{SM} expectation by \combinationsigSM\ standard deviations.
This enhanced branching ratio triggered the \ac{HEP} community to interpret this result under different new physics scenarios~\cite{Fridell:2023ssf,PhysRevD.109.075008, Gabrielli:2024wys}. Such reinterpretations, however, are only approximate as not all relevant information on the measurement was previously accessible to the public.

Generally, two distinct classes of new physics models can replicate the signature of a \BKnn decay.
These models naturally mimic the neutrino pair in the final state, as the neutrinos remain undetected at Belle~II. New physics scenarios can proceed via either three-body or two-body decays.
In three-body decays, the properties of new physics can be studied within the framework of the \ac{WET} (see \cref{sec:wet}), where new physics particles and force carriers have masses at or above the electroweak symmetry-breaking scale. Examples of such models include leptoquarks~\cite{PhysRevD.98.055003, Dorsner:2016wpm, Angelescu:2021lln,ParticleDataGroup:2024cfk} and heavy $Z'$ bosons~\cite{PhysRevD.104.053007, ParticleDataGroup:2024cfk, Leike:1998wr, Buras:2012jb}.
Two-body decays would signal the presence of light new physics, which lies below the scale of electroweak symmetry breaking and is not described by the \ac{WET}. Explicit examples include axions~\cite{MartinCamalich:2020dfe,ParticleDataGroup:2024cfk, Peccei:1977hh, Peccei:1977ur}, axion-like particles~\cite{ParticleDataGroup:2024cfk, PhysRevD.102.015023,Guerrera:2022ykl,Bruggisser:2023npd} or other dark-sector mediators~\cite{PhysRevD.101.095006,Datta:2022zng,Abdughani:2023dlr,Berezhnoy:2023rxx}.

The primary goals of this work are threefold: to publish the model-agnostic likelihood for the \BKnn analysis, which consists of the full likelihood from the \BKnn result~\cite{Belle-II:2023esi} and the joint number densities used in this reinterpretation~\cite{hepdata.166082,hepdata.146803} (for details see \cref{sec:BelleIIdata}); to reinterpret the result in the framework of the \ac{WET}, providing constraints on relevant Wilson coefficients; and to reinterpret the result in the context of a two-body light new physics model.
The published model-agnostic likelihood allows the broader scientific community to test alternative theoretical models in a statistically rigorous way. Importantly, this work sets a template for future Belle~II measurements that are suitable for reinterpretation and reflects the collaboration's commitment to publishing model-agnostic likelihoods as a means to maximize the scientific impact and reusability of its results.

This chapter is structured as follows:
In \cref{sec:wet paper analysis} I give an overview of the \BKnn analysis~\cite{Belle-II:2023esi}, including the data used and the analysis strategy.
How the reinterpretation method of \cref{sec:method} is applied in the context of the \BKnn analysis is described in \cref{sec:wet paper reinterpretation method}.

\section{The Belle~II \texorpdfstring{\BKnn}{B+->K+nunubar} analysis}
\label{sec:wet paper analysis}
In this section I provide an overview of the Belle~II analysis of the \BKnn decay~\cite{Belle-II:2023esi}, which serves as the basis for the reinterpretation presented in this work.

\subsection{Data and simulated samples}
\label{sec:analysis-data}
The Belle~II analysis of \BKnn decays~\cite{Belle-II:2023esi}, originating from an ${e^+ e^- \to \Upsilon(4S) \to B^+B^-}$ process, was conducted using an integrated luminosity of 362~fb$^{-1}$ of on-resonance data~\cite{Abudinén_2020}, recorded between 2019 and 2022 at a centre-of-mass energy of $\sqrt{s} \approx 10.58$~GeV, corresponding to the $\Upsilon(4S)$ resonance. This dataset contains $(387\pm6)\cdot 10^6$ $B^+B^-$ pairs~\cite{PhysRevD.72.032005}. An off-resonance sample of 42~fb$^{-1}$ was also recorded at $\sqrt{s} \approx 10.52$~GeV, which is used to constrain the background contributions from continuum processes.

The analysis was designed using \ac{MC} simulated data for the signal and seven background categories: decays of charged and neutral $B \bar{B}$ mesons; continuum processes ($u \bar{u}$, $d \bar{d}$, $s \bar{s}$, $c \bar{c}$); and low-multiplicity $\tau^+ \tau^-$ backgrounds.
These simulated data were used to estimate the efficiency of the selection criteria, train multivariate classifiers, and build the statistical model.
Detector simulation and reconstruction was performed using the \texttt{basf2} framework~\cite{basf2} interfaced with \texttt{GEANT4}~\cite{GEANT4:2002zbu}, as described in \cref{sec:computing}.

\subsection{Background suppression}
\label{sec:analysis-background-suppression}
This measurement was performed using two different reconstruction methods: the more sensitive \ac{ITA} and the more conventional \ac{HTA}, targeting nearly orthogonal datasets. 
In the \ac{HTA}, the non-signal $B$ meson of the $B^+B^-$ pair was reconstructed in specific hadronic decay modes. Subsequently, the signal $B$ meson is reconstructed from the remaining particles in the event. Requiring a full reconstruction of one $B$ meson ensures a low background, at the cost of lower signal efficiency.
In contrast, in the \ac{ITA}, the signal was identified by reconstructing the $K^{+}$ track, summing all remaining particles in the event, and then inferring the missing energy. The relaxed requirement on the reconstruction of the non-signal $B$ meson allows for a higher signal efficiency at the cost of increased background contamination.

In addition to a basic preselection, the \ac{ITA} used two \acp{BDT}, $\mathrm{BDT}_1$ and $\mathrm{BDT}_2$, to separate the signal from the background. 

A first binary classifier, \BDT1, is designed as a first-level filter after event selection. The most powerful discriminant is the difference between the \ac{ROE} energy in the centre-of-mass frame and half the total collision energy, $\Delta E_{\mathrm{ROE}} = E_{\mathrm{ROE}} - \sqrt{s}/2$, which is typically negative for signal events due to neutrinos but positive for background events containing additional reconstructed particles~\cite{Belle-II:2023esi}. 

The second classifier, \BDT2, is used for the final event selection, which is trained on events with \BDT1$>0.9$. This corresponds to a signal (background) selection efficiency of 34\% (1.5\%), using 35 input variables.
For \BDT2, the most discriminating variable is the cosine of the angle between the momentum of the signal-kaon candidate and the thrust axis of the \ac{ROE} energy in the centre-of-mass frame. This has a uniform distribution for the signal and a peaking shape for the jet-like continuum background~\cite{Belle-II:2023esi}.

The \ac{HTA} used one \acp{BDT}, $\mathrm{BDTh}$, to suppress the background. The $\mathrm{BDTh}$ input variable providing the highest discriminating power is $E_{extra}$, which is the sum of the energy deposited in the \ac{ECL} that cannot be directly associated with the reconstructed $B$-mesons. It is primarily expected to be zero or small for correctly reconstructed signal events, but can be larger for background events~\cite{Belle-II:2023esi}.

\subsection{Signal region definition}
\label{sec:analysis-signal-region}
For both the \ac{ITA} and \ac{HTA} methods, a \histfactory statistical model (see \cref{sec:histfactory} and Reference~\cite{histfactory}) was used. This was implemented in the \texttt{pyhf} framework (see \cref{sec:pyhf} and References~\cite{pyhf, Heinrich:2021gyp}), incorporating all systematic uncertainties. The \ac{ITA} analysis used $4 \times 3$ bins in the $\eta\left(\mathrm{BDT}_2\right) \times q^2_{\rm rec}$ space, where $\eta\left(\mathrm{BDT}_2\right)$ is the transformed $\mathrm{BDT}_2$ output, 
\begin{equation}
    \eta\left(\mathrm{BDT}_2\right) = 1 - \int^1_{\mathrm{BDT}_2} db  ~ \xi(b) \, ,
\label{eq:eta-bdt2}
\end{equation}
such that each bin in $\mathrm{BDT}_2$ represents a 2\% efficiency quantile for the selection. In this way, the distribution of $\eta(\BDT2)$ for simulated signal events is uniform. Here, $\xi(b)$ is the signal selection efficiency for the $\mathrm{BDT}_2$ output $b$. $q^2_{\rm rec}$ is the reconstructed $q^{2}$, defined as 
\begin{equation}
    q^2_{\rm rec} = \frac{s}{4} + M_K^2 - \sqrt{s} E_K^* \, ,
    \label{eq:q2rec}
\end{equation}
assuming the signal \B meson to be at rest in the $e^{+}e^{-}$ centre-of-mass frame.
Here, $M_K$ is the nominal kaon mass and $E_K^*$ is the reconstructed energy of the kaon in the collision centre-of-mass frame.
The signal regions for the \ac{ITA} were defined as $\mathrm{BDT}_1 > 0.9$ and $\eta\left(\mathrm{BDT}_2\right) > 0.92$, which maximizes the signal significance as expected by simulation.
The $q^{2}_{\mathrm{rec}}$ bin boundaries are given by $[-1.0, 4.0, 8.0, 25.0]\ \gev^{2}$, and the bin boundaries in $\eta(\BDT2)$ are $[0.92, 0.94, 0.96, 0.98, 1.00]$.
The bin with $\eta\left(\mathrm{BDT}_2\right) > 0.98$ provides the largest signal sensitivity.
The signal region for the \ac{HTA} was defined as $\eta\left(\mathrm{BDTh}\right) > 0.4$, where $\eta\left(\mathrm{BDTh}\right)$ is defined analogously to \cref{eq:eta-bdt2} but using the $\mathrm{BDTh}$ output. The \ac{HTA} used $6$ equally spaced bins in the $\eta\left(\mathrm{BDTh}\right)$ space~\cite{Belle-II:2023esi}.

\begin{figure}
    \centering
    \includegraphics[width=\linewidth]{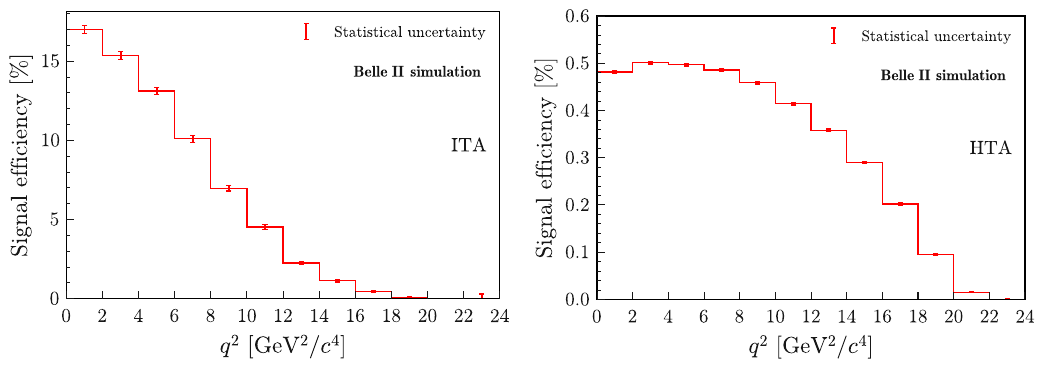}
    \caption{The signal-selection efficiencies for the \ac{ITA} (\textit{left}) and \ac{HTA} (\textit{right}) as a function of $q^2$. Error bars indicate the statistical uncertainty. Figures taken from Reference~\cite{Belle-II:2023esi}.}
    \label{fig:knunu-efficiency}
\end{figure}

The \ac{ITA} and \ac{HTA} $q^{2}$ resolutions are approximately $1\ \text{GeV}^{2}$ and $0.3\ \rm{GeV}^{2}$, respectively, leading to differences in the $q^{2}_{\rm rec}$ and $q^{2}$ distributions. Final-state radiation has a negligible effect on the presented results compared to other sources of uncertainties. This is discussed further in \cref{sec:fsr}.
The \ac{ITA} includes both a signal and a control region, resulting in a total of 24 reconstruction bins. Signal region bins are populated with events that pass all selection criteria in the on-resonance data. Control region bins are filled with events that pass the same selection in the off-resonance data. The control region improves constraints on the background contributions from continuum processes.
The \ac{HTA} signal region contains only events that pass all selection criteria in the on-resonance data.
Signal-selection efficiencies for simulated events, as a function of $q^2$ are shown in \cref{fig:knunu-efficiency} for the \ac{ITA} and \ac{HTA} methods. The \ac{ITA} has a higher signal efficiency than the \ac{HTA}, but the \ac{HTA} has a lower background contamination. In both cases, the signal efficiency decreases with increasing $q^2$.

\subsection{Signal extraction result}
\begin{figure}
    \centering
    \includegraphics[width=\linewidth]{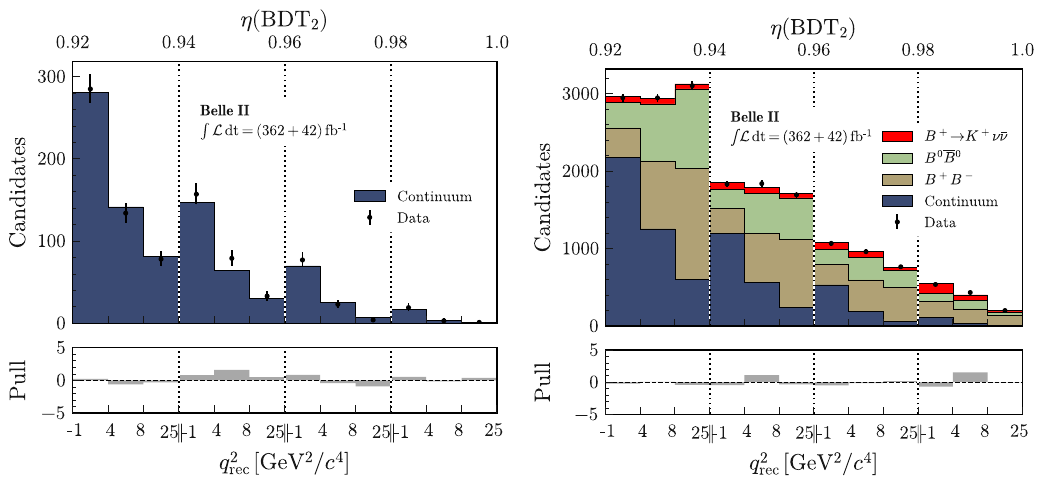}
    \caption{The best-fit yields of the \ac{ITA} analysis as a function of $\eta(\mathrm{BDT}_2) \times q^2_{\rm rec}$. The off-resonance fit is shown on the left, the on-resonance fit on the right. Pull distributions are shown in the lower panels.
    Figure taken from Reference~\cite{Belle-II:2023esi}.}
    \label{fig:ita-bf-paper}
\end{figure}

\begin{figure}
    \centering
    \includegraphics[width=0.55\linewidth]{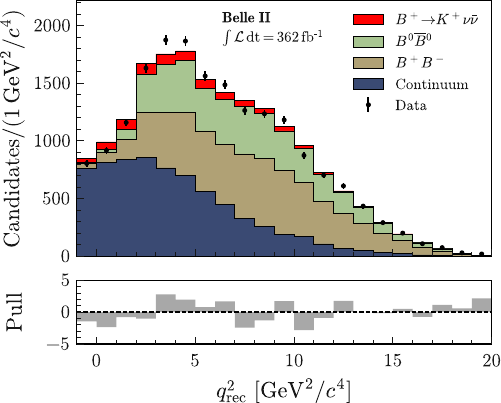}
    \caption{The best-fit projection of the \ac{ITA} analysis onto the $q^2_{\rm rec}$ variable.  Pull distributions are shown in the lower panels.
    Figure taken from Reference~\cite{Belle-II:2023esi}.}
    \label{fig:ita-q2-projection}
\end{figure}

The best-fit yields of the \ac{ITA} analysis are shown in \cref{fig:ita-bf-paper}. A maximum likelihood fit yields a signal strength relative to the \ac{SM} expectation of~\cite{Belle-II:2023esi}
\begin{equation}
    \mu_{\rm ITA} = 5.4 \pm 1.0~\text{(stat.)} \pm 1.1~\text{(syst.)} = 5.4 \pm 1.5 \, .
\end{equation}
This translates into a branching ratio of~\cite{Belle-II:2023esi}
\begin{equation}
    \mathcal{B}(\BKnn)_{\rm ITA} = (2.7 \pm 0.5~\text{(stat.)} \pm 0.5~\text{(syst.)}) \cdot 10^{-5} \, .
\end{equation}
A projection of the best-fit point onto the $q^2_{\rm rec}$ variable is shown in \cref{fig:ita-q2-projection}. This is done by computing a ratio of post- to pre-fit yields in the reconstruction binning for each of the samples and applying these weights to all events in the respective bins. Interestingly, one can see an excess in the data in the region of $ 3 < q^2_{\rm rec} < 7\ \text{GeV}^2$. I will discuss this further in \cref{sec:reinterpretation knunu bkx}.

\begin{figure}
    \centering
    \includegraphics[width=0.55\linewidth]{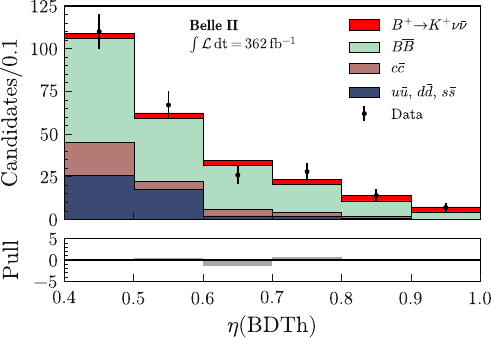}
    \caption{The best-fit yields of the \ac{HTA} analysis as a function of $\eta(\mathrm{BDT}_h)$. Pull distributions are shown in the lower panels.
    Figure taken from Reference~\cite{Belle-II:2023esi}.}
    \label{fig:hta-bf-paper}
\end{figure}

The best-fit yields of the \ac{HTA} analysis are shown in \cref{fig:hta-bf-paper}. 
A maximum likelihood fit yields a signal strength relative to the \ac{SM} expectation of~\cite{Belle-II:2023esi}
\begin{equation}
    \mu_{\rm HTA} = 2.2 ^{+1.8}_{-1.7}~\text{(stat.)}^{+1.6}_{-1.1}~\text{(syst.)} = 2.2 ^{+2.4}_{-2.0} \, .
\end{equation}
This translates into a branching ratio of~\cite{Belle-II:2023esi}
\begin{equation}
    \mathcal{B}(\BKnn)_{\rm HTA} = (1.1 ^{+0.9}_{-0.8}~\text{(stat.)} ^{+0.8}_{-0.5}~\text{(syst.)}) \cdot 10^{-5} \, .
\end{equation}

The \ac{ITA} and \ac{HTA} likelihoods were combined into one likelihood, accounting for the correlations between the systematic uncertainties in the two methods (for details, see section XIV of Reference~\cite{Belle-II:2023esi}).
The \textit{combined} likelihood includes 231 nuisance parameters $\boldsymbol{\chi}$ in addition to one \ac{POI}, the signal strength $\mu$. A maximum likelihood fit yields a signal strength of~\cite{Belle-II:2023esi}
\begin{equation}
    \mu = 4.6 \pm 1.0~\text{(stat.)} \pm 0.9~\text{(syst.)} = 4.6 \pm 1.3 \, ,
\end{equation}
which corresponds to a branching ratio of~\cite{Belle-II:2023esi}
\begin{equation}
    \mathcal{B}(\BKnn) = \combinationBFdetailed \, .
\end{equation}
This result has a significance of 3.5 standard deviations over the background-only hypothesis and a significance of 2.7 standard deviations over the \ac{SM} hypothesis. The combined likelihood serves as the basis for reinterpretation in this study.


\section{Joint number densities for the Belle~II \texorpdfstring{\BKnn}{B+->K+nunubar} analysis}
\label{sec:wet paper reinterpretation method}
To construct a model-agnostic likelihood for the \BKnn analysis~\cite{Belle-II:2023esi}, I follow the method introduced in \cref{sec:method}.
More concretely, since the \BKnn analysis is performed on binned data, I use the discrete reinterpretation method of \cref{sec:method paper reweighting-method-discrete}.

Recall that the method requires the definition of the kinematic variables $z$, the reconstruction variables $x$, the null and alternative distributions $\sigma_0$ and $\sigma_1$, and the joint number density $\nu_{0,xz}$, which is the basis of the model-agnostic likelihood.

Since the decay of the pseudoscalar $B$ meson is isotropic in its rest frame, the only experimentally accessible kinematic \ac{d.o.f.} is the squared dineutrino invariant mass, $q^2$, as mentioned in \cref{sec:knunu-theory}. Hence, the only kinematic variable in this case is $z=q^2$. 

The reconstruction variables are different for the \ac{ITA} and \ac{HTA} methods, as discussed in \cref{sec:wet paper analysis}. In the \ac{ITA}, the reconstruction variables are $x = \eta(\mathrm{BDT}_2) \times q^2_{\rm rec}$. In the \ac{HTA}, the reconstruction variable is $x = \eta(\mathrm{BDTh})$.

The null distribution $\sigma_0(q^2)$ is the \ac{SM} prediction for the \BKnn decay, which is presented in Reference~\cite{Parrott:2022zte}, with form factors taken from Reference~\cite{Parrott:2022rgu}. 

In this work, I will discuss two different cases for the alternative distribution $\sigma_1(q^2)$: one two-body and one three-body final-state scenario. In \cref{sec:reinterpretation knunu wet}, I will discuss the reinterpretation of the \BKnn measurement in the framework of the \ac{WET}. In \cref{sec:reinterpretation knunu bkx}, I will discuss the reinterpretation of the \BKnn measurement in a light new physics scenario, where the alternative distribution is constructed with the contribution of a two-body decay \BKX, with $X$ being a new light particle that decays invisibly. This manifests itself as a narrow resonance peak in the $q^2$ distribution, which is not present in the \ac{SM} prediction.


To obtain the joint number density $\nu_{0,xq^2}$ for both \ac{ITA} and \ac{HTA}, $10^7$ (\ac{ITA}) and $5 \cdot 10^7$ (\ac{HTA}) simulated \ac{SM} signal events from the \BKnn analysis~\cite{Belle-II:2023esi} are used, to which all selection criteria are applied. These samples include information on the generated and reconstructed squared momenta, $q^2$ and $q^2_{\rm rec}$, as well as the classifier responses $\eta(\mathrm{BDT}_2)$ and $\eta(\mathrm{BDTh})$. 

The granularity of the $q^2$ binning for $\nu_{0, x q^2}$ is determined by the differences between the null and the anticipated alternative distributions.
The usual guideline of ensuring sufficiently many events in each bin does not apply here, since the binning serves only as a temporary partitioning for the application of kinematic weights. In principle, finer binning is preferable; however, beyond the intrinsic resolution of the analysis, no improvement is gained from extremely fine binning.
The \ac{WET} predicts a broad distribution in $q^2$ and hence would not require fine binning. However, the \BKX scenario predicts a narrow resonance peak in the $q^2$ distribution, indicating a significant shape alteration. This requires finer binning to capture the shape of the distribution.
Consequently, one should adopt very fine binning in $q^2$, which will also facilitate future reinterpretations. I have therefore chosen to divide the $q^2$ range into 100 bins. In \cref{sec:q2 binning}, I present a comparative study of different binning choices, which shows that the choice of 100 bins is sufficient to capture the shape of the $q^2$ distribution in both cases.

I add an additional bin in the range ${-1 < q^2 < 0~\text{GeV}^2}$ to account for all events with unphysical or unassigned values. For the \ac{ITA} analysis, no such events are present, while in the \ac{HTA} analysis, $0.16\%$ of the events fall into this category. The origin of these unphysical values is a problem in the \ac{MC} truth matching, where an incorrect four-momentum is used to calculate $q^2$. Excluding these events has no impact on the inferred signal strength and only minimal effect on the best-fit values of the new physics parameters considered in this study (see \cref{sec:unphysical q2}). I therefore keep these events in the analysis for consistency. The corresponding bin remains unweighted in the reinterpretation process.

\begin{figure}
    \centering
    \includegraphics[width=0.8\linewidth]{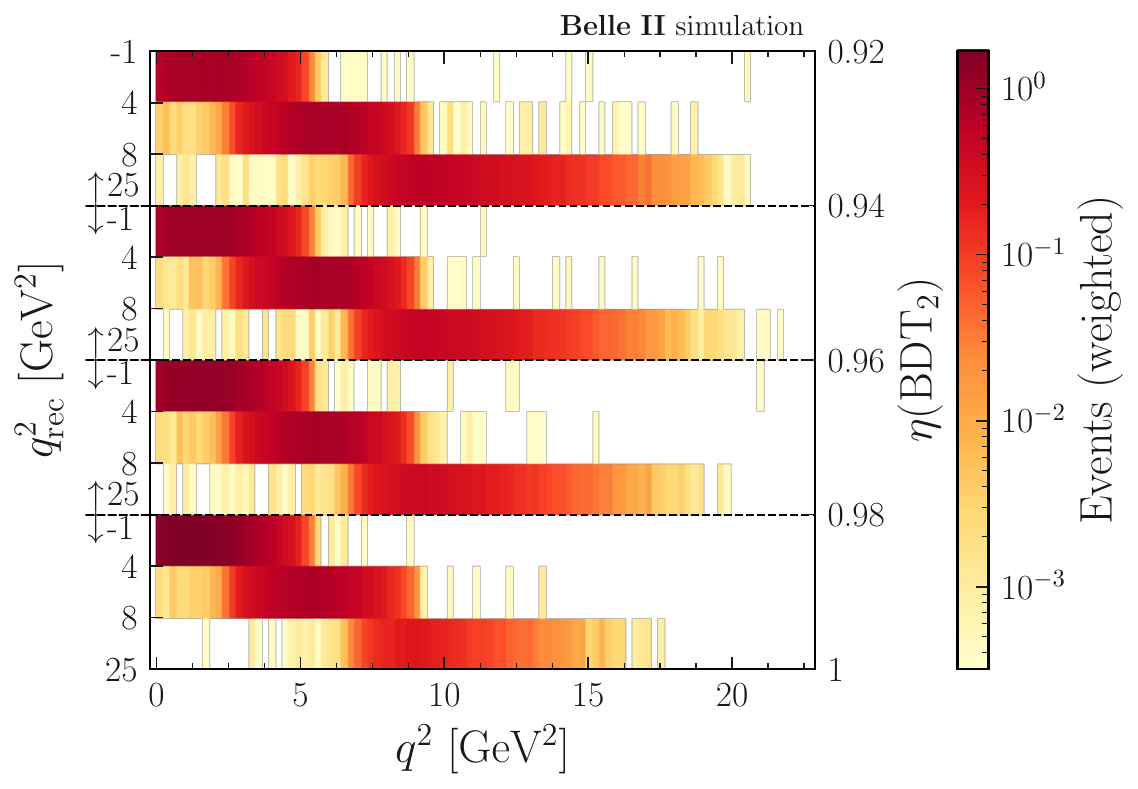}
    \includegraphics[width=0.8\linewidth]{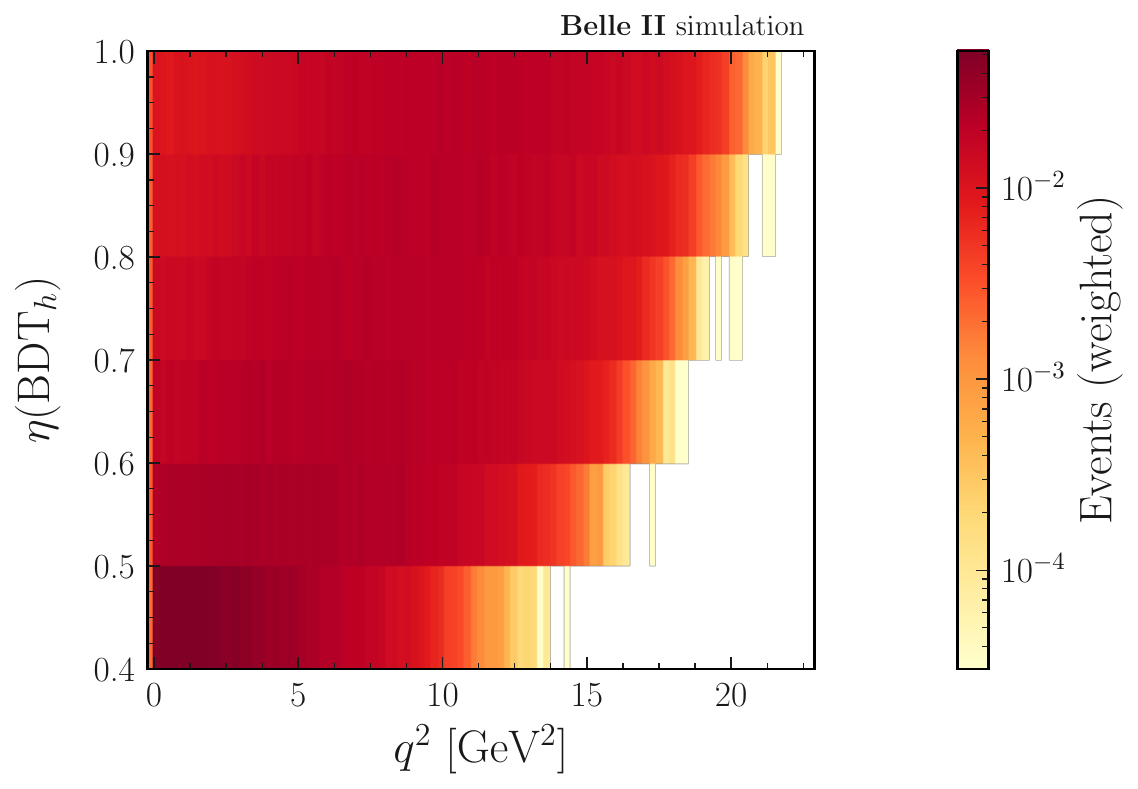}
    \caption{The \ac{ITA} (\textit{top}) and \ac{HTA} (\textit{bottom}) binned joint number densities. The horizontal axis corresponds to the generated $q^2$. The vertical axis represents the binning used in the \BKnn analysis~\cite{Belle-II:2023esi}.
    The heatmap shows the weighted signal events.}
    \label{fig:knunu-joint-nr-dens}
\end{figure}

The joint number densities, $\nu_{0,xq^2}$, for the \ac{ITA} and \ac{HTA} are shown in \cref{fig:knunu-joint-nr-dens}. Together with the \BKnn likelihood and the corresponding null distribution, one can perform a reinterpretation of the \BKnn measurement in terms of alternative theoretical models, as described in the following chapters.

A key difference between the \ac{ITA} and \ac{HTA} number densities is the correlation between the $q^2$ distributions in the kinematic bins and the reconstruction bins. 
In the \ac{ITA} analysis, which uses binning in $q^2_{\mathrm{rec}}$, this correlation is significantly higher than in the \ac{HTA} case, where it remains low. 
A strong correlation, as in the \ac{ITA} analysis, improves sensitivity to variations in the kinematic shape, since changes in the $q^2$ distribution affect the individual reconstruction bins in a more distinctive way.
I will discuss the implications of this in \cref{sec:wet paper results}.


%% file: chapters/knunu-wet.tex

\footnote{
The material presented in this chapter has been submitted to PRD~\cite{belle2pub88}. While the content largely follows that of the referenced paper, editorial modifications have been made, including restructuring of the presentation, rephrasing of portions of the text, and expansion of selected explanations. 
As this project is the result of collaborative work of the Belle~II collaboration, individual contributions of the proponents of this work, and further collaboration members are documented in \cref{sec:wet-contributions}.
}
This chapter presents the first-ever reinterpretation of the Belle~II \BKnn measurement in terms of the \ac{WET}. It demonstrates how the model-agnostic likelihood method developed in \cref{sec:method} can be used to extract new physics insights from experimental data beyond the original analysis scope.

The \ac{WET} provides a systematic, model-independent framework for parametrizing potential new physics contributions to the $b \to s \nu \bar \nu$ transition through Wilson coefficients, which quantify deviations from the \ac{SM} prediction. By reinterpreting the Belle~II data within this framework, this work presents the first-ever measurement of the $b \to s \nu \bar \nu$ \ac{WET} Wilson coefficients from experimental data with proper treatment of experimental uncertainties and kinematic shape information.\footnote{Excluding ``naive'' reinterpretation approaches that neglect kinematic shape information.}

A central question addressed in this chapter is whether the observed excess can be attributed solely to an enhancement in the vector sector (\ac{SM}-like kinematics), or whether additional contributions from scalar or tensor operators are required to optimally describe the data. This distinction is crucial, as different patterns of Wilson coefficients point to different classes of new physics models.

The analysis employs Bayesian inference to derive posterior distributions and credible intervals for the Wilson coefficients. To validate the results, goodness-of-fit tests and model comparison studies assess the compatibility of theoretical hypotheses with the observed data.

This chapter is structured as follows: 
The theoretical framework and parametrization details for the \ac{WET} predictions are summarized in \cref{sec:wet paper theory}.
The main results, including the marginal posterior distributions and credible intervals for the Wilson coefficients, are presented in \cref{sec:wet paper results}.
Statistical validation through goodness-of-fit tests is provided in \cref{sec:wet paper goodness-of-fit}, followed by a model comparison study in \cref{sec:wet paper model comparison}.
Finally, \cref{sec:wet paper discussion} summarizes the main results.


\section{Weak Effective Theory as alternative hypothesis}
\label{sec:wet paper theory}
This reinterpretation study is performed within the framework of the \ac{WET}, discussed in detail in \cref{sec:wet}. The prediction for the differential decay width of the \BKnn decay is given by \cref{eq:width}, which extends the \ac{SM} prediction by including additional operator contributions beyond the \ac{SM} vector operator.

The differential decay width is parametrized in terms of three combinations of Wilson coefficients (\cref{eq:BToKnunu-wc-sensitivity}):
\begin{equation}
    |C_{\mathrm{VL}}+C_{\mathrm{VR}}|, \quad |C_{\mathrm{SL}}+C_{\mathrm{SR}}|, \quad |C_{\mathrm{TL}}|,
\end{equation}
corresponding to vector, scalar, and tensor operator contributions, respectively. Due to the sensitivity to only the absolute values of these sums, this analysis treats each combination as a non-negative real-valued parameter.

In this reinterpretation study, the form factors are parametrized following the \ac{BSZ} parametrization~\cite{Bharucha_2016}. The parametrization is truncated at second order in the series expansion, yielding a total of 8 hadronic parameters that fully describe the relevant form factors $f_{+}(q^2)$, $f_{0}(q^2)$, and $f_{T}(q^2)$ (see \cref{sec:hadronic-matrix-elements}).

The values and uncertainties of these 8 hadronic parameters are extracted from a joint theoretical prior \ac{PDF} comprised of the 2021 lattice world average based on results by the Fermilab/MILC and HPQCD collaborations~\cite{FlavourLatticeAveragingGroupFLAG:2021npn} and recent results by the HPQCD collaboration~\cite{Parrott:2022rgu}; see \cref{tab:hadronic parameters full}.
This joint prior accounts for correlations between the form factor parameters, which are taken into account through their covariance matrices. Notably, the \ac{SM} prediction used in the original \BKnn analysis~\cite{Belle-II:2023esi} was based solely on hadronic parameters from the HPQCD collaboration~\cite{Parrott:2022rgu}, and only the vector form factor $f_{+}(q^2)$ contributes to the \ac{SM} decay width (see \cref{sec:knunu-theory,sec:wet paper analysis}).

All theoretical predictions for the differential distributions are computed using the \texttt{EOS} software package in version 1.0.16~\cite{EOSAuthors:2021xpv,EOS:v1.0.16}.


\section{Marginal posterior and credible intervals}
\label{sec:wet paper results}
This section presents the Bayesian analysis of the Belle~II \BKnn data under the \ac{WET} hypothesis, deriving posterior distributions and credible intervals for the Wilson coefficients.
The analysis employs the model-agnostic likelihood described in \cref{sec:reinterpretation knunu mall}, and the Bayesian inference framework described in \cref{sec:histfactory-bayesian}, implemented using the \texttt{bayesian pyhf} package~\cite{Feickert:2023hhr}.  The posterior distribution is sampled using \ac{MCMC} methods implemented in the \texttt{PyMC} software~\cite{AbrilPla2023PyMCAM}.

\subsection{Statistical model and prior choices}
To obtain a marginal posterior for the \ac{WET} Wilson coefficients, the \BKnn statistical model is extended by introducing 11~additional parameters beyond those already present in the original analysis. These comprise three unconstrained \acp{POI},
\begin{equation}
    \boldsymbol{\eta} = [C_{\mathrm{VL}}+C_{\mathrm{VR}}, ~C_{\mathrm{SL}}+C_{\mathrm{SR}}, ~C_{\mathrm{TL}}],
    \label{eq:wet-pois}
\end{equation}
representing the Wilson coefficient combinations of interest, along with 8~nuisance parameters that encode the hadronic form factor uncertainties. 

The hadronic parameters are correlated according to their joint prior distribution from lattice \ac{QCD}. These 8~correlated parameters are decorrelated via eigendecomposition of their covariance matrix, transforming them into 8~independent Gaussian-distributed nuisance parameters (see \cref{sec:method paper implementation,sec:histfactory-correlations}).

Importantly, the original \BKnn statistical model already contained 3~nuisance parameters for the hadronic parameters entering the \ac{SM} prediction (corresponding to the vector form factor $f_{+}(q^2)$). These are removed from the model to avoid double counting when the full set of 8~form factor parameters is introduced for the \ac{WET} interpretation. After this modification, the complete statistical model contains 239~parameters: 3~Wilson coefficients of interest, 8~hadronic nuisance parameters, and 228~systematic nuisance parameters from the original \BKnn analysis.

For the Wilson coefficient parameters, uniform priors are adopted over the range $[0,~20]$. Uniform priors are chosen to avoid assigning preference to any part of the parameter space and because no non-linear transformation of the Wilson coefficients is anticipated in this study. The upper bound of 20 is chosen to be sufficiently large to encompass the full posterior distribution, as verified a posteriori. The robustness of the results with respect to alternative prior choices is examined in detail in \cref{sec:wet paper sensitivity}.

\subsection{Posterior sampling and marginalization}
The marginal posterior distribution for the Wilson coefficients is obtained as follows. First, the joint posterior of all 239~parameters is sampled using \ac{MCMC} methods. Subsequently, this joint posterior is marginalized over the 236~nuisance parameters (8~hadronic and 228~systematic parameters), leaving only the 3-dimensional posterior distribution over the Wilson coefficients.

The \ac{MCMC} sampling employs 8~independent Markov chains, each generating 50\,000 samples, for a total of 400\,000 posterior samples. Multiple independent chains enable convergence diagnostics through inter-chain comparison (see \cref{sec:posterior-validation}). Convergence diagnostics are reported in \cref{sec:wet-posterior-validation-results}, confirming that the chains have converged to the target posterior distribution and that the sampling is reliable.

To improve computational efficiency, the symmetry of the differential decay width expression in \cref{eq:width} under sign changes of the Wilson coefficients is exploited. The sampling is performed only in the octant where all three Wilson coefficients are positive ($C_{\mathrm{VL}}+C_{\mathrm{VR}} \geq 0$, $C_{\mathrm{SL}}+C_{\mathrm{SR}} \geq 0$, $C_{\mathrm{TL}} \geq 0$), and the resulting samples are then reflected across the symmetry axes to populate all octants.

\begin{figure}
    \centering
    \includegraphics[width=0.8\linewidth]{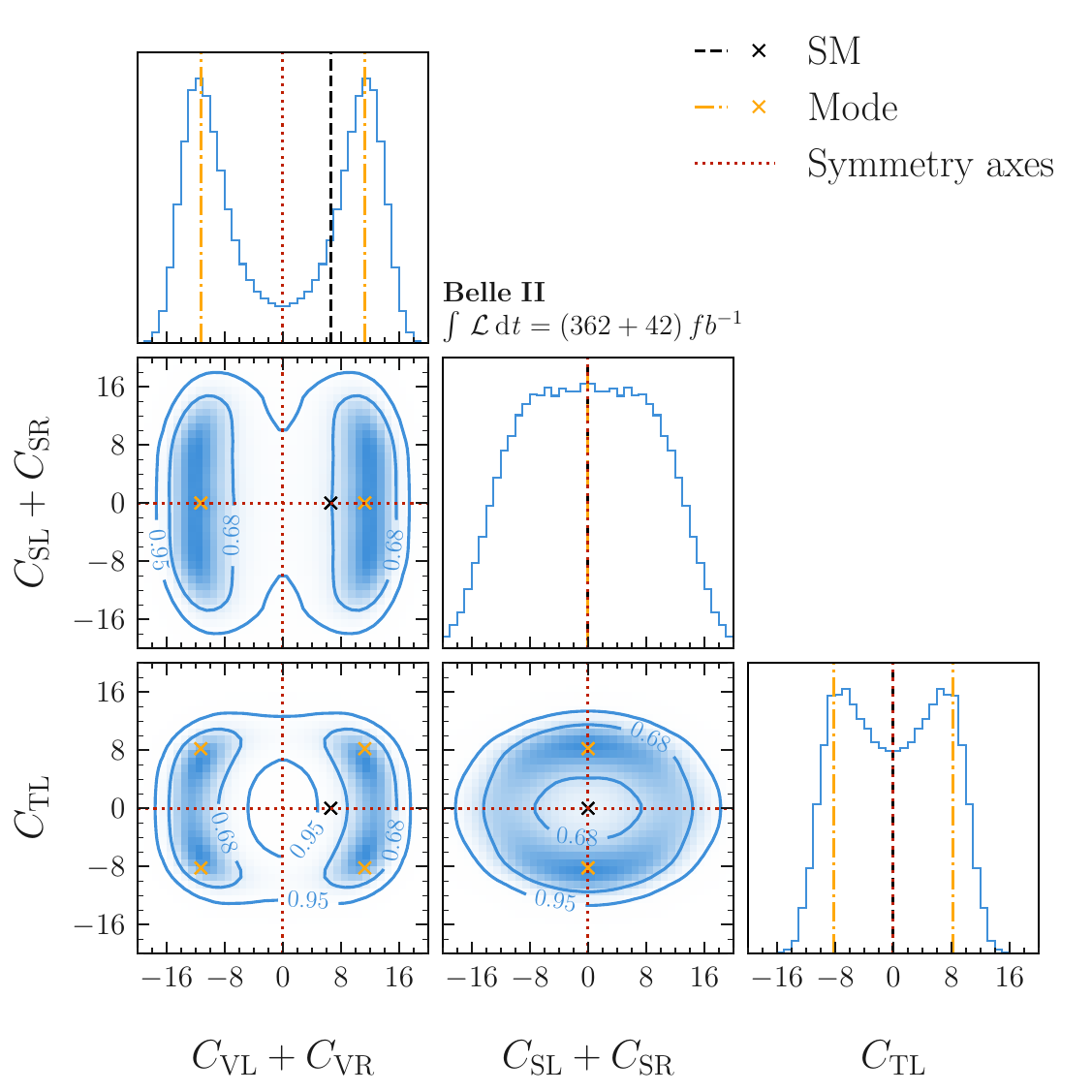}
    \caption{The marginalized posterior for the Wilson coefficients in \cref{eq:width}. The convention adopted is that $C_{\mathrm{VL}}+C_{\mathrm{VR}}$, $C_{\mathrm{SL}}+C_{\mathrm{SR}}$, and $C_{\mathrm{TL}}$ are real-valued.
    Diagonal and off-diagonal panels show the 1-dimensional and 2-dimensional sample density \acp{PDF} on a linear scale, respectively. The overall scale is omitted, as all relevant information is contained in the shape of the distribution. The contours indicate $68\%$ and $95\%$ credible intervals. The dashed black lines and cross mark the \ac{SM} point; the dash-dotted yellow lines and cross indicate the posterior mode; dotted red lines mark the symmetry axes used for sample symmetrization.
    }
    \label{fig:wet-posterior}
\end{figure}

The full set of 1- and 2-dimensional marginal posterior distributions and the resulting credible intervals at $68\%$ and $95\%$ probability is shown in \cref{fig:wet-posterior}.

\subsection{Credible intervals and posterior mode}
From the 1-dimensional marginal posterior distributions, I calculate the \acp{HDI} (see \cref{sec:credible-intervals}) at 68\% and 95\% probability for the Wilson coefficient combinations. The posterior mode and the credible intervals are summarized in \cref{tab:wet results}. 

\begin{table}[ht]
    \renewcommand{\arraystretch}{1.5}
    \caption{The mode of the posterior, and \ac{HDI} at 68\% and 95\% for the (sums of the) \ac{WET} Wilson coefficients in \cref{eq:width}, derived from the posterior in \cref{fig:wet-posterior}.}
    \centering
    \begin{tabularx}{\linewidth}{@{\extracolsep{\fill}}lYYY}
        \toprule \midrule
        \textbf{Parameters} & \textbf{Mode} & \textbf{68\% HDI} & \textbf{95\% HDI} \\
        \midrule
         $
         \begin{aligned}
            &|C_\mathrm{VL}+C_\mathrm{VR}|\\
            &|C_\mathrm{SL}+C_\mathrm{SR}|\\
            &|C_\mathrm{TL}|
        \end{aligned}
        $ & $
        \begin{aligned}
            11.3&\\
            0.0& \\
            8.2&
        \end{aligned}
        $ & $
         \begin{aligned}
            [7.8, ~14.6]&\\
            [0.0, ~9.6]& \\
            [2.3, ~9.6]&
        \end{aligned}
        $ & $
        \begin{aligned}
            [1.9, ~16.2]&\\
            [0.0, ~15.4]& \\
            [0.0, ~11.2]&
        \end{aligned}
        $ \\
        \midrule \bottomrule
    \end{tabularx}
    \label{tab:wet results}
\end{table}

\begin{figure}
    \centering
    \includegraphics[width=0.8\linewidth]{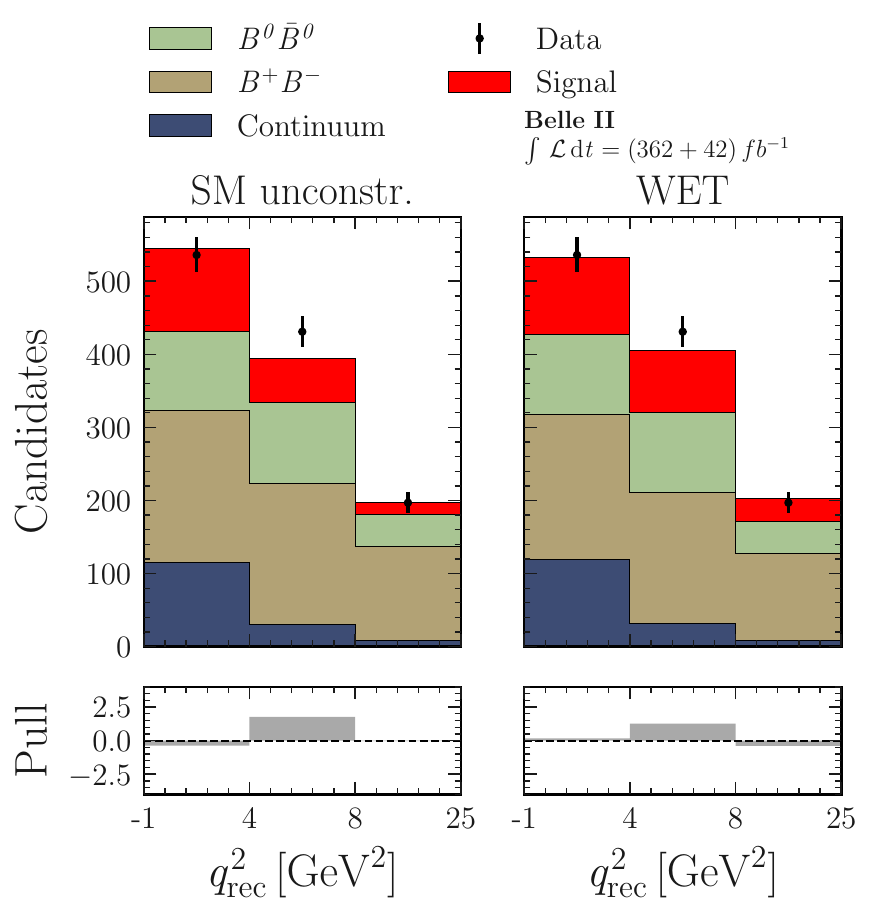}
    \caption{Observed and predicted best-fit yields in the \ac{HSR} of the analysis. The signal is shown for the unconstrained \BKnn \ac{SM} (\textit{left}) and the \ac{WET} (\textit{right})  predictions.
    The predicted background yields are shown individually for the neutral and charged \B-meson decays, and the summed five continuum categories. 
    Pulls are shown in the lower panels.
    }
    \label{fig:post-fit}
\end{figure}

Direct comparison of the observed data with the predicted yields at the posterior mode (see \cref{tab:wet results}) for the unconstrained \BKnn \ac{SM} ($\mu$ unconstrained) and the \ac{WET} is shown in \cref{fig:post-fit}, for the \ac{HSR} of the analysis (\ac{ITA}, $\eta(\BDT2)>0.98$). 
The \ac{WET} model provides a better fit to the data than the unconstrained \ac{SM} prediction, as indicated by the smaller pull values.

\subsection{Physical interpretation of the posterior}
The posterior distribution reveals a clear deviation from the \ac{SM} prediction in the vector sector, consistent with the original analysis result of a signal strength $\mu = 4.6 \pm 1.3$~\cite{Belle-II:2023esi}. This connection can be understood quantitatively as follows: Since the Wilson coefficients enter quadratically into the differential branching ratio (see \cref{eq:width}), the decay rate scales as $|C_{\gamma \delta}|^2$. For a pure vector contribution, the signal strength $\mu$ should satisfy
\begin{equation}
\mu = \frac{|C_\mathrm{VL}+C_\mathrm{VR}|^2}{|C_\mathrm{VL}^{SM}|^2},
\end{equation}
where $C_\mathrm{VL}^{SM} = 6.6$ is the \ac{SM} value. This implies an expected vector coefficient of
\begin{equation}
|C_\mathrm{VL}+C_\mathrm{VR}| \simeq \sqrt{4.6} \cdot 6.6 \simeq 14.2,
\end{equation}
assuming negligible contributions from scalar and tensor operators. The observed posterior mode of $11.3$ (see \cref{tab:wet results}) is slightly lower than this naive expectation, indicating that the data prefer a model in which additional operator contributions share some of the excess rate.

To verify this interpretation, I perform a constrained fit fixing $C_\mathrm{SL}+C_\mathrm{SR} = 0$ and $C_\mathrm{TL} = 0$, allowing only the vector coefficient to vary. This yields 
\begin{equation}
    C_\mathrm{VL}+C_\mathrm{VR} = 14.1 \pm 2.1,
\end{equation}
which is indeed close to the naive expectation. To quantify the consistency between this vector-only interpretation and the signal strength measurement, I evaluate the ratio
\begin{equation}
    \frac{|C_\mathrm{VL}+C_\mathrm{VR}|^2}{\mu \, |C_\mathrm{VL}^{SM}|^2} = 1.0 \pm 0.4,
\end{equation}
which equals unity within uncertainties, confirming the expected relationship.

Importantly, the posterior distribution reveals a preference for a non-zero tensor contribution, with the posterior mode at $C_\mathrm{TL} = 8.2$. This finding indicates that a pure \ac{SM}-like signal template does not provide the optimal description of the observed kinematic distributions. Instead, the data favour a linear combination of vector and tensor contributions, each with a distinct kinematic shape. This preference is detectable only because kinematic shape information is preserved in the reinterpretation procedure.

The marginal posterior and corresponding credible intervals derived independently from the \ac{ITA} and \ac{HTA} likelihoods are presented in \cref{sec:ita and hta posteriors}. This decomposition provides insight into which analysis drives the observed deviations from the \ac{SM}. The comparison reveals that the deviations in the vector and tensor sectors are features of the \ac{ITA}-based posterior. The \ac{HTA} plays a crucial complementary role by constraining the overall normalization. This normalization constraint is particularly important for the scalar and tensor sectors.

\subsection{Prior sensitivity}
\label{sec:wet paper sensitivity}
Bayesian inference inherently depends on the choice of prior distributions. To assess the robustness of the presented results and quantify the influence of prior choices, I perform a prior sensitivity analysis by comparing the baseline uniform priors to two physically motivated alternative prior specifications.

The first alternative employs \textit{truncated-normal priors} centred on the \ac{SM} expectation, representing a scenario in which deviations from the \ac{SM} are disfavoured a priori:
\begin{equation}
p\left( \eta_i \right) = 
\begin{cases}
    \mathcal{N}(\eta_i | \mu=C_i^{\mathrm{SM}}, \sigma=20) &~ \eta_i \geq 0 \\
    0 &~ \eta_i < 0
\end{cases},
\label{eq:wet priors normal}
\end{equation}
where $\eta_i$ correspond to the Wilson coefficient combinations in \cref{eq:wet-pois}, and ${C_i^{\mathrm{SM}} \in [6.6, 0.0, 0.0]}$ represents the \ac{SM} point. The standard deviation $\sigma=20$ is chosen to allow substantial deviations while maintaining a preference for \ac{SM}-like values.

The second alternative adopts \textit{uniform priors in the squared Wilson coefficients}, which translates to triangular priors for the Wilson coefficients themselves:
\begin{equation}
    p\left( \eta_i \right) \propto  \begin{cases}
        \eta_i &~ \eta_i \leq 30\\
        0 &~ \eta_i > 30
        \end{cases}.
    \label{eq:wet priors triangular}
\end{equation}
This choice is motivated by the fact that Wilson coefficients enter the decay rate quadratically (see \cref{eq:width}), so uniform priors on $|C_{\gamma\delta}|^2$ correspond to uniform priors on the decay rate enhancement. These priors favour larger values of the Wilson coefficients relative to the uniform priors.

The resulting posterior modes and credible intervals for both alternative prior choices are summarized in \cref{tab:wet results sensitivity}. The comparison reveals that the vector and tensor Wilson coefficients are the most robustly determined, with the posterior modes and credible intervals showing minimal sensitivity to the prior choice.

In contrast, the scalar Wilson coefficients exhibit the largest prior dependence, with the posterior mode shifting from $0.0$ (uniform and truncated-normal priors) to $8.9$ (triangular prior). This heightened sensitivity reflects the weak experimental constraints on the scalar sector due to the kinematic distribution peaking at high $q^2$ (see \cref{fig:knunu-theory}), where the experimental efficiency is low (see \cref{fig:knunu-efficiency}). When the data provide weak constraints, the posterior becomes more strongly influenced by the prior choice, as seen here.

\begin{table}[ht]
    \renewcommand{\arraystretch}{1.5}
    \caption{The posterior modes, and \acp{HDI} at 68\% and 95\% for the (sums of the) \ac{WET} Wilson coefficients in \cref{eq:width}, for alternative prior choices (cf. \cref{tab:wet results}). }
    \centering
    \begin{tabularx}{\linewidth}{@{\extracolsep{\fill}}llrYY}
        \toprule \midrule
        \textbf{Priors} & \textbf{Parameters} & \textbf{Mode} & \textbf{68\% HDI} & \textbf{95\% HDI} \\
        \midrule
        \cref{eq:wet priors normal} &
         $
         \begin{aligned}
            &|C_\mathrm{VL}+C_\mathrm{VR}|\\
            &|C_\mathrm{SL}+C_\mathrm{SR}|\\
            &|C_\mathrm{TL}|
        \end{aligned}
        $ & $
        \begin{aligned}
            11.4& \\
            0.0& \\
            7.7&
        \end{aligned}
        $ & $
         \begin{aligned}
            [9.0, 14.6]&\\
            [0.0, 9.2]& \\
            [1.5, 8.8]&
        \end{aligned}
        $ & $
        \begin{aligned}
            [2.2, 16.4]&\\
            [0.0, 14.7]& \\
            [0.0, 11.0]&
        \end{aligned}
        $
        \\
        \midrule
        \cref{eq:wet priors triangular} &
         $
         \begin{aligned}
            &|C_\mathrm{VL}+C_\mathrm{VR}|\\
            &|C_\mathrm{SL}+C_\mathrm{SR}|\\
            &|C_\mathrm{TL}|
        \end{aligned}
        $ & $
        \begin{aligned}
            11.6&\\
            8.9&\\
            7.2&
        \end{aligned}
        $ & $
         \begin{aligned}
            [8.2, ~14.0]&\\
            [4.6, ~12.6]& \\
            [3.9, ~9.6]&
        \end{aligned}
        $ & $
        \begin{aligned}
            [4.2, ~16.0]&\\
            [1.3, ~15.6]& \\
            [1.4, ~11.7]&
        \end{aligned}
        $\\
        \midrule \bottomrule
    \end{tabularx}
    \label{tab:wet results sensitivity}
\end{table}

\subsection{Naive reinterpretation}
\label{sec:wet paper naive reinterpretation}
The key advantage of the model-agnostic likelihood approach is that kinematic shape information is utilized for reinterpretation.
To quantify the impact of this feature and demonstrate its necessity, I perform a comparative ``naive'' reinterpretation that discards all kinematic shape information.

Technically, this naive approach is implemented by integrating the joint number densities over the entire kinematic range. This produces a single overall normalization constraint, discarding all differential information. This procedure is equivalent to performing inference using only the total number of observed signal events, without knowledge of how those events are distributed in $q^2$.

\begin{figure}
    \centering
    \includegraphics[width=0.8\textwidth]{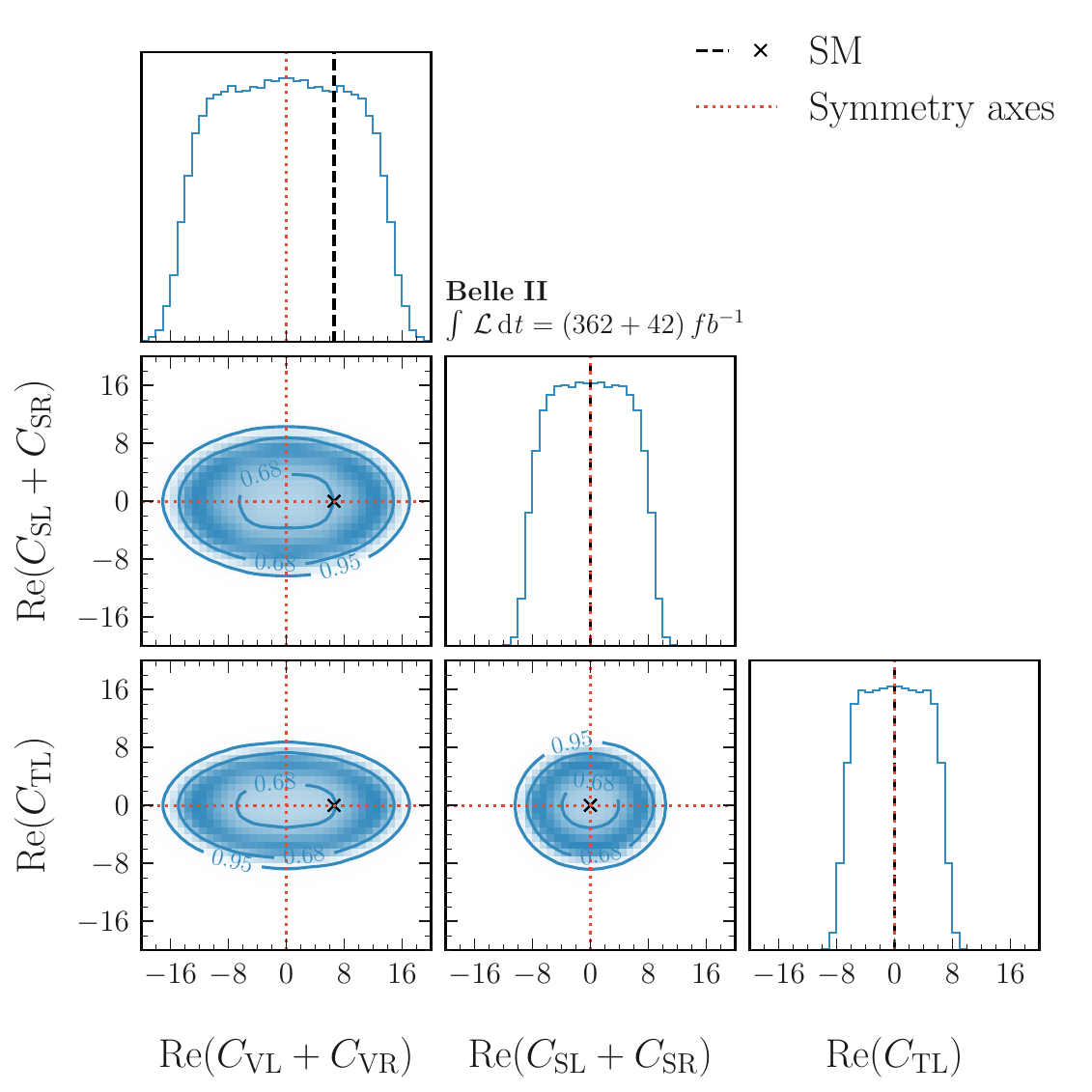}
    \caption{
    The marginalized posterior for the Wilson coefficients in \cref{eq:width}, based on a naive reinterpretation approach. The convention adopted is that $C_{\mathrm{VL}}+C_{\mathrm{VR}}$, $C_{\mathrm{SL}}+C_{\mathrm{SR}}$, and $C_{\mathrm{TL}}$ are real-valued.
    Diagonal and off-diagonal panels show the 1-dimensional and 2-dimensional sample density \acp{PDF} on a linear scale, respectively. The overall scale is omitted, as all relevant information is contained in the shape of the distribution. The contours indicate $68\%$ and $95\%$ credible intervals. The dashed black lines and cross mark the \ac{SM} point; dotted red lines mark the symmetry axes used for sample symmetrization.
    }
    \label{fig:wet-posterior-naive}
\end{figure}

The resulting marginalized posterior distribution from this naive approach is shown in \cref{fig:wet-posterior-naive}.

Within this naive framework, the resulting 95\% credible intervals are
${|C_\mathrm{VL}+C_\mathrm{VR}| < 14.4}$, ${|C_\mathrm{SL}+C_\mathrm{SR}|< 8.5}$, and ${|C_\mathrm{TL}|< 7.1}$.
Comparing these naive bounds to the full kinematic reinterpretation results in \cref{tab:wet results} (95\% HDI: $[1.9, 16.2]$, $[0.0, 15.4]$, $[0.0, 11.2]$, respectively) reveals systematic biases introduced by neglecting shape information.

The most significant difference can be observed in the vector sector, where the naive upper limit of $14.4$ underestimates the full result's upper bound of $16.2$. In addition, the naive approach yields only an upper limit, failing to establish a lower bound, whereas the full analysis provides a 95\% credible interval starting at $1.9$.

For the scalar and tensor sectors, the naive bounds appear tighter than the full analysis bounds. However, this apparent improvement is illusory --- it reflects biased central values rather than genuine enhanced precision. Any combination of vector, scalar, and tensor contributions that yields the observed total excess rate is equally plausible in the naive framework.

These systematic biases and loss of discriminating power highlight the importance of preserving kinematic shape information in reinterpretation studies.


\section{Goodness-of-fit}
\label{sec:wet paper goodness-of-fit}
Beyond obtaining credible intervals for the Wilson coefficients, it is important to validate that the \ac{WET} model provides a good description of the observed data.

I perform goodness-of-fit tests for several competing hypotheses: the \ac{WET} model with three free Wilson coefficients, the unconstrained \ac{SM} where only the signal strength $\mu$ is allowed to vary, and a background-only hypothesis that assumes no \BKnn signal contribution. The unconstrained \ac{SM} considers a signal which exhibits \ac{SM}-like kinematics, but with an overall free normalization (equivalent to a vector-only contribution of the \ac{WET}).
A detailed summary of these models and their defining characteristics is provided in \cref{tab:models}. The statistical framework for the goodness-of-fit test is described in \cref{sec:goodness-of-fit}.

\begin{table}[ht]
    \renewcommand{\arraystretch}{1.5}
    \caption{A summary of the models under consideration, their corresponding theoretical predictions, references for the hadronic parameters, and the number of hadronic parameters included in each model is provided.}
    \centering
    \begin{tabularx}{\linewidth}{@{\extracolsep{\fill}}llYYY}
        \toprule \midrule
        \textbf{Model} & \textbf{Prediction} & \textbf{Free param.} &\multicolumn{2}{c}{\textbf{Hadronic param.}} \\
        & & & \textbf{Ref.} & \textbf{Nr.} \\
        \midrule
        \ac{WET} & \cref{eq:width} & $C_{\mathrm{VL}}+C_{\mathrm{VR}}$, $C_{\mathrm{SL}}+C_{\mathrm{SR}}$, $C_{\mathrm{TL}}$ & \cite{FlavourLatticeAveragingGroupFLAG:2021npn,Parrott:2022rgu} &  8\\ \addlinespace \addlinespace
        \ac{SM} unconstrained & \makecell[l]{\cref{eq:width}, \\ $C_{\rm VL} \geq 0$, \\ $C_{\gamma \delta}=0$ otherwise} & $\mu$ &\cite{Parrott:2022rgu} & 3\\ \addlinespace \addlinespace
        \ac{SM} constrained & \makecell[l]{\cref{eq:width}, \\ $C_{\rm VL}=6.6$,\\ $C_{\gamma \delta}=0$ otherwise, \\ 5\% norm. unc.} & -- & \cite{Parrott:2022rgu} & 3\\ \addlinespace \addlinespace
        background-only (BKG) & No signal & -- & -- & --\\
        \midrule \bottomrule
    \end{tabularx}
    \label{tab:models}
\end{table}

The goodness-of-fit $P$-value is calculated from (see \cref{sec:goodness-of-fit})
\begin{equation}
        P_{\rm gof} = \int_{t_{\rm obs}}^\infty dt_{\rm gof} ~ p(t_{\rm gof}), \quad
        t_{\rm gof} = -2 \ln \frac{p(\boldsymbol{n}, \boldsymbol{a} \mid \hat{\boldsymbol{\eta}}, \hat{ \boldsymbol{\chi}})}{p_{\rm sat}(\boldsymbol{n}, \boldsymbol{a} \mid \bar{\boldsymbol{\chi}})},
        \label{eq:gof-pvalue-wet}
\end{equation}
where $p(\boldsymbol{n}, \boldsymbol{a} \mid \hat{\boldsymbol{\eta}}, \hat{ \boldsymbol{\chi}})$ is the likelihood evaluated at the best-fit point for the model under consideration, with $\hat{\boldsymbol{\eta}}$ denoting the fitted \acp{POI} (Wilson coefficients or signal strength) and $\hat{\boldsymbol{\chi}}$ the fitted nuisance parameters. The saturated likelihood $p_{\rm sat}(\boldsymbol{n}, \boldsymbol{a} \mid \bar{\boldsymbol{\chi}})$ represents a perfect fit where the expected event rates in each bin of the \histfactory likelihood~\cite{histfactory} are set exactly equal to the observed data, with all constraint terms maximized at $\bar \chi$.

The \ac{PDF} of the test statistic $p(t_{\rm gof})$ is obtained from fits to toy data sampled from
$
    p(\boldsymbol{n}, \boldsymbol{a} \mid \hat{\boldsymbol{\eta}}, \hat{ \boldsymbol{\chi}})
$~\cite{Cranmer:2014lly}.
The goodness-of-fit $P$-values are calculated as the fraction of toys with $t > t_{\rm obs}$, where $t_{\rm obs}$ is the value of the test statistic for the observed data.

\begin{figure}
    \centering
    \includegraphics[width=0.6\textwidth]{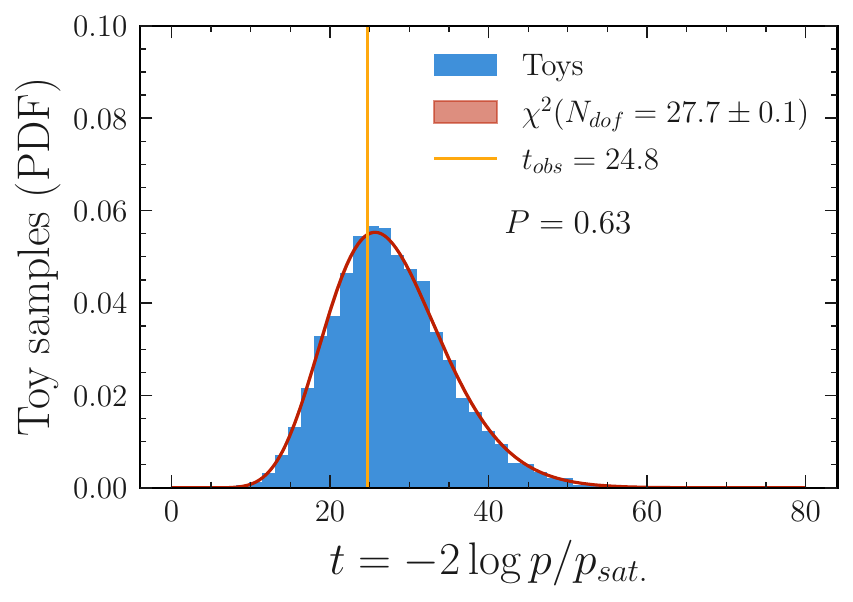}
    \caption{
    The distribution of the goodness-of-fit test statistic for the \ac{WET} model, obtained from 10\,000 toys. The value obtained from a fit to the data is shown as vertical yellow line. 
    The asymptotic $\chisq$-distribution (red band) corresponds to the best-fit $N_{dof}$, which is shown with the corresponding uncertainty band.
    }
    \label{fig:chi2-wet}
\end{figure}

For the \ac{WET} model, the distribution of the goodness-of-fit test statistic obtained from 10\,000 toy experiments is shown in \cref{fig:chi2-wet}. Under standard asymptotic assumptions, one expects this distribution to follow a $\chisq$-distribution with $N_{\rm dof} = 30 - 3 = 27$ \ac{d.o.f.} (30 bins in the analysis minus 3 free Wilson coefficients). However, a fit to the distribution from toys yields $N_{\rm dof}=27.7\pm0.1$, showing a small but statistically significant deviation.

This deviation from the asymptotic expectation can be understood from the structure of the \ac{WET} model: the Wilson coefficients can only contribute positively to the decay rate (they enter quadratically in \cref{eq:width}), and the accessible parameter space is therefore bounded at zero. Such boundary effects violate the conditions required for Wilks' theorem, which guarantees asymptotic $\chisq$ behaviour only when parameters are unconstrained. The presence of such boundaries modifies the test statistic distribution, as discussed in Reference~\cite{Bernlochner:2022oiw}. For this reason, I use the distribution from toys rather than the asymptotic approximation to calculate $P$-values.

Test statistic distributions for the unconstrained \ac{SM} and the background-only model are reported in \cref{sec:goodness-of-fit-wet}.

The goodness-of-fit $P$-values for all models are summarized in \cref{tab:bayes factors}. Both the \ac{WET} ($P_{\rm gof}=0.63$) and the unconstrained \ac{SM} ($P_{\rm gof}=0.58$) show good fits to the data. The background-only model yields a lower $P$-value of $0.13$, indicating a less favourable fit to the data.

\section{Model comparison}
\label{sec:wet paper model comparison}
Model comparison quantifies the \textit{relative} support that the data provide for competing hypotheses. I perform both Bayesian and frequentist model comparisons to assess the relative performance of the \ac{WET} and unconstrained \ac{SM} hypotheses against two reference models: the background-only hypothesis (BKG), and the constrained \ac{SM} (see \cref{tab:models}).

The constrained \ac{SM} model (see \cref{tab:models}) fixes $\mu=1$ to the \ac{SM} prediction. To account for theoretical normalization uncertainties in the \ac{SM} prediction, this model includes a 5\% relative uncertainty, which combines in quadrature a $4.4\%$ uncertainty from the \ac{CKM} matrix elements $|V_{ts}^*V_{tb}|^2$ and a $2.3\%$ uncertainty from the Wilson coefficient calculation $|C_{\rm VL}^{\rm SM}|^2$~\cite{Parrott:2022zte}.

\subsection{Bayes factors}
Bayesian model comparison employs the Bayes factor, defined as the ratio of marginal likelihoods between two competing models (see \cref{sec:global-comparison}).

The Bayes factors are computed using sequential \ac{MC} sampling methods~\cite{AbrilPla2023PyMCAM,10.1093/gji/ggt180}, which provide numerical estimates of the marginal likelihoods for each model. The results are presented in \cref{tab:bayes factors}.

Both the \ac{WET} ($\log_{10} B_{\rm BKG} = 1.78$) and the unconstrained \ac{SM} ($\log_{10} B_{\rm BKG} = 2.02$) exceed Jeffreys's threshold~\cite{jeffreys1961theory} of $\log_{10} B > 1.5$ for a \textit{very strong} model preference (see \cref{tab:jeffreys-scale}).

Comparing both models to the constrained \ac{SM}, I find \textit{substantial} model preference ($\log_{10} B > 0.5$) for the unconstrained scenarios: $\log_{10} B_{\rm SM}^{\rm constr.} = 0.68$ for the \ac{WET} and $\log_{10} B_{\rm SM}^{\rm constr.} = 0.92$ for the unconstrained \ac{SM}.

Interestingly, the unconstrained \ac{SM} shows slightly stronger evidence than the \ac{WET}, suggesting that the data do not strongly require the additional \ac{d.o.f.} provided by the scalar and tensor Wilson coefficients --- the vector enhancement alone largely suffices to explain the observations.

\begin{table}[ht]
    \renewcommand{\arraystretch}{1.5}
    \caption{ 
        Bayes factors and goodness-of-fit $P$-values for the \ac{WET} and unconstrained \ac{SM}. The Bayes factors are reported relative to the background-only hypothesis ($B_{\rm BKG}$) and the constrained \BKnn \ac{SM} hypothesis ($B_{\rm SM}^{\rm constr.}$).
        }
    \centering
    \begin{tabularx}{\linewidth}{@{\extracolsep{\fill}}lYYY}
        \toprule \midrule
        \textbf{Model} & $\boldsymbol{\log_{10} B_{\rm BKG}}$ & $\boldsymbol{\log_{10} B_{\rm SM}^{\rm constr.}}$ & $\boldsymbol{P_{\rm gof}}$\\
        \midrule
        \ac{WET} & 1.78 & 0.68 & 0.63\\
        \ac{SM} unconstrained  & 2.02 & 0.92 & 0.58\\
        background-only  & 0.00 & -1.08 & 0.13\\
        \midrule \bottomrule
    \end{tabularx}
    \label{tab:bayes factors}
\end{table}

\subsection{Frequentist hypothesis test}
To complement the Bayesian model comparison and provide a frequentist perspective, I perform hypothesis tests that quantify the significance of the signal hypotheses relative to the background-only hypothesis.

The tests are performed using the $P$-value defined as
\begin{equation}
        P = \int_{t_{obs}}^\infty dt ~ p(t), \quad
        t = -2 \ln \frac{p(\boldsymbol{n}, \boldsymbol{a} \mid \boldsymbol{\eta} = \boldsymbol{0}, \hat{\hat{ \boldsymbol{\chi}}})}{p(\boldsymbol{n}, \boldsymbol{a} \mid \hat{\boldsymbol{\eta}}, \hat{ \boldsymbol{\chi}})},
        \label{eq:hypothesis-test-pvalue}
\end{equation}
where $p(\boldsymbol{n}, \boldsymbol{a} \mid \hat{\boldsymbol{\eta}}, \hat{ \boldsymbol{\chi}})$ is the likelihood at the best-fit point with all parameters free to vary, and $p(\boldsymbol{n}, \boldsymbol{a} \mid \boldsymbol{\eta} = \boldsymbol{0}, \hat{\hat{ \boldsymbol{\chi}}})$ is the likelihood with Wilson coefficients or signal strength fixed to zero (background-only hypothesis) but nuisance parameters still optimized at $\hat{\hat{ \boldsymbol{\chi}}}$.

The distribution of the test statistic $p(t)$ is determined
by generating toy datasets from the background-only hypothesis $p(\boldsymbol{n}, \boldsymbol{a} \mid \boldsymbol{\eta} = \boldsymbol{0}, \hat{\hat{ \boldsymbol{\chi}}})$ and computing the test statistic for each toy. The $P$-value is the fraction of toys yielding test statistics larger than the one observed, $t > t_{\rm obs}$.

\begin{figure}
    \centering
    \includegraphics[width=0.6\textwidth]{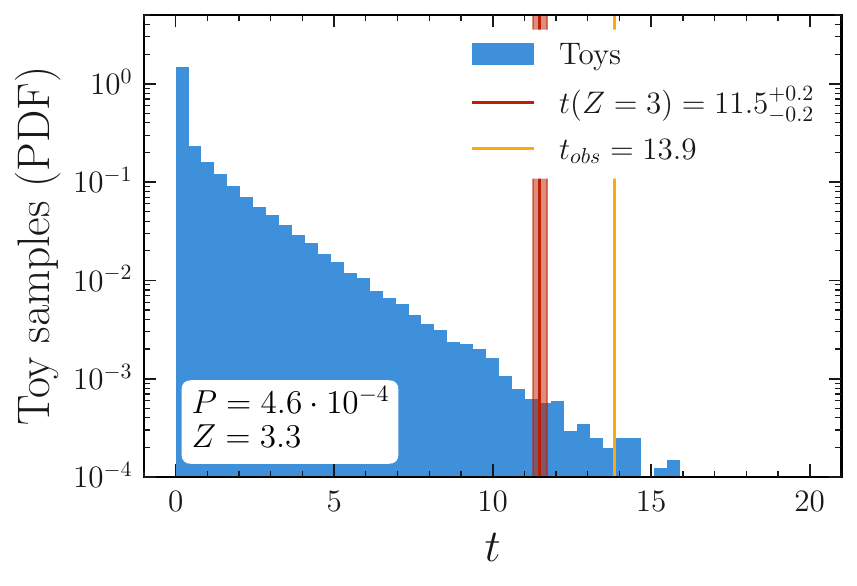}
    \caption{
        Distribution of the test statistic for the \ac{WET} hypothesis over the background-only hypothesis, obtained from 100\,000 toy experiments. 
        The value from the fit to data is shown as a yellow line. 
        The vertical red line indicates the test statistic corresponding to a significance of $Z=3$, with the shaded band denoting its uncertainty.
    }
    \label{fig:hypo-wet}
\end{figure}

For the \ac{WET} hypothesis, the hypothesis test yields a $P$-value of $P = 4.6 \cdot 10^{-4}$, corresponding to a significance of $Z = 3.3\sigma$ with respect to the background-only hypothesis. The distribution of the test statistic obtained from 100\,000 toy experiments is shown in \cref{fig:hypo-wet}, where the observed value $t_{\rm obs} = 13.9$ falls in the far tail of the distribution.

As discussed in \cref{sec:wet paper goodness-of-fit}, the test statistic distribution for the \ac{WET} does not follow the asymptotic $\chisq$-distribution due to the positivity constraints on Wilson coefficients. The distribution from toys shows that the test statistic value corresponding to $Z = 3.0\sigma$ significance is $t(Z=3) = 11.5^{+0.2}_{-0.2}$, which is notably smaller than the observed $t_{\rm obs} = 13.9$. This confirms that the observed excess represents a significant deviation from the background-only hypothesis.

\begin{figure}
    \centering
    \includegraphics[width=0.6\textwidth]{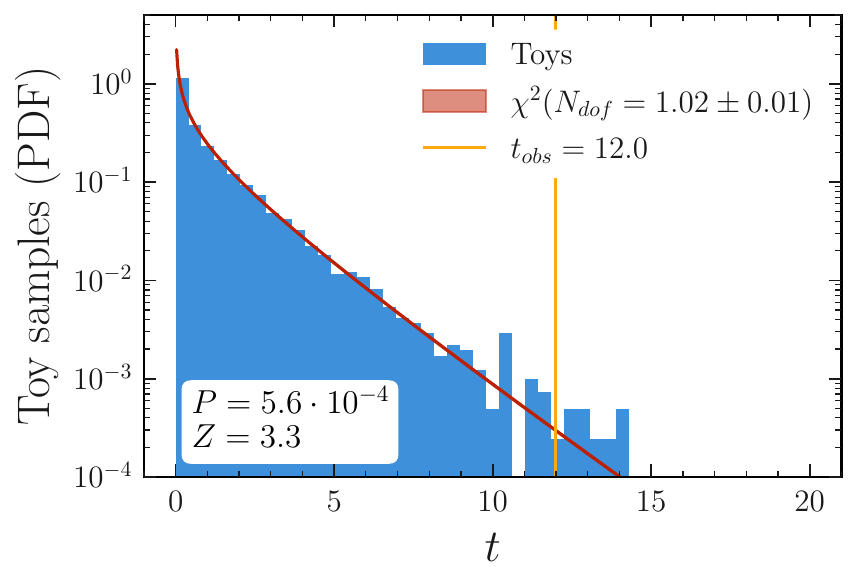}
    \caption{
        Distribution of the test statistic for the unconstrained \ac{SM} hypothesis over the background-only hypothesis, obtained from 10\,000 toy experiments. 
        The value from the fit to data is shown as a vertical yellow line.
        The asymptotic $\chisq$-distribution (red band) corresponds to the best-fit $N_{dof}$, which is shown with the corresponding uncertainty band.
    }
    \label{fig:hypo-sm}
\end{figure}

For comparison, the unconstrained \ac{SM} hypothesis yields a $P$-value of $P = 5.6 \cdot 10^{-4}$, corresponding to a significance of $Z = 3.3\sigma$. The test statistic distribution from 10\,000 toy experiments is shown in \cref{fig:hypo-sm}. In this case, the distribution closely follows a $\chisq$-distribution with $N_{\rm dof} = 1.02 \pm 0.01$ \ac{d.o.f.}, as expected for a single free parameter. The agreement with the asymptotic expectation validates the use of asymptotic formulae for this simpler hypothesis. The observed test statistic is $t_{\rm obs} = 12.0$.

The significances of $Z = 3.3\sigma$ for both the \ac{WET} and unconstrained \ac{SM} hypotheses indicate that both models provide comparable descriptions of the deviation from the background-only hypothesis. The slightly smaller $P$-value for the \ac{WET} suggests a marginally better fit, consistent with the posterior preference for tensor contributions discussed earlier, though the difference is not substantial.

One can compare these results to the original Belle~II publication~\cite{Belle-II:2023esi}, which reported a significance of $Z = 3.5\sigma$. That value was obtained using the approximation $Z = \sqrt{t_{\rm obs}}$, valid for an asymptotic one-sided test with one \ac{d.o.f.} (corresponding to an upper-limit test statistic as defined in Reference~\cite{Cowan:2010js}). The difference between $3.5\sigma$ and the present $3.3\sigma$ arises mainly from the fact that the present analysis uses the likelihood ratio test statistic, rather than an upper limit test statistic. The updated form-factor parameters also play a minor role. 
Nonetheless, the conclusions remain consistent.

\section{Discussion}
\label{sec:wet paper discussion}
This chapter presents the first comprehensive reinterpretation of the Belle~II \BKnn measurement~\cite{Belle-II:2023esi} in terms of the \ac{WET} framework, demonstrating the power of the model-agnostic likelihood method developed in \cref{sec:method}. Starting from the published analysis likelihood and joint number densities, I have constructed full Bayesian posterior distributions and derived credible intervals for the $b\to s \nu \bar \nu$ \ac{WET} Wilson coefficients. This represents the first time that experimental data have been used to constrain this set of coefficients with proper treatment of uncertainties and kinematic shape information.

The principal findings are summarized in \cref{tab:wet results}. The posterior mode for the Wilson coefficient combinations is located at ${(|C_\mathrm{VL}+C_\mathrm{VR}|,\, |C_\mathrm{SL}+C_\mathrm{SR}|,\, |C_\mathrm{TL}|) = (11.3,\, 0.0,\, 8.2)}$, with 95\% credible intervals of $[1.9,\, 16.2]$, $[0.0,\, 15.4]$, and $[0.0,\, 11.2]$, respectively.

The data require an enhancement in the vector sector beyond the \ac{SM} expectation of $C_{\rm VL}^{\rm SM} = 6.6$. The 95\% credible interval excludes the \ac{SM} value, and the posterior sharply peaks at $C_{\rm VL}+C_{\rm VR} \approx 11.3$, consistent with the large signal strength $\mu = 4.6 \pm 1.3$ reported in the original analysis.

The data exhibit a preference for an additional tensor contribution, with the posterior mode at $C_{\rm TL} = 8.2$. The preserved kinematic shape information allows for the identification of this preferred admixture of vector and tensor contributions.

The scalar Wilson coefficient remains weakly constrained, with the posterior consistent with zero but allowing values up to $|C_{\rm SL}+C_{\rm SR}| \sim 15$. This weak constraint stems from the unfavourable overlap of a kinematic distribution that peaks at high $q^2$ (see \cref{fig:knunu-theory}) with the analysis efficiency, which is low for high-$q^2$ signals (see \cref{fig:knunu-efficiency}).

Goodness-of-fit tests and model comparisons support the reliability of these conclusions. The \ac{WET} model provides a good absolute fit to the data ($P_{\rm gof} = 0.63$), comparable to the unconstrained \ac{SM} ($P_{\rm gof} = 0.58$). The background-only hypothesis provides a less favourable fit to the data ($P_{\rm gof} = 0.13$). This confirms that the \ac{WET} model adequately describes the data.

Bayesian model comparison via Bayes factors shows \textit{very strong} preference ($\log_{10} B_{\rm BKG} = 1.78$ for \ac{WET}, $2.02$ for unconstrained \ac{SM}) for signal-plus-background models over the background-only hypothesis, and a \textit{substantial} preference ($\log_{10} B_{\rm SM}^{\rm constr.} = 0.68$ for \ac{WET}, $0.92$ for unconstrained \ac{SM}) over the constrained \ac{SM} prediction. Interestingly, the unconstrained \ac{SM} shows slightly stronger preference than the \ac{WET}, suggesting that the additional complexity of scalar and tensor operators is not strongly required by the data.

Frequentist hypothesis tests support these findings, yielding a $3.3\sigma$ significance for both models against the background-only hypothesis. The significances confirm that both the \ac{WET} and unconstrained \ac{SM} provide strong evidence for new physics contributions. The \ac{WET}'s marginally smaller $P$-value ($4.6 \cdot 10^{-4}$ vs.\ $5.6 \cdot 10^{-4}$) reflects its slightly improved fit from the tensor contribution.

The comparison to the naive reinterpretation in \cref{sec:wet paper naive reinterpretation} illustrates the value of preserving kinematic shape information. The naive approach, which discards all differential distribution information, yields biased credible intervals and fails to identify the preference for a tensor contribution. This comparison validates, that inference results can depend critically on the kinematic shape information, which reinterpretation approaches based on published summary results cannot utilize.


%% file: chapters/knunu-bkx.tex
Having demonstrated the successful application of the model-agnostic reinterpretation method to extract \ac{WET} Wilson coefficients in the previous chapter, this chapter explores an alternative theoretical scenario that addresses a specific feature in the experimental data. The \ac{WET} reinterpretation in \cref{sec:reinterpretation knunu wet} assumed heavy new physics integrated out at the electroweak scale. Here I investigate the possibility of light new physics contributing to the \BKnn decay.

In the \BKnn decay, the final-state neutrinos are not directly detected but inferred from missing energy and momentum in the detector. This does not only hold for \ac{SM} neutrinos --- any light, weakly-interacting particle $X$ that escapes detection would produce such an experimental signature~\cite{MartinCamalich:2020dfe,ParticleDataGroup:2024cfk, Peccei:1977hh, Peccei:1977ur, ParticleDataGroup:2024cfk, PhysRevD.102.015023,Guerrera:2022ykl,Bruggisser:2023npd, PhysRevD.101.095006,Datta:2022zng,Abdughani:2023dlr,Berezhnoy:2023rxx}.

The recent Belle~II \BKnn measurement~\cite{Belle-II:2023esi} observed an excess of events compared to \ac{SM} expectations, with a non-\ac{SM}-like distribution in $q^2_{\rm rec}$. The enhancement was observed in the region ${3~\mathrm{GeV}^2< q^2_{\rm rec} < 7~\mathrm{GeV}^2}$ (see \cref{fig:ita-q2-projection,sec:wet paper analysis}). This would correspond to a resonance near $m_X \sim 2~\text{GeV}$.

Simplified studies of light new physics scenarios in the context of the Belle~II result have been published in References~\cite{Fridell:2023ssf,PhysRevD.109.075008, Gabrielli:2024wys}. These analyses provide initial insights, but they rely on simplified background templates and neglect the full experimental systematic uncertainties.

This chapter addresses these limitations by performing the first reinterpretation of the Belle~II \BKnn measurement that includes the full experimental statistical model. Using the model-agnostic likelihood approach, I derive Bayesian credible intervals and frequentist confidence limits on both the new physics branching ratio $\mathcal{B}(\BKX)$ and the mass $m_X$ of the hypothetical light particle. The analysis is performed for several choices of the resonance width $\Gamma_X$.

This chapter is structured as follows: 
In \cref{sec:bkx-model} I introduce the \BKX model as an alternative hypothesis, and give the parametrized model used for reinterpretation.
In \cref{sec:bkx-results} I present the marginal posterior and credible intervals for the model parameters.
In \cref{sec:mass-scan} I present a frequentist mass scan, deriving upper limits on the branching ratio $\mathcal{B}(\BKX)$ as a function of the mass $m_X$.
The goodness-of-fit test is shown in \cref{sec:goodness-of-fit-bkx}, followed by the model comparison in \cref{sec:model-comparison-bkx}. Finally, I conclude with a discussion of the results in \cref{sec:bkx-summary}.

\section{The \texorpdfstring{\BKX}{B+->K+X} model as alternative hypothesis}
\label{sec:bkx-model}
To test the hypothesis of a light invisible resonance contributing to the \BKnn signal, I construct a theoretical model that combines the \ac{SM} prediction with a resonant two-body contribution.

The presence of an unknown two-body final state \BKX would manifest itself as a sharp resonant peak in the $q^2$ spectrum at the squared mass of the new particle, $q^2 = m_X^2$. To model this resonance in a general way, I employ a relativistic Breit-Wigner \ac{PDF},
\begin{equation}
\begin{aligned}
    p_{\rm BW}(q^2| m_X, \Gamma_X) &= \frac{k}{(q^2 - m_X^2)^2 + m_X^2 \Gamma_X^2}, \\
    k &= \frac{m_X \Gamma_X}{\pi / 2 + \arctan(m_X/\Gamma_X)},
\end{aligned}
\end{equation}
where $m_X$ is the mass of the resonance, $\Gamma_X$ is its decay width, and $k$ is a normalization constant.

Including both the \ac{SM} contribution from the $b \to s \nu \bar \nu$ transitions and the new physics resonance, the total differential branching ratio is modelled as
\begin{equation}
    \frac{d \mathcal{B}}{d q^2} = \mu\frac{d \mathcal{B}_{SM}}{d q^2} + \mu_X \sigma_X ~ p_{\rm BW}(q^2| m_X, \Gamma_X),
    \label{eq:bkx-width}
\end{equation}
where $\mu$ is a scale factor for the \ac{SM} contribution, and $\mu_X$ controls the strength of the new physics contribution. The factor $\sigma_X = 10^{-6}$ is introduced for numerical stability. The new physics branching ratio is given by $\mathcal{B}(\BKX) = \mu_X \cdot 10^{-6}$.

The \ac{SM} prediction $d \mathcal{B}_{SM}/d q^2$ is computed using the theoretical framework described in \cref{sec:knunu-theory}, with hadronic form factor parameters from the lattice \ac{QCD} calculation by the HPQCD collaboration~\cite{Parrott:2022rgu}. This is the same \ac{SM} baseline used in the original Belle~II analysis~\cite{Belle-II:2023esi}.

The model defined in \cref{eq:bkx-width} contains five free parameters:
\begin{itemize}
    \item $\mu$: \ac{SM} scale factor
    \item $\mu_X$: New physics scale (related to $\mathcal{B}(\BKX)$ by a factor $\sigma_X = 10^{-6}$)
    \item $m_X$: Resonance mass
    \item $\Gamma_X$: Resonance width
\end{itemize}

In this study, the \ac{SM} contribution is treated as a background. Hence, the \ac{SM} scale factor is constrained around unity, $\mu=1.0$, with a $5\%$ normalization uncertainty added to all other systematic uncertainties. This $5\%$ uncertainty accounts for a $4.4\%$ uncertainty on the \ac{CKM} matrix elements $|V_{ts}^*V_{tb}|^2$, and a $2.3\%$ uncertainty on the Wilson coefficient $|C_{\rm VL}^{\rm SM}|^2$~\cite{Parrott:2022zte}.

The primary \acp{POI} for this analysis are the new physics scale $\mu_X$ and the resonance mass $m_X$.
Rather than treating the width $\Gamma_X$ as a free parameter, I fix it to three representative values: $\Gamma_X = 0.1, 0.5, 1.0~\text{GeV}$. 
Different theoretical models for the particle $X$ predict different widths. By considering multiple widths, I remain relatively model-agnostic.
In addition, the analysis has a limited momentum transfer resolution, which makes widths below $\Gamma_X \lesssim 0.1~\text{GeV}$ indistinguishable. Therefore, $\Gamma_X = 0.1~\text{GeV}$ represents the narrow-width limit accessible to this analysis. All chosen widths lead to extremely short lifetimes of $X$, which is consequently assumed to decay invisibly.

\begin{figure}
    \centering
    \includegraphics[width=0.8\textwidth]{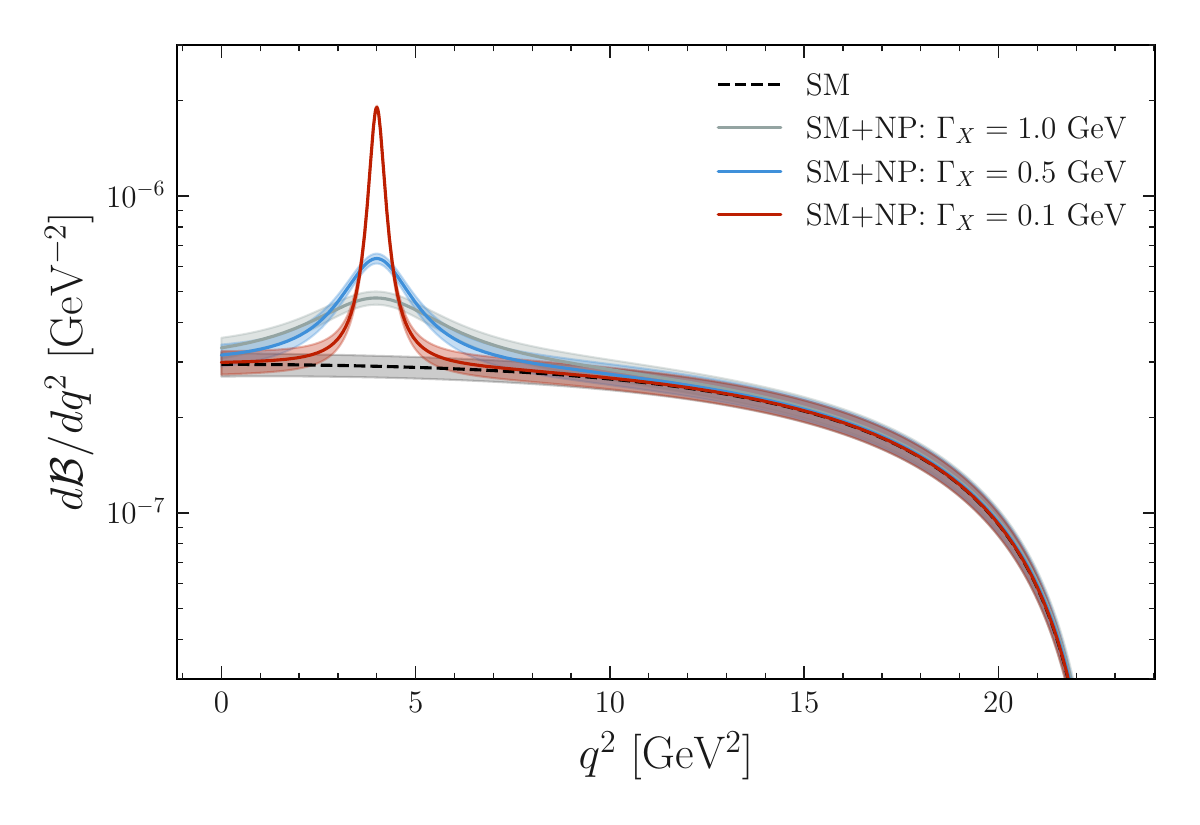}
    \caption{The predicted differential branching ratio from \cref{eq:bkx-width} for three choices of resonance width, shown for the exemplary parameter point $\mu_X=1$ and $m_X=2~\text{GeV}$. The bands represent the combined theoretical uncertainty from hadronic form factor parameters and the $5\%$ normalization uncertainty on the \ac{SM} contribution.} 
    \label{fig:lightNP-theory}
\end{figure}

The predicted differential branching ratio for these three width choices is illustrated in \cref{fig:lightNP-theory} for an exemplary parameter point with $\mu_X=1$ and $m_X = 2~\text{GeV}$.

\begin{figure}
    \centering
    \includegraphics[width=0.8\textwidth]{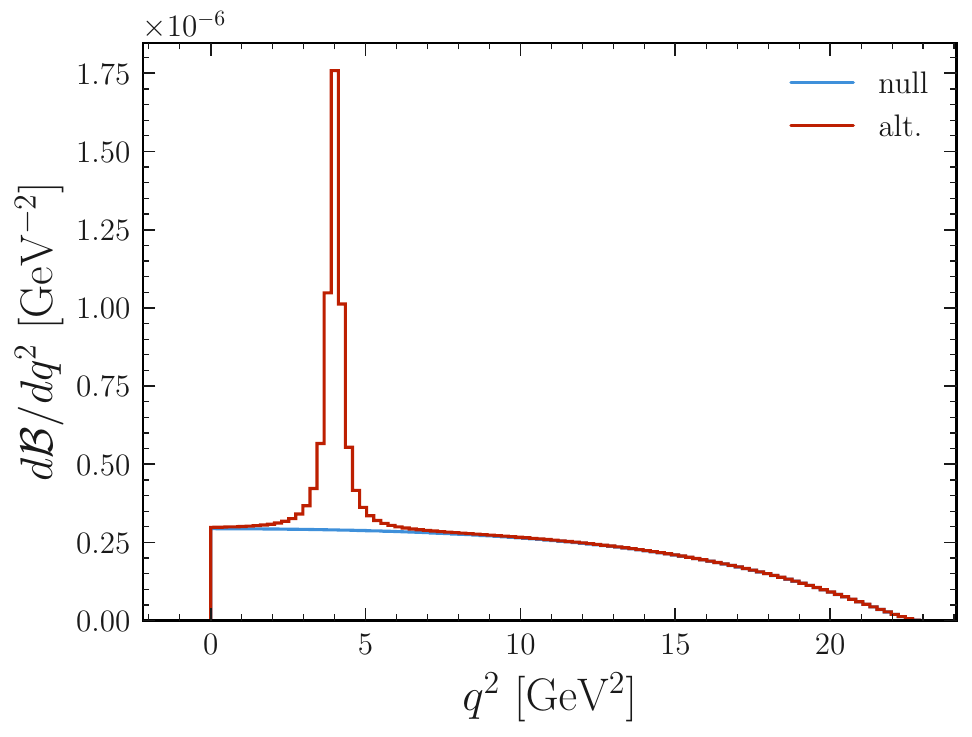}
    \caption{Comparison of bin-integrated distributions for the null hypothesis (blue, \ac{SM} only~\cite{Parrott:2022zte}) and alternative hypothesis (red, \ac{SM} plus \BKX resonance) according to \cref{eq:bkx-width}. The alternative distribution is shown for $\mu_{X}=1$, $m_X = 2~\text{GeV}$, and $\Gamma_X = 0.1~\text{GeV}$.}
    \label{fig:lightNP-dists}
\end{figure}

\Cref{fig:lightNP-dists} shows the binned distributions for the null hypothesis and the alternative hypothesis for the same exemplary parameter point.

\section{Marginal posterior and credible intervals}
\label{sec:bkx-results}
This section presents the Bayesian analysis of the Belle~II \BKnn data under the \BKX hypothesis, deriving posterior distributions and credible intervals for the model parameters. The analysis employs the model-agnostic likelihood described in \cref{sec:reinterpretation knunu mall}, and the Bayesian inference framework described in \cref{sec:histfactory-bayesian}, implemented using the \texttt{bayesian pyhf} package~\cite{Feickert:2023hhr}.  The posterior distribution is sampled using \ac{MCMC} methods implemented in the \texttt{PyMC} software~\cite{AbrilPla2023PyMCAM}.

\subsection{Statistical model and prior choices}
The objective is to obtain the marginal posterior distributions for the \acp{POI},
\begin{equation}
    \boldsymbol{\eta} = [\mu_X, ~m_X],
\end{equation}
for each of the three fixed resonance widths $\Gamma_X=0.1, ~0.5, ~1.0 ~\text{GeV}$. To achieve this, I extend the baseline \BKnn statistical model (which contains 232 parameters) with the two \acp{POI} $\mu_X$ and $m_X$. The complete statistical model contains 234 parameters: the 2 \acp{POI}, 1 constrained \ac{SM} scale parameter, and 231 additional nuisance parameters (the \ac{SM} scale is considered a nuisance parameter here).

Prior distributions for the \acp{POI} are chosen to be uniform over the ranges specified in \cref{tab:bkx priors}. Uniform priors are chosen to avoid assigning preference to any part of the parameter space and because no non-linear transformation of the parameters is anticipated in this study. Ranges are chosen to cover the full posterior, as verified a posteriori. The robustness of the results with respect to alternative prior choices is examined in detail in \cref{sec:bkx-sensitivity}.

\begin{table}[ht]
    \renewcommand{\arraystretch}{1.5}
    \caption{Prior ranges for the new physics scale $\mu_X$ and resonance mass $m_X$, specified for each of the three resonance widths considered.}
    \centering
    \begin{tabularx}{\linewidth}{@{\extracolsep{\fill}}lYYY}
        \toprule \midrule
        $\Gamma_X=$ & $0.1~\text{GeV}$ & $0.5~\text{GeV}$ & $1.0~\text{GeV}$ \\
        \midrule
         $
         \begin{aligned}
            &\mu_X \\
            &m_X~[\text{GeV}]
        \end{aligned}
        $ & $
        \begin{aligned}
            [0.0,~24.0]&\\
            [1.5,~3.0]&
        \end{aligned}
        $ & $
         \begin{aligned}
            [0.0,~32.0]&\\
            [1.5,~3.2]&
        \end{aligned}
        $ & $
        \begin{aligned}
            [0.0,~42.0]&\\
            [1.5,~3.4]&
        \end{aligned}
        $
        \\
        \midrule \bottomrule
    \end{tabularx}
    \label{tab:bkx priors}
\end{table}

\subsection{Posterior sampling and marginalization}
The marginal posterior distributions for $\mu_X$ and $m_X$ are obtained as follows. First, the joint posterior of all 234~parameters is sampled using \ac{MCMC} methods. Subsequently, this joint posterior is marginalized over the 232~nuisance parameters, leaving only the 2-dimensional posterior distribution over the \acp{POI}, $(\mu_X, m_X)$.

Each analysis (one per value of $\Gamma_X$) employs 8~independent Markov chains, each generating 50\,000 samples, for a total of 400\,000 posterior samples per width. Multiple independent chains enable convergence diagnostics through inter-chain comparison (see \cref{sec:posterior-validation}). Convergence diagnostics are reported in \cref{sec:bkx-posterior-validation-results}, confirming that the chains have converged to the target posterior distribution and that the sampling is reliable.

\begin{figure}
    \centering
    \includegraphics[width=0.8\textwidth]{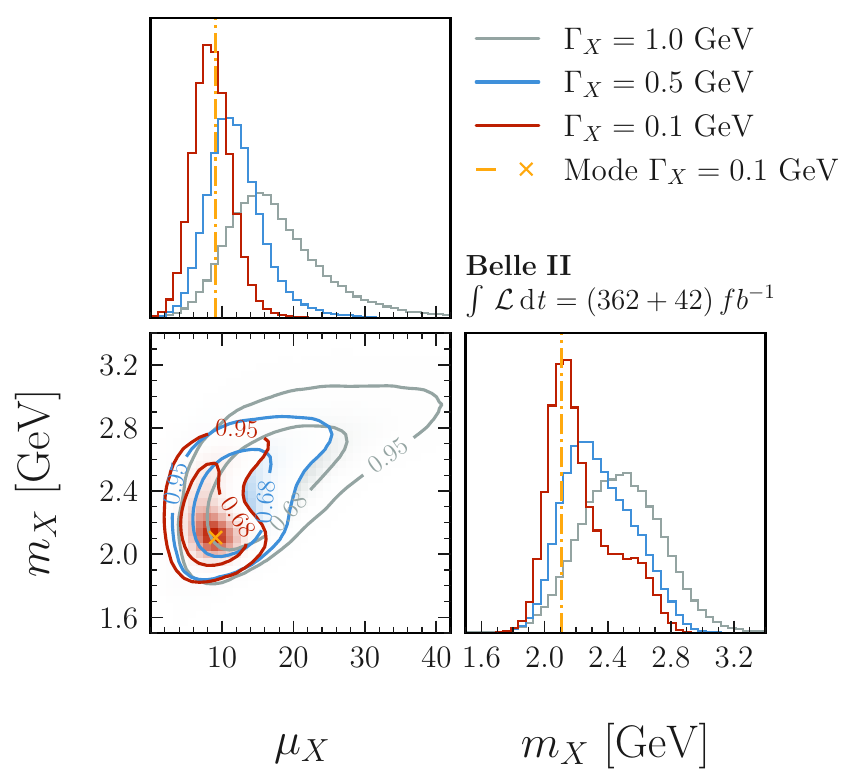}
    \caption{Marginalized posterior distributions for the \BKX model parameters from \cref{eq:bkx-width}, shown for the three resonance widths $\Gamma_X = 0.1, 0.5, 1.0~\text{GeV}$.
    Diagonal and off-diagonal panels show the 1-dimensional and 2-dimensional sample density \acp{PDF} on a linear scale, respectively. The overall scale is omitted, as all relevant information is contained in the shape of the distribution. The contours indicate $68\%$ and $95\%$ credible intervals. The dash-dotted yellow lines and yellow cross indicate the posterior mode.
    }
    \label{fig:bkx-posterior}
\end{figure}

The full set of 1- and 2-dimensional marginal posterior distributions is shown in \cref{fig:bkx-posterior} for all three widths. The contours enclose $68\%$ and $95\%$ credible regions.

\subsection{Credible intervals and posterior mode}
From the 1-dimensional marginal posterior distributions, I calculate the \acp{HDI} (see \cref{sec:credible-intervals}) at 68\% and 95\% probability for the model parameters. The posterior mode and the credible intervals are summarized in \cref{tab:bkx results}. 

\begin{table}[ht]
    \renewcommand{\arraystretch}{1.5}
    \caption{Posterior mode and \acp{HDI} at 68\% and 95\% credible levels for the \BKX model parameters from \cref{eq:bkx-width}, derived from the posterior distributions in \cref{fig:bkx-posterior}. The new physics branching ratio is $\mathcal{B}(\BKX) = \mu_X \cdot 10^{-6}$. Results are shown for three choices of resonance width.}
    \centering
    \begin{tabularx}{\linewidth}{@{\extracolsep{\fill}}lYYYY}
        \toprule \midrule
        \textbf{Param.} & \textbf{$\boldsymbol{\Gamma_X}~[\text{GeV}]$} &\textbf{Mode} & \textbf{68\% HDI} & \textbf{95\% HDI} \\
        \midrule
         $
            \mu_X
        $ & $
        \begin{aligned}
            &0.1\\
            &0.5\\
            &1.0
        \end{aligned}
        $ & $
        \begin{aligned}
            9.2&\\
            11.1&\\
            15.3&
        \end{aligned}
        $ & $
         \begin{aligned}
            [5.8, ~11.0]&\\
            [7.8, ~15.0]&\\
            [9.6, ~21.8]&
        \end{aligned}
        $ & $
        \begin{aligned}
            [3.4, ~14.0]&\\
            [4.2, ~20.2]&\\
            [5.5, ~32.1]&
        \end{aligned}
        $
        \\[0.7cm]
        $
            m_X~[\text{GeV}]
        $ & $
        \begin{aligned}
            &0.1\\
            &0.5\\
            &1.0
        \end{aligned}
        $ & $
        \begin{aligned}
            &2.1\\
            &2.2\\
            &2.4
        \end{aligned}
        $ & $
         \begin{aligned}
            [2.0,~2.3]&\\
            [2.1,~2.5]&\\
            [2.2,~2.7]&
        \end{aligned}
        $ & $
        \begin{aligned}
            [1.9,~2.7]&\\
            [2.0,~2.8]&\\
            [2.0,~3.0]&
        \end{aligned}
        $ \\
        \midrule \bottomrule
    \end{tabularx}
    \label{tab:bkx results}
\end{table}

Direct comparison of the observed data with the predicted yields at the posterior mode (see \cref{tab:bkx results}) for the unconstrained \BKnn \ac{SM} and \BKX ($\Gamma_X=0.1~\text{GeV}$) is shown in \cref{fig:post-fit}, for the \ac{HSR} of the analysis (\ac{ITA}, $\eta(\BDT2)>0.98$). 
The \BKX model provides a significantly better fit to the data than the unconstrained \ac{SM} prediction, as indicated by the smaller pull values.

\begin{figure}
    \centering
    \includegraphics[width=0.8\linewidth]{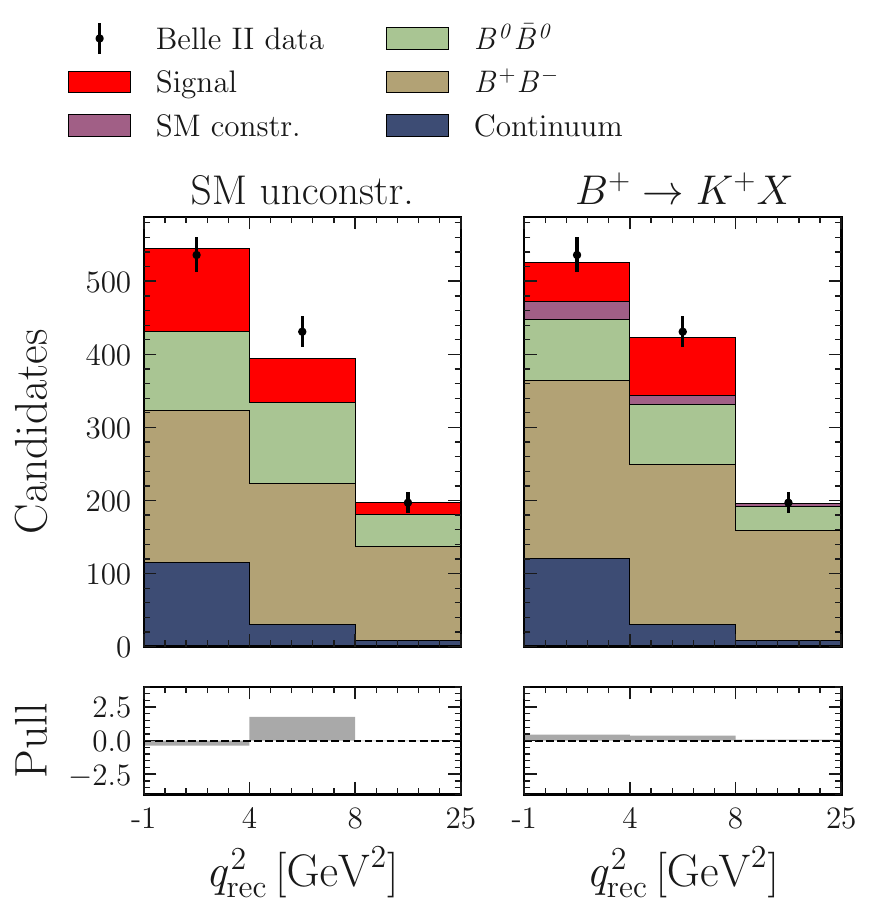}
    \caption{
        Observed and predicted best-fit yields in the \ac{HSR} of the analysis. The signal is shown for the unconstrained \BKnn \ac{SM} (\textit{left}) and the \BKX ($\Gamma_X=0.1~\text{GeV}$) (\textit{right}) predictions. In the latter case, the constrained \BKnn \ac{SM} is separated from the \BKX contribution.
        The predicted background yields are shown individually for the neutral and charged \B-meson decays, and the summed five continuum categories. 
        Pulls are shown in the lower panels.}
    \label{fig:post-fit-bkx}
\end{figure}

\subsection{Physical interpretation of the posterior}
For all three widths, the mass parameter $m_X$ has a sharp peak in the range $2.1-2.4~\text{GeV}$, with an extended tail toward higher masses. The peak location is consistent across the different width models, indicating that the data prefer a resonance in this mass range regardless of the assumed width. This result agrees well with the region where an excess in data were observed in Reference~\cite{Belle-II:2023esi} and \cref{fig:ita-q2-projection}.

These findings are consistent with the analyses in References~\cite{PhysRevD.109.075008,Fridell:2023ssf}, which identified best-fit masses around $m_X \approx 2~\text{GeV}$ using simplified treatments of the experimental data.

The new physics scale parameter $\mu_X$ shows a clear deviation from zero for all widths, indicating that the data prefer a model with a non-zero \BKX contribution. The posterior mode increases with resonance width: $\mu_X = 9.2$ for $\Gamma_X = 0.1~\text{GeV}$, $\mu_X = 11.1$ for $\Gamma_X = 0.5~\text{GeV}$, and $\mu_X = 15.3$ for $\Gamma_X = 1.0~\text{GeV}$. This is expected because broader resonances distribute the same total rate over a larger $q^2$ range, requiring a larger overall normalization to match the observed excess in any given bin.

Translating to branching ratios, the narrowest resonance ($\Gamma_X = 0.1~\text{GeV}$) yields $\mathcal{B}(\BKX) = 9.2 \cdot 10^{-6}$ at the posterior mode, with a 95\% credible interval of $[3.4, 14.0] \cdot 10^{-6}$. For comparison, the \ac{SM} prediction for the total \BKnn branching ratio is $\mathcal{B}_{\rm SM} = (4.97\pm 0.37) \cdot 10^{-6}$~\cite{Parrott:2022zte}. The preferred \BKX branching ratio is thus roughly twice the \ac{SM} expectation.


\subsection{Prior sensitivity}
\label{sec:bkx-sensitivity}
Bayesian inference inherently depends on the choice of prior distributions. To assess the robustness of the presented results and quantify the influence of prior choices, I perform a prior sensitivity analysis by comparing the baseline uniform priors to two physically motivated alternative prior specifications.

For simplicity, I perform the sensitivity study for the $B \to K X$ model in \cref{eq:bkx-width} for a width of ${\Gamma_X = 0.1~\text{GeV}}$ only.

The first alternative employs a \textit{truncated-normal prior} for $\mu_{X}$ and a uniform prior for $m_X$ (as in \cref{tab:bkx priors}),
\begin{equation}
\begin{aligned}
    p\left( \mu_{X}\right) &= 
    \begin{cases}
        \mathcal{N}(\mu_{X} | \mu=0, \sigma=20) &~ \mu_{X} \geq 0 \\
        0 &~ \mu_{X} < 0
    \end{cases},\\
    p\left( m_X \right) &= \mathcal{U}([1.5, 3.0]~\text{GeV}).
    \end{aligned}
\label{eq:bkx priors normal}
\end{equation}
Here $\mathcal{U}$ corresponds to a uniform distribution.

The second alternative adopts a uniform prior for $\mu_{X}$ (as in \cref{tab:bkx priors}) and a \textit{uniform prior in the squared mass} $m_X^2$, translating to a triangular prior in the mass,
\begin{equation}
\begin{aligned}
    p\left( \mu_{X}\right) &= 
    \mathcal{U}([0, 24]),\\
    p\left( m_X \right) &\propto  \begin{cases}
        m_X &~ m_{X} \leq 4.8~\text{GeV}\\
        0 &~ m_{X} > 4.8~\text{GeV}
    \end{cases}.
    \end{aligned}
\label{eq:bkx priors triangular}
\end{equation}

The resulting posterior modes and credible intervals for both alternative prior choices are summarized in \cref{tab:bkx results sensitivity}. The comparison reveals that the mass $m_X$ is most robustly determined, with the posterior mode ($m_X=2.1~\text{GeV}$ for all priors) and credible intervals showing minimal sensitivity to the prior choice. The new physics scale $\mu_X$ exhibits a small shift in the posterior mode from $9.2$ (uniform prior) to $9.0$ (truncated-normal and triangular priors), and the credible intervals demonstrate similar robustness against changes in the priors.

\begin{table}[ht]
    \renewcommand{\arraystretch}{1.5}
    \caption{The posterior modes, and \acp{HDI} at 68\% and 95\% for the \BKX model in \cref{eq:bkx-width}, for alternative priors (cf. \cref{tab:bkx results}). These results are based on the $\Gamma_X = 0.1~\text{GeV}$ model.}
    \centering
    \begin{tabularx}{\linewidth}{@{\extracolsep{\fill}}llYYY}
        \toprule \midrule
        \textbf{Priors} & \textbf{Param.} &\textbf{Mode} & \textbf{68\% HDI} & \textbf{95\% HDI} \\
        \midrule
        \cref{eq:bkx priors normal} & $
        \begin{aligned}
            &\mu_X\\
            &m_X~[\text{GeV}]
        \end{aligned}
        $ & $
        \begin{aligned}
            &9.0\\
            &2.1
        \end{aligned}
        $ & $
         \begin{aligned}
            [5.7, ~10.8]&\\
            [2.0,~2.4]&
        \end{aligned}
        $ & $
        \begin{aligned}
            [3.4, ~13.8]&\\
            [1.9,~2.7]&
        \end{aligned}
        $\\
         \midrule
        \cref{eq:bkx priors triangular}& $
        \begin{aligned}
            &\mu_X\\
            &m_X~[\text{GeV}]
        \end{aligned}
        $ & $
        \begin{aligned}
            9.0&\\
            2.1&
        \end{aligned}
        $ & $
         \begin{aligned}
            [5.7, ~11.0]&\\
            [2.0,~2.4]&
        \end{aligned}
        $ & $
        \begin{aligned}
            [3.0, ~14.4]&\\
            [1.9,~2.7]&
        \end{aligned}
        $\\
        \midrule \bottomrule
    \end{tabularx}
    \label{tab:bkx results sensitivity}
\end{table}

\section{Frequentist mass scan}
\label{sec:mass-scan}
This section presents a frequentist scan over the resonance mass, deriving 95\% confidence-level upper limits on the branching ratio $\mathcal{B}(\BKX)$ as a function of $m_X$.

For each fixed value of $m_X$, I determine the 95\% confidence-level upper limit on $\mu_X$ using the modified frequentist test statistic for setting upper limits~\cite{Cowan:2010js}:
\begin{equation}
    q(\mu_X) = 
    \begin{cases}
        t(\mu_X) & \hat{\mu}_X \leq \mu_X,\\[1ex]
        0 & \hat{\mu}_X > \mu_X,
    \end{cases}
\end{equation}
where $t(\mu_X)$ is the likelihood ratio test statistic defined in \cref{eq:nested-likelihood-ratio-test-statistic}. This test statistic excludes values of $\mu_X$ below the best-fit value $\hat{\mu}_X$ from contributing. This modification is appropriate for upper limit setting because observing a signal strength larger than the hypothesized value should not be considered evidence against the hypothesis~\cite{Cowan:2010js}.

The confidence limit is determined using the $CL_s$ criterion of \cref{eq:cls}~\cite{read2002},
\begin{equation}
    CL_s = \frac{P(q_{\rm obs} \mid s+b)}{P(q_{\rm obs} \mid b)}.
\end{equation}
The $CL_s$ method is more conservative than using $P(q_{\rm obs} \mid s+b)$ alone. The 95\% confidence-level upper limit is defined as the value of $\mu_X$ for which $CL_s = 0.05$.

\begin{figure}
    \centering
    \includegraphics[width=0.8\linewidth]{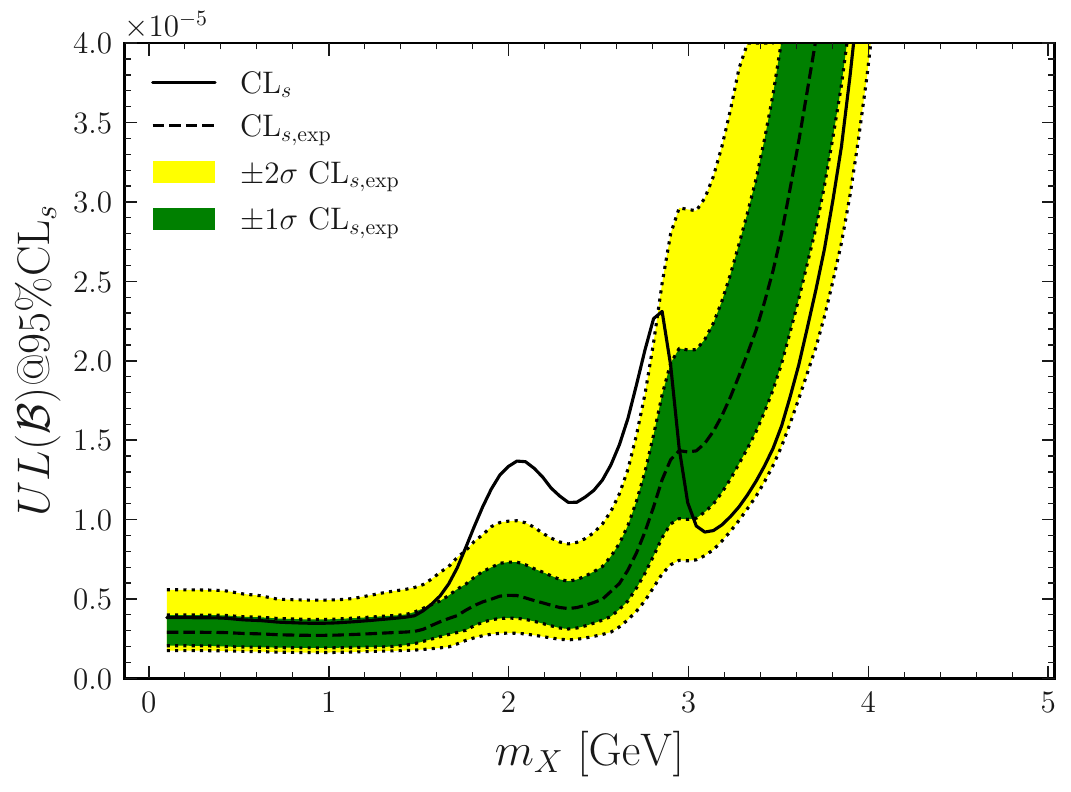}
    \caption{Frequentist 95\% confidence-level upper limit on $\mathcal{B}(\BKX)$ as a function of resonance mass $m_X$ for $\Gamma_X = 0.1~\text{GeV}$, determined using the $CL_s$ criterion with $CL_s \leq 0.05$. The solid black line shows the observed upper limit from data. The dashed black line shows the expected limit under the \ac{SM} hypothesis, with green and yellow bands indicating the $\pm 1\sigma$ and $\pm 2\sigma$ ranges, respectively.}
    \label{fig:mass-scan}
\end{figure}

The resulting 95\% confidence-level upper limit on $\mathcal{B}(\BKX)$ as a function of $m_X$ is shown in \cref{fig:mass-scan} for the narrow resonance case $\Gamma_X = 0.1~\text{GeV}$. 

A pronounced excess of the observed limit relative to the expected limit is found in the mass range $m_X \in [1.5, 2.9]~\text{GeV}$. In this region, the observed limit lies significantly above the expected limit. The excess is most prominent around $m_X \sim 2.1~\text{GeV}$. The location and extent of the excess region are consistent with the 95\% credible intervals on $m_X$ reported in \cref{tab:bkx results}.

For $m_X > 3.5~\text{GeV}$, both the observed and expected limits rise steeply, indicating reduced sensitivity. This loss of sensitivity arises because the analysis efficiency decreases rapidly at high $q^2$ (see \cref{fig:knunu-efficiency}).

The observed limit exhibits a localized increase near $m_X \sim 2~\text{GeV}$ and a change in slope around $m_X \sim 3~\text{GeV}$. The location of these features correspond to the analysis bin boundaries in $q^2_{\rm rec}$. This behaviour is expected, as the sensitivity is reduced when the new physics contribution is distributed across two bins rather than fully contained within a single bin.

\subsection{Injection study validation}
To validate the reliability of the mass scan results and quantify the expected performance in different regions of parameter space, I perform an injection study. In this study, toy data are generated from the \BKX model \ac{PDF} at specific parameter points $(\mu_X, m_X)$, and fitting is applied to recover these injected values. The results are presented in detail in \cref{sec:bkx-injection}.

The injection study reveals that parameter recovery is good in the well-constrained mass range $1.0~\text{GeV} < m_X < 2.5~\text{GeV}$. Here the fitted values recover the injected truth within statistical uncertainties.

For $m_X > 3~\text{GeV}$, the reduced sensitivity makes parameter determination challenging. In addition, the analysis employs coarse binning in $q^2_{\rm rec}$ with boundaries at $[-1, 4, 8, 25]~\text{GeV}^2$. This binning is less ideal for resolving narrow resonances.

\section{Goodness-of-fit}
\label{sec:goodness-of-fit-bkx}
Beyond determining credible intervals and confidence limits, it is important to validate that the \BKX hypothesis provides a good description of the observed data. The goodness-of-fit $P$-values are evaluated analogously to the \ac{WET} model in \cref{eq:gof-pvalue-wet}, using a toy study with fluctuated data.

\begin{figure}
    \centering
    \includegraphics[width=0.6\textwidth]{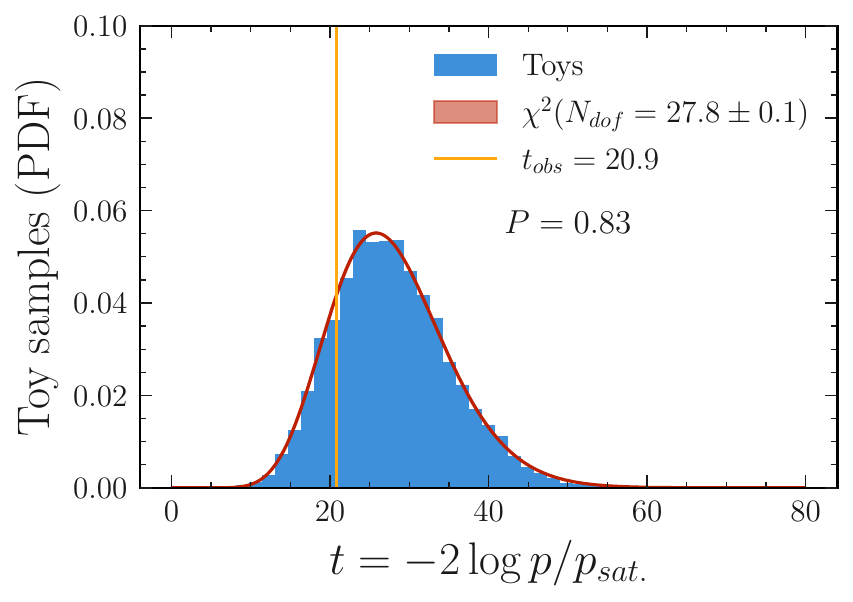}
    \includegraphics[width=0.6\textwidth]{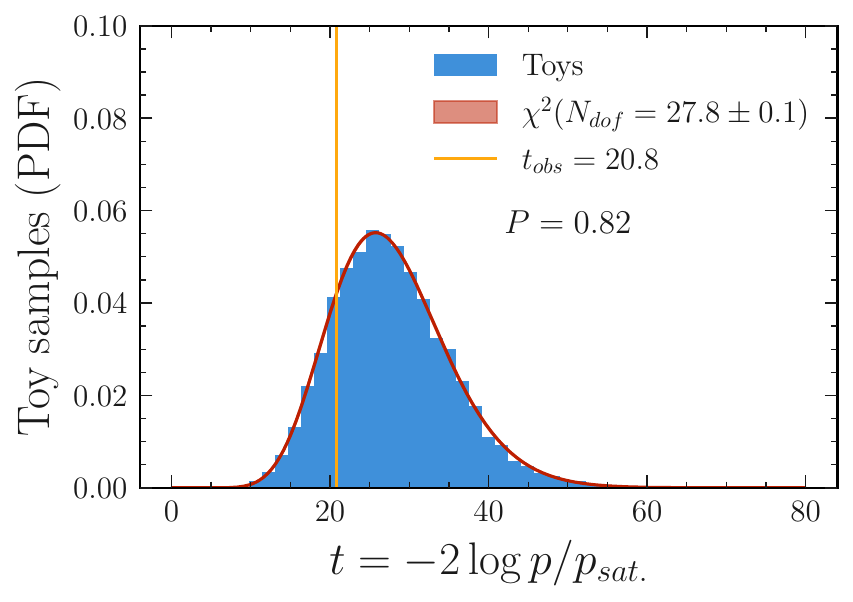}
    \includegraphics[width=0.6\textwidth]{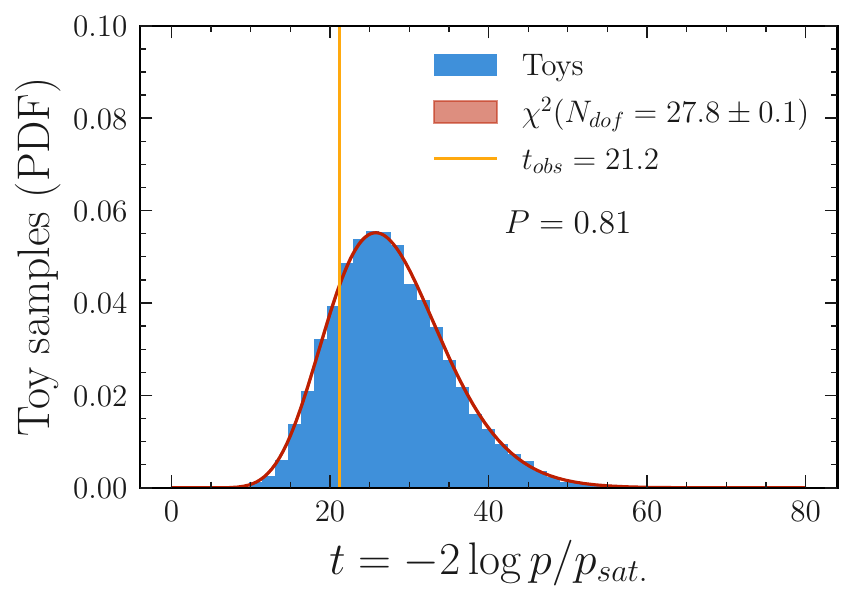}
    \caption{
    The distribution of the goodness-of-fit test statistic for the \BKX model with resonance widths $\Gamma_X = 0.1, 0.5, 1.0~\text{GeV}$ (\textit{top/middle/bottom}), each obtained from 10\,000 toys. The values obtained from a fit to the data are shown as vertical yellow lines.
    The asymptotic $\chisq$-distributions (red band) correspond to the best-fit $N_{dof}$, which are shown with the corresponding uncertainty band.
    }
    \label{fig:chi2-dm-sm}
\end{figure}

The distribution of the goodness-of-fit test statistic under the \BKX hypothesis is determined from 10\,000 toy experiments. 
The distributions of $t_{\rm gof}$ for the three resonance widths are shown in \cref{fig:chi2-dm-sm}. Under standard asymptotic assumptions, $t_{\rm gof}$ should follow a $\chisq$-distribution with $N_{\rm dof} = 30-2=28$ \ac{d.o.f.}, where $N_{\rm bins} = 30$ is the total number of bins in the analysis and $N_{\rm POI} = 2$ is the number of free \acp{POI} ($\mu_X$ and $m_X$). This yields an expectation of $N_{\rm dof} = 28$. Fits to the distributions from toys yield $N_{\rm dof}=27.8\pm0.1$ for all three widths, showing a small deviation but overall good agreement with the expectation. The present deviation arises likely due to slightly non-Gaussian distributions of the \acp{POI}.

The goodness-of-fit $P$-values for all three resonance widths are summarized in \cref{tab:bayes factors bkx}. These high $P$-values indicate excellent agreement between the \BKX models and the observed data --- about 80\% of toy experiments under the \BKX hypothesis would yield worse fits than what is observed in the real data.

The consistency of the $P$-values across different widths demonstrates that the fit quality is robust. This insensitivity to $\Gamma_X$ arises because of the coarse binning of the analysis ($q^2_{\rm rec}$ bins of $[-1, 4, 8, 25]~\text{GeV}^2$). As long as resonances fall in the same bin, they produce a similar distribution in $q^2_{\rm rec}$.

Comparing the \BKX goodness-of-fit results to those from the \ac{WET} analysis in \cref{sec:wet paper goodness-of-fit}, the \ac{WET} model yielded $P_{\rm gof} = 0.63$, and the unconstrained \ac{SM} yielded $P_{\rm gof} = 0.58$.
The superior goodness-of-fit for the \BKX models indicates that the resonance hypothesis provides a better explanation for the observed data. This advantage arises because the \BKX model fits well to the localized data excess in the $q^2_{\rm rec}$ distribution around $3-7~\text{GeV}^2$ by construction.

\section{Model comparison}
\label{sec:model-comparison-bkx}
Model comparison quantifies the \textit{relative} support that the data provide for competing hypotheses. I perform both Bayesian and frequentist model comparisons to assess the relative performance of the \BKX hypothesis against two reference models: the background-only hypothesis (BKG), the constrained \BKnn \ac{SM}, where $\mu$ is fixed to the \ac{SM} prediction.

\subsection{Bayes factors}
Bayesian model comparison employs the Bayes factor, defined as the ratio of marginal likelihoods between two competing models (see \cref{sec:global-comparison}).

The Bayes factors are computed using sequential \ac{MC} sampling methods~\cite{AbrilPla2023PyMCAM,10.1093/gji/ggt180}, which provide numerical estimates of the marginal likelihoods for each model. The Bayes factors for all three resonance widths are summarized in \cref{tab:bayes factors bkx}.

Comparing to the constrained \ac{SM}, the Bayes factors are $\log_{10} B_{\rm SM}^{\rm constr.} = 1.71$ for $\Gamma_X = 0.1~\text{GeV}$, $1.83$ for $\Gamma_X = 0.5~\text{GeV}$, and $1.82$ for $\Gamma_X = 1.0~\text{GeV}$. These values all exceed Jeffreys's threshold~\cite{jeffreys1961theory} of $\log_{10} B > 1.5$ for a \textit{very strong} model preference (see \cref{tab:jeffreys-scale}).

Comparing to the background-only hypothesis, the Bayes factors are $\log_{10} B_{\rm BKG} = 2.82$ for $\Gamma_X = 0.1~\text{GeV}$, $2.93$ for $\Gamma_X = 0.5~\text{GeV}$, and $2.92$ for $\Gamma_X = 1.0~\text{GeV}$. These values all exceed the threshold of $\log_{10} B > 2.0$ for a \textit{decisive} model preference.

Comparing the Bayes factors across different new physics hypotheses (see \cref{tab:bayes factors} for \ac{WET} and unconstrained \ac{SM} results), the \BKX models show the strongest overall preference.
The \ac{WET} model yields $\log_{10} B_{\rm BKG} = 1.78$, and the unconstrained \ac{SM} yields $\log_{10} B_{\rm BKG} = 2.02$, both lower than the \BKX results. 
This hierarchy reflects, that the excess in data around $q^2_{\rm rec} \sim 3-7~\text{GeV}^2$ is better described by a resonance than by smooth modifications to the kinematic distribution.

The Bayes factor inherently accounts for the look-elsewhere effect through the integration over the full parameter space in the marginal likelihood (see \cref{eq:marginal-likelihood}). The look-elsewhere effect is the reduction of global significance compared to local significance, arising because testing many possible mass values increases the chance of observing a random fluctuation that looks signal-like~\cite{Gross:2010qma}.
The strong Bayes factors obtained here thus represent global evidence for the \BKX hypothesis, not just local significance at the best-fit mass.

\begin{table}[ht]
    \renewcommand{\arraystretch}{1.5}
    \caption{Bayes factors and goodness-of-fit $P$-values for the \BKX models with three different resonance widths. The Bayes factors are reported relative to the background-only hypothesis ($B_{\rm BKG}$) and the constrained \BKnn \ac{SM} hypothesis ($B_{\rm SM}^{\rm constr.}$).}
    \centering
    \begin{tabularx}{\linewidth}{@{\extracolsep{\fill}}llYYY}
        \toprule \midrule
        \textbf{Model} & $\Gamma_X$ & $\boldsymbol{\log_{10} B_{\rm BKG}}$ & $\boldsymbol{\log_{10} B_{\rm SM}^{\rm constr.}}$ & $\boldsymbol{P_{\rm gof}}$\\
        \midrule
        \BKX & $0.1~\text{GeV}$ & 2.82 & 1.71 & 0.83\\
        \BKX & $0.5~\text{GeV}$ & 2.93 & 1.83 & 0.82\\
        \BKX & $1.0~\text{GeV}$ & 2.92 & 1.82 & 0.81\\
        \midrule \bottomrule
    \end{tabularx}
    \label{tab:bayes factors bkx}
\end{table}

\subsection{Frequentist hypothesis test}
To complement the Bayesian model comparison and provide a frequentist perspective, I perform hypothesis tests that quantify the significance of the signal hypotheses relative to the background-only hypothesis.

The tests were conducted using the $P$-value defined in \cref{eq:hypothesis-test-pvalue}. The distribution of the test statistic $p(t)$ is determined through toy experiments.

\begin{figure}
    \centering
    \includegraphics[width=0.6\textwidth]{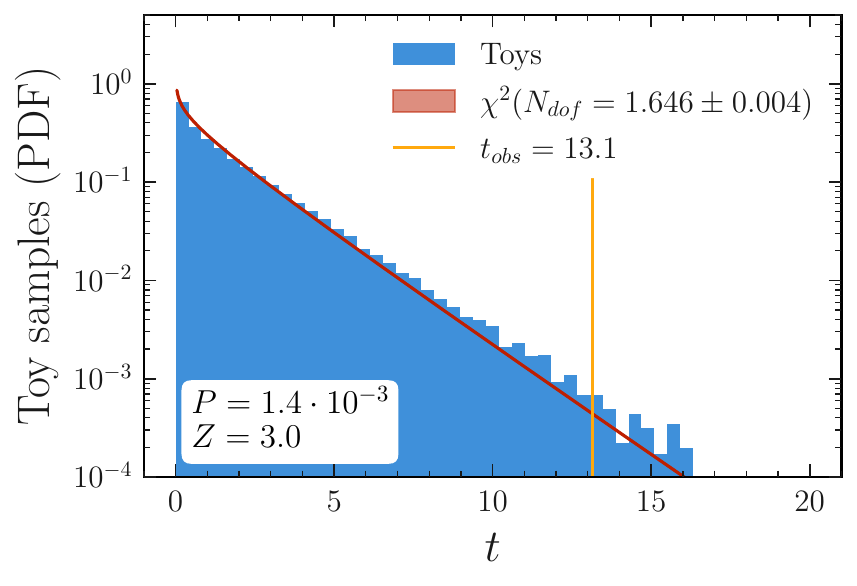}
    \caption{
        Distribution of the test statistic for the \BKX hypothesis ($\Gamma_X = 0.1~\text{GeV}$) versus the constrained \ac{SM} hypothesis, obtained from 100\,000 toy experiments. 
        The value from the fit to data is shown as vertical yellow line.
        The asymptotic $\chisq$-distribution (red band) corresponds to the best-fit $N_{dof}$, which is shown with the corresponding uncertainty band.
    }
    \label{fig:hypo-dm-sm}
\end{figure}

For the \BKX model with $\Gamma_X = 0.1~\text{GeV}$, the hypothesis test yields a $P$-value of $P = 1.4 \cdot 10^{-3}$, corresponding to a significance of $Z = 3.0\sigma$ with respect to the constrained \ac{SM} hypothesis. The distribution of the test statistic from 100\,000 toy experiments is shown in \cref{fig:hypo-dm-sm}, where the observed value $t_{\rm obs} = 13.1$ falls in the far tail of the distribution.

This significance is notably higher than the $Z = 2.7\sigma$ quoted in the original Belle~II publication~\cite{Belle-II:2023esi} for the excess relative to the \ac{SM}. The increased significance when testing the \BKX hypothesis reflects the fact that the shape of the excess in data matches the \BKX prediction better than a simple overall enhancement.

For a model with two free parameters ($\mu_X$ and $m_X$), the asymptotic expectation is a $\chi^2$ distribution with $N_{\rm dof} = 2$. Fitting to the toy distribution yields $N_{\rm dof} = 1.646 \pm 0.004$, showing a statistically significant deviation from the asymptotic expectation. This potentially arises from boundary effects ($m_X$ must be positive) introducing non-Gaussian features in the test statistic distribution. The obtained $P$-value is based on the toy distribution and hence reliable.

The hypothesis test as performed here does account for the look-elsewhere effect. The test statistic distribution is generated from toys sampled under the constrained \BKnn \ac{SM} hypothesis. Each toy is fitted allowing both $\mu_X$ and $m_X$ to vary freely. Therefore, this procedure incorporates the penalty for searching over different mass values. In the toys, random fluctuations can mimic resonances at various masses. The test statistic distribution reflects how often such fluctuations occur when fitting with a free mass parameter.

\section{Discussion}
\label{sec:bkx-summary}
This chapter presents a comprehensive reinterpretation of the Belle~II \BKnn measurement under the hypothesis of a light invisible resonance \BKX. Using the model-agnostic likelihood method of \cref{sec:method}, I have derived Bayesian credible intervals and frequentist confidence limits for the mass and branching ratio of this potential contribution.

The principal findings are summarized in \cref{tab:bkx results}.
The data exhibit a strong preference for a resonance mass in the range of $2.1-2.4~\text{GeV}$. This finding is approximately independent of the three tested widths ($\Gamma_X = 0.1, 0.5, 1.0~\text{GeV}$), with posterior modes at $m_X = 2.1, 2.2, 2.4~\text{GeV}$, respectively. The 95\% credible intervals are $[1.9, 2.7]~\text{GeV}$ for $\Gamma_X = 0.1~\text{GeV}$, broadening slightly to $[2.0, 3.0]~\text{GeV}$ for $\Gamma_X = 1.0~\text{GeV}$.
This preferred mass range corresponds to $q^2 = m_X^2 \sim 4-6~\text{GeV}^2$, in agreement with the localized excess observed in the Belle~II data around $3~\text{GeV}^2 < q^2_{\rm rec} < 7~\text{GeV}^2$~\cite{Belle-II:2023esi}.

For the narrowest resonance ($\Gamma_X = 0.1~\text{GeV}$), the branching ratio posterior mode is $\mathcal{B}(\BKX) = 9.2 \cdot 10^{-6}$ with a 95\% credible interval of $[3.4, 14.0] \cdot 10^{-6}$. This branching ratio is approximately twice the \ac{SM} prediction for the total \BKnn rate, $\mathcal{B}_{\rm SM} = (4.97 \pm 0.37) \cdot 10^{-6}$~\cite{Parrott:2022zte}.

Both the mass and branching ratio results are consistent with the findings of References~\cite{Fridell:2023ssf,PhysRevD.109.075008, Gabrielli:2024wys}. These References identify best-fit masses around $m_X \approx 2~\text{GeV}$ using simplified treatments of the Belle~II data. The key advance of this analysis is the correct treatment of experimental uncertainties through the model-agnostic likelihood framework. Previous studies necessarily rely on simplified assumptions, and neglected systematic uncertainties.

Goodness-of-fit tests and model comparisons support the reliability of these conclusions.
All three \BKX models provide excellent fits to the data, with goodness-of-fit $P$-values of $P_{\rm gof} \sim 0.81-0.83$. These values exceed those obtained for the \ac{WET} model ($P_{\rm gof} = 0.63$) and the unconstrained \ac{SM} ($P_{\rm gof} = 0.58$) presented in \cref{sec:wet paper goodness-of-fit}. This indicates that the \BKX models fit the data significantly better.

Bayesian model comparison via Bayes factors shows \textit{decisive} preference ($\log_{10} B_{\rm BKG} = 2.82-2.93$) for the \BKX models over the background-only hypothesis, and a \textit{very strong} preference ($\log_{10} B_{\rm SM}^{\rm constr.} = 1.71-1.83$) over the constrained \ac{SM}.
These Bayes factors are the largest observed among all tested hypotheses (\ac{WET}, unconstrained \ac{SM}, \BKX).

The frequentist hypothesis test yields a significance of $Z = 3.0\sigma$ for the \BKX model ($\Gamma_X = 0.1~\text{GeV}$) against the constrained \ac{SM} hypothesis. This represents a global significance that accounts for the look-elsewhere effect through the freedom to fit the resonance mass. The $3.0\sigma$ significance exceeds the $2.7\sigma$ reported in the original Belle~II publication~\cite{Belle-II:2023esi}. This reflects the improved fit quality when the kinematic structure of the excess is explicitly modelled.

However, despite the high statistical significance and the strong Bayes factor, caution is advised before claiming evidence for \BKX. Two factors are the reason for this. First, the Belle~II analysis was optimized for measuring the \ac{SM} \BKnn branching ratio, not for searching for narrow resonances. Second, the \BKX model was motivated by features observed in the unblinded data.\footnote{Particle physics analyses are designed using simulated data. Experimental data are examined only after the analysis procedure is finalized, to avoid bias.} Therefore, the present findings should be regarded as motivation for dedicated resonance searches in future analyses rather than as evidence for new physics.

This comprehensive reinterpretation demonstrates that a light invisible resonance with $m_X \sim 2~\text{GeV}$ and $\mathcal{B}(\BKX) \sim 10^{-5}$ provides an excellent description of the Belle~II \BKnn data, significantly outperforming the \ac{WET} model and the \ac{SM}.

%% file: chapters/errors_syst.tex
\def\errSysnominalBzeroToKshort {\ensuremath 2.75}
\def\errBrnominalBzeroToKshort {\ensuremath 5.64}
\def\errSysnominalBzeroToKstarZero {\ensuremath 1.49}
\def\errBrnominalBzeroToKstarZero {\ensuremath 13.49}
\def\errSysnominalBplusToKstarPlus {\ensuremath 2.19}
\def\errBrnominalBplusToKstarPlus {\ensuremath 21.41}
\def\errSysnominalBplusToKplus {\ensuremath 1.24}
\def\errBrnominalBplusToKplus {\ensuremath 5.49}
\def\errSysnormsBBBzeroToKshort {\ensuremath 1.38}
\def\errSysnormsBBBzeroToKstarZero {\ensuremath 0.60}
\def\errSysnormsBBBplusToKstarPlus {\ensuremath 1.20}
\def\errSysnormsBBBplusToKplus {\ensuremath 0.64}
\def\errSysnormsCBzeroToKshort {\ensuremath 1.64}
\def\errSysnormsCBzeroToKstarZero {\ensuremath 0.53}
\def\errSysnormsCBplusToKstarPlus {\ensuremath 0.58}
\def\errSysnormsCBplusToKplus {\ensuremath 0.22}
\def\errSysMCstatsBzeroToKshort {\ensuremath 1.14}
\def\errSysMCstatsBzeroToKstarZero {\ensuremath 0.59}
\def\errSysMCstatsBplusToKstarPlus {\ensuremath 0.95}
\def\errSysMCstatsBplusToKplus {\ensuremath 0.47}
\def\errSysKLeffBzeroToKshort {\ensuremath 0.02}
\def\errSysKLeffBzeroToKstarZero {\ensuremath 0.34}
\def\errSysKLeffBplusToKstarPlus {\ensuremath 0.19}
\def\errSysKLeffBplusToKplus {\ensuremath 0.27}
\def\errSysKaonIDBzeroToKshort {\ensuremath 0.02}
\def\errSysKaonIDBzeroToKstarZero {\ensuremath 0.07}
\def\errSysKaonIDBplusToKstarPlus {\ensuremath 0.03}
\def\errSysKaonIDBplusToKplus {\ensuremath 0.11}
\def\errSysgammaEMBzeroToKshort {\ensuremath 0.20}
\def\errSysgammaEMBzeroToKstarZero {\ensuremath 0.08}
\def\errSysgammaEMBplusToKstarPlus {\ensuremath 0.46}
\def\errSysgammaEMBplusToKplus {\ensuremath 0.01}
\def\errSysgammaHBzeroToKshort {\ensuremath 0.01}
\def\errSysgammaHBzeroToKstarZero {\ensuremath 0.06}
\def\errSysgammaHBplusToKstarPlus {\ensuremath 0.71}
\def\errSysgammaHBplusToKplus {\ensuremath 0.29}
\def\errSysTrackingBzeroToKshort {\ensuremath 0.03}
\def\errSysTrackingBzeroToKstarZero {\ensuremath 0.00}
\def\errSysTrackingBplusToKstarPlus {\ensuremath 0.21}
\def\errSysTrackingBplusToKplus {\ensuremath 0.11}
\def\errSysBcountingBzeroToKshort {\ensuremath 0.04}
\def\errSysBcountingBzeroToKstarZero {\ensuremath 0.04}
\def\errSysBcountingBplusToKstarPlus {\ensuremath 0.04}
\def\errSysBcountingBplusToKplus {\ensuremath 0.03}
\def\errSysLuminosityBzeroToKshort {\ensuremath 0.01}
\def\errSysLuminosityBzeroToKstarZero {\ensuremath 0.00}
\def\errSysLuminosityBplusToKstarPlus {\ensuremath 0.01}
\def\errSysLuminosityBplusToKplus {\ensuremath 0.01}
\def\errSysLeadingBFBzeroToKshort {\ensuremath 0.16}
\def\errSysLeadingBFBzeroToKstarZero {\ensuremath 0.20}
\def\errSysLeadingBFBplusToKstarPlus {\ensuremath 0.45}
\def\errSysLeadingBFBplusToKplus {\ensuremath 0.21}
\def\errSysDstarstarBzeroToKshort {\ensuremath 0.15}
\def\errSysDstarstarBzeroToKstarZero {\ensuremath 0.31}
\def\errSysDstarstarBplusToKstarPlus {\ensuremath 0.40}
\def\errSysDstarstarBplusToKplus {\ensuremath 0.14}
\def\errSysnnbarBzeroToKshort {\ensuremath 0.16}
\def\errSysnnbarBzeroToKstarZero {\ensuremath 0.08}
\def\errSysnnbarBplusToKstarPlus {\ensuremath 0.16}
\def\errSysnnbarBplusToKplus {\ensuremath 0.20}
\def\errSysBDTtwoefficiencyBzeroToKshort {\ensuremath 0.05}
\def\errSysBDTtwoefficiencyBzeroToKstarZero {\ensuremath 0.06}
\def\errSysBDTtwoefficiencyBplusToKstarPlus {\ensuremath 0.08}
\def\errSysBDTtwoefficiencyBplusToKplus {\ensuremath 0.03}
\def\errSysKplusKLKLBzeroToKshort {\ensuremath 0.02}
\def\errSysKplusKLKLBzeroToKstarZero {\ensuremath 0.01}
\def\errSysKplusKLKLBplusToKstarPlus {\ensuremath 0.05}
\def\errSysKplusKLKLBplusToKplus {\ensuremath 0.48}
\def\errSysKplusKSKLBzeroToKshort {\ensuremath 0.00}
\def\errSysKplusKSKLBzeroToKstarZero {\ensuremath 0.00}
\def\errSysKplusKSKLBplusToKstarPlus {\ensuremath 0.01}
\def\errSysKplusKSKLBplusToKplus {\ensuremath 0.02}
\def\errSysKzKzKzBzeroToKshort {\ensuremath 0.49}
\def\errSysKzKzKzBzeroToKstarZero {\ensuremath 0.00}
\def\errSysKzKzKzBplusToKstarPlus {\ensuremath 0.00}
\def\errSysKzKzKzBplusToKplus {\ensuremath 0.00}
\def\errSysKstarKzKzBzeroToKshort {\ensuremath 0.02}
\def\errSysKstarKzKzBzeroToKstarZero {\ensuremath 0.22}
\def\errSysKstarKzKzBplusToKstarPlus {\ensuremath 0.01}
\def\errSysKstarKzKzBplusToKplus {\ensuremath 0.01}
\def\errSysDtoKLBzeroToKshort {\ensuremath 0.04}
\def\errSysDtoKLBzeroToKstarZero {\ensuremath 0.02}
\def\errSysDtoKLBplusToKstarPlus {\ensuremath 0.17}
\def\errSysDtoKLBplusToKplus {\ensuremath 0.13}
\def\errSysOffreslumiBzeroToKshort {\ensuremath 0.06}
\def\errSysOffreslumiBzeroToKstarZero {\ensuremath 0.11}
\def\errSysOffreslumiBplusToKstarPlus {\ensuremath 0.23}
\def\errSysOffreslumiBplusToKplus {\ensuremath 0.10}
\def\errSysBDTcallBzeroToKshort {\ensuremath 0.51}
\def\errSysBDTcallBzeroToKstarZero {\ensuremath 0.41}
\def\errSysBDTcallBplusToKstarPlus {\ensuremath 0.46}
\def\errSysBDTcallBplusToKplus {\ensuremath 0.09}
\def\errSysPizeffBzeroToKshort {\ensuremath 0.01}
\def\errSysPizeffBzeroToKstarZero {\ensuremath 0.01}
\def\errSysPizeffBplusToKstarPlus {\ensuremath 0.03}
\def\errSysPizeffBplusToKplus {\ensuremath 0.01}
\def\errSysKshortStatBzeroToKshort {\ensuremath 0.05}
\def\errSysKshortStatBzeroToKstarZero {\ensuremath 0.00}
\def\errSysKshortStatBplusToKstarPlus {\ensuremath 0.01}
\def\errSysKshortStatBplusToKplus {\ensuremath 0.00}
\def\errSysKshortSystBzeroToKshort {\ensuremath 0.28}
\def\errSysKshortSystBzeroToKstarZero {\ensuremath 0.01}
\def\errSysKshortSystBplusToKstarPlus {\ensuremath 0.02}
\def\errSysKshortSystBplusToKplus {\ensuremath 0.01}
\def\errSysSignalFFBzeroToKshort {\ensuremath 0.06}
\def\errSysSignalFFBzeroToKstarZero {\ensuremath 0.13}
\def\errSysSignalFFBplusToKstarPlus {\ensuremath 0.13}
\def\errSysSignalFFBplusToKplus {\ensuremath 0.06}
\def\errSysFakeYfourSBzeroToKshort {\ensuremath 0.01}
\def\errSysFakeYfourSBzeroToKstarZero {\ensuremath 0.14}
\def\errSysFakeYfourSBplusToKstarPlus {\ensuremath 0.14}
\def\errSysFakeYfourSBplusToKplus {\ensuremath 0.06}
\def\errSysFakeDplusBzeroToKshort {\ensuremath 0.00}
\def\errSysFakeDplusBzeroToKstarZero {\ensuremath 0.01}
\def\errSysFakeDplusBplusToKstarPlus {\ensuremath 0.01}
\def\errSysFakeDplusBplusToKplus {\ensuremath 0.00}
\def\errSysFakeDzeroBzeroToKshort {\ensuremath 0.01}
\def\errSysFakeDzeroBzeroToKstarZero {\ensuremath 0.08}
\def\errSysFakeDzeroBplusToKstarPlus {\ensuremath 0.04}
\def\errSysFakeDzeroBplusToKplus {\ensuremath 0.02}
\def\errSysKoneNormBzeroToKshort {\ensuremath 0.03}
\def\errSysKoneNormBzeroToKstarZero {\ensuremath 0.14}
\def\errSysKoneNormBplusToKstarPlus {\ensuremath 0.09}
\def\errSysKoneNormBplusToKplus {\ensuremath 0.01}
\def\errSysKoneCompositionBzeroToKshort {\ensuremath 0.01}
\def\errSysKoneCompositionBzeroToKstarZero {\ensuremath 0.06}
\def\errSysKoneCompositionBplusToKstarPlus {\ensuremath 0.02}
\def\errSysKoneCompositionBplusToKplus {\ensuremath 0.00}
\def\numberOfNuisanceParams {\ensuremath 338}
\def\numberOfBFNuisanceParams {\ensuremath 113}
\def\bgYieldSRKplus {\ensuremath 15455 }
\def\bgYieldSRKshort {\ensuremath 14516 }
\def\bgYieldSRKstarZ {\ensuremath 39101 }
\def\bgYieldSRKstarP {\ensuremath 39186 }
\def\bgYieldHSRKplus {\ensuremath  896 }
\def\bgYieldHSRKshort {\ensuremath  736 }
\def\bgYieldHSRKstarZ {\ensuremath 1331 }
\def\bgYieldHSRKstarP {\ensuremath  708 }
\def\sgYieldSRKplus {\ensuremath  141 }
\def\sgYieldSRKshort {\ensuremath   60 }
\def\sgYieldSRKstarZ {\ensuremath  164 }
\def\sgYieldSRKstarP {\ensuremath  111 }
\def\sgYieldHSRKplus {\ensuremath   35 }
\def\sgYieldHSRKshort {\ensuremath   15 }
\def\sgYieldHSRKstarZ {\ensuremath   33 }
\def\sgYieldHSRKstarP {\ensuremath   19 }

\def\errIsoCentPS {\ensuremath 1.13}
\def\errIsoBrBplusToKplus {\ensuremath 5.03}
\def\errIsoBrBzeroToKshort {\ensuremath 2.32}
\def\errIsoCentV {\ensuremath 1.30}
\def\errIsoBrBplusToKstarPlus {\ensuremath 12.70}
\def\errIsoBrBzeroToKstarZero {\ensuremath 11.74}

\def\errOneCent {\ensuremath 0.82}
\def\errOneBrBzeroToKshort {\ensuremath 1.68}
\def\errOneBrBzeroToKstarZero {\ensuremath 7.44}
\def\errOneBrBplusToKstarPlus {\ensuremath 8.05}
\def\errOneBrBplusToKplus {\ensuremath 3.65}

%% file: chapters/errors_syst_pyhf.tex
\def\BrnominalBzeroToKshort {2.05}
\def\BrnominalBzeroToKstarZero {9.05}
\def\BrnominalBplusToKstarPlus {9.79}
\def\BrnominalBplusToKplus {4.44}

\def\errSysnominalBzeroToKshortPYHF {2.75}
\def\errSysnominalBzeroToKstarZeroPYHF {1.49}
\def\errSysnominalBplusToKstarPlusPYHF {2.17}
\def\errSysnominalBplusToKplusPYHF {1.23}

\def\errSysnominalISOmuPSPYHF {1.13}
\def\errSysnominalISOmuVPYHF {1.30}

\def\errSysnominalALLmuPYHF {0.82}

\pgfmathsetmacro{\errBrnominalBzeroToKshortPYHF} {\BrnominalBzeroToKshort * \errSysnominalBzeroToKshortPYHF}
\pgfmathsetmacro{\errBrnominalBzeroToKstarZeroPYHF} {\BrnominalBzeroToKstarZero * \errSysnominalBzeroToKstarZeroPYHF}
\pgfmathsetmacro{\errBrnominalBplusToKstarPlusPYHF} {\BrnominalBplusToKstarPlus * \errSysnominalBplusToKstarPlusPYHF}
\pgfmathsetmacro{\errBrnominalBplusToKplusPYHF} {\BrnominalBplusToKplus * \errSysnominalBplusToKplusPYHF}

\pgfmathsetmacro{\errBrnominalISOBzeroToKshortPYHF} {\BrnominalBzeroToKshort * \errSysnominalISOmuPSPYHF}
\pgfmathsetmacro{\errBrnominalISOBzeroToKstarZeroPYHF} {\BrnominalBzeroToKstarZero * \errSysnominalISOmuVPYHF}
\pgfmathsetmacro{\errBrnominalISOBplusToKstarPlusPYHF} {\BrnominalBplusToKstarPlus * \errSysnominalISOmuVPYHF}
\pgfmathsetmacro{\errBrnominalISOBplusToKplusPYHF} {\BrnominalBplusToKplus * \errSysnominalISOmuPSPYHF}

\pgfmathsetmacro{\errBrnominalALLBzeroToKshortPYHF} {\BrnominalBzeroToKshort * \errSysnominalALLmuPYHF}
\pgfmathsetmacro{\errBrnominalALLBzeroToKstarZeroPYHF} {\BrnominalBzeroToKstarZero * \errSysnominalALLmuPYHF}
\pgfmathsetmacro{\errBrnominalALLBplusToKstarPlusPYHF} {\BrnominalBplusToKstarPlus * \errSysnominalALLmuPYHF}
\pgfmathsetmacro{\errBrnominalALLBplusToKplusPYHF} {\BrnominalBplusToKplus * \errSysnominalALLmuPYHF}

%% file: chapters/knunu-update.tex
\footnote{
The material presented in this chapter is the result of collaborative work within the Belle~II collaboration. Individual contributions of the proponents of this work and further collaboration members are documented in \cref{sec:update-contributions}. My personal contributions are the theoretical predictions in \cref{sec:update theory} and the statistical inference in \cref{sec:signal_extraction}.
Furthermore, this is a work-in-progress project, and the presented results are subject to change.
}
In the previous chapters, I have demonstrated how powerful reinterpretation can be. For reinterpretation in terms of the \ac{WET}, the underlying theory only allows us to probe sums of Wilson coefficients rather than individual ones for \BKnunu decays. This is evident from \cref{eq:width}.
I have also shown that the \ac{WET} prediction for \BKsnunu decays is sensitive to the same set of Wilson coefficients but has a different structure, as shown in \cref{eq:widthKs}.
Hence, a combined analysis of \BKnunu and \BKsnunu enables probing the magnitude of all five contributing Wilson coefficients, as discussed in \cref{sec:method paper ksnunu}.

In this chapter, I present the work-in-progress Belle~II analysis of the four channels \BpKpnn, \BzKshnn, \BzKstnn, and \BpKstnn in combination. 
This study is not only interesting as a comparison with the existing \BKnn analysis, but also enables very powerful reinterpretation studies, as shown in \cref{sec:method paper ksnunu}.

This chapter is structured as follows:
\Cref{sec:update theory} presents the theoretical predictions for all four decay channels.
\Cref{sec:outline} provides an overview of the analysis strategy, extending from the published \BpKpnn measurement to a simultaneous four-channel analysis.
The background suppression techniques are described in \cref{sec:background_suppression}.
The signal region definition is covered in \cref{sec:SR_def}.
\Cref{sec:signal_extraction} describes the statistical framework for simultaneous signal extraction, including uncorrelated, isospin-averaged, and fully correlated scenarios.
Finally, \cref{sec:discussion_update} discusses the expected sensitivity improvements and implications for reinterpretation studies.

\section{Theory predictions}
\label{sec:update theory}
As should be evident by now, theory predictions form the foundation of any analysis. This analysis is based on the \ac{SM} prediction, which is obtained from \cref{eq:width,eq:widthKs} at the \ac{SM} Wilson coefficient point given in \cref{eq:WET-SM-point}.

\begin{figure}[hbt]
    \centering
    \includegraphics[width=0.49\linewidth]{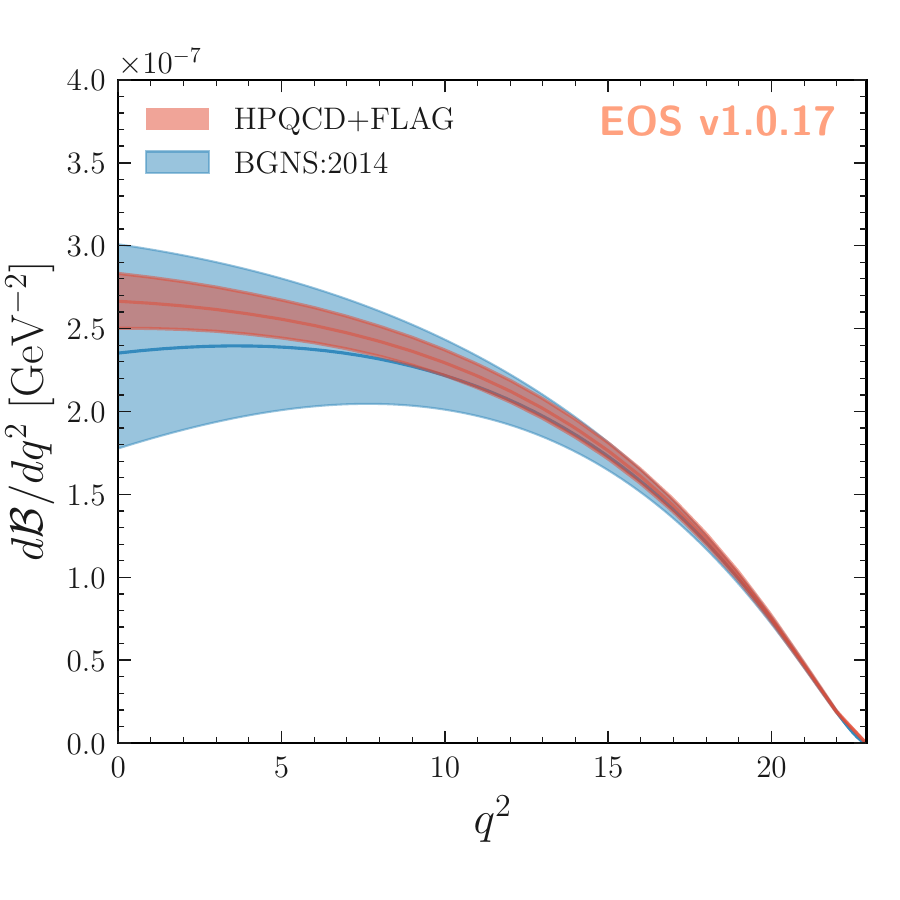}
    \includegraphics[width=0.49\linewidth]{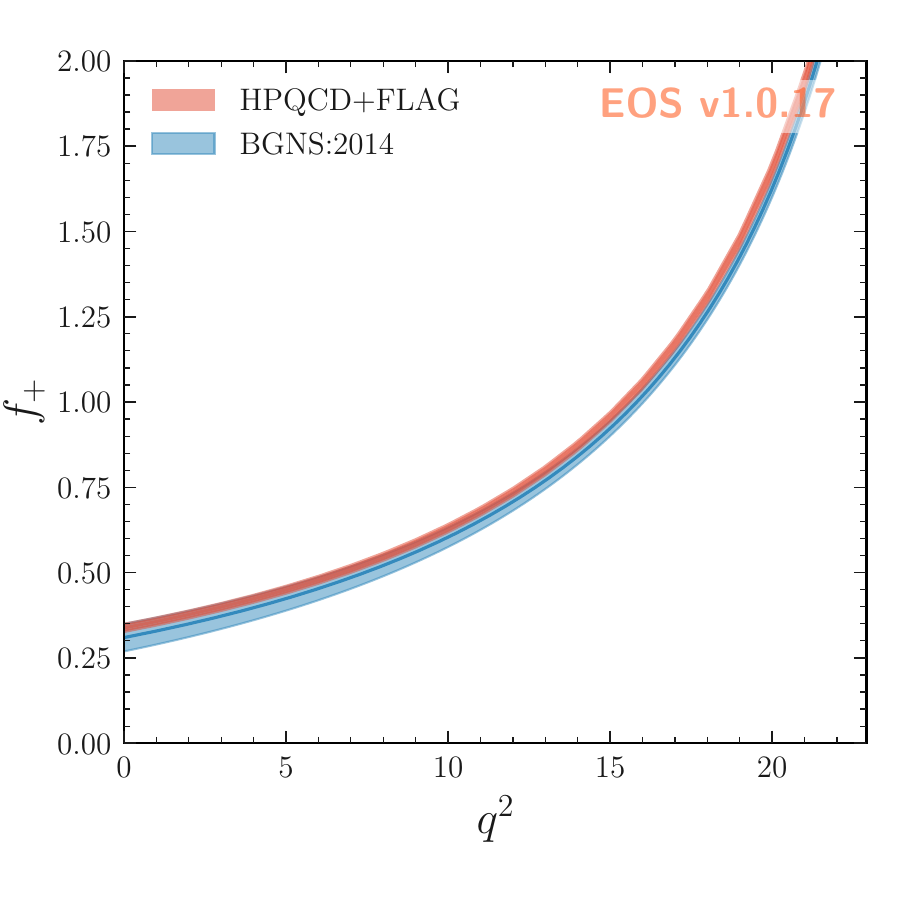}
    \caption{The \BKnunu \ac{SM} differential branching ratio (left) and $f_+$ form factor (right), based on the hadronic parameters from Reference~\cite{Buras:2014fpa} (labelled as \texttt{BGNS:2014}) and from References~\cite{FlavourLatticeAveragingGroupFLAG:2021npn,Parrott:2022rgu} (labelled as \texttt{HPQCD+FLAG}). The bands represent the uncertainty in the hadronic parameters.}
    \label{fig:hadronic parameters comparison K}
\end{figure}

The signal \ac{MC} for the \BpKpnn and \BzKshnn channels is based on the hadronic parameters from Reference~\cite{Buras:2014fpa}. To base the analysis on the most recent findings, the samples are reweighted to hadronic parameters obtained from a joint theoretical prior \ac{PDF} composed of the 2021 lattice world average, based on results by the Fermilab/MILC and HPQCD collaborations~\cite{FlavourLatticeAveragingGroupFLAG:2021npn,Parrott:2022rgu}, see \cref{tab:hadronic parameters full}. The reduction of uncertainties in the hadronic parameters is significant, as shown in \cref{fig:hadronic parameters comparison K}.

\begin{figure}[hbt]
    \centering
    \includegraphics[width=0.49\linewidth]{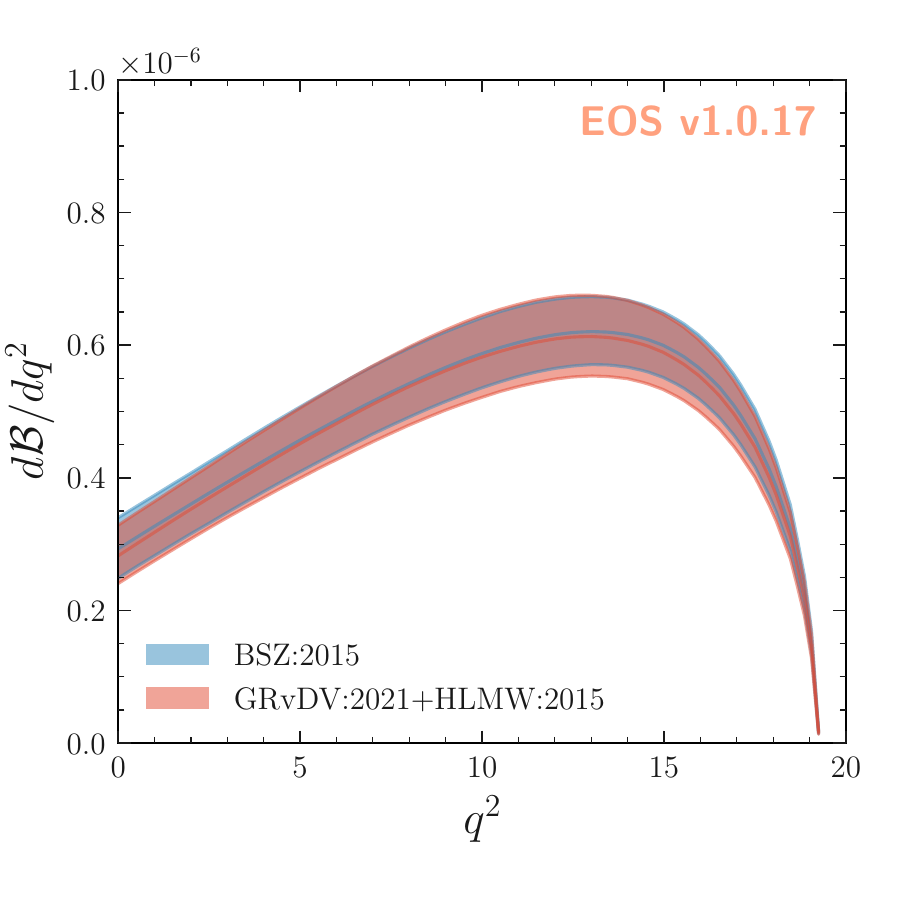}
    \includegraphics[width=0.49\linewidth]{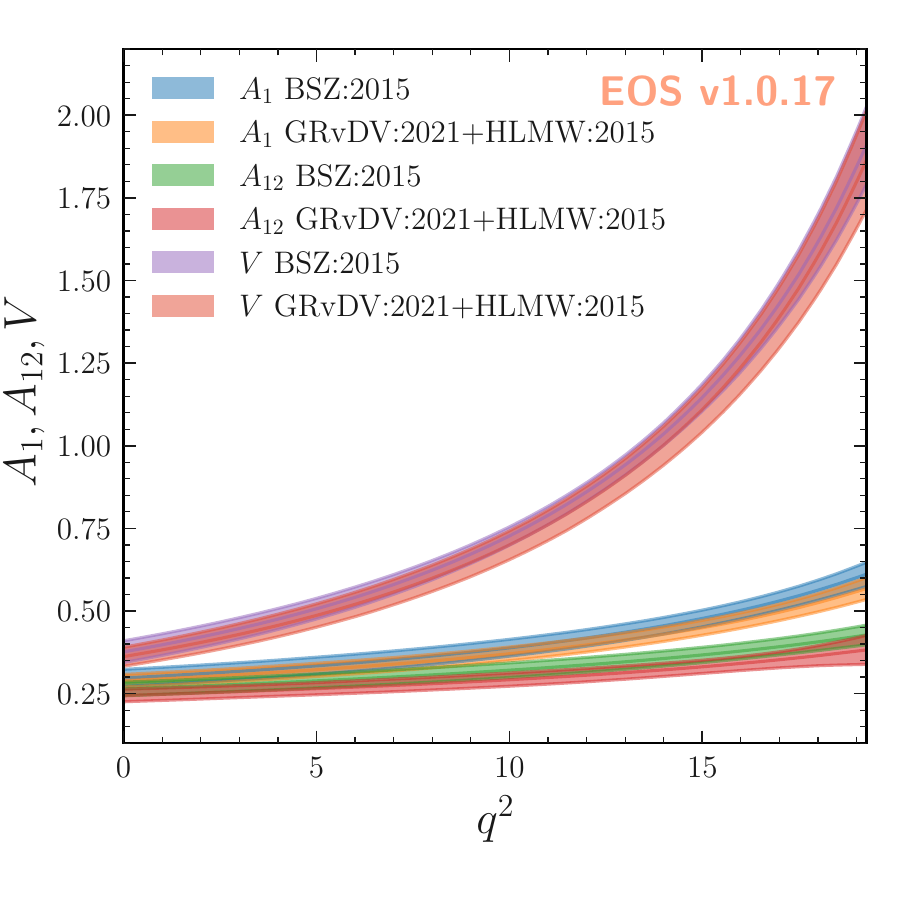}
    \caption{The \BKsnunu \ac{SM} differential branching ratio (left) and $A_1, A_{12}, V$ form factors (right), based on the hadronic parameters from Reference~\cite{Bharucha_2016} (labelled as \texttt{BSZ:2015}) and from References~\cite{Gubernari:2020eft,Horgan:2015vla} (labelled as \texttt{GRvDV:2021+HLMW:2015}). The bands represent the uncertainty in the hadronic parameters.}
    \label{fig:hadronic parameters comparison Ks}
\end{figure}

The signal \ac{MC} for the \BzKstnn and \BpKstnn channels is based on the hadronic parameters from Reference~\cite{Bharucha_2016}. In this case, the samples are reweighted to hadronic parameters obtained from a joint theoretical prior \ac{PDF} based on results from References~\cite{Gubernari:2020eft,Horgan:2015vla}, see \cref{tab:hadronic parameters kstar full update}. Here there is no improvement in the uncertainties, as shown in \cref{fig:hadronic parameters comparison Ks}. The result is nevertheless more reliable due to the combination of multiple sources.

The predictions are obtained using the \EOS software, discussed in \cref{sec:eos}. The relevant input parameters are summarized in \cref{tab:sm parameters}. 
The product $ |V_{tb}V_{ts}^*| $ is evaluated using \ac{CKM} unitarity from the value of $|V_{cb}|$, which can be determined using exclusive or inclusive methods~\cite{ParticleDataGroup:2024cfk}.
The predictions reported here are based on the product ${|V_{tb}V_{ts}^*| = 0.039\pm0.001}$ from exclusive modes~\cite{Becirevic:2023aov,ParticleDataGroup:2024cfk,FlavourLatticeAveragingGroupFLAG:2021npn}.

Theoretical uncertainties arise from two main sources: uncertainties in the \ac{CKM} matrix elements and the hadronic form factors, with small additional contributions from other parametric dependencies and higher-order corrections.
The discrepancies in the determinations of the matrix element $|V_{cb}|$ using exclusive or inclusive methods contribute a 6\% uncertainty~\cite{ParticleDataGroup:2024cfk}.
Predictions for decays involving a pseudoscalar in the final state contain a 4\% form factor uncertainty, such that the \ac{CKM} uncertainties dominate.
Conversely, predictions for decays with a vector meson in the final state have a dominant 10\% form factor uncertainty.

The \ac{SM} predictions for the total branching ratios of the four decays are summarized in \cref{tab:predbr}. 

\begin{table}[ht]
  \caption{\ac{SM} branching ratios for \BKnunu decays used in this work (as described in the text), compared to predictions from Reference~\cite{Becirevic:2023aov}. The uncertainties are separated into contributions from hadronic form factors (FF) and \ac{CKM} elements (CKM). Note that Reference~\cite{Becirevic:2023aov} uses slightly different input parameters. In addition, it assumes that \BKsnunu form factor parameters are uncorrelated, leading to larger form factor uncertainties.}
  \centering
  \begin{tabularx}{\linewidth}{@{\extracolsep{\fill}}lYYY}
    \toprule \midrule
    Decay &  \multicolumn{2}{c}{$\mathcal{B_{\rm SM}}\cdot 10^{6}$}\\
    & This work & Reference~\cite{Becirevic:2023aov}\\
    \midrule
    \BpKpnn  & $4.32 \pm 0.16 ({\rm FF}) \pm 0.26 ({\rm CKM})$ & $4.37 \pm 0.15({\rm FF}) \pm 0.26 ({\rm CKM})$   \\
    \BzKshnn & $2.00 \pm 0.07 ({\rm FF}) \pm 0.12 ({\rm CKM})$ & $2.02 \pm 0.07({\rm FF}) \pm 0.12 ({\rm CKM})$   \\
    \BpKstnn & $9.35 \pm 0.94 ({\rm FF}) \pm 0.56 ({\rm CKM})$ & $9.73 \pm 1.37({\rm FF}) \pm 0.58 ({\rm CKM})$  \\
    \BzKstnn & $8.65 \pm 0.87 ({\rm FF}) \pm 0.57 ({\rm CKM})$ & $8.97 \pm 1.22({\rm FF}) \pm 0.54 ({\rm CKM})$   \\
    \midrule \bottomrule
  \end{tabularx}
  \label{tab:predbr}
\end{table}



The long-distance double-charged current contribution $B \to (\tau \to K \nu_\tau) \bar \nu_\tau$ is treated as a background in this analysis. The contribution is normalized based on experimental measurements of $B^+ \to \tau^+ \nu_\tau$ and $\tau^+ \to K^{(*)+} \bar{\nu}_\tau$ decays~\cite{ParticleDataGroup:2024cfk}.



\section{Analysis outline}
\label{sec:outline}

The analysis is designed as an extension of the published \BpKpnn\ measurement described in \cref{sec:wet paper analysis}, using the same Belle~II data sample (see \cref{sec:analysis-data}). The \BpKpnn\ analysis itself is largely unchanged, with the selected data sample in the signal region left unmodified, and the simulated samples also remain unchanged. However, the relative contributions of the leading backgrounds are updated to reflect the latest knowledge, ensuring consistency with the other channels. In contrast to the published version, where only the uncertainties were reweighted, both the central values and uncertainties of the background contributions are now updated. This leads to an overall reduction in background and reduces the expected uncertainty for the \BpKpnn channel compared to the published result.
The signal-extraction machinery is updated to determine the four channels of interest simultaneously, accounting for their mutual cross-influences. Additional updates include newly developed control channels.

Conversely, the measurements of \BzKshnn, \BpKstnn, and \BzKstnn\ are designed to follow the \BpKpnn\ analysis (see \cref{sec:wet paper analysis}) as closely as possible, unless deviations provide significant benefits. The main update is the use of improved simulated samples that incorporate dedicated data to model beam-background activity. This enables the inclusion of additional variables to discriminate signal from background and to fine-tune the reconstruction. It was verified that these updates have only a minor impact on the \BpKpnn\ measurement.

The analysis begins with the reconstruction of charged and neutral particles, followed by the selection of signal kaon candidates in events with one or more $K$ mesons. Subsequently, relevant quantities to discriminate between signal and background processes are computed using the kaon candidates together with the remaining particles in the event. These quantities are used in \acp{BDT}, which are optimized and trained on simulated samples. A signal region is then defined, and a single $K$-meson candidate is selected. 

Up to this point, the analysis is performed for each channel individually, while accounting for correlated systematic uncertainties. With the signal-region samples prepared for all four channels, a simultaneous binned profile-likelihood sample-composition fit is performed on data using a \histfactory statistical model (\cref{sec:histfactory}). This fit uses simulated samples to provide predictions and determines the branching ratios of the four decay modes along with the rates of background processes. The fit incorporates systematic uncertainties arising from detector and physics-modelling imperfections as nuisance parameters. To validate the modelling of signal and background processes in simulation, several control channels are employed.

\section{Background suppression}
\label{sec:background_suppression}
Simulated signal and background events are used to train \acp{BDT} that suppress background. Several inputs are considered, including general event-shape variables as well as variables characterizing the signal candidates and the kinematic properties of the \ac{ROE}.

This analysis uses two consecutive \acp{BDT}, as described in \cref{sec:analysis-background-suppression}.
The \BDT1 classifier is constructed using the same input variables for the \BzKshnn and \BzKstnn channels as the published \BpKpnn channel. For the \BpKstnn channel, the classifier uses a different set of 11 variables, which additionally contains kinematic properties of the \Kstarp decay products. 
In the \BzKshnn and \BzKstnn channels, the most powerful discriminant is the difference between the \ac{ROE} energy in the centre-of-mass frame and half the total collision energy, $\Delta E_{\mathrm{ROE}}$.
For the \BpKstnn channel, the modified Fox-Wolfram moments~\cite{Belle:2003fgr} provide the strongest separation power.

The second classifier, \BDT2, is used for the final event selection, which is trained on events with \BDT1$>0.9$. 
For the \BzKshnn channel, the most discriminating variable in \BDT2 is the cosine of the angle between the momentum of the signal \KS candidate and its vertex displacement vector.
For the \BzKstnn channel, the most powerful variable is the invariant mass of the \Kp\pim pair of the \Kstarz candidate.
For the \BpKstnn channel, the most discriminating variable is $\Delta E_{\mathrm{ROE}}$.

\section{Signal region definition}
\label{sec:SR_def}
Using the simulated signal sample, the \BDT2 variable is mapped to the complement of the integrated signal-selection efficiency, as in \cref{eq:eta-bdt2}.
The \ac{SR} is defined by requiring $\BDT1 > 0.9$ for all channels. In addition, channel-dependent requirements are applied to $\eta(\BDT2)$: $\eta(\BDT2) > 0.92$ for \BzKshnn, $\eta(\BDT2) > 0.95$ for \BzKstnn, and $\eta(\BDT2) > 0.97$ for \BpKstnn.

The \ac{SR} is further divided into $4\times3$, $5\times3$ and $6\times3$ bins in the $\eta(\BDT2)\times q^2_{\mathrm{rec}}$ space for the channels \BzKshnn, \BzKstnn and \BpKstnn, respectively. The $q^{2}_{\mathrm{rec}}$ bin boundaries are the same for all channels and are given by $[-1.0, 4.0, 8.0, 25.0]~\gev^{2}$. The bin boundaries in $q^2_{\mathrm{rec}}$ are chosen to follow the theoretical predictions from Reference~\cite{Buras:2014fpa} while ensuring a sufficient number of expected signal events in each bin.

For the \BzKshnn channel, the bin boundaries in $\eta(\BDT2)$ are $[0.92, 0.94, 0.96, 0.98, 1.00]$. For the \BzKstnn channel, the bin boundaries in $\eta(\BDT2)$ are $[0.95, 0.96, 0.97, 0.98, 0.99, 1.00]$. For the \BpKstnn channel, the bin boundaries in $\eta(\BDT2)$ are $[0.97, 0.975, 0.98, 0.985, 0.99, 0.995, 1.00]$. 
The binning choice and the lower boundary for $\eta(\BDT2)$ are optimized by minimizing the expected statistical uncertainty on the \ac{POI}. 
The last bin of $\eta(\BDT2)$ provides the dominant sensitivity to the signal and is therefore referred to as the \ac{HSR}.

\begin{figure}[ht]
    \centering
    \includegraphics[width=0.49\linewidth]{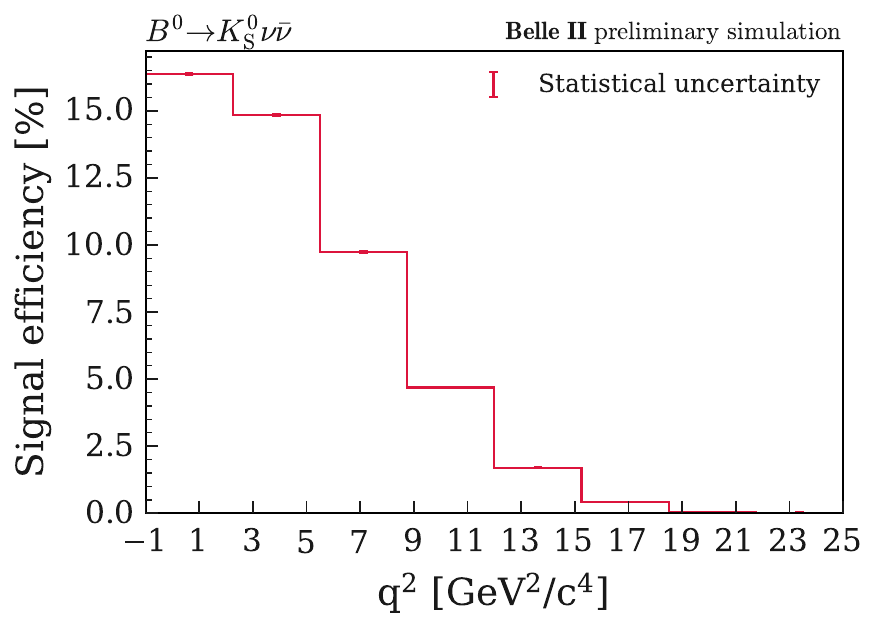}\\
    \includegraphics[width=0.49\linewidth]{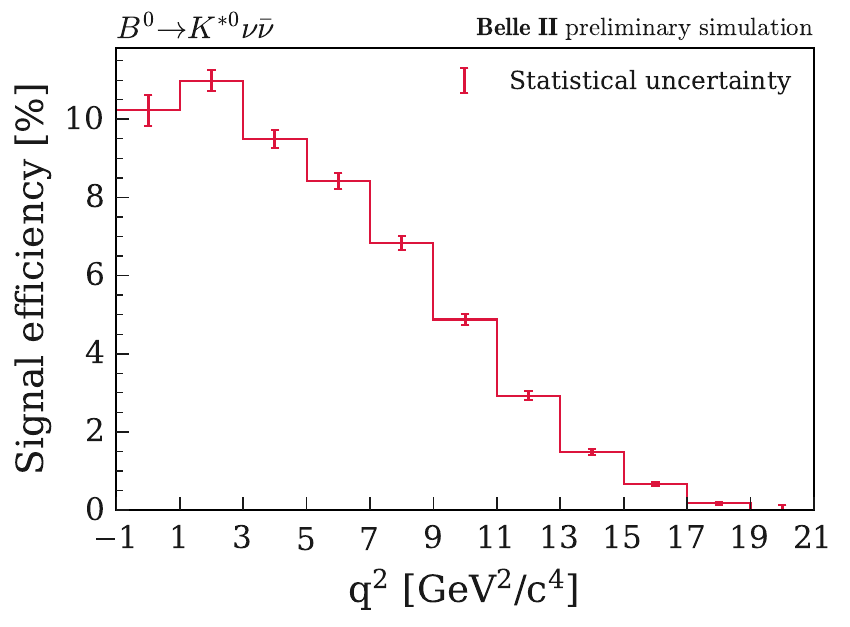}
    \includegraphics[width=0.49\linewidth]{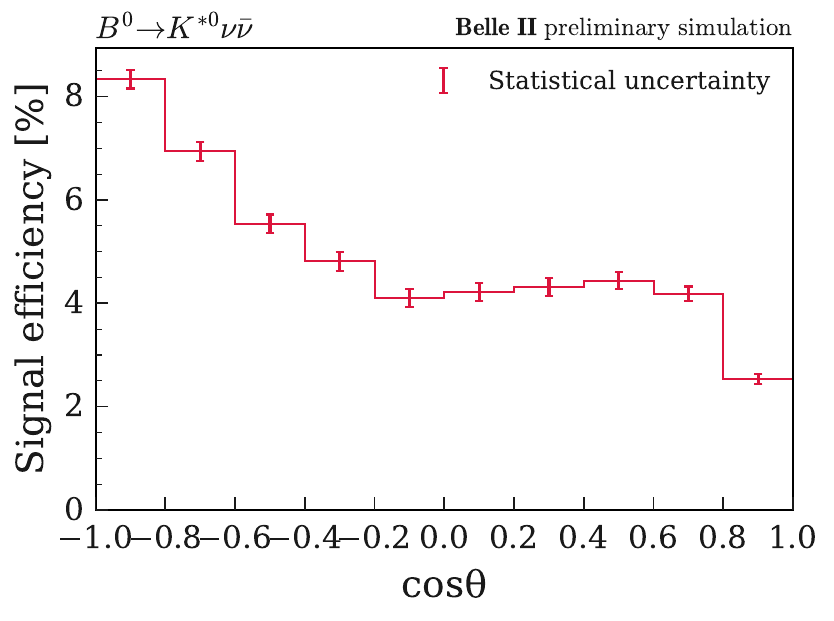}
    \includegraphics[width=0.49\linewidth]{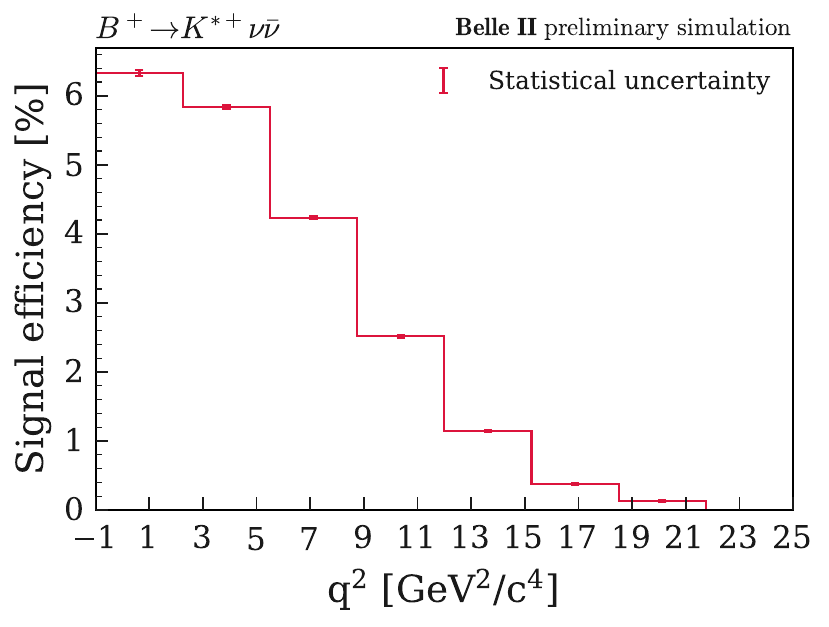}
    \includegraphics[width=0.49\linewidth]{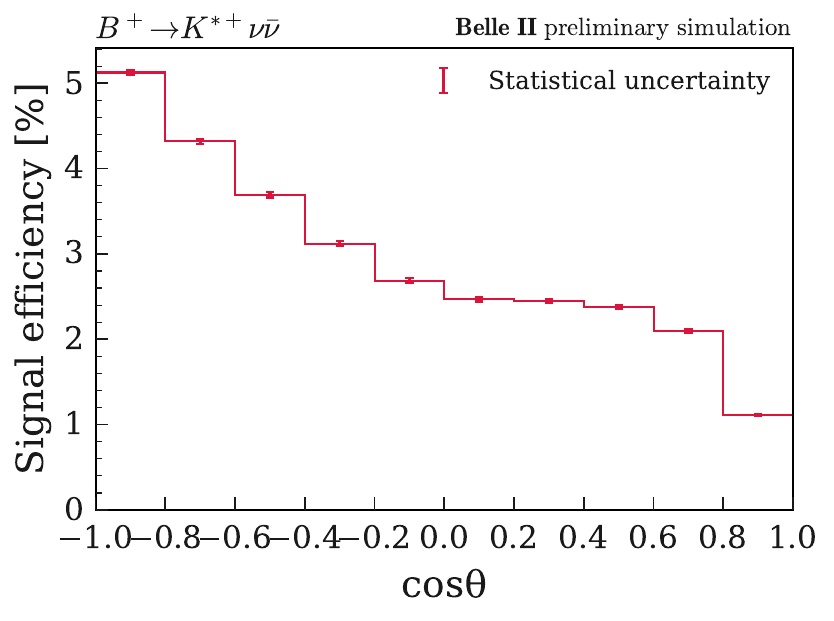}
 \caption{Signal efficiency as a function of $q^2$ and $\cos\theta$ for all channels, determined from the signal region selection. The $\cos\theta$ distributions are provided for vector mesons only.}
    \label{fig:effvsq2th}
\end{figure}

The signal-selection efficiency in the \ac{SR} for the \BzKshnn, \BpKstnn, and \BzKstnn decays is shown in \cref{fig:effvsq2th}. The efficiency varies significantly as a function of the kinematic variables $q^2$ and the helicity angle $\cos \theta$, which is defined as the angle between the \Kstar boost vector and the kaon direction in the \Kstar rest frame.



\section{Signal extraction}
\label{sec:signal_extraction}
A \histfactory statistical model is constructed to extract the signal strengths of the four channels \BpKpnn, \BpKstnn, \BzKshnn, and \BzKstnn simultaneously. Inference is performed with \pyhf (\cref{sec:pyhf}), across the four channels, using both on-resonance $\Upsilon(4S)$ data and off-resonance data to constrain the continuum background.



The systematic uncertainties affect normalizations and may also influence the shape of the templates.
The three dominant sources are:
\begin{itemize}
\item
The yields of the seven individual background categories are allowed to vary independently in the fit. In each case, a Gaussian constraint (\texttt{normsys} modifier, see \cref{sec:histfactory-modifiers}) is added to the fit, centred at the expectation based on simulation and with a standard deviation corresponding to 50\% of the central value. The conservative 50\% value is motivated by a global normalization difference between the off-resonance data and continuum simulation for the \BpKpnn channel. 
The uncertainty is modelled by using seven nuisance parameters for each of the four channels leading to in total 28 parameters in the fit.
\item
The systematic uncertainty due to the limited size of simulated samples (\texttt{staterror} modifier, see \cref{sec:histfactory-modifiers}), which also includes small contributions of uncorrelated uncertainty arising from covariance matrix decomposition for \ac{PID}, \KS, and \piz efficiency is modelled by an independent nuisance parameter for each analysis bin (114 nuisance parameters in total).
This is the second-largest uncertainty of the analysis, ranging between 0.5 and 1.2.
\item
The branching ratios of decay modes contributing to more than $80\%$ of the background $B$ decays in the \ac{SR} are allowed to vary according to their known uncertainties~\cite{ParticleDataGroup:2024cfk}. They are described by 113 nuisance parameters (\texttt{histosys} modifier, see \cref{sec:histfactory-modifiers}). 
\end{itemize}
To account for all systematic sources, including the ones above, but also smaller contributions, a total of 338 nuisance parameters, along with the four signal strengths $\mu$, are varied in the fit.

The fitting proceeds in two steps: A first fit is performed using \texttt{scipy}~\cite{2020SciPy-NMeth} as the minimizer. Then, a second fit is performed using \texttt{Minuit (Migrad)}~\cite{James:1994vla} as the minimizer with the result of the first fit as the starting point. The covariance matrices for the \acp{POI} and nuisance parameters are obtained using the \texttt{Hesse} algorithm. This two-step approach significantly increases the robustness of the fit. 

\begin{figure}[hbt]
    \centering
    \includegraphics[width=0.45\linewidth]{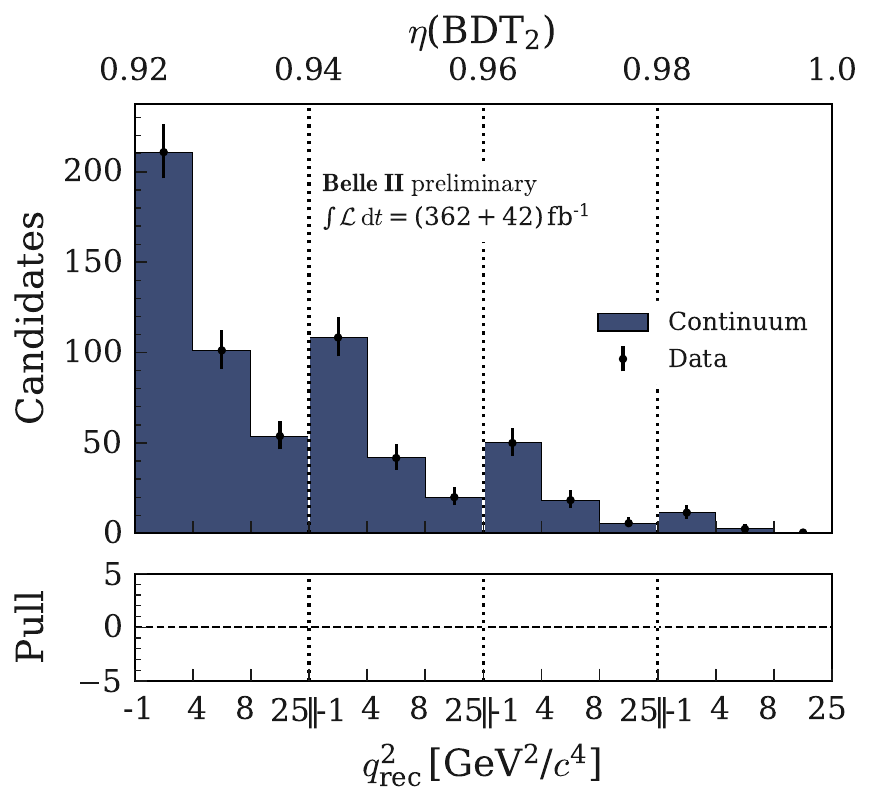}
    \includegraphics[width=0.45\linewidth]{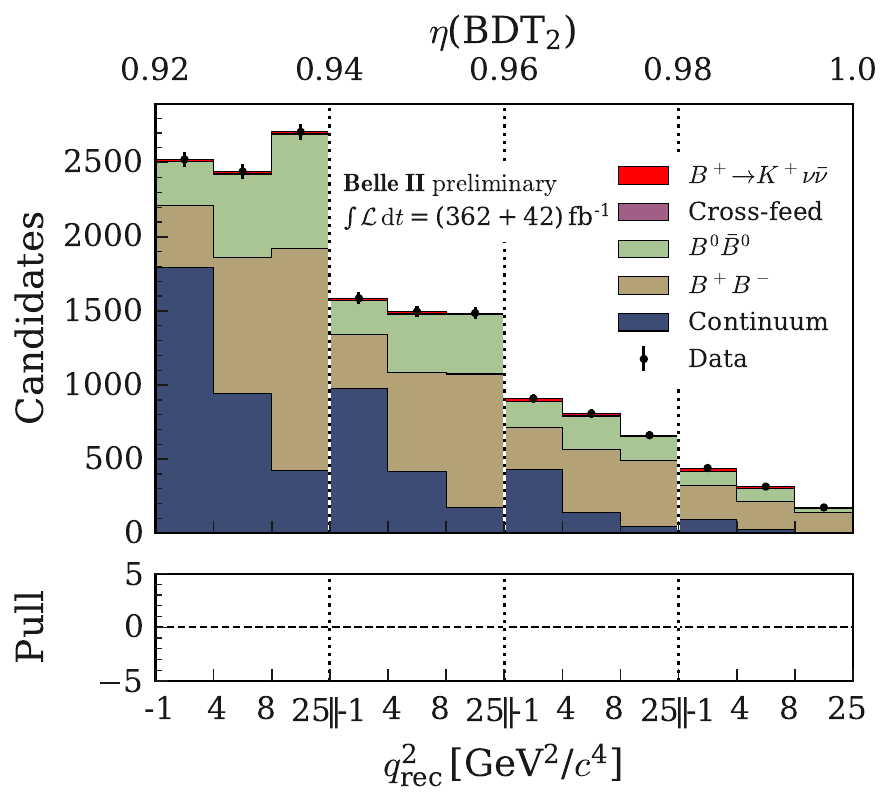}
    \includegraphics[width=0.45\linewidth]{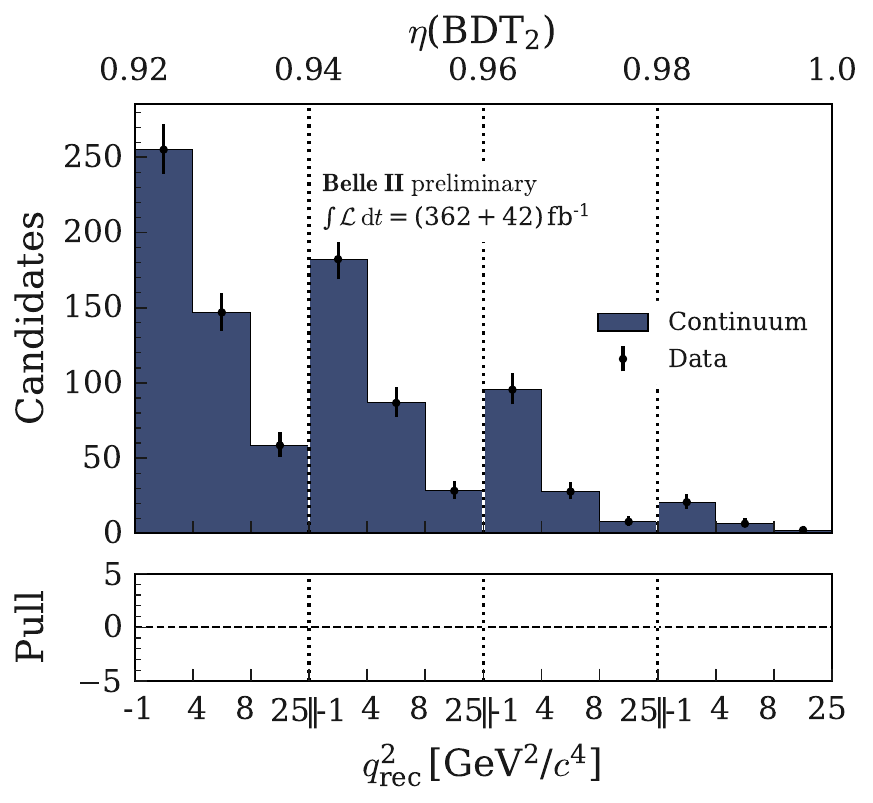}
    \includegraphics[width=0.45\linewidth]{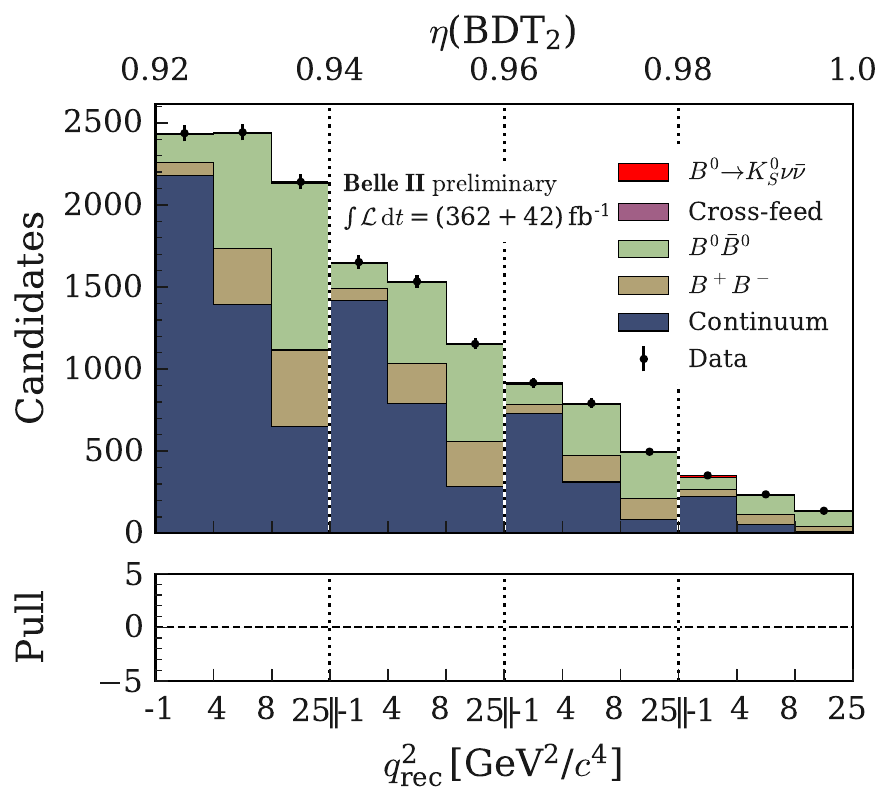}
    \caption{Post fit distribution from the Asimov fit of the \BpKpnn and \BzKshnn channels. The off-resonance region is shown on the \textit{left} and the on-resonance region on the \textit{right}. For each signal channel, the cross-feed indicates the events from the other three signal channels combined.}
    \label{fig:postfit_asimov1}
\end{figure}

\begin{figure}[hbt]
    \centering
    \includegraphics[width=0.45\linewidth]{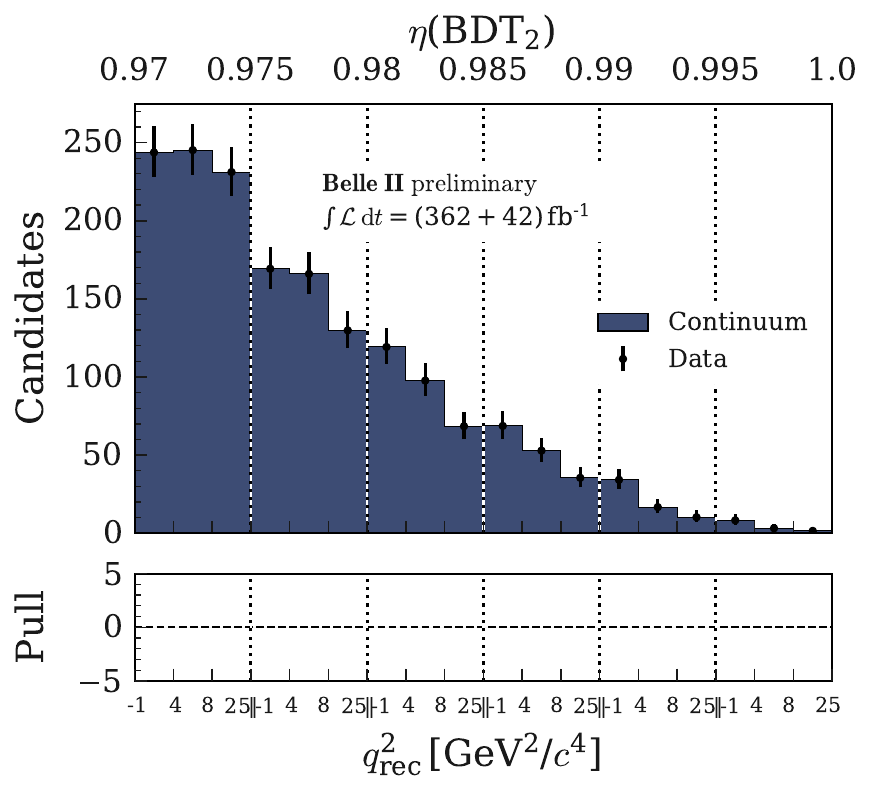}
    \includegraphics[width=0.45\linewidth]{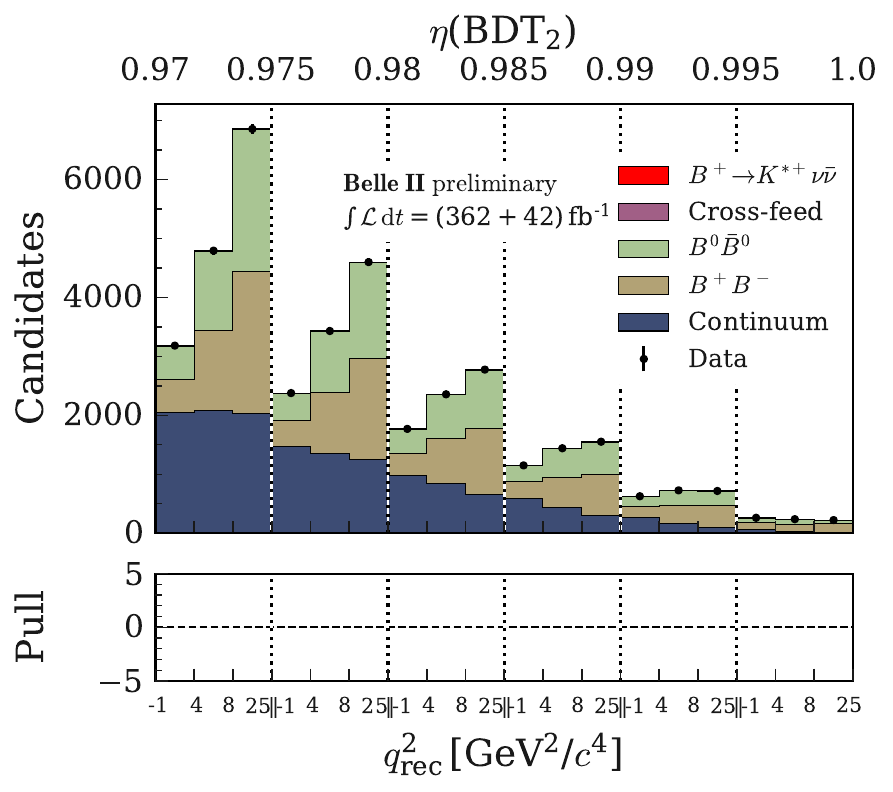}
    \includegraphics[width=0.45\linewidth]{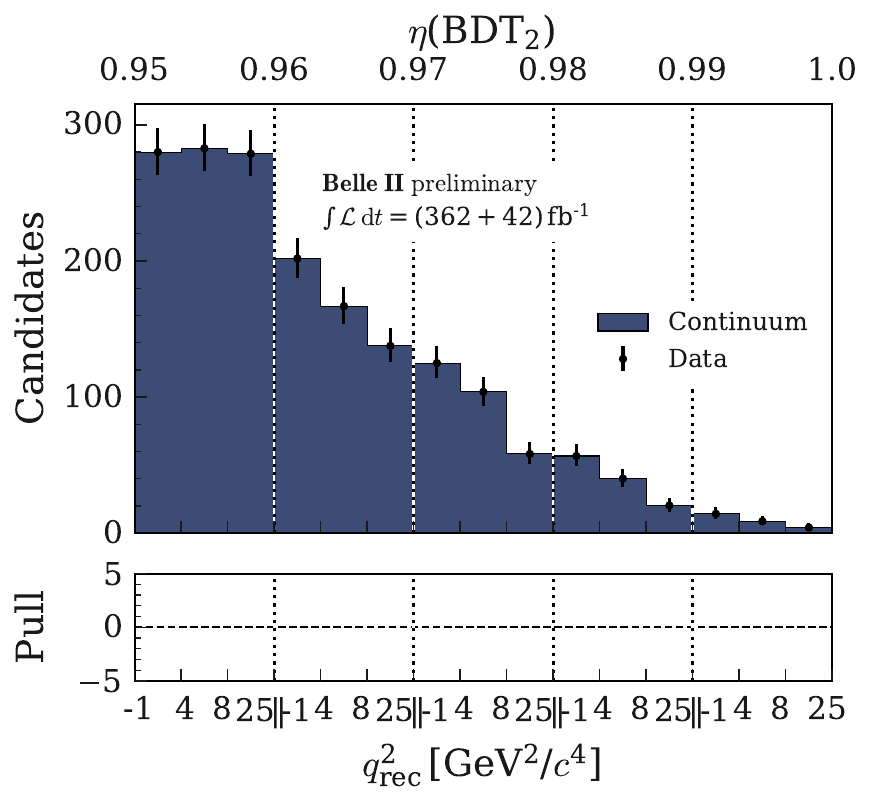}
    \includegraphics[width=0.45\linewidth]{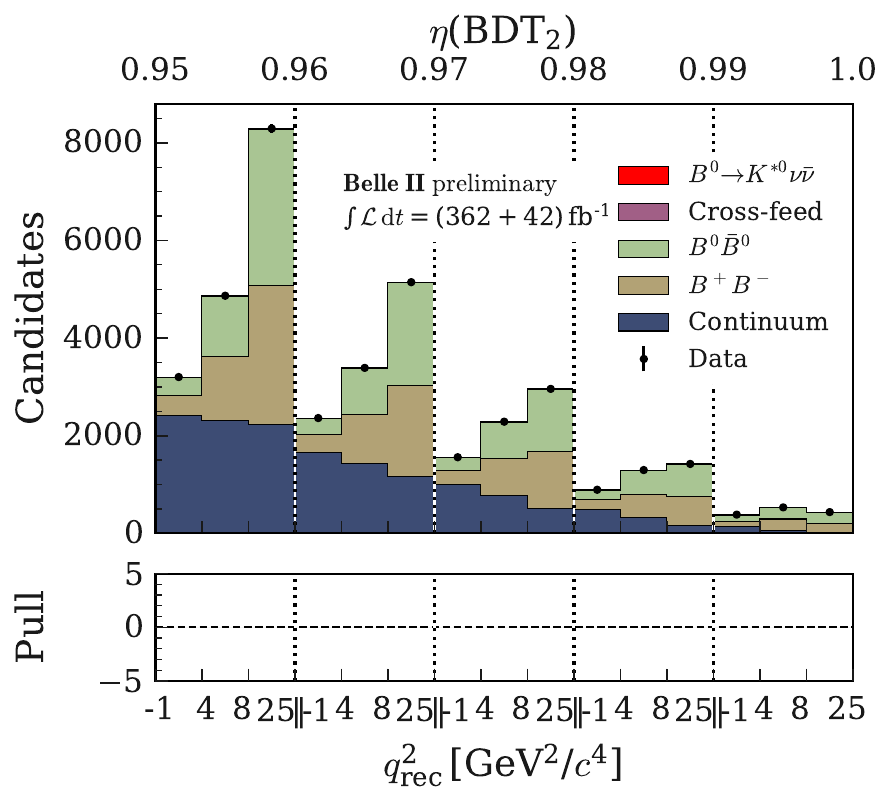}
    \caption{Post fit distribution from the Asimov fit of the \BpKstnn and \BzKstnn channels. The off-resonance region is shown on the \textit{left} and the on-resonance region on the \textit{right}. For each signal channel, the cross-feed indicates the events from the other three signal channels combined.}
    \label{fig:postfit_asimov2}
\end{figure}

The fit has been performed on the Asimov dataset, defined as the dataset with the number of observed events equivalent to the number of expected events for every bin. The Asimov dataset is obtained as the sum of all process templates (signals and backgrounds).
The information gained from this procedure consists of estimates of the uncertainties and correlations of the fit parameters. The post-fit distributions are shown in \cref{fig:postfit_asimov1,fig:postfit_asimov2}.

The fit results in terms of uncertainty on the signal strength multiplier, $\sigma_\mu$, and uncertainty on the \ac{SM} branching ratio ($\sigma_{\mathcal B}=\sigma_{\mu}\mathcal B_{SM}$) are shown in \cref{tab:fit_Asimov_err}. The world's best results~\cite{Belle:2017oht,BelleNote1413} are also reported for reference (obtained by Belle with a semileptonic tag on the full dataset).\footnote{Numerical branching ratio uncertainties are only found in the Belle internal note~\cite{BelleNote1413}.} The expected sensitivity is comparable to the world's best results in all channels.

\begin{table}[ht]
    \renewcommand{\arraystretch}{1.5}
    \caption{Fit uncertainties on \acp{POI} in terms of signal strength $\mu$ and branching ratio $\mathcal B$. The world's best results are taken from Reference~\cite{Belle:2017oht,BelleNote1413} (The Belle~II result from Reference~\cite{Belle-II:2023esi} is excluded).}
    \centering
    \begin{tabularx}{\linewidth}{@{\extracolsep{\fill}}lYYYY}
        \toprule \midrule
        Channel & $\sigma_\mu$ & $\sigma_{\mathcal B} \cdot 10^{-6}$ & World's best $\sigma_{\mathcal B} \cdot 10^{-6}$ & $\pm\sigma_\mu$ profiled \\
        \midrule
        \BpKpnn  & \errSysnominalBplusToKplusPYHF & \pgfmathprintnumber[fixed, precision=2]{\errBrnominalBplusToKplusPYHF} & 5.7 & + 1.24 - 1.24\\
        \BzKshnn & \errSysnominalBzeroToKshortPYHF & \pgfmathprintnumber[fixed, precision=2]{\errBrnominalBzeroToKshortPYHF} & 6.5 & + 2.90 - 2.82\\
        \BpKstnn & \errSysnominalBplusToKstarPlusPYHF & \pgfmathprintnumber[fixed, precision=2]{\errBrnominalBplusToKstarPlusPYHF} & 17 & + 2.26 - 2.18\\
        \BzKstnn & \errSysnominalBzeroToKstarZeroPYHF &  \pgfmathprintnumber[fixed, precision=2]{\errBrnominalBzeroToKstarZeroPYHF} & 11 & + 1.55 - 1.48\\
        \midrule \bottomrule
    \end{tabularx}
    \label{tab:fit_Asimov_err}
\end{table}

\begin{figure}[hbt]
    \centering
    \includegraphics[width=0.8\linewidth]{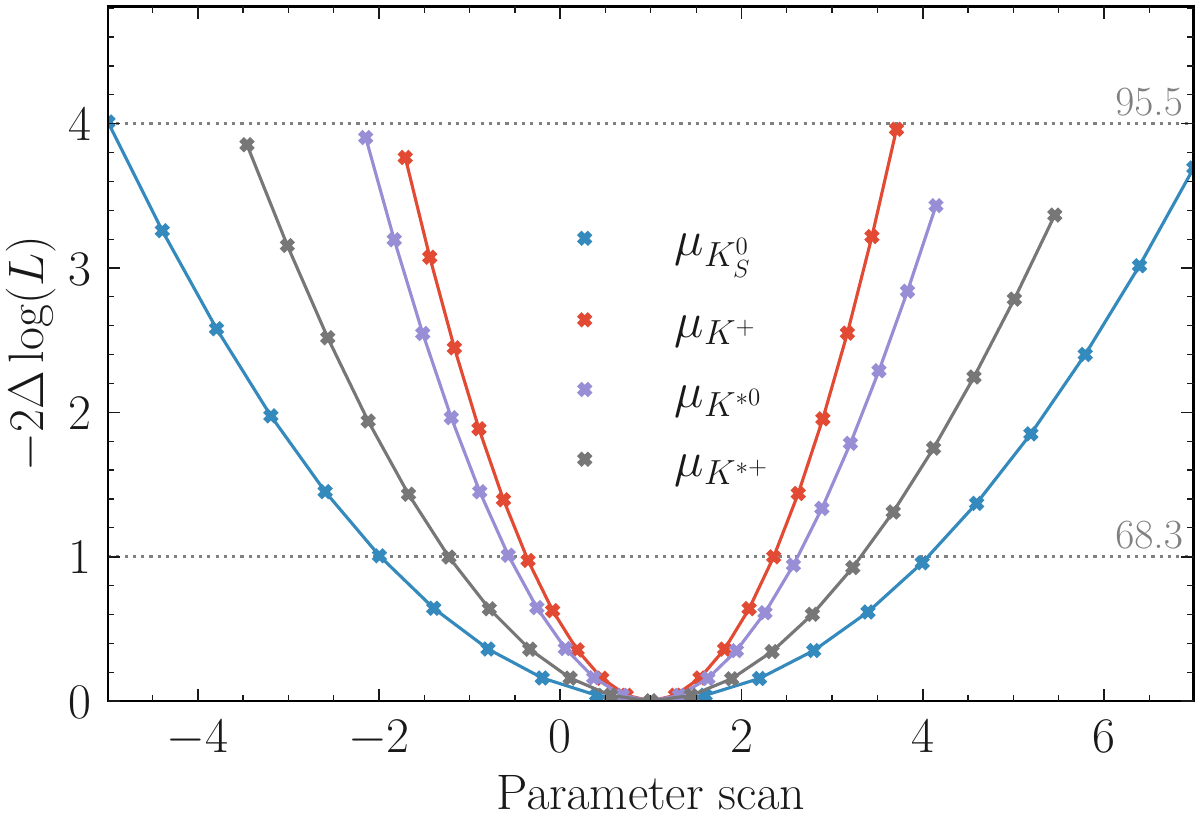}
    \caption{Profiled likelihood scan of the Asimov fit for the four signal channels.}
    \label{fig:fit_likelihood_scan}
\end{figure}

As a test for sensitivity and symmetry of the parameter distributions, 1-dimensional profiled likelihood scans for each of the four signal strengths are performed on Asimov data.
The results are shown in \cref{fig:fit_likelihood_scan}. 
The negative-log-likelihood curve is highly symmetric in all cases, which indicates expected asymptotic behaviour.
From the profiled likelihood scan, the parameter uncertainties can be calculated through the relative change in the value of the negative log likelihood compared to the best-fit point~\cite{Cowan1998},
\begin{equation}
    -2 \ln L(\hat \mu \pm \sigma_\mu) + 2 \ln L(\hat \mu) = 1.
    \label{eq:profiled-uncertainty}
\end{equation}
The resulting profiled uncertainties are shown in \cref{tab:fit_Asimov_err}.

\begin{figure}[hbt]
    \centering
    \includegraphics[width=\linewidth]{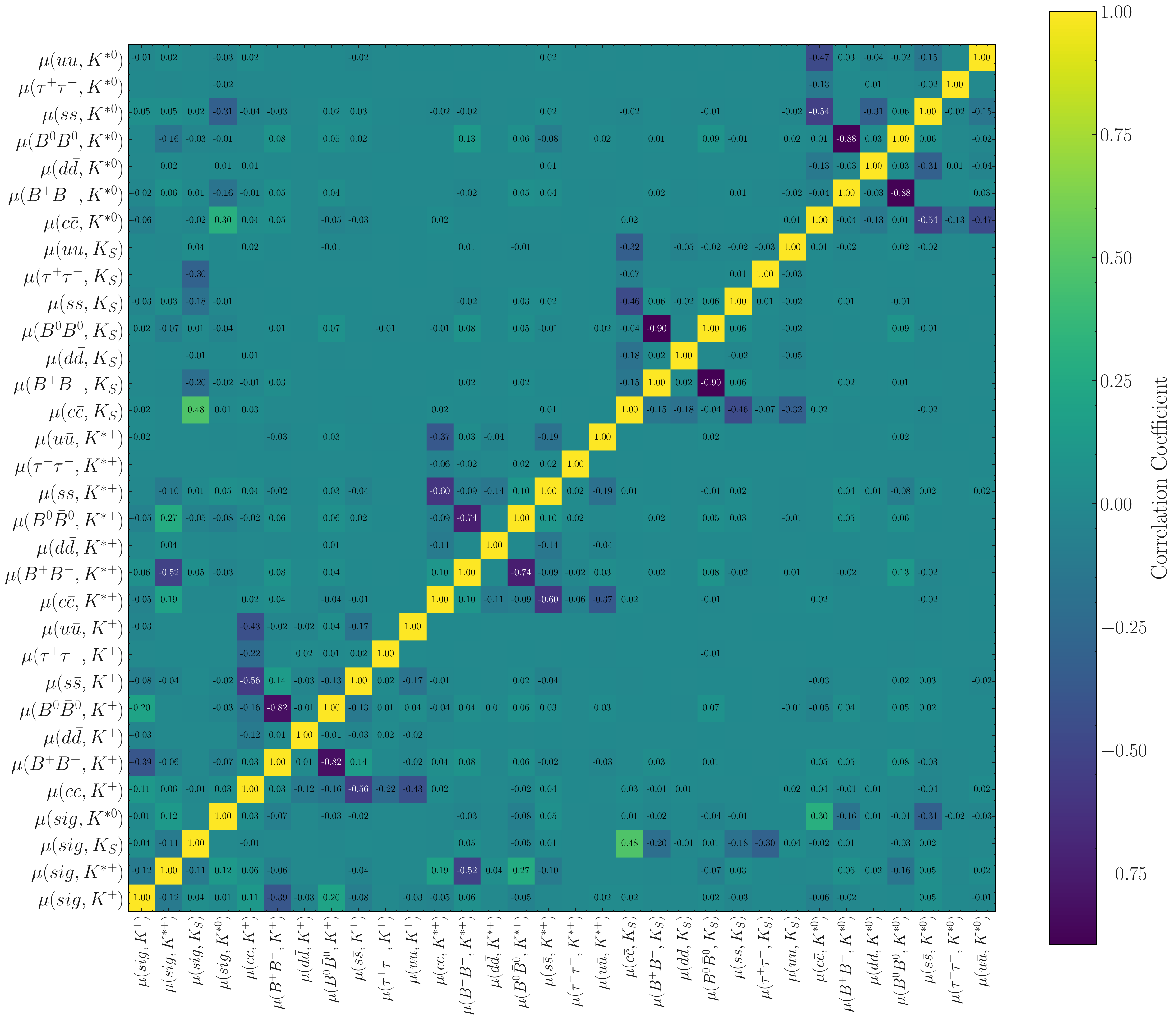}
    \caption{Correlation matrix of the uncorrelated model resulting from the Asimov fit. Only correlations of the normalization parameters are reported.}
    \label{fig:fit_asimov_correlation_matrix}
\end{figure}

In \cref{fig:fit_asimov_correlation_matrix}, the correlation matrix of the Asimov fit is shown. Most importantly, there is only a small correlation between the four signal strengths. The largest correlation between the signal and background components is found for the $B^+ B^-$ background in the \BpKstnn channel.

To investigate by how much a particular systematic source increases the uncertainty on $\mu$, this particular source is removed from the fit, and the uncertainty on $\mu$ of the modified fit is subtracted from that of the nominal fit in quadrature.
The results are reported in \cref{tab:systImportance}.
The background normalization uncertainties have the largest contribution to $\sigma_{\mu}$. Overall, the normalizations of \BBbar backgrounds play a more important role. For the \BzKshnn decay, in contrast, the normalization uncertainty of the continuum background is more important.
The systematic uncertainty due to the limited size of simulated samples is the second-largest uncertainty of the analysis.
The branching ratio uncertainty of the leading $B$ backgrounds is the third-largest uncertainty.

\begin{table}[ht]
    \renewcommand{\arraystretch}{1.5}
    \caption{The impact of systematic uncertainties, estimated using the Asimov dataset.}
    \centering
    \begin{tabularx}{\linewidth}{lYYYY}
        \toprule \midrule
        & \multicolumn{4}{c}{Uncertainty on $\mu$} \\
        Source & {\BpKpnn} & {\BzKshnn} &{\BpKstnn} & {\BzKstnn} \\
        \midrule
        $B\bar{B}$ norm. & \errSysnormsBBBplusToKplus & \errSysnormsBBBzeroToKshort  & \errSysnormsBBBplusToKstarPlus & \errSysnormsBBBzeroToKstarZero   \\
        Continuum norm. & \errSysnormsCBplusToKplus & \errSysnormsCBzeroToKshort  & \errSysnormsCBplusToKstarPlus & \errSysnormsCBzeroToKstarZero   \\
        \ac{MC} stats        & \errSysMCstatsBplusToKplus & \errSysMCstatsBzeroToKshort  & \errSysMCstatsBplusToKstarPlus & \errSysMCstatsBzeroToKstarZero   \\
        Leading $\mathcal{B}$      & \errSysLeadingBFBplusToKplus & \errSysLeadingBFBzeroToKshort  & \errSysLeadingBFBplusToKstarPlus & \errSysLeadingBFBzeroToKstarZero   \\
        \midrule \bottomrule
    \end{tabularx}
    \label{tab:systImportance}
\end{table}

\subsection{Isospin average}

\pgfmathsetmacro{\errSysISOratioPS}{round((\errSysnominalBplusToKplusPYHF / \errSysnominalISOmuPSPYHF - 1)*100)}
\pgfmathsetmacro{\errSysISOratioV}{round((\errSysnominalBzeroToKstarZeroPYHF / \errSysnominalISOmuVPYHF - 1)*100)} 

\begin{figure}[hbt]
    \centering
    \includegraphics[width=\linewidth]{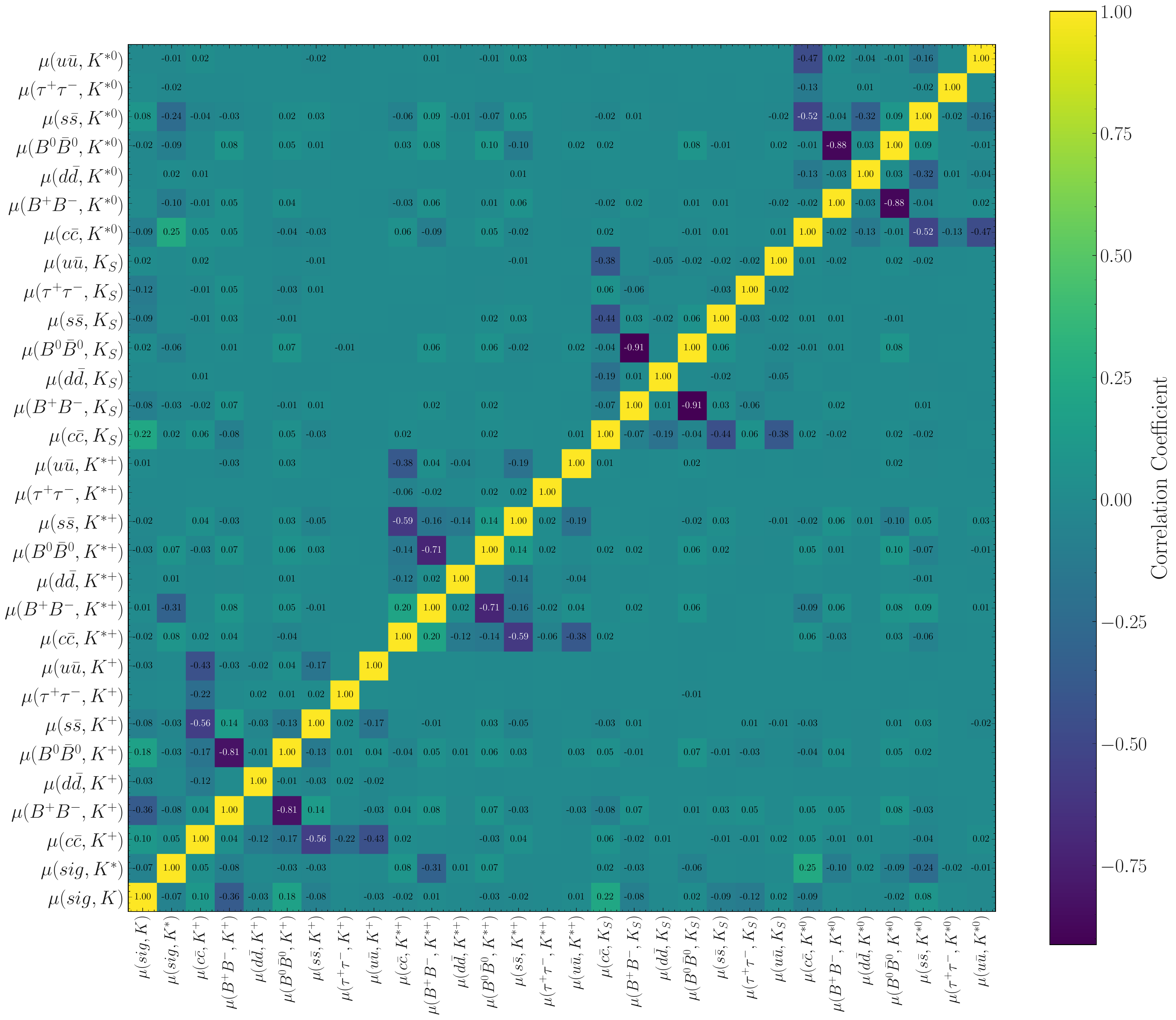}
    \caption{Correlation matrix of the isospin correlated model resulting from the Asimov fit. Only correlations of the normalization parameters are reported.}
    \label{fig:fit_asimov_IsoCorr_correlation_matrix}
\end{figure}

The isospin average combines the two $K$ and the two \Kstar modes, respectively.
In this case, a single \ac{POI} $\mu_{B\to K\nu\bar \nu}$ is used for the $K^+$ and $K_S^0$ modes, and a second $\mu_{B\to K^*\nu\bar \nu}$ is used for the $K^*+$ and $K^{*0}$ modes. The resulting uncertainties are $\sigma(\mu_{B\to K\nu\bar \nu}) = $~\errSysnominalISOmuPSPYHF~and $\sigma(\mu_{\BKstnn}) = $~\errSysnominalISOmuVPYHF. This corresponds to a~\errSysISOratioPS\% and a~\errSysISOratioV\% improvement in accuracy compared to using the \BpKpnn and \BzKstnn decays alone. 
The resulting branching ratio uncertainties are reported in \cref{tab:fit_Asimov_IsoCorr_err}. 
The correlation matrix is reported in \cref{fig:fit_asimov_IsoCorr_correlation_matrix}. There is only a weak correlation between the two signal strengths. The largest correlation between the signal and background components is found between $\mu_{B\to K\nu\bar \nu}$ and the $B^+ B^-$ background in the \BKnn channel.

\begin{table}[ht]
    \renewcommand{\arraystretch}{1.5}
    \caption{Fit uncertainties on \acp{POI} in terms of signal strength $\mu$ and branching ratio $\mathcal B$. The world's best results are taken from Reference~\cite{Belle:2017oht,BelleNote1413} (The Belle~II result from Reference~\cite{Belle-II:2023esi} is excluded).}
    \centering
    \begin{tabularx}{\linewidth}{@{\extracolsep{\fill}}lYYYY}
        \toprule \midrule
        Channel & $\sigma_\mu$ & $\sigma_{\mathcal B} \cdot 10^{-6}$ & World's best $\sigma_{\mathcal B} \cdot 10^{-6}$ & $\pm\sigma_\mu$ profiled \\
        \midrule
        \BpKpnn  & \errSysnominalISOmuPSPYHF & \pgfmathprintnumber[fixed, precision=2]{\errBrnominalISOBplusToKplusPYHF} & 5.7 & + 1.14 - 1.15\\
        \BzKshnn & \errSysnominalISOmuPSPYHF & \pgfmathprintnumber[fixed, precision=2]{\errBrnominalISOBzeroToKshortPYHF} & 6.5 & + 1.14 - 1.15\\
        \BpKstnn & \errSysnominalISOmuVPYHF & \pgfmathprintnumber[fixed, precision=2]{\errBrnominalISOBplusToKstarPlusPYHF} & 17 & + 1.34 - 1.29\\
        \BzKstnn & \errSysnominalISOmuVPYHF &  \pgfmathprintnumber[fixed, precision=2]{\errBrnominalISOBzeroToKstarZeroPYHF} & 11 & + 1.34 - 1.29\\
        \midrule \bottomrule
    \end{tabularx}
    \label{tab:fit_Asimov_IsoCorr_err}
\end{table}

\begin{figure}[hbt]
    \centering
    \includegraphics[width=0.8\linewidth]{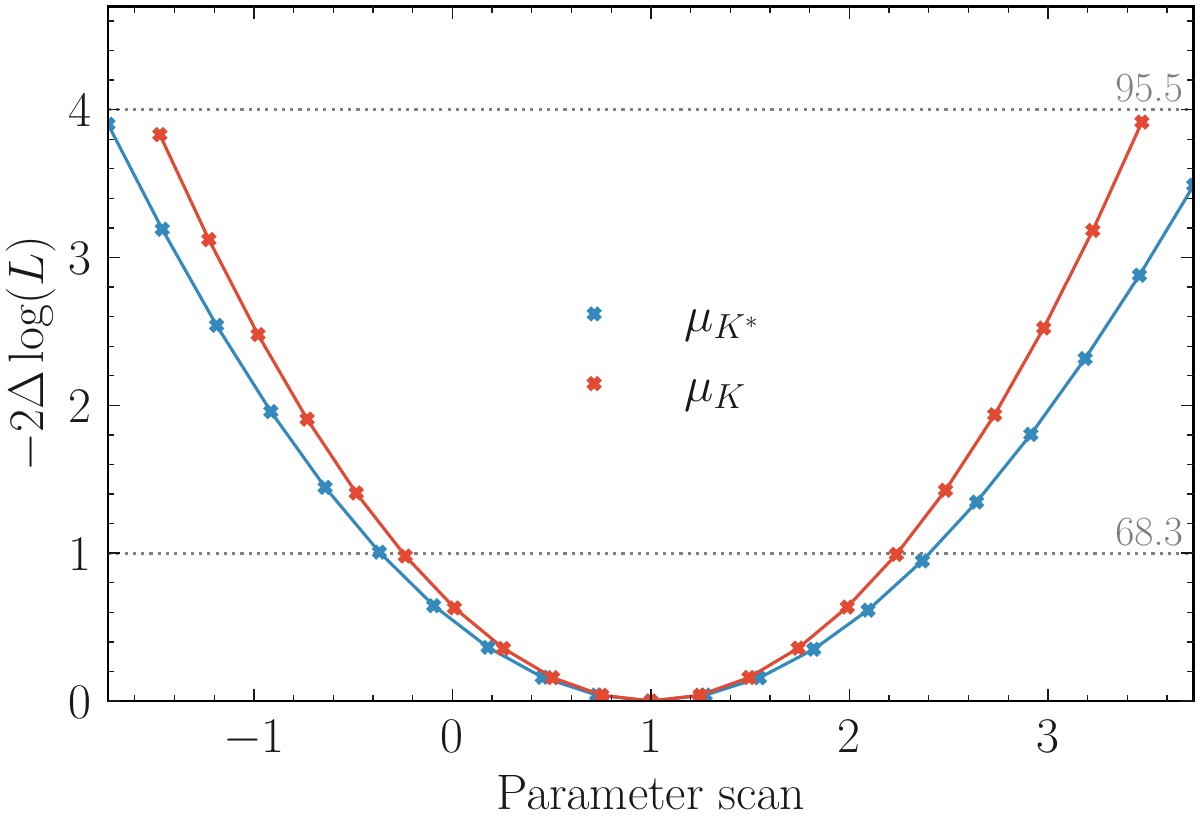}
    \caption{Profiled likelihood scans for the pseudoscalar \ac{POI} $\mu_{K}={\mu(B\to K\nu\bar \nu)}$ and vector \ac{POI} ${\mu_{K^*}=\mu(B\to K^*\nu\bar \nu)}$ of the isospin averaged model, fit on Asimov data.}
    \label{fig:fit_likelihood_scan_isospin}
\end{figure}

The results of 1-dimensional profiled likelihood scans for each of the two parameters, performed on Asimov data, are shown in \cref{fig:fit_likelihood_scan_isospin}. 
The negative-log-likelihood curve is highly symmetric in all cases, which indicates the expected parabolic behaviour in the asymptotic regime. 
The uncertainties resulting from the profiled likelihood scan are shown in \cref{tab:fit_Asimov_IsoCorr_err}.

\subsection{4-channel correlation}\label{sec:fit_4channels}

\pgfmathsetmacro{\errSysALLratio}{round((\errSysnominalBplusToKplusPYHF / \errSysnominalALLmuPYHF - 1)*100)}

\begin{figure}[hbt]
    \centering
    \includegraphics[width=\linewidth]{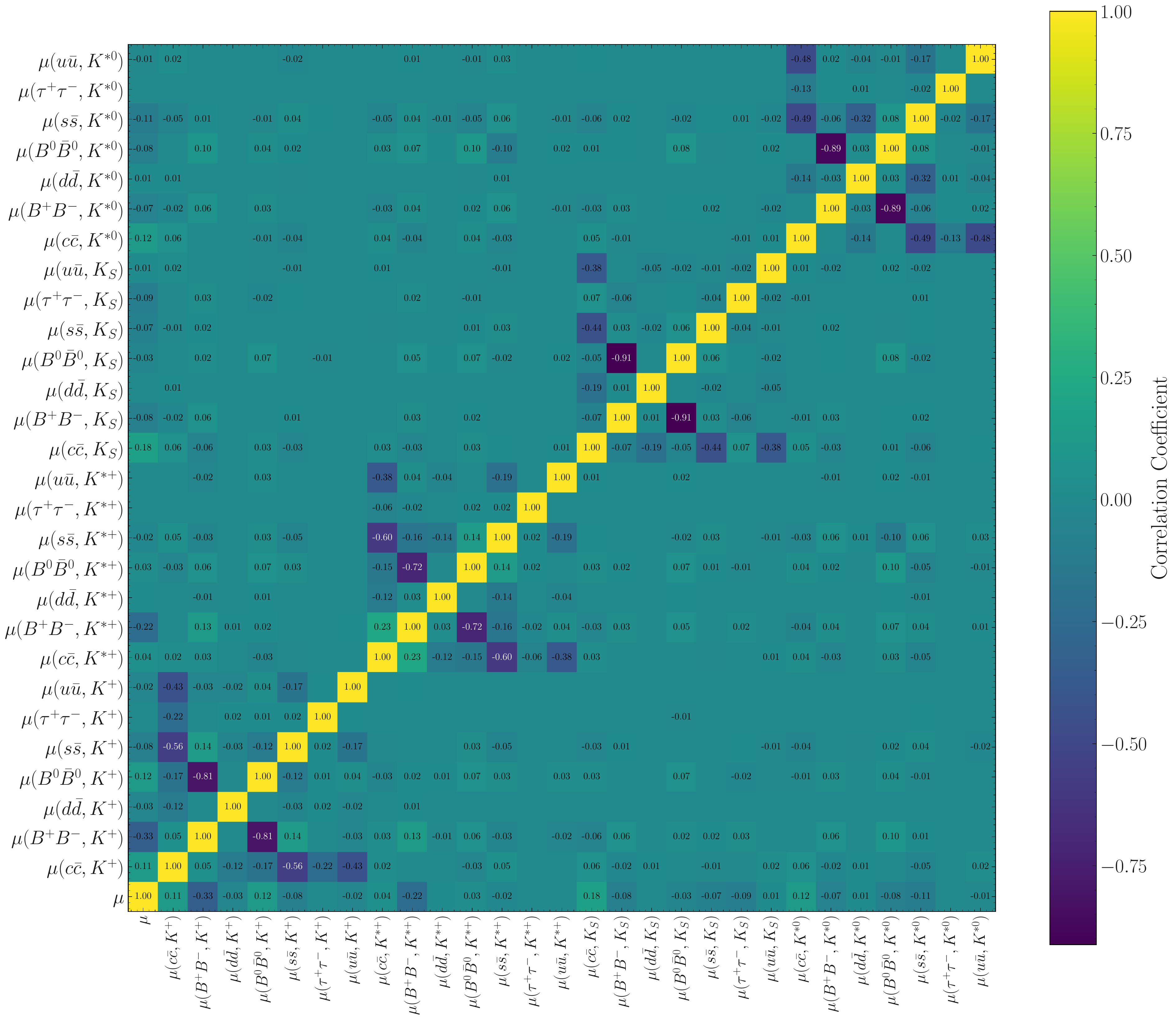}
    \caption{Correlation matrix of the fully correlated model resulting from the Asimov fit. Only correlations of the normalization parameters are reported.}
    \label{fig:fit_asimov_allCorr_correlation_matrix}
\end{figure}

Lastly, an analysis is performed assuming full correlation between the four $K$ modes; thus, a single \ac{POI} $\mu_{\BKnunu}$ is used for all modes. The resulting uncertainty is $\sigma(\mu_{\BKnunu}) = $ \errSysnominalALLmuPYHF. This corresponds to a~\errSysALLratio\% improvement in accuracy compared to using the \BpKpnn decay alone. 
The resulting branching ratio uncertainties are reported in \cref{tab:fit_Asimov_allCorr_err}. 
The correlation matrix is reported in \cref{fig:fit_asimov_allCorr_correlation_matrix}. The largest correlation between the signal and background components is found for the charged $B^+ B^-$ background in the \BKnn channel.

\begin{figure}[hbt]
    \centering
    \includegraphics[width=0.8\linewidth]{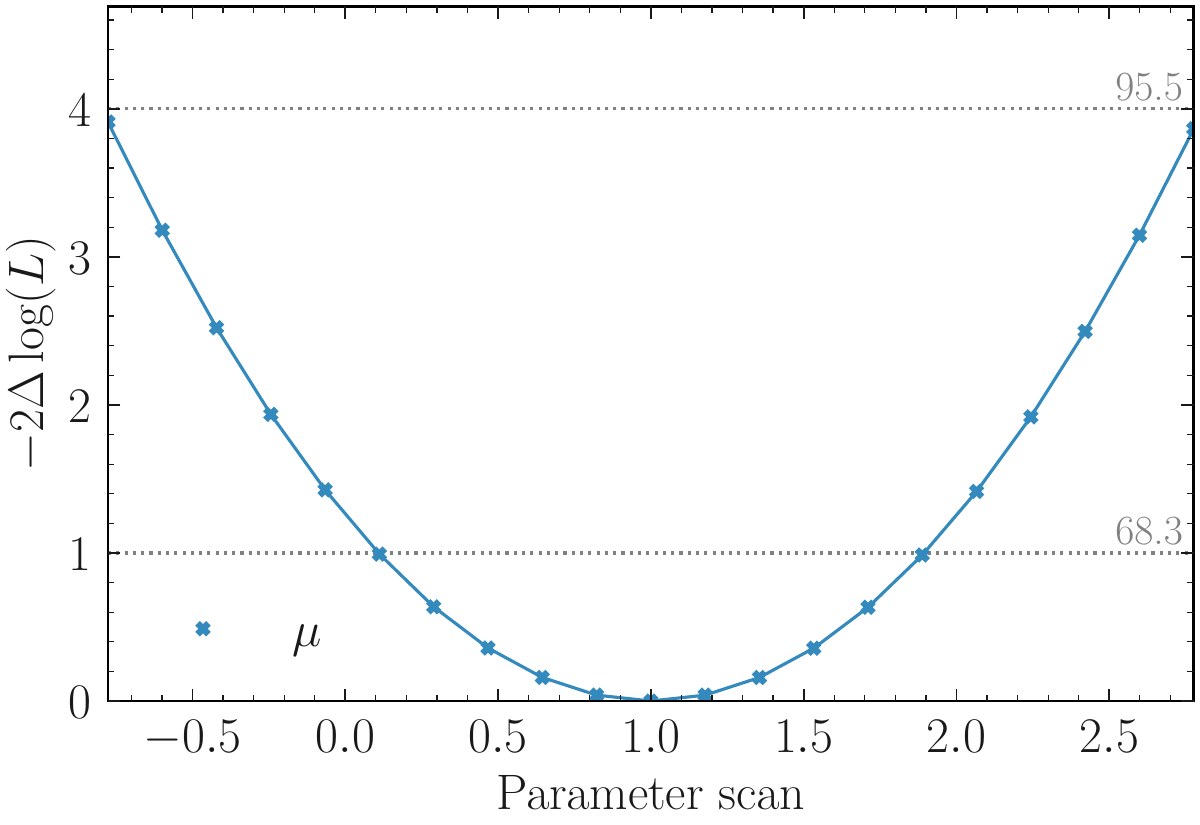}
    \caption{Profiled likelihood scan for the fully correlated signal strength across all channels, fit on Asimov data.}
    \label{fig:fit_likelihood_scan_all}
\end{figure}

The results of 1-dimensional profiled likelihood scans for the single parameter, performed on Asimov data, are shown in \cref{fig:fit_likelihood_scan_all}. 
The negative-log-likelihood curve is highly symmetric, which indicates the expected parabolic behaviour in the asymptotic regime. 
The uncertainties resulting from the profiled likelihood scan are shown in \cref{tab:fit_Asimov_allCorr_err}.

\begin{table}[ht]
    \renewcommand{\arraystretch}{1.5}
    \caption{Fit uncertainties on \acp{POI} in terms of signal strength $\mu$ and branching ratio $\mathcal B$. The world's best results are taken from Reference~\cite{Belle:2017oht,BelleNote1413} (The Belle~II result from Reference~\cite{Belle-II:2023esi} is excluded).}
    \centering
    \begin{tabularx}{\linewidth}{@{\extracolsep{\fill}}lYYYY}
        \toprule \midrule
        Channel & $\sigma_\mu$ & $\sigma_{\mathcal B} \cdot 10^{-6}$ & World's best $\sigma_{\mathcal B} \cdot 10^{-6}$ & $\pm\sigma_\mu$ profiled \\
        \midrule
        \BpKpnn  & \errSysnominalALLmuPYHF & \pgfmathprintnumber[fixed, precision=2]{\errBrnominalALLBplusToKplusPYHF} & 5.7 & + 0.83 - 0.83\\
        \BzKshnn & \errSysnominalALLmuPYHF & \pgfmathprintnumber[fixed, precision=2]{\errBrnominalALLBzeroToKshortPYHF} & 6.5 & + 0.83 - 0.83\\
        \BpKstnn & \errSysnominalALLmuPYHF & \pgfmathprintnumber[fixed, precision=2]{\errBrnominalALLBplusToKstarPlusPYHF} & 17 & + 0.83 - 0.83\\
        \BzKstnn & \errSysnominalALLmuPYHF &  \pgfmathprintnumber[fixed, precision=2]{\errBrnominalALLBzeroToKstarZeroPYHF} & 11 & + 0.83 - 0.83\\
        \midrule \bottomrule
    \end{tabularx}
    \label{tab:fit_Asimov_allCorr_err}
\end{table}

\section{Discussion}
\label{sec:discussion_update}

This work-in-progress analysis demonstrates the significant advantages of a simultaneous four-channel approach to \BKnunu measurements at Belle~II. 

The expected uncertainties presented in \cref{tab:fit_Asimov_err,tab:fit_Asimov_IsoCorr_err,tab:fit_Asimov_allCorr_err} show that Belle~II can achieve sensitivity competitive with the world's best results using a significantly smaller dataset than the one of the Belle experiment. This demonstrates the power of combining multiple decay modes and the superior performance of the Belle~II detector and analysis techniques. The systematic uncertainties are well controlled, with background normalization being the dominant source, followed by \ac{MC} statistics and branching ratio uncertainties of leading backgrounds.

The simultaneous measurement of four $B \to K^{(*)} \nu\bar{\nu}$ channels opens new opportunities for model-agnostic reinterpretation studies. As demonstrated in \cref{sec:method paper ksnunu}, the combination of pseudoscalar (\BpKpnn, \BzKshnn) and vector (\BpKstnn, \BzKstnn) final states enables the determination of the magnitudes of individual Wilson coefficients rather than just their sums. The different kinematic structures of these decays provide complementary sensitivity to various \ac{BSM} scenarios. The statistical framework constructed here provides an ideal foundation for reinterpreting these results in terms of new physics models.

%% file: chapters/conclusion.tex
This thesis addresses a fundamental challenge in \ac{HEP}: how to accelerate the discovery of new physics through efficient and accurate reinterpretation of experimental results. The core idea is to reinterpret measurements designed for one theoretical hypothesis in terms of alternative frameworks. I develop, validate, and apply a novel reinterpretation method that preserves statistical rigour while dramatically reducing computational costs compared to a full reanalysis.

The key contribution is a model-agnostic reinterpretation method based on kinematic reweighting. This approach addresses limitations of existing techniques by enabling direct inference on theoretical parameters while correctly accounting for changes in kinematic distributions. Compared to current methods, it offers computational efficiency, statistical accuracy, accessibility, and broad applicability.

The method has been validated through toy examples and then applied to real experimental data. I focus on the Belle~II $B^+ \to K^+ \nu \bar{\nu}$ measurement, which shows a $2.7\sigma$ tension with \ac{SM} predictions. I demonstrate reinterpretations in two theoretical contexts:

\textbf{Weak Effective Theory}: I have obtained constraints on the \ac{WET} Wilson coefficients relevant to $b \to s \nu \bar{\nu}$ processes, providing model-independent bounds on new physics contributions above the electroweak scale. These results represent both a successful application of the reinterpretation method and novel scientific contributions. The presented results are the first direct experimental constraints on the $b\to s \nu \bar{\nu}$ \ac{WET} Wilson coefficients.

\textbf{$\boldsymbol{\BKX}$ two-body decays}: I apply the framework to a two-body light new physics scenario, demonstrating the versatility of the method in constraining specific \ac{BSM} hypotheses with different kinematic signatures. The results demonstrate significantly improved compatibility with the experimental data compared to the \ac{SM}. This illustrates how the method can explore different kinematic signatures and guide future investigations.

This work addresses a bottleneck in particle physics today. The complexity of experimental analyses and the proliferation of theoretical models create a mismatch between the pace of theoretical development and experimental validation. By enabling rapid, accurate reinterpretation of existing measurements, the method presented here has several important implications:

\textbf{Enhanced scientific impact}: The resources invested in particle physics experiments can yield broader scientific impact when their results can be efficiently applied to test multiple theoretical hypotheses beyond those originally considered. Measurements with open source analysis logic are more likely to be cited and reused in future works, increasing their scientific impact~\cite{Butterworth:2929207}.

\textbf{Democratization of experimental data}: By making reinterpretation accessible to theorists without requiring deep experimental expertise or computational resources, the method lowers barriers to scientific collaboration and accelerates the theory-experiment dialogue.

\textbf{Extension of scientific legacy}: Experimental measurements maintain their relevance and utility even as new theoretical frameworks emerge, ensuring that today's measurements can continue contributing to tomorrow's discoveries.

The reproducibility and distributability aspects of the method enable powerful combined analyses that can dramatically reduce uncertainties.

The foundation established in this work opens several avenues for future development:

\textbf{Method extensions}: The reweighting framework can be extended to handle more complex scenarios. Alternative approaches to joint number density construction can be explored to improve performance and robustness. The method can be applied within the context of other inference tools and techniques.

\textbf{Broader applications}: While this work has focused on Belle~II measurements, the method applies equally to results from other experiments. As of today, colleagues at the LHCb experiment are applying the method and corresponding tool for a number of analyses (see e.g.\ \cite{colonna2025redist}).

\textbf{Global fits}:  The statistical models generated through this approach can be integrated into global fitting frameworks and combined with constraints from other experiments, providing much tighter bounds on theoretical parameters than any single measurement.

The reinterpretation method presented in this thesis contributes to a more sustainable and productive model for scientific progress. The approach advances open science initiatives by making experimental results more accessible and reusable. This directly supports the \ac{FAIR} data principles that are increasingly recognized as essential for modern scientific practice~\cite{Wilkinson:2016myn}.

The $2.7\sigma$ tension observed in the Belle~II $B^+ \to K^+ \nu \bar{\nu}$ measurement illustrates why such advances are important. The ability to rapidly and accurately test multiple theoretical models against the same experimental data is a powerful accelerator of scientific progress.

As the field of particle physics continues to scale towards more data and higher energies, methods that enhance what we can learn from existing measurements become increasingly valuable. 
The combination of improved experimental techniques, more sophisticated theoretical frameworks, and advanced statistical methods provides a foundation for such enhanced interpretation.
This work contributes one piece to this larger endeavour, helping to ensure that experimental particle physics continues to deliver important scientific discoveries.

%% file: chapters/publications.tex
\section*{Papers}

\begin{refsection}
\nocite{Gartner:2024muk}
\printbibliography[heading=none,env=notnumbered]
\end{refsection}

\section*{Proceedings}

\begin{refsection}
\nocite{colonna2025redist}
\printbibliography[heading=none,env=notnumbered]
\end{refsection}

\section*{In publication}
\begin{refsection}
\nocite{belle2pub88}
\printbibliography[heading=none,env=notnumbered]
\end{refsection}

%% file: chapters/appendix-theory.tex
\section{Group generators}
\label{sec:generators}

The Standard Model gauge group $\mathrm{G}_{SM} = \mathrm{SU}(3)_C \times \mathrm{SU}(2)_W \times \mathrm{U}(1)_Y$ is characterized by its generators, which determine the structure of gauge interactions and the number of gauge bosons. Here we follow the conventions of References~\cite{Schwartz_2013,Peskin:1995ev}.

\subsection{U(1) generators}
The Abelian group $\mathrm{U}(1)$ has a single generator, which can be taken as the identity operator. For $\mathrm{U}(1)_Y$ (hypercharge), transformations take the form
\begin{equation}
    \psi \to e^{i \alpha Y} \psi,
\end{equation}
where $\alpha$ is a real parameter and $Y$ is the hypercharge quantum number. This gives rise to a single gauge boson $B_\mu$.

\subsection{SU(2) generators}
The $\mathrm{SU}(2)_W$ group has $2^2 - 1 = 3$ generators, conventionally chosen as
\begin{equation}
    T_i = \frac{\sigma_i}{2}, \quad i = 1, 2, 3,
\end{equation}
where $\sigma_i$ are the Pauli matrices:
\begin{equation}
    \sigma_1 = \begin{pmatrix} 0 & 1 \\ 1 & 0 \end{pmatrix}, \quad
    \sigma_2 = \begin{pmatrix} 0 & -i \\ i & 0 \end{pmatrix}, \quad
    \sigma_3 = \begin{pmatrix} 1 & 0 \\ 0 & -1 \end{pmatrix}.
\end{equation}

These generators satisfy the $\mathrm{SU}(2)$ algebra
\begin{equation}
    [T_i, T_j] = i \varepsilon_{ijk} T_k,
\end{equation}
where $\varepsilon_{ijk}$ is the totally antisymmetric Levi-Civita symbol. The generators are normalized according to
\begin{equation}
    \mathrm{Tr}(T_i T_j) = \frac{1}{2} \delta_{ij}.
\end{equation}

Transformations under $\mathrm{SU}(2)_W$ take the form
\begin{equation}
    \psi \to \exp\left(i \alpha_i T_i\right) \psi = \exp\left(i \frac{\alpha_i \sigma_i}{2}\right) \psi,
\end{equation}
where $\alpha_i$ are three real parameters. This generates three gauge bosons $W_\mu^i$.

\subsection{SU(3) generators}
The $\mathrm{SU}(3)_C$ group has $3^2 - 1 = 8$ generators, conventionally chosen as
\begin{equation}
    T^a = \frac{\lambda^a}{2}, \quad a = 1, 2, \ldots, 8,
\end{equation}
where $\lambda^a$ are the Gell-Mann matrices:
\begin{align}
    \lambda_1 &= \begin{pmatrix} 0 & 1 & 0 \\ 1 & 0 & 0 \\ 0 & 0 & 0 \end{pmatrix}, &
    \lambda_2 &= \begin{pmatrix} 0 & -i & 0 \\ i & 0 & 0 \\ 0 & 0 & 0 \end{pmatrix}, &
    \lambda_3 &= \begin{pmatrix} 1 & 0 & 0 \\ 0 & -1 & 0 \\ 0 & 0 & 0 \end{pmatrix}, \\
    \lambda_4 &= \begin{pmatrix} 0 & 0 & 1 \\ 0 & 0 & 0 \\ 1 & 0 & 0 \end{pmatrix}, &
    \lambda_5 &= \begin{pmatrix} 0 & 0 & -i \\ 0 & 0 & 0 \\ i & 0 & 0 \end{pmatrix}, &
    \lambda_6 &= \begin{pmatrix} 0 & 0 & 0 \\ 0 & 0 & 1 \\ 0 & 1 & 0 \end{pmatrix}, \\
    \lambda_7 &= \begin{pmatrix} 0 & 0 & 0 \\ 0 & 0 & -i \\ 0 & i & 0 \end{pmatrix}, &
    \lambda_8 &= \frac{1}{\sqrt{3}} \begin{pmatrix} 1 & 0 & 0 \\ 0 & 1 & 0 \\ 0 & 0 & -2 \end{pmatrix}.
\end{align}

The $\mathrm{SU}(3)$ generators satisfy the commutation relations
\begin{equation}
    [T^a, T^b] = i f^{abc} T^c,
\end{equation}
where $f^{abc}$ are the totally antisymmetric structure constants of $\mathrm{SU}(3)$. The structure constants are normalized according to
\begin{equation}
    f^{ade} f^{bce} = \delta^{ab} N,
\end{equation}
where $N = 3$ for $\mathrm{SU}(3)$. This normalization ensures that the structure constants satisfy the Jacobi identity and are consistent with the trace normalization of the generators.

The generators are normalized according to
\begin{equation}
    \mathrm{Tr}(T^a T^b) = \frac{1}{2} \delta^{ab}.
\end{equation}

Transformations under $\mathrm{SU}(3)_C$ take the form
\begin{equation}
    q \to \exp\left(i \alpha_a T^a\right) q = \exp\left(i \frac{\alpha_a \lambda^a}{2}\right) q,
\end{equation}
where $\alpha_a$ are 8 real parameters and $q$ represents a quark field in the fundamental representation. This generates 8 gauge bosons $A_\mu^a$ (gluons).

\section{Dirac matrices}
\label{sec:gamma-matrices}

The Dirac matrices $\gamma^\mu$ are fundamental objects in the theory of relativistic fermions~\cite{Schwartz_2013,Peskin:1995ev}. They satisfy the Clifford algebra
\begin{equation}
    \{\gamma^\mu, \gamma^\nu\} = 2 g^{\mu\nu} \mathbb{1},
\end{equation}
where $g^{\mu\nu} = \text{diag}(1, -1, -1, -1)$ is the Minkowski metric tensor and $\mathbb{1}$ is the identity matrix. The anticommutator is defined as $\{A, B\} = AB + BA$.

In the chiral (Weyl) representation, the gamma matrices are given by
\begin{equation}
    \gamma^0 = \begin{pmatrix} 0 & \mathbb{1} \\ \mathbb{1} & 0 \end{pmatrix}, \quad
    \gamma^i = \begin{pmatrix} 0 & \sigma^i \\ -\sigma^i & 0 \end{pmatrix},
\end{equation}
where $i = 1, 2, 3$ and $\sigma^i$ are the Pauli matrices defined in \cref{sec:generators}.

The fifth gamma matrix is defined as
\begin{equation}
    \gamma^5 = i \gamma^0 \gamma^1 \gamma^2 \gamma^3 = \begin{pmatrix} -\mathbb{1} & 0 \\ 0 & \mathbb{1} \end{pmatrix}.
\end{equation}
This matrix satisfies $(\gamma^5)^2 = \mathbb{1}$ and anticommutes with all other gamma matrices:
\begin{equation}
    \{\gamma^5, \gamma^\mu\} = 0.
\end{equation}

The projection operators for left- and right-handed fermions are constructed using $\gamma^5$:
\begin{equation}
    P_L = \frac{1 - \gamma^5}{2}, \quad P_R = \frac{1 + \gamma^5}{2},
\end{equation}
satisfying $P_L + P_R = \mathbb{1}$, $P_L^2 = P_L$, $P_R^2 = P_R$, and $P_L P_R = P_R P_L = 0$.

The charge conjugation matrix $C$ satisfies
\begin{equation}
    C \gamma^\mu C^{-1} = -(\gamma^\mu)^T, \quad C \gamma^5 C^{-1} = (\gamma^5)^T,
\end{equation}
where $T$ denotes matrix transposition. In the chiral representation,
\begin{equation}
    C = \begin{pmatrix} i\sigma^2 & 0 \\ 0 & -i\sigma^2 \end{pmatrix}.
\end{equation}

The slash notation is commonly used for contractions with four-vectors:
\begin{equation}
    \slashed{p} = \gamma^\mu p_\mu = p_0 \gamma^0 - \vec{p} \cdot \vec{\gamma}.
\end{equation}

\section{\texorpdfstring{$B \to K^*$}{B->K*} form factors}
\label{sec:form-factors}
For the $B \to K^*$ transitions, vector operators in \cref{eq:operators} give rise to four independent form factors~\cite{Felkl:2021uxi,Gubernari:2018wyi}:
\begin{align}
    \langle K^*(k,\epsilon) | \bar{s} \gamma^\mu b | B(p) \rangle &= \frac{2 V(q^2)}{M_B + M_{K^*}} \varepsilon^{\mu\nu\rho\sigma} \epsilon^*_\nu p_\rho k_\sigma, \\
    \langle K^*(k,\epsilon) | \bar{s} \gamma^\mu \gamma_5 b | B(p) \rangle &= i \epsilon^{\mu*} (M_B + M_{K^*}) A_1(q^2) \nonumber \\
    &\quad - i (p+k)^\mu (\epsilon^* \cdot q) \frac{A_2(q^2)}{M_B + M_{K^*}} \nonumber \\
    &\quad - i q^\mu (\epsilon^* \cdot q) \frac{2 M_{K^*}}{q^2} \left[ A_3(q^2) - A_0(q^2) \right],
    \label{eq:vector-axial-matrix-element}
\end{align}
where
\begin{align}
    A_3(q^2) &= \frac{M_B + M_{K^*}}{2 M_{K^*}} A_1(q^2) - \frac{M_B - M_{K^*}}{2 M_{K^*}} A_2(q^2), \\
    A_3(0) &= A_0(0),
\end{align}
and $\varepsilon^{\mu\nu\rho\sigma}$ is the totally antisymmetric Levi-Civita symbol in four dimensions, and $V(q^2)$, $A_0(q^2)$, $A_1(q^2)$, and $A_2(q^2)$ are the $B \to K^*$ form factors.

For the scalar operators, the non-zero matrix element is~\cite{Felkl:2021uxi,Gubernari:2018wyi}
\begin{align}
    \langle K^*(k,\epsilon) | \bar{s} \gamma_5 b | B(p) \rangle = i (\epsilon^* \cdot q) \frac{2 M_{K^*}}{m_b - m_s} A_0(q^2).
    \label{eq:vector-scalar-matrix-element}
\end{align}

The tensor operator matrix elements are~\cite{Felkl:2021uxi,Gubernari:2018wyi}
\begin{align}
    \langle K^*(k,\epsilon) | \bar{s} \sigma^{\mu\nu} b | B(p) \rangle &= 2 \varepsilon^{\mu\nu\rho\sigma} \epsilon^*_\nu p_\sigma k_\rho T_1(q^2) \\
    \langle K^*(k,\epsilon) | \bar{s} \sigma^{\mu\nu} \gamma_5 b | B(p) \rangle &= i \left[ \epsilon^{\mu*} (p + k)^\nu - \epsilon^{\nu*} (p + k)^\mu \right] T_2(q^2) \nonumber \\
    &\quad + i (\epsilon^* \cdot q) \left[q^\mu (p+k)^\nu - q^\nu (p+k)^\mu\right] \frac{T_3(q^2)}{M_B^2 - M_{K^*}^2},
    \label{eq:vector-tensor-matrix-element}
\end{align}
where $T_1(q^2)$, $T_2(q^2)$, and $T_3(q^2)$ are the tensor form factors for $B \to K^*$ transitions, with the constraint $T_1(0) = T_2(0)$.

It is common to replace the form factors $A_2(q^2)$ and $T_3(q^2)$ with the linear combinations~\cite{Felkl:2021uxi,Gubernari:2018wyi}
\begin{align}
    A_{12}(q^2) &= \frac{(M_B + M_{K^*})^2 (M_B^2 - M_{K^*}^2 - q^2) A_1(q^2) - \lambda(q^2, M_B, M_{K^*}) A_2(q^2)}{16 M_B M_{K^*}^2 (M_B + M_{K^*})}, 
    \label{eq:a12} \\
    T_{23}(q^2) &= \frac{(M_B^2 - M_{K^*}^2) (M_B^2 + 3 M_{K^*}^2 - q^2) T_2(q^2) - \lambda(q^2, M_B, M_{K^*}) T_3(q^2)}{8 M_B M_{K^*}^2 (M_B - M_{K^*})}, 
    \label{eq:t23}
\end{align}
where $\lambda(q^2, M_B, M_{K^*})$ is the Källén function.

Scalar currents do not contribute to $B \to K^*$ transitions, by virtue of the symmetries of \ac{QCD}. In particular, the invariance of the strong interaction under $C$, $P$, and \ac{CP} transformations implies that these $B \to K^*$ matrix elements vanish.

\section{\texorpdfstring{\BKsnunu}{B->K*nunubar} hadronic form factor parameters}
\label{sec:kstar form factors}

The theoretical predictions for $B \to K^* \nu \bar{\nu}$ decays discussed in \cref{sec:ksnunu-theory} rely on precise determinations of the $V(q^2)$, $A_{0}(q^2)$, $A_{1}(q^2)$, $A_{12}(q^2)$, $T_1(q^2)$, $T_2(q^2)$ and $T_{23}(q^2)$ hadronic form factors that parametrize the non-perturbative \ac{QCD} matrix elements. The values for the corresponding $19$ hadronic parameters obtained from the Gaussian likelihood provided in Reference~\cite{Gubernari:2023puw} and are reported in \cref{tab:hadronic parameters kstar full}. The values obtained from a joint theoretical prior \ac{PDF} based on results from References~\cite{Gubernari:2020eft,Horgan:2015vla} are reported in \cref{tab:hadronic parameters kstar full update}.

\begin{sidewaystable}
    \caption{The hadronic parameters used to parametrize the $V(q^2)$, $A_{0}(q^2)$, $A_{1}(q^2)$, $A_{12}(q^2)$, $T_1(q^2)$, $T_2(q^2)$ and $T_{23}(q^2)$ form factors in the \ac{BSZ} parametrization~\cite{Bharucha_2016}, based on results of Reference~\cite{Gubernari:2023puw}. The table shows the mean values (first row) and the covariance matrix (upper triangular part).}
    \centering
    \scriptsize
    \setlength{\tabcolsep}{3pt}
    \begin{tabular}{ccccccccccccccccccc}
    \toprule
    $v_0$ & $v_1$ & $v_2$ & $a_0^0$ & $a_1^0$ & $a_2^0$ & $a_0^1$ & $a_1^1$ & $a_2^1$ & $a_{12}^1$ & $a_{12}^2$ & $t_0^1$ & $t_1^1$ & $t_2^1$ & $t_1^2$ & $t_2^2$ & $t_{23}^0$ & $t_{23}^1$ & $t_{23}^2$ \\
    \midrule
        0.3628 & -1.0893 & 2.8029 & 0.3421 & -1.1474 & 2.3728 & 0.2893 & 0.4599 & 1.2242 & 0.5512 & 0.5781 & 0.3192 & -0.9619 & 2.0931 & 0.5932 & 1.6795 & 0.6212 & 0.9733 & 1.8213 \\
    \midrule
    0.0014 & -0.0008 & -0.0633 & 0.0008 & 0.0039 & -0.0001 & 0.0135 & 0.0506 & 0.0012 & -0.0030 & -0.0630 & 0.0006 & 0.0011 & 0.0130 & 0.0040 & 0.0139 & 0.0011 & 0.0015 & -0.0199 \\
     & 0.0660 & 0.3765 & -0.0010 & -0.0010 & -0.0195 & 0.0219 & 0.1498 & -0.0034 & 0.0147 & 0.1876 & -0.0009 & -0.0033 & -0.0733 & -0.0083 & -0.0549 & -0.0025 & -0.0039 & 0.0500 \\
     & & 7.8174 & -0.0242 & -0.1684 & -0.0881 & -0.5806 & -1.5977 & -0.0509 & 0.0928 & 4.0359 & -0.0087 & -0.1056 & -1.9914 & -0.1922 & -1.5282 & -0.0424 & 0.0235 & 1.3014 \\
     & & & 0.0009 & 0.0036 & -0.0038 & 0.0069 & 0.0273 & 0.0014 & -0.0032 & -0.0635 & 0.0008 & 0.0003 & -0.0059 & 0.0042 & 0.0044 & 0.0011 & 0.0033 & -0.0102 \\
     & & & & 0.0286 & 0.0496 & 0.0571 & 0.2767 & 0.0061 & -0.0119 & -0.3353 & 0.0035 & 0.0039 & 0.0102 & 0.0192 & 0.0344 & 0.0052 & 0.0130 & -0.0528 \\
     & & & & & 0.5464 & 0.0785 & 0.5791 & -0.0011 & -0.0094 & 0.0201 & -0.0024 & 0.0085 & 0.1791 & -0.0114 & 0.0269 & -0.0014 & -0.0106 & -0.0011 \\
     & & & & & & 0.2434 & 1.2378 & 0.0097 & -0.0156 & -0.6045 & 0.0042 & 0.0162 & 0.2251 & 0.0387 & 0.1792 & 0.0093 & 0.0013 & -0.2289 \\
     & & & & & & & 7.0963 & 0.0371 & -0.0532 & -2.2307 & 0.0170 & 0.0732 & 0.9817 & 0.1582 & 0.7220 & 0.0370 & -0.0092 & -0.9501 \\
     & & & & & & & & 0.0025 & -0.0069 & -0.1353 & 0.0014 & 0.0007 & -0.0065 & 0.0071 & 0.0098 & 0.0019 & 0.0055 & -0.0183 \\
     & & & & & & & & & 0.0464 & 0.4091 & -0.0033 & -0.0008 & 0.0285 & -0.0161 & -0.0127 & -0.0048 & -0.0140 & 0.0571 \\
     & & & & & & & & & & 10.012 & -0.0612 & -0.0618 & -0.1433 & -0.3813 & -0.8975 & -0.0937 & -0.2439 & 1.2014 \\
     & & & & & & & & & & & 0.0010 & -0.0005 & -0.0189 & 0.0038 & -0.0045 & 0.0012 & 0.0039 & -0.0056 \\
     & & & & & & & & & & & & 0.0205 & 0.1149 & 0.0087 & 0.0599 & 0.0002 & 0.0031 & -0.0219 \\
     & & & & & & & & & & & & & 2.6624 & -0.0004 & 1.0087 & -0.0053 & -0.1341 & -0.2744 \\
     & & & & & & & & & & & & & & 0.0451 & 0.1751 & 0.0057 & 0.0138 & -0.1063 \\
     & & & & & & & & & & & & & & & 1.5341 & 0.0075 & -0.0664 & -0.5565 \\
     & & & & & & & & & & & & & & & & 0.0022 & 0.0039 & -0.0454 \\
     & & & & & & & & & & & & & & & & & 0.0998 & 0.4790 \\
     & & & & & & & & & & & & & & & & & & 5.9109 \\
    \midrule
    \bottomrule
    \end{tabular}
    \label{tab:hadronic parameters kstar full}
\end{sidewaystable}

\begin{sidewaystable}
    \caption{The hadronic parameters used to parametrize the $V(q^2)$, $A_{0}(q^2)$, $A_{1}(q^2)$, $A_{12}(q^2)$, $T_1(q^2)$, $T_2(q^2)$ and $T_{23}(q^2)$ form factors in the \ac{BSZ} parametrization~\cite{Bharucha_2016}, based on results of References~\cite{Gubernari:2020eft,Horgan:2015vla}. The table shows the mean values (first row) and the covariance matrix (upper triangular part).}
    \centering
    \scriptsize
    \setlength{\tabcolsep}{3pt}
    \begin{tabular}{ccccccccccccccccccc}
    \toprule
    $v_0$ & $v_1$ & $v_2$ & $a_0^0$ & $a_1^0$ & $a_2^0$ & $a_0^1$ & $a_1^1$ & $a_2^1$ & $a_{12}^1$ & $a_{12}^2$ & $t_0^1$ & $t_1^1$ & $t_2^1$ & $t_1^2$ & $t_2^2$ & $t_{23}^0$ & $t_{23}^1$ & $t_{23}^2$ \\
    \midrule
        0.3631 & -1.0903 & 2.7817 & 0.2894 & 0.4597 & 1.2199 & 0.5422 & 0.5465 & 0.3415 & -1.1442 & 2.4375 & 0.3193 & -0.9574 & 2.1215 & 0.5971 & 1.7001 & 0.6216 & 0.9705 & 1.8025 \\
    \midrule
    0.0008 & -0.0013 & -0.0348 & 0.0004 & 0.0018 & -0.0013 & 0.0017 & 0.0067 & 0.0003 & -0.0002 & -0.0012 & 0.0005 & 0.0000 & -0.0085 & 0.0026 & 0.0029 & 0.0006 & 0.0024 & 0.0019 \\
     & 0.0284 & 0.1068 & -0.0002 & 0.0019 & -0.0058 & 0.0033 & 0.0118 & -0.0004 & 0.0043 & -0.0242 & -0.0002 & 0.0003 & 0.0030 & -0.0006 & 0.0000 & -0.0004 & -0.0008 & 0.0090 \\
     & & 4.0579 & -0.0115 & -0.0790 & 0.1297 & -0.0983 & -0.3947 & -0.0043 & -0.0306 & 0.3915 & -0.0139 & 0.0009 & 0.2689 & -0.0745 & -0.0657 & -0.0168 & -0.0617 & 0.0177 \\
     & & & 0.0004 & 0.0012 & -0.0031 & 0.0018 & 0.0059 & 0.0003 & 0.0000 & -0.0035 & 0.0003 & 0.0000 & -0.0062 & 0.0018 & 0.0022 & 0.0004 & 0.0017 & 0.0023 \\
     & & & & 0.0174 & 0.0550 & 0.0261 & 0.1422 & 0.0011 & 0.0049 & -0.0235 & 0.0014 & 0.0001 & -0.0247 & 0.0076 & 0.0094 & 0.0018 & 0.0073 & 0.0053 \\
     & & & & & 0.5522 & 0.0782 & 0.6026 & -0.0007 & 0.0046 & 0.1593 & -0.0011 & -0.0001 & 0.0382 & -0.0074 & -0.0093 & -0.0004 & -0.0055 & -0.0454 \\
     & & & & & & 0.1186 & 0.6613 & 0.0047 & 0.0167 & -0.1270 & 0.0014 & -0.0009 & -0.0272 & 0.0064 & 0.0047 & 0.0020 & 0.0081 & 0.0030 \\
     & & & & & & & 4.3665 & 0.0118 & 0.0759 & 0.0016 & 0.0053 & -0.0024 & -0.0999 & 0.0231 & 0.0110 & 0.0064 & 0.0280 & 0.0263 \\
     & & & & & & & & 0.0007 & 0.0004 & -0.0195 & 0.0002 & -0.0001 & -0.0041 & 0.0011 & 0.0011 & 0.0003 & 0.0013 & 0.0004 \\
     & & & & & & & & & 0.0380 & 0.1255 & -0.0001 & 0.0006 & 0.0053 & 0.0003 & 0.0038 & -0.0003 & -0.0004 & 0.0008 \\
     & & & & & & & & & & 3.0766 & -0.0014 & 0.0074 & 0.0218 & -0.0026 & 0.0059 & -0.0050 & -0.0131 & 0.0332 \\
     & & & & & & & & & & & 0.0004 & -0.0004 & -0.0110 & 0.0019 & -0.0004 & 0.0005 & 0.0017 & 0.0023 \\
     & & & & & & & & & & & & 0.0189 & 0.0704 & 0.0043 & 0.0198 & -0.0003 & 0.0047 & -0.0073 \\
     & & & & & & & & & & & & & 1.6254 & -0.0675 & 0.1173 & -0.0063 & -0.0624 & 0.0737 \\
     & & & & & & & & & & & & & & 0.0331 & 0.1339 & 0.0021 & 0.0143 & 0.0048 \\
     & & & & & & & & & & & & & & & 0.9558 & 0.0026 & 0.0103 & 0.0404 \\
     & & & & & & & & & & & & & & & & 0.0012 & 0.0017 & -0.0264 \\
     & & & & & & & & & & & & & & & & & 0.0957 & 0.5292 \\
     & & & & & & & & & & & & & & & & & & 5.7139 \\
    \midrule
    \bottomrule
    \end{tabular}
    \label{tab:hadronic parameters kstar full update}
\end{sidewaystable}

%% file: chapters/appendix-stat.tex
\section{Probability density functions}
\label{sec:pdfs}
For a measurement outcome depending on a continuous variable $x$, the probability to find $x$ in an infinitesimal range around a value $x'$ is given by the \ac{PDF} $p(x)$~\cite{Cowan1998},
\begin{equation}
    P(x \in [x', x'+dx]) = p(x') dx.
    \label{eq:pdf}
\end{equation}
The \ac{PDF} is normalized to 1 over the whole range of $x$, $[a,b]$,
\begin{equation}
    \int_a^b dx \ p(x)  = 1.
\end{equation}
\acp{PDF} play a fundamental role in statistical analyses. The fundamental goal of any data analysis is to gather information about certain parameters, given large amounts of data. For example, if one attempts to measure the mass of a particle, what is actually obtained is a \ac{PDF} of the mass that indicates how likely certain values are. Ideally, this \ac{PDF} is a narrow peak around the true value of the mass. The width of the \ac{PDF} then defines the uncertainty on the result.

In particle physics, data are often binned in histograms. Consider a set of $N$ measurements, $x_1, \ldots, x_N$. The $x$-range can be subdivided into $m$ intervals of width $\Delta x_i$ for $i = 1,\ldots, m$, and one can count how many events $n_i$ are in each of those intervals. The value of the \ac{PDF} is then given by normalizing the area under the histogram to 1.
For a specific bin $i$ this is given by
\begin{equation}
    p_i = \frac{n_i}{\sum_i n_i \Delta x_i}.
    \label{eq:histogram}
\end{equation}
The probability of finding an event in a certain bin $i$ is then
\begin{equation}
    P(x \in \Delta x_i) = p_i \Delta x_i.
\end{equation}
In the continuous limit $m \to \infty$, the system returns to \cref{eq:pdf}.
From \cref{eq:histogram}, it can be easily seen that the \ac{PDF} is correctly normalized, since
\begin{equation}
    \sum_i p_i \Delta x_i = 1.
\end{equation}
Hence, there is a mapping between continuous and binned probability models,
\begin{equation}
    \int p(x) dx \leftrightarrow \sum p_i \Delta x_i,
\end{equation}
which can be applied to any of the following definitions in this text.

\subsection{Cumulative distribution functions}
\label{sec:cdfs}
The \ac{CDF} is the probability of a measurement less than or equal to $x$~\cite{Cowan1998},
\begin{equation}
    F(x) = \int_{-\infty}^x p(x') dx'.
\end{equation}
A \textit{quantile} of $\alpha$, $x_\alpha$, is then defined such that the value $F(x_\alpha) = \alpha$, where $0\leq \alpha \leq 1$. The quantile is then given by
\begin{equation}
    x_\alpha = F^{-1}(\alpha).
\end{equation}







\subsection{Transformation properties}
\label{sec:pdf-transformation}
Consider now the transformation properties of \acp{PDF}. Suppose there is a random variable $x$ with \ac{PDF} $p_x(x)$, and a transformation is applied to $x$ such that a new variable $y$ is related to $x$ by the equation
\begin{equation}
    y = g(x),
\end{equation}
where $g$ is a differentiable and invertible function. The goal is to find the \ac{PDF} of $y$, denoted $p_y(y)$, in terms of the \ac{PDF} of $x$.

By the law of total probability, the probability that $y$ lies in the range $[y', y' + dy]$ is the same as the probability that $x$ lies in the range $[x', x' + dx]$, where $x' = g^{-1}(y')$. Therefore, the following holds~\cite{Cowan1998}:
\begin{equation}
    P(y \in [y', y'+dy]) = P(x \in [x', x' + dx]).
\end{equation}
Since $P(x \in [x', x' + dx]) = p_x(x') \, dx$, it follows that:
\begin{equation}
    p_y(y') \, dy = p_x(x') \, dx.
\end{equation}
Substituting $dx = \left| \frac{dx}{dy} \right| dy$ gives:
\begin{equation}
    p_y(y') = p_x(g^{-1}(y')) \left| \frac{dx}{dy} \right|.
\end{equation}
Thus, the \ac{PDF} of $y$ is related to the \ac{PDF} of $x$ by the transformation rule:
\begin{equation}
    p_y(y) = p_x(g^{-1}(y)) \left| \frac{d}{dy} g^{-1}(y) \right|.
\end{equation}

\subsection{Expectation values}
\label{sec:expectation-values}
The \textit{expectation value} of a random variable $x$ distributed according to the \ac{PDF} $p(x)$ is given by~\cite{Cowan1998}
\begin{equation}
    E[x] = \int_{-\infty}^{\infty} xp(x)dx,
    \label{eq:expectation-value}
\end{equation}
which is not a function of $x$, but depends only on the shape of $p(x)$. If $p(x)$ is concentrated in one region, $E[x]$ corresponds to likely observed values of $x$. If $p(x)$ has separated peaks, then values of $x$ corresponding to $E[x]$ can actually be unlikely values to observe.

The \textit{variance} is defined as~\cite{Cowan1998}
\begin{equation}
    V[x] = E[(x-E[x])^2] = E[x^2] - E[x]^2,
    \label{eq:variance}
\end{equation}
which measures the spread of $x$ around its mean.
The \textit{standard deviation} is
\begin{equation}
    \sigma[x] = \sqrt{V[x]}.
\end{equation}
In the case of multiple variables, the \textit{covariance} is defined as
\begin{equation}
    cov[x_1,x_2] = E[(x_1-E[x_1])(x_2-E[x_2])] = E[x_1x_2] - E[x_1]E[x_2],
\end{equation}
which encodes information on the correlation of the two variables.
The covariance matrix is symmetric by construction and the diagonal elements correspond to $cov[x_i,x_i] = \sigma[x_i]^2$.
To have a dimensionless measure of correlation, one can define the \textit{correlation matrix}
\begin{equation}
    \rho_{ij} = \frac{cov[x_i, x_j]}{\sigma[x_i]\sigma[x_j]}.
\end{equation}
The range of any coefficient in the correlation matrix is bounded by $-1 \leq \rho_{ij} \leq 1$. 

A way to think about the meaning of the correlation matrix is as follows: a positive correlation coefficient $\rho_{12}>0$ corresponds to ``if I measure $x_1 > E[x_1]$, then it is more likely to obtain $x_2 > E[x_2]$''. Conversely, a negative correlation coefficient $\rho_{12}<0$ corresponds to ``if I measure $x_1 > E[x_1]$, then it is more likely to obtain $x_2 < E[x_2]$''.

\section{Conjugate priors}
\label{sec:conjugate-priors}
A \textit{conjugate prior} is a prior distribution that, when combined with a specific likelihood function, results in a posterior distribution of the same family as the prior. This property simplifies the process of Bayesian updating, as the posterior can be easily computed and interpreted.

The constraint likelihoods in the \histfactory statistical model are either normal or Poisson \acp{PDF}. Consider the two cases:

\paragraph{Normal constraints} Most modifiers have normal constraint terms of the form
\begin{equation}
    \mathcal{N}(a | \chi, \sigma_{\chi}) \ .
\end{equation}
In these cases, a suitable choice of conjugate prior is another normal \ac{PDF},
\begin{equation}
    \mathcal{N}(\chi | \chi', \sigma_{\chi}') \ .
\end{equation}
Since the product of two normal distributions is another normal distribution~\cite{gelman2013bayesian}, the posterior becomes
\begin{equation}
    p(\chi) = \mathcal{N}(\chi | \bar \chi, \bar \sigma) \propto \mathcal{N}(a | \chi, \sigma) \mathcal{N}(\chi | \chi', \sigma_{\chi}') \ ,
\end{equation}
with
\begin{equation}
    \bar \chi = \bar \sigma^2 \left( \frac{\chi'}{\sigma_{\chi}'^2} + \frac{a}{\sigma_{\chi}^2}\right) ~, \qquad \bar \sigma^2 = \frac{\sigma_{\chi}^2 \sigma_{\chi}'^2}{\sigma_{\chi}^2 +\sigma_{\chi}'^2} \ .
\end{equation}
In the limit $\sigma_{\chi} \ll \sigma_{\chi}'$, 
\begin{equation}
    \bar \chi \to a \ , \qquad \bar \sigma \to \sigma_{\chi} \ ,
\end{equation}
the posterior is (auxiliary) data-driven and not prior-driven.

\paragraph{Poisson constraints} For the uncorrelated shape modifier, the individual constraint terms have the form 
\begin{equation}
    \pois(a|a\chi).
\end{equation}
The conjugate prior for the Poisson distribution is the Gamma distribution \cite{gelman2013bayesian},
\begin{equation}
    p_{\Gamma}(\theta|\alpha, \beta) = \frac{\beta^{\alpha}}{\Gamma(\alpha)}\theta^{\alpha-1} e^{-\beta \theta}.
\end{equation}
Combining the above two equations and setting $\theta=a\chi$, the posterior is
\begin{equation}
    p(a \chi ) = p_{\Gamma}(a \chi | \alpha + a, \beta+1) \propto \pois(a|a\chi)p(a \chi |\alpha, \beta).
\end{equation}
With the parameter transformation property (see \cref{sec:pdf-transformation})
\begin{equation}
    p(\chi) = \frac{d(a\chi)}{d \chi} p(a\chi) = a p(a\chi),
\end{equation}
the posterior for $\chi$ is simply another Gamma distribution,
\begin{equation}
    p( \chi ) = p_{\Gamma}(\chi | \bar \alpha, \bar \beta) = a ~  p_{\Gamma}(a \chi | \bar \alpha, \bar \beta / a),
\end{equation}
with
\begin{equation}
    \bar \alpha = \alpha + a \ , \qquad \bar \beta = a(\beta+1) \ .
\end{equation}
In the limit of $\alpha \to 0$ and $\beta \to 0$,
\begin{equation}
    p( \chi ) \propto \pois(a | a \chi),
\end{equation}
such that the posterior is (auxiliary) data-driven and not prior-driven.

\section{Proof of the Cramér-Rao bound}
\label{sec:cramer-rao-bound}
The \textit{Cramér-Rao bound} is an important result, which sets a lower bound on the variance of any estimator. The variance of an estimator $\estimator_i \in \estimator$ with a sampling distribution $p(\estimator_i | \params_0)$ is given by,
\begin{equation}
    \boldsymbol{V}_i[\estimator] = \int d^m \hat \theta \ (\estimator_i - \boldsymbol{E}_i[\estimator])^2 \ p(\estimator_i | \params_0) \ .
\end{equation}
The Cramér-Rao bound sets a lower limit on the variance in terms of the Fisher information and the bias~\cite{James:2006zz},
\begin{equation}
    \boldsymbol{V}_i[\estimator] \geq \frac{\partial \boldsymbol{E}_i[\estimator] / \partial \theta_i}{\boldsymbol{I}_{ii}(\estimator)} = \frac{(1+\partial \boldsymbol{b}_i / \partial \theta_i)^2}{\boldsymbol{I}_{ii}(\estimator)} \ .
    \label{eq:cramer-rao-bound}
\end{equation}

This proof only considers a single estimator, but it is easily extendable to multidimensional estimators, $\estimator$. 
Consider a scalar estimator $\hat\theta(\data)$ with a sampling distribution $p(\hat \theta | \theta_0)$. 
The conditions under which \cref{eq:cramer-rao-bound} holds are 
\begin{itemize}
    \item the range of $\data$ is independent of $\theta$,
    \item $\partial /\partial \theta_0$ commutes with $\int d x$ or equivalently $\int d \hat \theta$.
\end{itemize}
Starting with the normalization condition of any \ac{PDF},
\begin{equation}
    \int d \hat \theta \ p(\hat \theta | \theta_0) = 1 \ .
\end{equation}
Differentiating with respect to $\theta_0$ gives
\begin{equation}
    \int d \hat \theta \ \frac{\partial p(\hat \theta | \theta_0)}{\partial \theta_0} = \int d \hat \theta \ \frac{\partial \ln p(\hat \theta | \theta_0)}{\partial \theta_0} p(\hat \theta | \theta_0) = 0 \ .
    \label{eq:cramer-rao-step1}
\end{equation}
Furthermore, differentiating $E[\hat \theta]$ with respect to $\theta_0$ gives
\begin{equation}
    \frac{\partial E[\hat \theta]}{\partial \theta_0} = \frac{\partial}{\partial \theta_0} \int d \hat \theta \ \hat \theta p(\hat \theta | \theta_0) = \int d \hat \theta \ \hat \theta \frac{\partial \ln p(\hat \theta | \theta_0)}{\partial \theta_0} p(\hat \theta | \theta_0)\ .
    \label{eq:cramer-rao-step2}
\end{equation}
Since $E[\hat \theta]$ is a constant, one can multiply \cref{eq:cramer-rao-step1} with it to obtain zero and subtract it from \cref{eq:cramer-rao-step2}. This gives
\begin{equation}
    \frac{\partial E[\hat \theta]}{\partial \theta_0} = \int d \hat \theta \ \left[\hat \theta - E[\hat \theta]\right] \frac{\partial \ln p(\hat \theta | \theta_0)}{\partial \theta_0} p(\hat \theta | \theta_0)
\end{equation}
An application of Schwarz' inequality yields
\begin{equation}
    \int d \hat \theta \ \left[\hat \theta - E[\hat \theta]\right]^2 \ \int d \hat \theta \ \left[\frac{\partial \ln p(\hat \theta | \theta_0)}{\partial \theta_0}\right]^2 p(\hat \theta | \theta_0) \geq \left(\frac{\partial E[\hat \theta]}{\partial \theta_0}\right)^2
\end{equation}
The first expression is the definition of the variance $V[\hat \theta]$. The second expression is the Fisher information (\cref{eq:fisher-information})
\begin{equation}
    \int d \hat \theta \ \left[\frac{\partial \ln p(\hat \theta | \theta_0)}{\partial \theta_0}\right]^2 p(\hat \theta | \theta_0) = E\left[\left(\frac{\partial \ln p(\hat \theta | \theta_0)}{\partial \theta_0}\right)^2\right] = I(\hat \theta).
\end{equation}
Hence, this gives 
\begin{equation}
    V[\hat \theta] \geq \frac{\partial E[\hat \theta] / \partial \theta_0}{I(\hat \theta)} = \frac{(1+\partial b / \partial \theta_0)^2}{I(\hat \theta)} \ ,
\end{equation}
where the last step comes from the definition of the bias. This proves \cref{eq:cramer-rao-bound}~\cite{James:2006zz,Young_Smith_2005}.

\section{Fisher information}
\label{sec:fisher-information}
The Fisher information is a key quantity in statistical inference that quantifies the amount of information a random variable $x$ carries about an unknown parameter $\theta$~\cite{James:2006zz}.

For a given \ac{PDF} $p(x|\theta)$, the Fisher information $I(\theta)$ is defined as the expected value of the squared derivative of the log-likelihood function with respect to the parameter $\theta$,
\begin{equation}
    I(\theta) = E\left[ \left( \frac{\partial}{\partial \theta} \log p(x|\theta) \right)^2 \right] = \int dx \  \left( \frac{\partial}{\partial \theta} \log p(x|\theta) \right)^2 p(x|\theta) \ .
    \label{eq:fisher-information}
\end{equation}
For multiple parameters, the Fisher information takes the matrix form
\begin{equation}
    \boldsymbol{I}_{ij}(\params) = E\left[ \left(\frac{\partial}{\partial \theta_i} \log p(x|\params) \right) \left( \frac{\partial}{\partial \theta_j} \log p(x|\params)  \right) \right].
    \label{eq:fisher-information-multidim}
\end{equation}

The Fisher information provides a measure of the sensitivity of the likelihood function with respect to changes in $\theta$. A high Fisher information value indicates that small changes in $\theta$ result in large changes in the likelihood, implying that the data provide more information about $\theta$. Conversely, a low Fisher information value suggests that the likelihood is relatively insensitive to changes in $\theta$, indicating less information about the parameter.


\section{Proof of consistency of maximum likelihood estimators}
\label{sec:consistency-mle}
To prove the consistency of the \ac{MLE}, Jensen's inequality is applied to the strictly concave logarithm function~\cite{Young_Smith_2005}:
\begin{equation}
    E\left[ \ln\left(\frac{p(x|\params)}{p(x|\params_0)}\right)\right] < \ln\left(E\left[ \frac{p(x|\params)}{p(x|\params_0)}\right]\right) = 0 \quad \forall \ \params \neq \params_0 \ .
    \label{eq:mle-jensen-inequality}
\end{equation}
The last step follows from
\begin{equation}
    E\left[ \frac{p(x|\params)}{p(x|\params_0)}\right] = \int dx \ p(x |\params_0)  \frac{p(x|\params)}{p(x|\params_0)} = 1 \ .
\end{equation}
Hence, from \cref{eq:mle-jensen-inequality},
\begin{equation}
    E\left[ \ln p(x|\params_0)\right] > E\left[ \ln p(x|\params)\right] \quad \forall \ \params \neq \params_0 \ .
\end{equation}
Consider now two solutions to \cref{eq:mleestimator}, one inconsistent, $\estimator'$, and one consistent, $\estimator$. Then from the law of large numbers (\cref{eq:law-of-large-numbers}),
\begin{equation}
    \lim_{N \to \infty} \frac{1}{N} \sum_{i=1}^{N} \ln p(x_i|\estimator) = E[\ln p(x|\params_0)] > E[\ln p(x|\estimator')]
\end{equation}
Hence, the consistent solution will asymptotically correspond to the absolute maximum of the likelihood.
Note that this proof does not require differentiability of the likelihood. This means that the \ac{MLE} is consistent, even if not found through \cref{eq:mle-condition-diff}. However, if the likelihood is differentiable and the estimator found through \cref{eq:mle-condition-diff}, it is consistent, but not necessarily unique~\cite{James:2006zz,Young_Smith_2005}.

\section{Proof of asymptotic normality of the maximum likelihood estimator}
\label{sec:asymptotic-normality-mle}
To prove the asymptotic normality of the \ac{MLE}, the approach from Reference~\cite{Young_Smith_2005} is followed. The condition is enforced that the log-likelihood $\ln p(\data | \theta)$ is twice differentiable in a neighbourhood of $\theta_0$\footnote{This proof is presented for the case of one parameter $\theta$, but it is easily extendable to multiple parameters.}. By the property of consistency, just discussed above, there exists (at least) one parameter point $\hat \theta$ where 
\begin{equation}
    \partial_\theta \ln p(\data | \theta)|_{\hat \theta} = \frac{\partial \ln p(\data | \theta)}{\partial \theta} \bigg|_{\hat \theta} = 0 \ ,
\end{equation}
 such that $\lim_{N \to \infty}\hat \theta = \theta_0$.

Expanding the negative log-likelihood around a point $\theta^*$ which lies between $\theta_0$ and $\hat \theta$,
\begin{equation}
    -\partial_\theta \ln p(\data | \theta)|_{\theta_0} = \partial_\theta \ln p(\data | \theta)|_{\hat \theta} - \partial_\theta \ln p(\data | \theta)|_{\theta_0} = (\hat \theta - \theta_0)\partial_\theta^2 \ln p(\data | \theta)|_{\theta^*} \ .
\end{equation}
Hence,
\begin{equation}
    \hat \theta - \theta_0 = - \frac{\partial_\theta \ln p(\data | \theta)|_{\theta_0}}{\partial_\theta^2 \ln p(\data | \theta)|_{\theta^*}} \ ,
\end{equation}
which can be written as 
\begin{equation}
    \sqrt{N I(\theta_0)} (\hat \theta - \theta_0) = 
    \left(\frac{\partial_\theta \ln p(\data | \theta)|_{\theta_0}}{\sqrt{N I(\theta_0)}} \right) 
    \left(\frac{\partial_\theta^2 \ln p(\data | \theta)|_{\theta_0}}{\partial_\theta^2 \ln p(\data | \theta)|_{\theta^*}}\right) 
    \left(\frac{\partial_\theta^2 \ln p(\data | \theta)|_{\theta_0}}{N I(\theta_0)}\right)^{-1} \ .
    \label{eq:asymptotic-normality-step1}
\end{equation}
For the first term on the right-hand side of \cref{eq:asymptotic-normality-step1} the central limit theorem (\cref{eq:central-limit-thm}) can be applied. It is known that
\begin{equation}
    E_{\theta_0}\left[\partial_\theta \ln p(\data | \theta)\right] = E\left[\partial_\theta \ln p(\data | \theta)\right]|_{\theta_0} = 0 \ ,
\end{equation}
and
\begin{equation}
    V_{\theta_0}\left[\partial_\theta \ln p(\data | \theta)\right] = E_{\theta_0}\left[\left(\partial_\theta \ln p(\data | \theta)\right)^2\right]|_{\theta_0} = I(\theta_0) \ .
\end{equation}
Hence, it follows that
\begin{equation}
    \frac{\partial_\theta \ln p(\data | \theta)|_{\theta_0}}{\sqrt{N I(\theta_0)}} = \frac{\sqrt{N}}{\sqrt{I(\theta_0)}} \sum_i \partial_\theta \ln p(x_i | \theta) \to^d \mathcal{N}(0,1) \ .
\end{equation}

The second term on the right-hand side of \cref{eq:asymptotic-normality-step1} approaches unity in the large sample limit,
\begin{equation}
    \frac{\partial_\theta^2 \ln p(\data | \theta)|_{\theta_0}}{\partial_\theta^2 \ln p(\data | \theta)|_{\theta^*}} \to 1 \ ,
\end{equation}
by the consistency condition $\hat \theta \to \theta_0$ and the fact that $\theta^*$ lies between $\theta_0$ and $\hat \theta$.

The third term on the right-hand side of \cref{eq:asymptotic-normality-step1} also approaches unity in the large sample limit, by the law of large numbers (\cref{eq:law-of-large-numbers}).

Hence, it follows that 
\begin{equation}
    \sqrt{N I(\theta_0)} (\hat \theta - \theta_0) \to^d \mathcal{N}(0,1) \ ,
\end{equation}
which is equivalent to \cref{eq:asymptotic-normality}.

\section{Proof of the Neyman-Pearson lemma}
\label{sec:proof-npl}
To prove the Neyman-Pearson lemma, the method of Lagrange multipliers can be adapted~\cite{Young_Smith_2005}. The goal is to maximize the power of the test, $1-\beta(t)$, subject to the constraint $\alpha(t) \leq \alpha$. 
This constraint ensures that the Type I error probability does not exceed the chosen significance level $\alpha$.

The Lagrangian to maximize is
\begin{align*}
    L(t_{cut}, \lambda) & = (1-\beta(t_{cut})) - \frac{1}{\lambda}(\alpha(t_{cut}) - \alpha)                                                 \\
                        & = \int_{t_{cut}}^{\infty} p(t|H_1)dt - \frac{1}{\lambda} \left(\int_{t_{cut}}^{\infty} p(t|H_0)dt - \alpha\right),
\end{align*}
for a positive-valued $\lambda$.
This can be written as a function of $\data$, integrating over a rejection region $\data \in \mathcal{R}$,
\begin{align*}
    L(\data, \lambda) & = \int_{\mathcal{R}} p(\data|H_1) d^n \data - \frac{1}{\lambda} \left(\int_{\mathcal{R}} p(\data|H_0) d^n \data - \alpha\right) \\
                      & = \int_{\mathcal{R}} \left(p(\data|H_1) - \frac{1}{\lambda} p(\data|H_0)\right) d^n \data - \frac{1}{\lambda} \alpha.
\end{align*}
To maximize this integral, the rejection region must be defined such that
\begin{equation}
    \mathcal{R}(\lambda) = \left\{ \data \mid p(\data|H_1) > \frac{1}{\lambda} p(\data|H_0) \right\},
\end{equation}
or, rearranging, this defines a rejection region where
\begin{equation}
    \frac{p(\data|H_0)}{p(\data|H_1)} < \lambda.
\end{equation}
The value of $\lambda$ is determined by the upper bound of the constraint $\alpha(t) \leq \alpha$,
\begin{equation}
    [\alpha(\data)]_{\mathcal{R}(\lambda)} = \alpha.
\end{equation}
For the proof of the Neyman-Pearson lemma, consider now an alternative region $\mathcal{R}'$. Since $p(\data|H_1) > \frac{1}{\lambda} p(\data|H_0)$ in all of $\mathcal{R}$, any deviation to $\mathcal{R}'$ will contain parts where $p(\data|H_1) < \frac{1}{\lambda} p(\data|H_0)$. Hence, the inequality can be defined as
\begin{equation}
    \left[(1-\beta(\data)) - \frac{1}{\lambda}\alpha(\data)\right]_{\mathcal{R}} \geq \left[(1-\beta(\data)) - \frac{1}{\lambda}\alpha(\data)\right]_{\mathcal{R'}}.
\end{equation}
Rearranging this inequality gives
\begin{equation}
    [1-\beta(\data)]_{\mathcal{R}} - [1-\beta(\data)]_{\mathcal{R'}} \geq
    \frac{1}{\lambda}\left([\alpha(\data)]_{\mathcal{R}} - [\alpha(\data)]_{\mathcal{R'}}\right) \geq 0,
\end{equation}
where the last equality uses the fact that $[\alpha(\data)]_{\mathcal{R}} = \alpha$ and $[\alpha(\data)]_{\mathcal{R'}} \leq \alpha$. Hence, it follows that
\begin{equation}
    [1-\beta(\data)]_{\mathcal{R}} \geq [1-\beta(\data)]_{\mathcal{R'}}
\end{equation}
showing that the rejection region defined by the likelihood ratio results in the most powerful test. This proves the Neyman-Pearson lemma~\cite{Young_Smith_2005}.

\section{Likelihood ratio test statistic for normally distributed data}
\label{sec:likelihood-ratio-test-statistic-normal}
This section shows the asymptotic distribution of the likelihood ratio test statistic for the case of normally distributed data. This proves to be useful for Wilks' theorem~\cite{wilks} and goodness-of-fit testing.

For example, for a Gaussian likelihood of $N$ data points $\data$ with means $\boldsymbol{\mu}$ and variances $\boldsymbol{\sigma}^2$,
\begin{equation}
    p(\data | \boldsymbol{\mu}, \boldsymbol{\sigma}^2) = \prod_{i=0}^N \mathcal{N}(x_i | \mu_i, \sigma_i^2) = \prod_{i=0}^N \frac{1}{\sqrt{2 \pi \sigma_i^2}} e^{-\frac{(x_i - \mu_i)^2}{2\sigma_i^2}} \ .
\end{equation}
The likelihood ratio test statistic is then simply
\begin{equation}
    t = \sum_{i=0}^N \frac{(x_i - \mu_i)^2}{\sigma_i^2} \ .
\end{equation}
One might recognize this term as the classic $\chisq$ variable.
So what is the \ac{PDF} of the test statistic in this case? For one data-point
\begin{equation}
    p(x_i | \mu_i, \sigma_i^2) = \frac{1}{\sqrt{2 \pi \sigma_i^2}} e^{-\frac{(x_i - \mu_i)^2}{2\sigma_i^2}}.
\end{equation}
The distribution of the test statistic is
\begin{equation}
    p(t_i) = p(x_i | \mu_i, \sigma_i^2) \left|\frac{dx_i}{d t_i}\right| = \frac{1}{\sqrt{2 \pi t_i}} e^{-t_i/2},
\end{equation}
where an extra factor of 2 comes from the change in integration limits. This is the $\chisq$-distribution for 1 \ac{d.o.f.}. The characteristic function\footnote{The Fourier transform of the \ac{PDF}.} of this \ac{PDF} is
\begin{equation}
    \phi_i(k) = \int_0^\infty d t_i \ p(t_i) e^{ikt_i} = (1-2ik)^{-1/2}.
\end{equation} 
Under the assumption of independent data points, the characteristic function of the sum $t = \sum_{i=0}^N t_i$ is the product of the individual characteristic functions
\begin{equation}
    \phi(k) = \prod_{i=0}^N \phi_i(k) = (1-2ik)^{-N/2},
\end{equation}
which is the characteristic function of a $\chisq$-distribution with $N$ \ac{d.o.f.}. If the system parametrizes $\mu(x | \params)$ with $m$ parameters and infers these parameters from the data, this results in losing $m$ independent pieces of information and the resulting test statistic \ac{PDF} will be a $\chisq$-distribution with $N-m$ \ac{d.o.f.}.
This is an important example, since from the central limit theorem in \cref{eq:central-limit-thm} it is known that in the asymptotic limit, all parameters become normally distributed~\cite{Cowan1998}. 

The above calculation proves useful in two cases: 
\begin{itemize}
    \item \textbf{Wilks' theorem}: For one \ac{POI}, in the large sample limit where the Wald approximation of \cref{eq:wald-test-statistic} holds and $\hat \mu$ is normally distributed, the test statistic $t$ is distributed as a $\chisq$-distribution with 1 \ac{d.o.f.}~\cite{wilks}.
    \item \textbf{Goodness-of-fit test}: In this case, the asymptotic distribution of the test statistic is a $\chisq$-distribution with $N-m$ \ac{d.o.f.}~\cite{Cowan1998}.
\end{itemize}

\section{Convergence of random variables}
\label{sec:convergence-random-variables}
In the context of parameter estimation and statistical inference, two foundational results in probability theory are particularly important: the \textit{Law of large numbers} and the \textit{Central limit theorem}. These theorems provide insight into the behaviour of sample-based estimators in the asymptotic limit of large sample sizes. They are key for proving fundamental properties of \textit{maximum likelihood estimators} (see \cref{sec:method-of-maximum-likelihood}).

\paragraph{Law of large numbers} Consider a sample $\data = (x_1, \ldots, x_N)$ drawn from a \ac{PDF} $p(x)$. The law of large numbers states that for any function $a(x)$ with finite variance $V[a(x)]$, the sample average converges to the expectation value in the limit of large $N$~\cite{James:2006zz},
\begin{equation}
    \lim_{N\to \infty} \frac{1}{N} \sum_{i=1}^{N} a(x_i) = E[a(x)] = \int dx \ a(x) p(x).
    \label{eq:law-of-large-numbers}
\end{equation}

\paragraph{Central limit theorem} This theorem states that for a sample $\data = (x_1, \ldots, x_N)$ of independent and identically distributed random variables with finite mean $\mu = E[x_i]$ and variance $\sigma^2 = V[x_i]$, $\frac{\sqrt{N}}{\sigma}(\bar{x} - \mu)$ converges in distribution to the standard normal distribution $\norm(0, 1)$ as $N$ becomes large,
\begin{equation}
    \frac{\sqrt{N}}{\sigma}(\bar{x} - \mu) \to^d \norm(0, 1)
    \label{eq:central-limit-thm}
\end{equation}
where $\bar{x} = \frac{1}{N} \sum_{i=1}^{N} x_i$ is the sample mean~\cite{Cowan1998,Young_Smith_2005}. This theorem forms the basis for the use of normal approximations in hypothesis testing and confidence interval construction. 

The proof of the central limit theorem can be shown through the use of characteristic functions. The characteristic function $\phi_x(k)$ of a \ac{PDF} $p(x)$ is the expectation value of $e^{ikx}$, which is the Fourier transform of the \ac{PDF}~\cite{James:2006zz,Cowan1998},
\begin{equation}
    \phi_x(k) = E_x\left[e^{ikx}\right] = \int dx ~ e^{ikx} p(x).
    \label{eq:characteristic-fn}
\end{equation}

Consider the definition of the variable
\begin{equation}
Z_N = \frac{\bar{x}-\mu}{\sigma/\sqrt{N}} = \sum_{i=1}^{N} \frac{x_i-\mu}{\sqrt{N}\sigma} = \sum_{i=1}^{N} \frac{1}{\sqrt{N}} Y_i,  
\end{equation}
with $Y_i = (x_i-\mu)/\sigma$. The expectation value of $Y_i$ is $E_x[Y_i]=0$ by the law of large numbers (\cref{eq:law-of-large-numbers}), and $E_x[Y_i^2] = 1$. The proof proceeds by showing that the characteristic function of $Z_N$ is equivalent to the one of a standard normal, with mean 0 and variance 1, in the limit of $N \to \infty$.

The characteristic function of $Z_N$ is
\begin{equation}
    \phi_{Z_N}(k) = \prod_{i=1}^{N} \phi_{Y_i/\sqrt{N}}(k)= \prod_{i=1}^{N} \phi_{Y_i}\left(\frac{k}{\sqrt{N}}\right),
\end{equation}
using the property $\phi_{ax}(k) = \phi_x(ak)$ for some constant $a$.
Since all $x_i$ are equally distributed,
\begin{equation}
    \phi_{Z_N}(k) = \left(\phi_{Y_i}\left(\frac{k}{\sqrt{N}}\right)\right)^N.
\end{equation}

The characteristic function $\phi_{Y_i}(k/\sqrt{N})$ can be expanded for large $N$ as
\begin{equation}
    \phi_{Y_i}\left(\frac{k}{\sqrt{N}}\right) = \sum_{m=0}^{\infty} \frac{k^m}{m!} \frac{\partial^m \phi_{Y_i}(k/\sqrt{N}) }{\partial (k/\sqrt{N})^m} \bigg|_{k/\sqrt{N}=0} = 1 - \frac{k^2}{2N} + o\left(\frac{k^2}{N}\right),
\end{equation}
where $o(k^2/N)$ refers to all terms which approach zero faster than $k^2/N$ as $N \to \infty$.

Neglecting all higher order terms, this gives
\begin{equation}
    \lim_{N \to \infty} \phi_{Z_N}(k) = \lim_{N \to \infty}\left(1 - \frac{k^2}{2N} \right) = e^{-\frac{k^2}{2}},
\end{equation}
which is the characteristic function of a standard normal distribution~\cite{James:2006zz,Cowan1998}.


\section{Posterior validation}
\label{sec:posterior-validation}
The reliability of Bayesian posterior inference critically depends on the convergence and mixing properties of the \ac{MCMC} sampling. The quality of \ac{MCMC} chains is validated following established best practices~\cite{Vehtari_2021,gelman2013bayesian, roy2019convergence}.

\subsection{Potential scale reduction factor}
The potential scale reduction factor $\hat{R}$ quantifies convergence by comparing between-chain and within-chain variance~\cite{gelman2013bayesian}. The improved split-$\hat{R}$ diagnostic~\cite{Vehtari_2021} is used, which splits each chain in half to increase sensitivity to non-stationarity. For parameter $\theta$, split-$\hat{R}$ is defined as:
\begin{equation}
\hat{R} = \sqrt{\frac{\widehat{\text{Var}}^+(\theta|\data)}{W}}
\end{equation}
where $\widehat{\text{Var}}^+(\theta|\data) = \frac{n-1}{n}W + \frac{1}{n}B$ combines within-chain variance $W$ and between-chain variance $B$ and $n$ is the number of draws per chain. Values $\hat{R} < 1.01$ indicate excellent convergence, while $\hat{R} \geq 1.1$ suggests insufficient sampling~\cite{Vehtari_2021}.

\subsection{Effective sample size}
The effective sample size (ESS) accounts for autocorrelation within chains, quantifying how many independent samples the correlated \ac{MCMC} output is equivalent to. Both bulk ESS (for central quantiles) and tail ESS (for extreme quantiles) are computed~\cite{Vehtari_2021}. Following recommendations, ESS $> 400$ is required for reliable inference, with ESS $> 1000$ preferred for high precision~\cite{Vehtari_2021}.

\subsection{Visual diagnostics}
\paragraph{Trace plots} display parameter evolution across iterations, providing assessment of chain behaviour. Well-mixed chains exhibit rapid fluctuation around a stable mean without systematic trends or chain separation~\cite{roy2019convergence}.

\paragraph{Rank plots} offer sensitivity for detecting mixing problems. For each parameter, all samples across chains are ranked, then histograms of ranks per chain are plotted. Uniform rank distributions indicate proper mixing, while systematic patterns reveal convergence issues~\cite{Vehtari_2021}.

%% file: chapters/appendix-method.tex
\section{Individual contributions}
\label{app:method-contributions}

The work presented in \cref{sec:method} is a joint effort of Lorenz Gärtner (L. G.), Nikolai Krug (N. K.; former Nikolai Hartmann), Lukas Heinrich (L. H.), Malin Horstmann (M. H.), Thomas Kuhr (T. K.), M\'{e}ril Reboud (M. R.), Slavomira Stefkova (S. S.), and Danny van Dyk (D. v. D.).

The initial idea for the method was developed by L. G. as an extension of the ``brief idea'' proposed by L. H. in Reference~\cite{cranmer_2017_1013926}.

The idea for the toy study was developed by L. G. and S. S., based on the plan to apply it to the \BKnn analysis at Belle~II~\cite{Belle-II:2023esi}.

The implementation of the method was done by L. G. with the guidance of N. K., while L. H. provided support with the custom modifier implementation. The code is available at this \href{https://github.com/lorenzennio/redist}{repository}~\cite{redist_v1.0.4}, and contributions can be traced back through the commit history. In the \texttt{examples} folder, one can find the study presented in this work.

Theoretical input is provided by M. R. and D. v. D., who also developed the \EOS package~\cite{EOSAuthors:2021xpv,EOS:v1.0.11} used to obtain the theoretical predictions. The implementation of \cref{eq:width} in \EOS was done by L. G. with the help of M. R. and D. v. D.

The statistical inference is performed using \pyhf~\cite{pyhf,Heinrich:2021gyp}, which was developed by L. H. The Bayesian \pyhf package~\cite{Feickert:2023hhr} used for the Bayesian inference in this work was developed by Matthew Feickert, L. H., and M. H.

T. K. provided general guidance and support throughout the project.

\section{Optimizing the kinematic binning}
\label{app:kinematic-binning}
To obtain suggestions on the number of bins to use for the kinematic variable(s), one can follow the following procedure:

For a large set of models covering your parameter space as thoroughly as possible, compute the expected yields for a finer and finer binning in the kinematic \ac{d.o.f.} (a new joint number density and new weights need to be computed every time). At each step, compute the difference to the results of the previous step and stop when reaching a predefined convergence condition. The maximum number of bins over all the looped models should give a good estimate of the number of kinematic bins to use.

\section{A recipe for application of this reinterpretation method}
\label{app:recipe}
To assist with easy application of this reinterpretation method to any analysis, I provide a simple 4-step guide on what needs to be done to reinterpret a result from \ac{HEP}. This guide focuses on the discrete approach of \cref{sec:method paper reweighting-method-discrete}.
\begin{enumerate}
    \item \textit{Samples}. Gather your post-reconstruction samples and ensure that they contain information on all kinematic \ac{d.o.f.} as well as the reconstruction variable.
    \item \textit{Null joint number density}. From these samples, build the null joint number density by binning samples in bins of the reconstruction variable times the kinematic \ac{d.o.f.} (see \cref{app:kinematic-binning} on how to optimize the kinematic binning).
    \item \textit{Weights}. Identify your null prediction used for producing the original \ac{MC} samples. Choose your alternative theoretical prediction(s) and ensure that the support of the null distribution covers the full range of the alternative distribution (this can also be done by setting an upper bound on the weights). Compute the weights as the ratio of the bin-integrated alternative to the bin-integrated null distribution (as in \cref{eq:binned-weights}).
    \item \textit{Inference}. Either by making use of the code in~\cite{redist_v1.0.4} or by implementing \cref{eq:reweight_discrete}, compute the expected yields given the alternative prediction, making use of the joint number density and the computed weights. Using either \pyhf~\cite{pyhf,Heinrich:2021gyp} or alternative tools, statistical inference can be used to compute results for the alternative theory. 
\end{enumerate}

%% file: chapters/appendix-knunu-mall.tex
\section{Individual contributions}
\label{sec:wet-contributions}
The work presented in \cref{sec:reinterpretation knunu mall,sec:reinterpretation knunu wet} is a joint effort by the Belle~II collaboration. The main proponents of this work are Lorenz Gärtner (L. G.), Nikolai Krug (N. K.; former Nikolai Hartmann), Thomas Kuhr (T. K.), and  Slavomira Stefkova (S. S.).

The initial idea for the project was developed by T. K. and S. S. The details of the analysis were discussed and refined by L. G., with guidance from N. K., T. K. and S. S. 

The method was implemented by L. G., with guidance from N. K., T. K. and S. S.

Valuable input on the analysis and the corresponding publication~\cite{belle2pub88} was provided by all members of the Belle~II collaboration. Particular gratitude goes to Swagato Banerjee, Gaetano de Marino, Sasha Glazov, Eldar Ganiev, Enrico Graziani, Romulus Godang, Bob Kowalewski, James Libby, Elisa Manoni, Mikihiko Nakao, Leo Piilonen, Soeren Prell, Yoshi Sakai, Alan Schwartz, Wei Shan, Diego Tonelli, Phillip Urquijo, Roberta Volpe, and Bruce Yabsley, who provided detailed comments and were particularly active during the review process of this work.

\section{Final-state radiation}
\label{sec:fsr}
\begin{figure}
    \centering
    \includegraphics[width=0.8\linewidth]{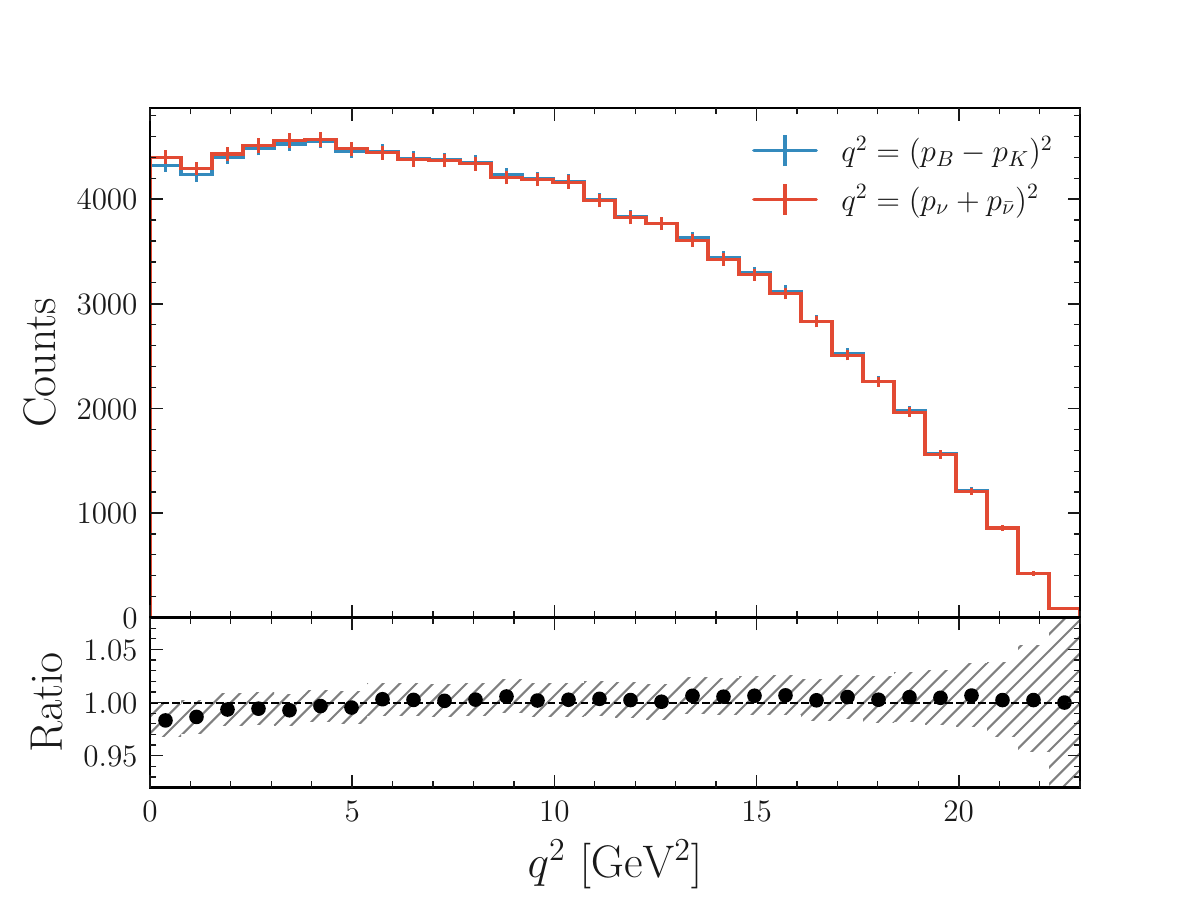}
    \caption{The generator level $q^2$ distributions for a set of 100\,000 simulated \ac{MC} samples, including final-state radiation.}
    \label{fig:q2dist}
\end{figure}

Final-state radiation potentially changes the $q^2$ spectrum, depending on how $q^2$ is defined. If no final-state radiation is present
\begin{equation}
    q^2 = (p_\nu +p_{\bar \nu})^2=(p_B-p_K)^2.
\end{equation}
Otherwise, the two definitions differ, $(p_\nu +p_{\bar \nu})^2 \neq (p_B-p_K)^2$, and consequently their distributions will change slightly. The comparison is shown in \cref{fig:q2dist}, where final-state radiation was included using \texttt{PHOTOS}~\cite{Barberio:1993qi}. Differences in the distributions are at most around 2\% in the lowest $q^2$ bins. The higher $q^2$ bins are less affected by final-state radiation.

As a comparison, the form factor uncertainties constitute an uncertainty of around 6\% in the low $q^2$ region. 
Given the small differences in the distributions, depending on the two definitions, I neglect this effect in the reinterpretation study.

\section{Kinematic $q^2$ binning}
\label{sec:q2 binning}
When creating the number densities, the binning of the kinematic variables is an important choice. To enable reinterpretation studies with narrow resonances, I consider relatively fine binnings and test options of 50, 100, and 250 bins in the kinematic range $q^2 \in [0, 22.9]~\text{GeV}^2$, corresponding to bin widths of $0.45$, $0.23$, and $0.09~\text{GeV}^2$, respectively. The resulting best-fit values of the \acp{POI} are summarised in \cref{tab:q2 binning}. For all three binning choices, the best-fit values agree within uncertainties. The uncertainty in $C_\mathrm{SL}+C_\mathrm{SR}$ is difficult to estimate due to limited sensitivity and is therefore expected to fluctuate. Since potential use cases may require rather fine binning, I adopt 100 bins for this reinterpretation study.

\begin{table}[ht]
    \renewcommand{\arraystretch}{1.5}
    \caption{Best-fit values for the \ac{WET} and \BKX reinterpretation, under different $q^2$ binning scenarios.}
    \centering
    \begin{tabularx}{\linewidth}{@{\extracolsep{\fill}}YlYYY}
        \toprule \midrule
        \textbf{Model} & \textbf{Parameters} & \textbf{50 $q^2$ bins} & \textbf{100 $q^2$ bins} & \textbf{250 $q^2$ bins} \\
        \midrule
         \ac{WET} & $
         \begin{aligned}
            &C_\mathrm{VL}+C_\mathrm{VR}\\
            &C_\mathrm{SL}+C_\mathrm{SR}\\
            &C_\mathrm{TL}
        \end{aligned}
        $ & $
        \begin{aligned}
            11.4 &\pm 3.2\\
            0.1  &\pm 15.8 \\
            8.4  &\pm 3.0
        \end{aligned}
        $ & $
         \begin{aligned}
            11.4 &\pm 3.2  \\
            0.1  &\pm 13.7 \\
            8.4  &\pm 3.0
        \end{aligned}
        $ & $
        \begin{aligned}
            11.4 &\pm 3.2  \\
            0.2  &\pm 10.0 \\
            8.4  &\pm 3.0
        \end{aligned}
        $
        \\
        \midrule
         \makecell{\BKX \\ $\Gamma_X=0.1$ GeV}& 
         $
         \begin{aligned}
            &\mu_X\\
            &m_X ~ [\text{GeV}]\\
        \end{aligned}
        $ & $
        \begin{aligned}
            9.1 &\pm 2.6\\
            2.1 &\pm 0.1\\
        \end{aligned}
        $ & $
         \begin{aligned}
            9.0 &\pm 2.6\\
            2.1 &\pm 0.1\\
        \end{aligned}
        $ & $
         \begin{aligned}
            9.0 &\pm 2.6\\
            2.1 &\pm 0.1\\
        \end{aligned}
        $
         \\
        \midrule \bottomrule
    \end{tabularx}
    \label{tab:q2 binning}
\end{table}

\section{Impact of removing unphysical $q^2$ values in the hadronic tagging analysis}
\label{sec:unphysical q2}
In \cref{sec:wet paper reinterpretation method} I discussed that 0.16\% of events in the \ac{HTA} analysis have an unphysical generator-level $q^2$ assigned. In \cref{tab:rm q2 best-fit} I summarise the best-fit values of the \acp{POI} with and without these unphysical samples. Since the impact on the best-fit point is negligible, I decide to keep the unphysical samples in the analysis without reweighting them in the reinterpretation process.

\begin{table}[h]
    \caption{Best-fit values for a model with all samples and one with the unphysical $q^2$ events removed.}
    \centering
    \begin{tabularx}{\linewidth}{@{\extracolsep{\fill}}YlYY}
        \toprule \midrule
        \textbf{Model} & \textbf{Parameters} & \textbf{All samples} & \textbf{Unphys. $q^2$ removed} \\
        \midrule
         \ac{SM} &
         $\mu$
         &
         $4.6 \pm 1.3$
         &
         $4.6 \pm 1.3$
         \\
        \midrule
         \ac{WET} & $
         \begin{aligned}
            &C_\mathrm{VL}+C_\mathrm{VR}\\
            &C_\mathrm{SL}+C_\mathrm{SR}\\
            &C_\mathrm{TL}
        \end{aligned}
        $ & $
         \begin{aligned}
            11.4 &\pm 3.2  \\
            0.1  &\pm 13.7 \\
            8.4  &\pm 3.0 
        \end{aligned}
        $ & $
        \begin{aligned}
            11.4 &\pm 3.2  \\
            0.1  &\pm 16.9 \\
            8.4  &\pm 3.0
        \end{aligned}
        $
        \\
        \midrule
         \makecell{\BKX \\ $\Gamma_X=0.1$ GeV}& 
         $
         \begin{aligned}
            &\mu_X\\
            &m_X~[\text{GeV}]\\
        \end{aligned}
        $ & $
         \begin{aligned}
            9.0 &\pm 2.6\\
            2.1 &\pm 0.1\\
        \end{aligned}
        $ & $
         \begin{aligned}
            9.0 &\pm 2.6\\
            2.1 &\pm 0.1\\
        \end{aligned}
        $
         \\
        \midrule \bottomrule
    \end{tabularx}
    \label{tab:rm q2 best-fit}
\end{table}

%% file: chapters/appendix-knunu-wet.tex
\section{Belle~II HEPData reinterpretation inventory}
\label{sec:BelleIIdata}

To enable reinterpretation under any new physics model with the model-agnostic likelihood~\cite{Gartner:2024muk}, the necessary information from Belle~II will be published on \href{https://www.hepdata.net/}{HEPData}~\cite{hepdata.166082,hepdata.146803}. The release will include the following components:

\begin{itemize}
\item The \ac{SM} \BKnn differential branching ratio as a function of $q^{2}$;
\item Signal selection efficiency as a function of $q^{2}$;
\item Binned joint number densities:
\begin{itemize}
    \item \ac{ITA}: x-axis: $q^{2}$, y-axis: $q^{2}_{\rm{rec}} \times \eta(\rm{BDT}_{2})$ (flattened), z-axis: events (weighted);
    \item \ac{HTA}: x-axis: $q^{2}$, y-axis: $\eta(\rm{BDTh})$ (flattened), z-axis: events (weighted);
\end{itemize}  
\item \texttt{pyhf} combined likelihood in JSON format:
\begin{itemize}
\item Containing templates for signal and background after all selections, binned in $q^{2}_{\rm{rec}} \times \eta(\rm{BDT}_{2})$ (\ac{ITA}) and $\eta(\rm{BDTh})$ (\ac{HTA});
\end{itemize}  
\item The code to reproduce the \ac{WET} reinterpretation results obtained in this analysis.
\end{itemize}

\section{Weak Effective Theory posterior validation}
\label{sec:wet-posterior-validation-results}
To validate the \ac{MCMC} sampling, I apply the diagnostics described in \cref{sec:posterior-validation} to the \ac{WET} posterior samples discussed in \cref{sec:wet paper results} (\cref{fig:wet-posterior}).
I find $\hat{R} = 1.000$ for all parameters, indicating complete convergence across the chains, with bulk ESS $> 2400$ and tail ESS $> 3100$ for all parameters, ensuring high-precision inference.

\begin{figure}
    \centering
    \includegraphics[width=\textwidth]{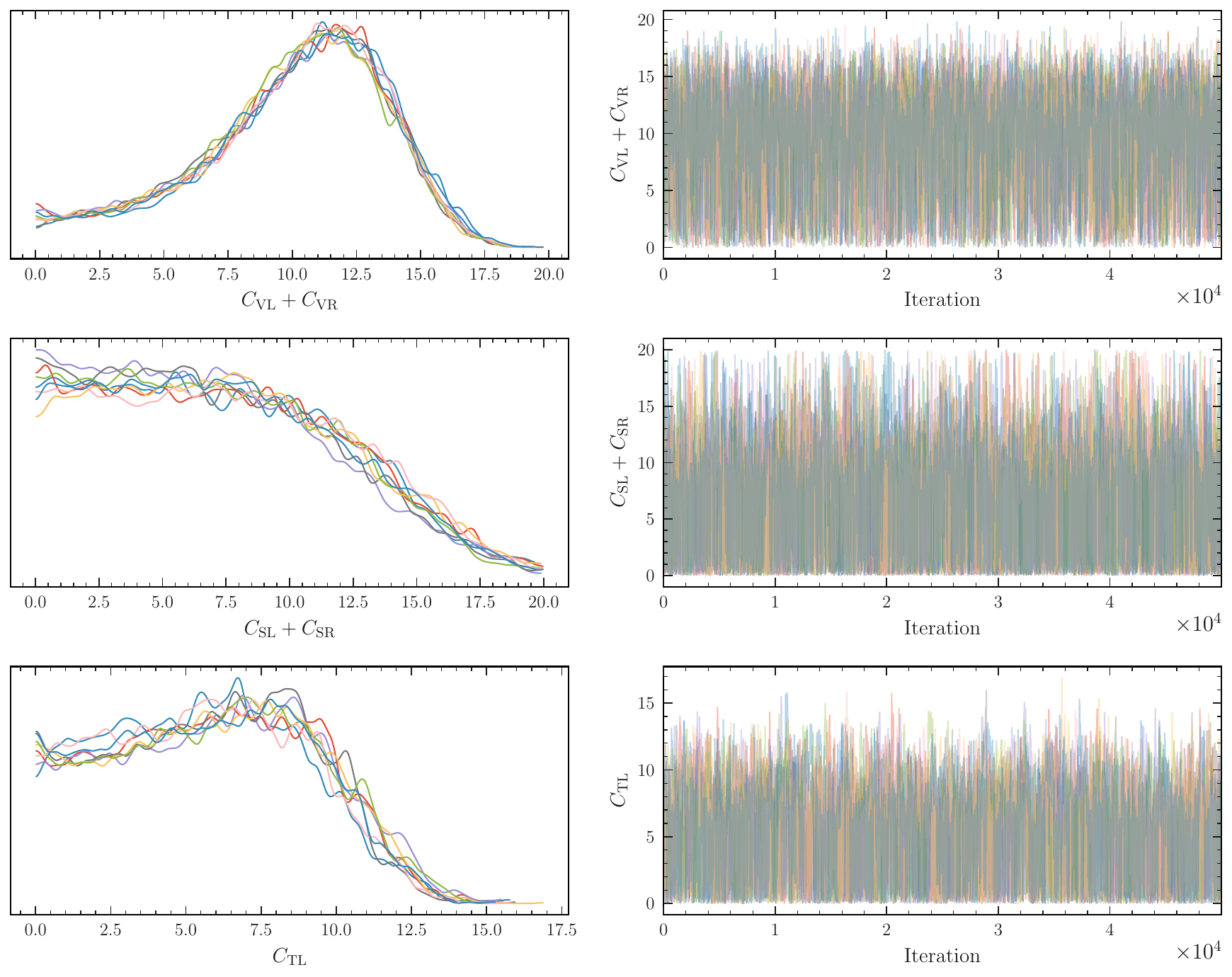}
    \caption{Chain-wise marginal posterior distributions (\textit{left}) and trace plots (\textit{right}) for the \ac{WET} Wilson coefficients in \cref{eq:width}, from the posterior discussed in \cref{sec:wet paper results} (\cref{fig:wet-posterior}). Each colour represents a different \ac{MCMC} chain. No trends or chain separation are visible.}
    \label{fig:trace-plots-wet} 
\end{figure}

\begin{figure}
    \centering
    \includegraphics[width=\textwidth]{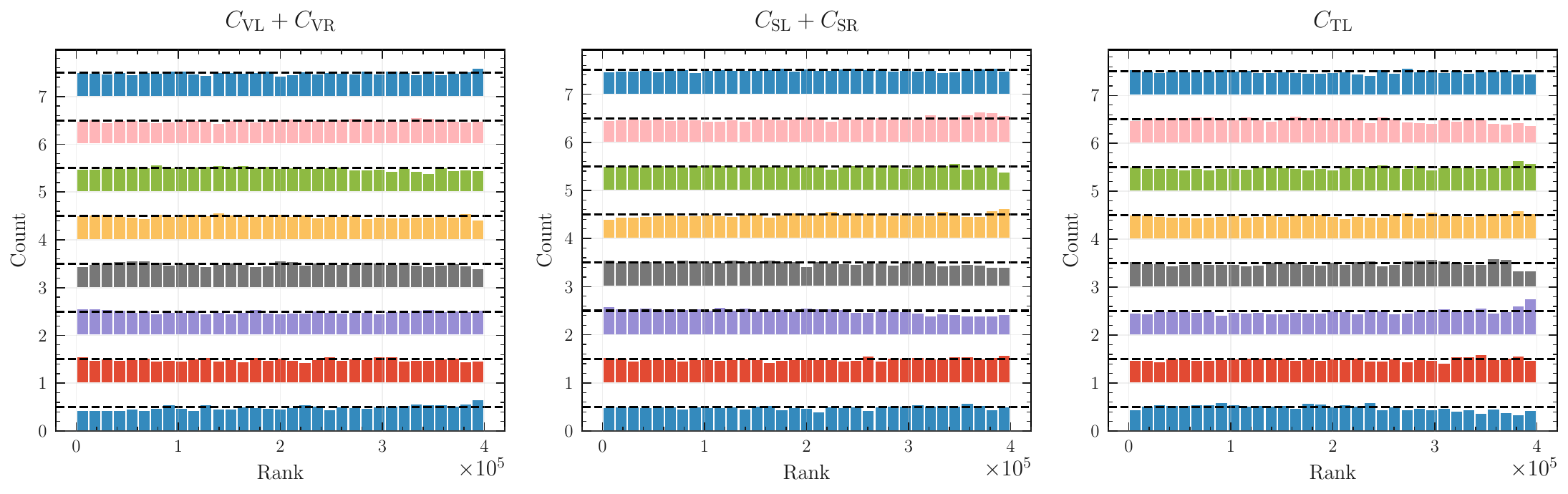}
    \caption{Rank plots for the \ac{WET} Wilson coefficients in \cref{eq:width}, from the posterior discussed in \cref{sec:wet paper results} (\cref{fig:wet-posterior}). Each colour represents a different \ac{MCMC} chain. The uniform distributions indicate good mixing.}
    \label{fig:rank-plots-wet}
\end{figure}

Trace plots (see \cref{fig:trace-plots-wet}) show ideal ``hairy caterpillar'' behaviour with rapid mixing and no visible trends. Rank plots (see \cref{fig:rank-plots-wet}) display uniform distributions across all chains, confirming good mixing. 

These results provide confidence that the posterior constraints presented in \cref{tab:wet results} are statistically robust and suitable for physics interpretation. The diagnostic validation ensures that reported credible intervals accurately reflect posterior uncertainty and that no additional sampling is required.


\section{Weak Effective Theory posteriors for the inclusive and hadronic tagging analyses}
\label{sec:ita and hta posteriors}
In \cref{sec:reinterpretation knunu wet}, I presented the marginal posterior of the \ac{WET} Wilson coefficients based on the combined \ac{ITA} and \ac{HTA} likelihood. Here, I present the marginal posteriors for the \ac{ITA} and \ac{HTA} analyses separately.

I adopt the same approach as in \cref{sec:wet paper results} but now only use the \ac{ITA} or \ac{HTA} likelihood, respectively. Priors are chosen to be uniform for all Wilson coefficients with support $[0, 20]$. I also use the same symmetrization technique as described in \cref{sec:wet paper results}. 

\subsection{Marginal posterior for the inclusive tagging analysis}
\label{sec:wet-ita}
\begin{figure}
    \centering
    \includegraphics[width=0.8\textwidth]{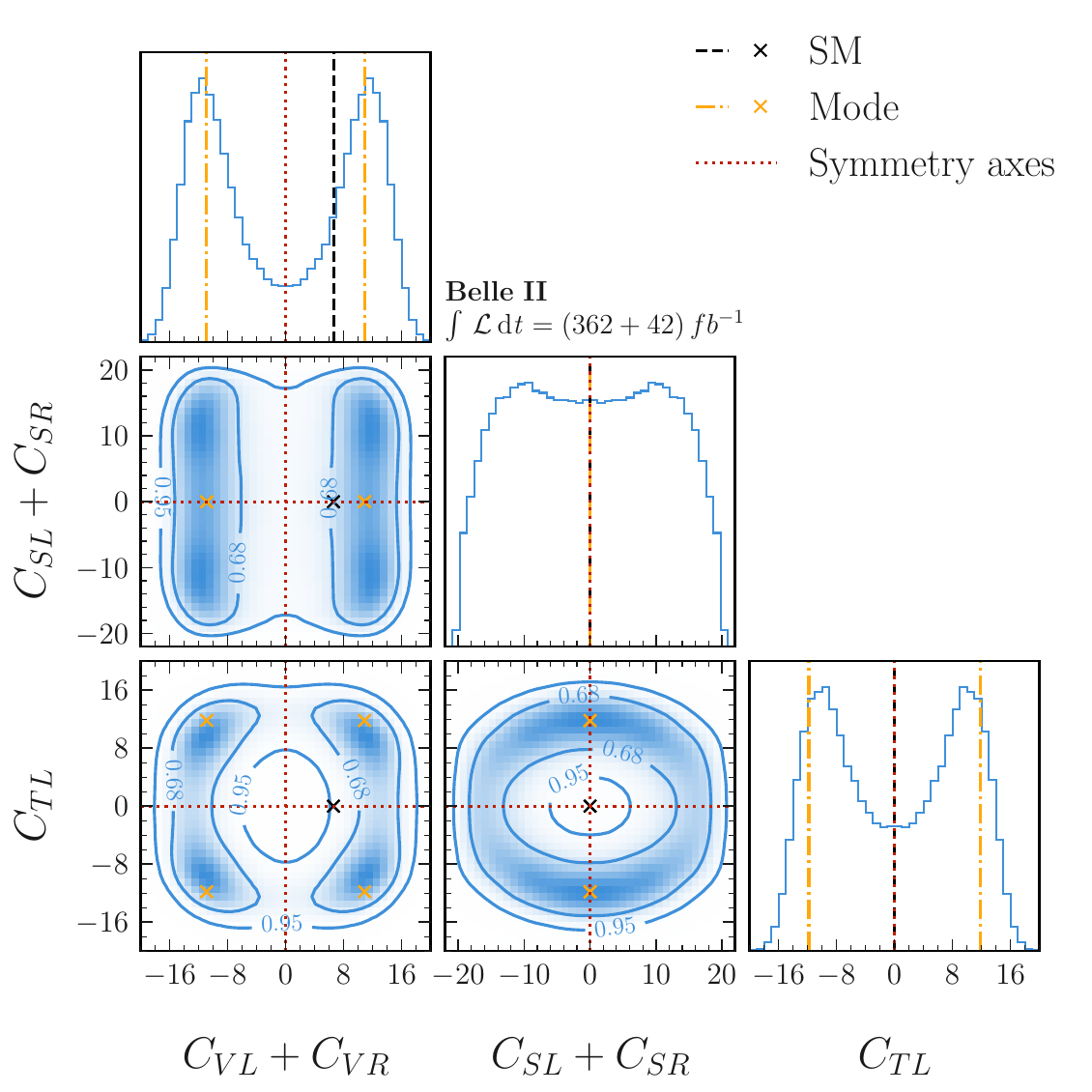}
    \caption{
        The marginalized posterior for the Wilson coefficients in \cref{eq:width}, based on the \ac{ITA} likelihood. This analysis adopts the convention that $C_{\mathrm{VL}}+C_{\mathrm{VR}}$, $C_{\mathrm{SL}}+C_{\mathrm{SR}}$, and $C_{\mathrm{TL}}$ are real-valued.
        Diagonal and off-diagonal panels show the 1-dimensional and 2-dimensional sample density \acp{PDF} on a linear scale, respectively. The overall scale is omitted, as all relevant information is contained in the shape of the distribution. The contours indicate $68\%$ and $95\%$ credible intervals. The dashed black lines and cross mark the \ac{SM} point; the dash-dotted yellow lines and cross indicate the posterior mode; dotted red lines mark the symmetry axes used for sample symmetrization.
        }
    \label{fig:ita-posterior}
\end{figure}

The 1- and 2-dimensional marginalizations of the posterior based on the \ac{ITA} likelihood and the resulting credible intervals at $68\%$ and $95\%$ probability are shown in \cref{fig:ita-posterior}. Posterior mode and credible intervals for the absolute values of the Wilson coefficients are shown in \cref{tab:wet results ita}. 

\begin{table}[ht]
    \renewcommand{\arraystretch}{1.5}
    \caption{The mode of the posterior, and \ac{HDI} at 68\% and 95\% for the (sums of the) \ac{WET} Wilson coefficients in \cref{eq:width}, derived from the posterior in \cref{fig:ita-posterior}.}
    \centering
    \begin{tabularx}{\linewidth}{@{\extracolsep{\fill}}lYYY}
        \toprule \midrule
        \textbf{Parameters} & \textbf{Mode} & \textbf{68\% HDI} & \textbf{95\% HDI} \\
        \midrule
         $
         \begin{aligned}
            &|C_\mathrm{VL}+C_\mathrm{VR}|\\
            &|C_\mathrm{SL}+C_\mathrm{SR}|\\
            &|C_\mathrm{TL}|
        \end{aligned}
        $ & $
        \begin{aligned}
            10.9&\\
            0.0 &\\
            11.8&
        \end{aligned}
        $ & $
         \begin{aligned}
            [7.0, ~14.7]&\\
            [1.9, ~14.3]& \\
            [4.9, ~13.6]&
        \end{aligned}
        $ & $
        \begin{aligned}
            &[1.0, ~16.3]&\\
            &[0.0, ~18.1]& \\
            &[0.0, ~14.5]&
        \end{aligned}
        $ \\
        \midrule \bottomrule
    \end{tabularx}
    \label{tab:wet results ita}
\end{table}

There is a clear deviation from the \ac{SM} prediction in the vector sector, as expected from the result of the original \ac{ITA} analysis, which reported a signal strength of $\mu = 5.4 \pm 1.5$~\cite{Belle-II:2023esi}. Since the Wilson coefficients enter quadratically into the differential branching ratio (see \cref{eq:width}), one expects the posterior of the dominant vector Wilson coefficient to peak approximately at
\begin{equation}
|C_\mathrm{VL}+C_\mathrm{VR}| \simeq \sqrt{5.4} \, C_\mathrm{VL}^{SM} \simeq 15.3,
\end{equation}
assuming negligible contributions from the other Wilson coefficients. The posterior indeed peaks below this value, with an additional non-zero contribution from the tensor sector.

Furthermore, the analysis finds that the posterior distribution peaks around a non-zero value for the tensor contribution. This indicates that a pure \ac{SM} signal template does not provide the best description of the data, and the combination with a different kinematic distribution is preferred.

\subsection{Marginal posterior for hadronic tagging analysis}
\label{sec:wet-hta}
\begin{figure}
    \centering
    \includegraphics[width=0.8\textwidth]{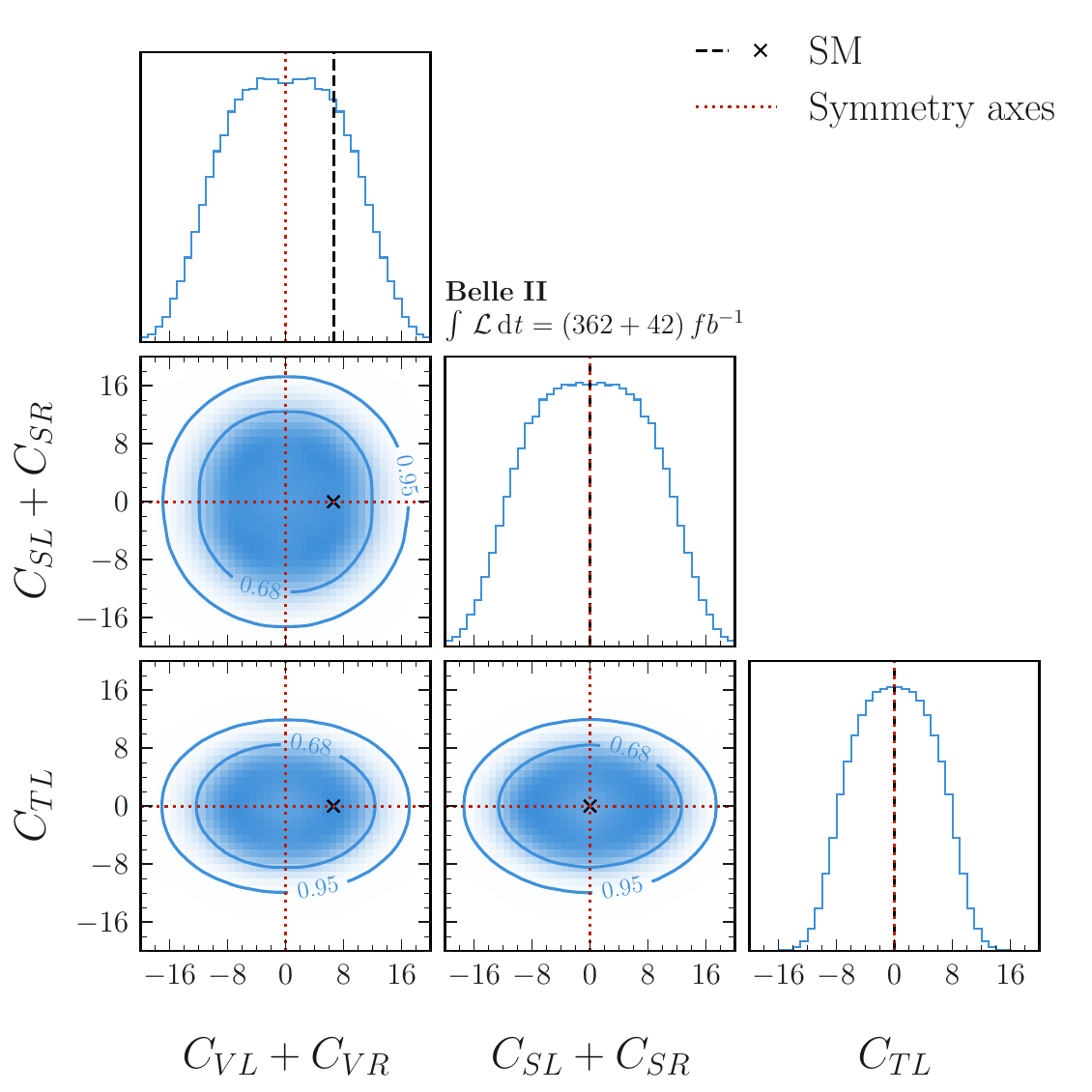}
    \caption{
        The marginalized posterior for the Wilson coefficients in \cref{eq:width}, based on the \ac{HTA} likelihood. This analysis adopts the convention that $C_{\mathrm{VL}}+C_{\mathrm{VR}}$, $C_{\mathrm{SL}}+C_{\mathrm{SR}}$, and $C_{\mathrm{TL}}$ are real-valued.
        Diagonal and off-diagonal panels show the 1-dimensional and 2-dimensional sample density \acp{PDF} on a linear scale, respectively. The overall scale is omitted, as all relevant information is contained in the shape of the distribution. The contours indicate $68\%$ and $95\%$ credible intervals. The dashed black lines and cross mark the \ac{SM} point; dotted red lines mark the symmetry axes used for sample symmetrization.
        }
    \label{fig:hta-posterior}
\end{figure}

The 1- and 2-dimensional marginalizations of the posterior based on the \ac{HTA} likelihood and the resulting credible intervals at $68\%$ and $95\%$ probability are shown in \cref{fig:hta-posterior}.
Credible intervals for the absolute values of the Wilson coefficients are shown in \cref{tab:wet results hta}. 

\begin{table}[ht]
    \renewcommand{\arraystretch}{1.5}
    \caption{The mode of the posterior, and \ac{HDI} at 68\% and 95\% for the (sums of the) \ac{WET} Wilson coefficients in \cref{eq:width}, derived from the posterior in \cref{fig:hta-posterior}.}
    \centering
    \begin{tabularx}{\linewidth}{@{\extracolsep{\fill}}lYY}
        \toprule \midrule
        \textbf{Parameters} & \textbf{68\% HDI} & \textbf{95\% HDI} \\
        \midrule
         $
         \begin{aligned}
            &|C_\mathrm{VL}+C_\mathrm{VR}|\\
            &|C_\mathrm{SL}+C_\mathrm{SR}|\\
            &|C_\mathrm{TL}|
        \end{aligned}
        $ & $
         \begin{aligned}
            [0.0, ~8.3]&\\
            [0.0, ~8.6]& \\
            [0.0, ~5.7]&
        \end{aligned}
        $ & $
        \begin{aligned}
            [0.0, ~14.1]&\\
            [0.0, ~14.3]& \\
            [0.0, ~9.7]&
        \end{aligned}
        $ \\
        \midrule \bottomrule
    \end{tabularx}
    \label{tab:wet results hta}
\end{table}

In this case, the posterior peaks at zero for all (sums of) Wilson coefficients. 
This can be attributed to the choice of the binning variable, $\eta(\mathrm{BDTh})$, which shows only a weak correlation with the kinematic variable $q^2$ (see \cref{sec:wet paper reinterpretation method}). 
Therefore, the posterior mode is not a sensible measure. 
Consequently, the \ac{HTA} analysis provides less constraining power on the individual Wilson coefficients than the \ac{ITA} analysis, where the binning variable $\eta(\BDT2)$ exhibits a stronger correlation with $q^2$ (see \cref{sec:wet paper reinterpretation method,fig:knunu-joint-nr-dens}).

Nonetheless, the \ac{HTA} likelihood provides complementary information to the \ac{ITA} likelihood, as it is sensitive to the overall normalization of the signal.

\section{Goodness-of-fit}
\label{sec:goodness-of-fit-wet}
Here, I report the goodness-of-fit tests for the \ac{SM} and background-only hypotheses. For details on the goodness-of-fit test, see \cref{sec:goodness-of-fit} and the paragraph around \cref{eq:gof-pvalue-wet}. Here, more details on the results reported in \cref{tab:bayes factors} are provided.

\begin{figure}
    \centering
    \includegraphics[width=0.6\textwidth]{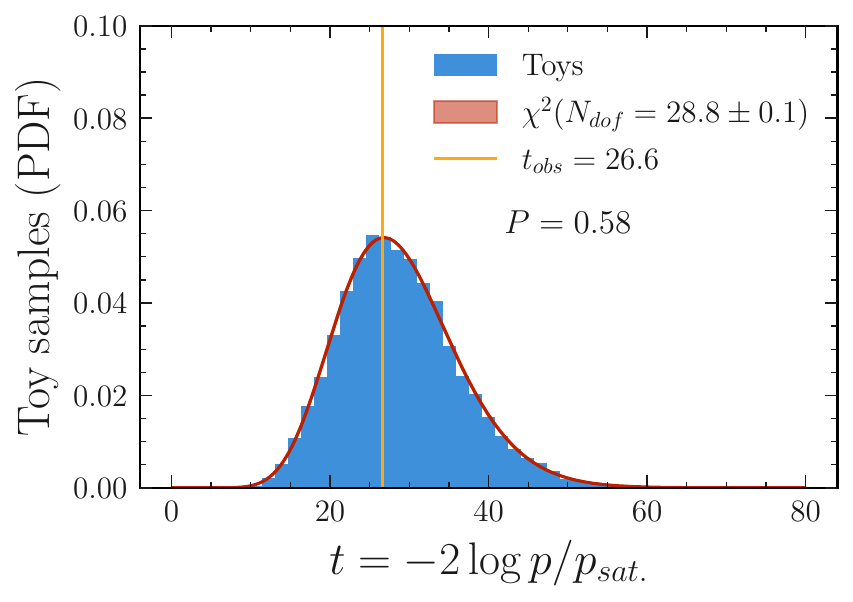}
    \includegraphics[width=0.6\textwidth]{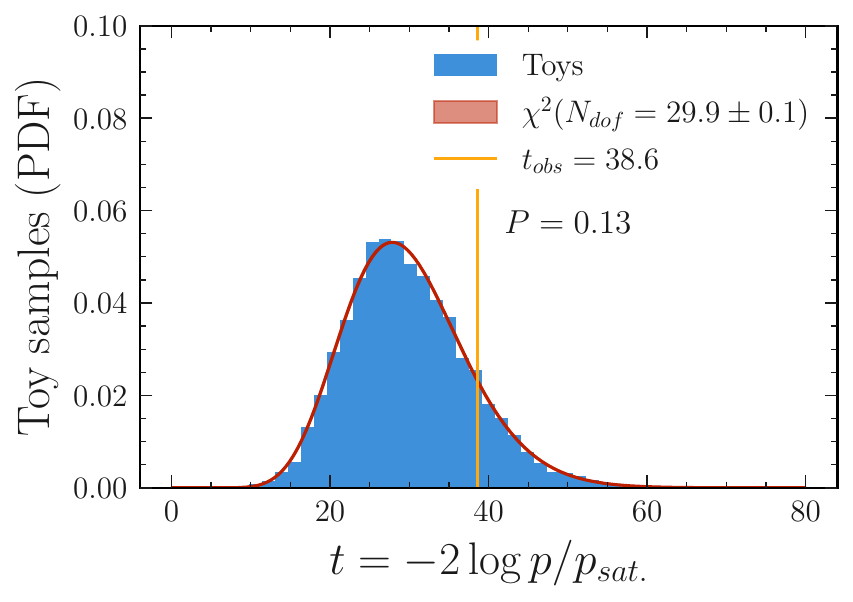}

    \caption{
        The distribution of the goodness-of-fit test statistic for the \ac{SM} (\textit{top}) and background-only (\textit{bottom}) hypotheses, each obtained from 10\,000 toys. The values obtained from a fit to the data are shown as vertical yellow lines.
        The asymptotic $\chisq$-distributions (red band) correspond to the best-fit $N_{dof}$, which are shown with the corresponding uncertainty band.
    }
    \label{fig:chi2-sm-bkg}
\end{figure}

The test statistic distributions are shown in \cref{fig:chi2-sm-bkg}. The expected $N_{\rm dof}$ is obtained when fitting to the toy test statistic values. For the \ac{SM} case, the expected $N_{\rm dof} = 29$, as there is only the signal strength as a free parameter and 30~analysis bins. For the background-only hypothesis, there are no free parameters, and hence the expected $N_{\rm dof}=30$.
The $P$-value for the \ac{SM} hypothesis, $P=0.58$, indicates a very good fit to the data. The $P$-value for the background-only hypothesis, $P=0.13$, indicates an acceptable fit to the data.

%% file: chapters/appendix-knunu-bkx.tex
\section{\texorpdfstring{\BKX}{B+->K+X} posterior validation}
\label{sec:bkx-posterior-validation-results}
To validate the \ac{MCMC} sampling, I apply the diagnostics described in \cref{sec:posterior-validation} to all \BKX posterior samples discussed in \cref{sec:bkx-results} (\cref{fig:bkx-posterior}).
For all three marginal posteriors, I find $\hat{R} = 1.000$ for all parameters, indicating complete convergence across the chains, with bulk ESS $> 5800$ and tail ESS $> 4200$ for all parameters, ensuring high-precision inference.

\begin{figure}
    \centering
    \includegraphics[width=\textwidth]{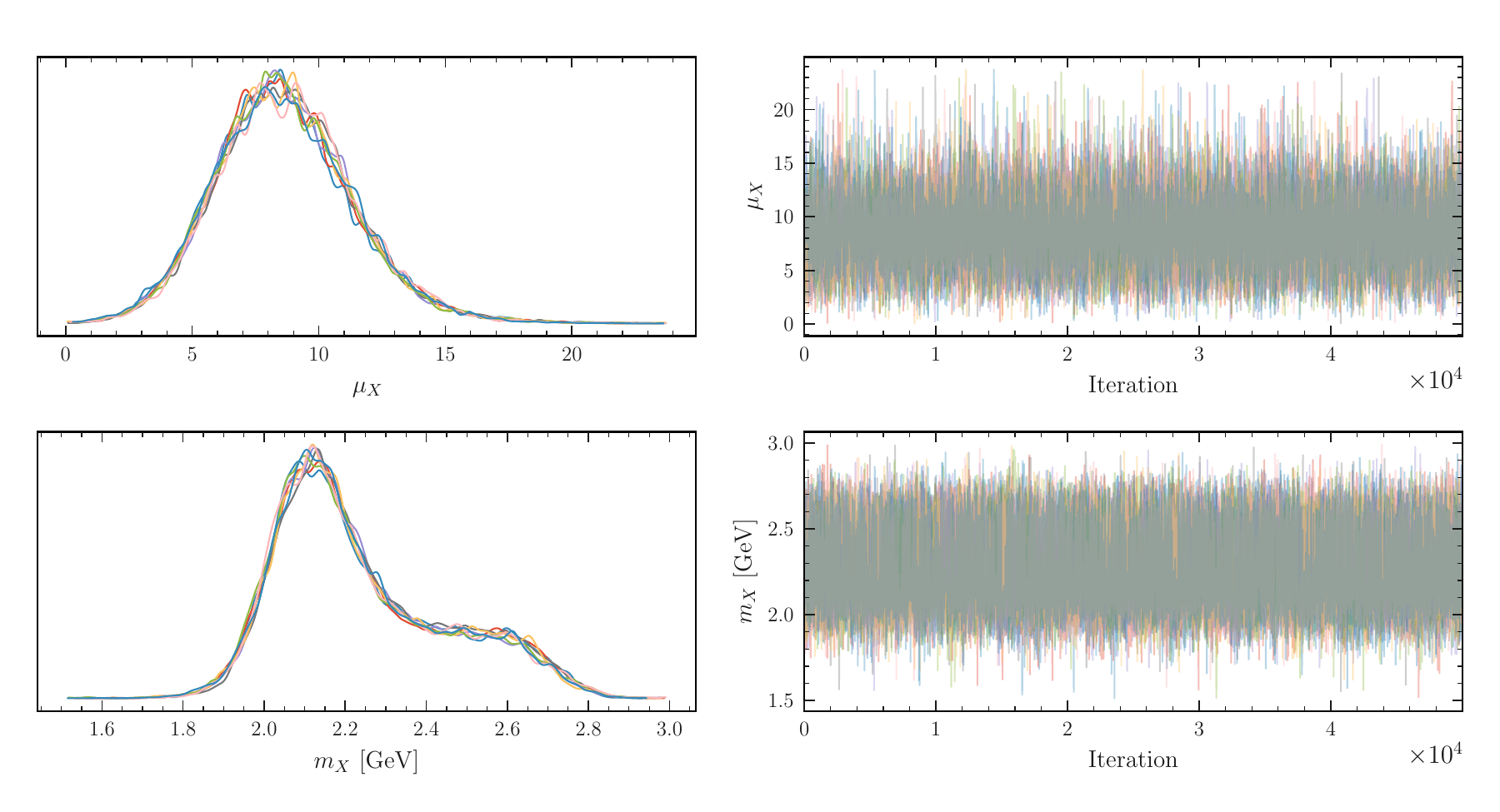}
    \caption{Chain-wise marginal posterior distributions (\textit{left}) and trace plots (\textit{right}) for the \BKX signal strength $\mu_X$ and mass $m_X$ in \cref{eq:bkx-width}, from the $\Gamma_X = 0.1~\text{GeV}$ posterior discussed in \cref{sec:bkx-results} (\cref{fig:bkx-posterior}). Each colour represents a different \ac{MCMC} chain. No trends or chain separation are visible.}
    \label{fig:trace-plots-bkx-01} 
\end{figure}

\begin{figure}
    \centering
    \includegraphics[width=\textwidth]{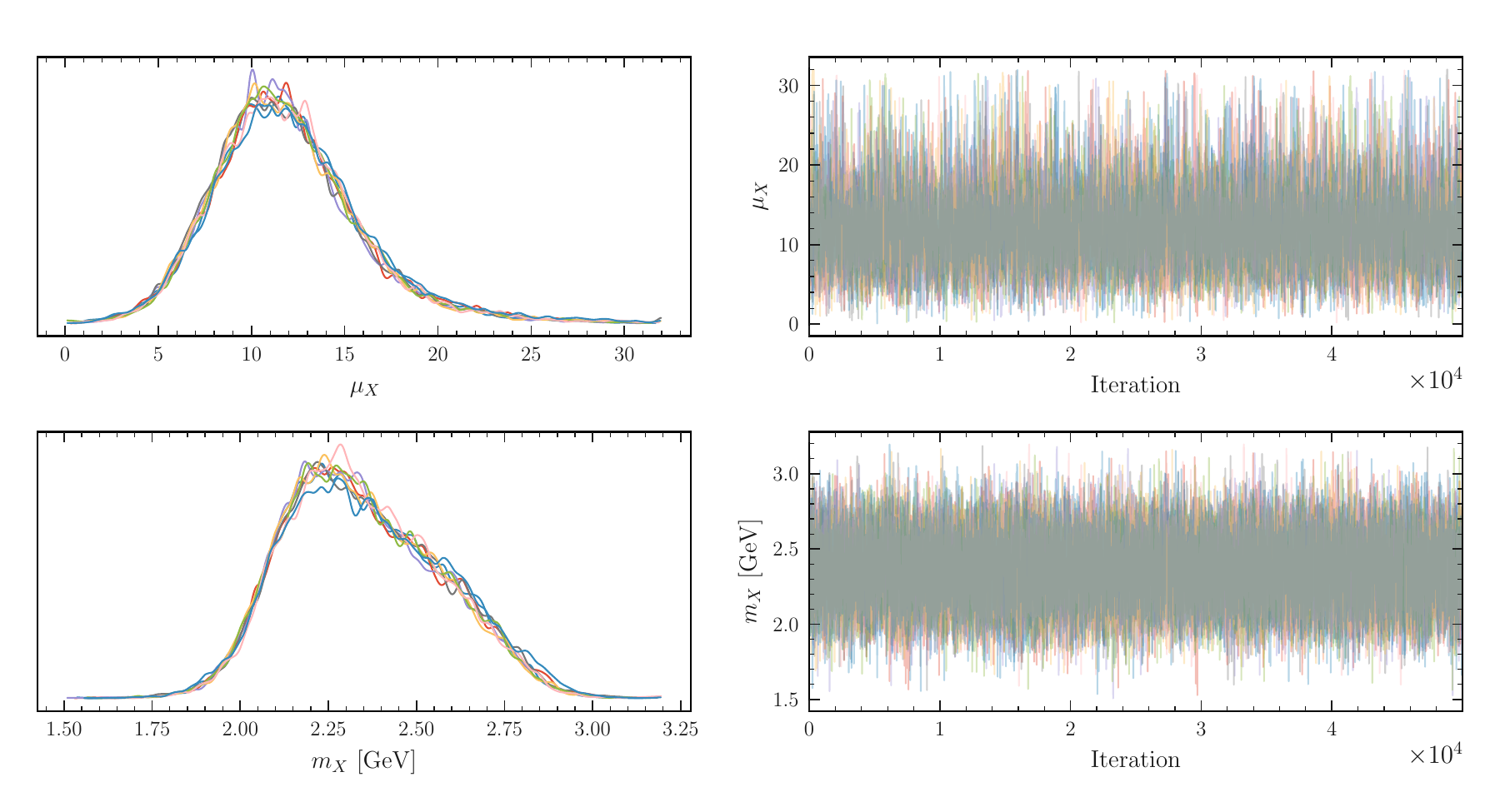}
    \caption{Chain-wise marginal posterior distributions (\textit{left}) and trace plots (\textit{right}) for the \BKX signal strength $\mu_X$ and mass $m_X$ in \cref{eq:bkx-width}, from the $\Gamma_X = 0.5~\text{GeV}$ posterior discussed in \cref{sec:bkx-results} (\cref{fig:bkx-posterior}). Each colour represents a different \ac{MCMC} chain. No trends or chain separation are visible.}
    \label{fig:trace-plots-bkx-05} 
\end{figure}

\begin{figure}
    \centering
    \includegraphics[width=\textwidth]{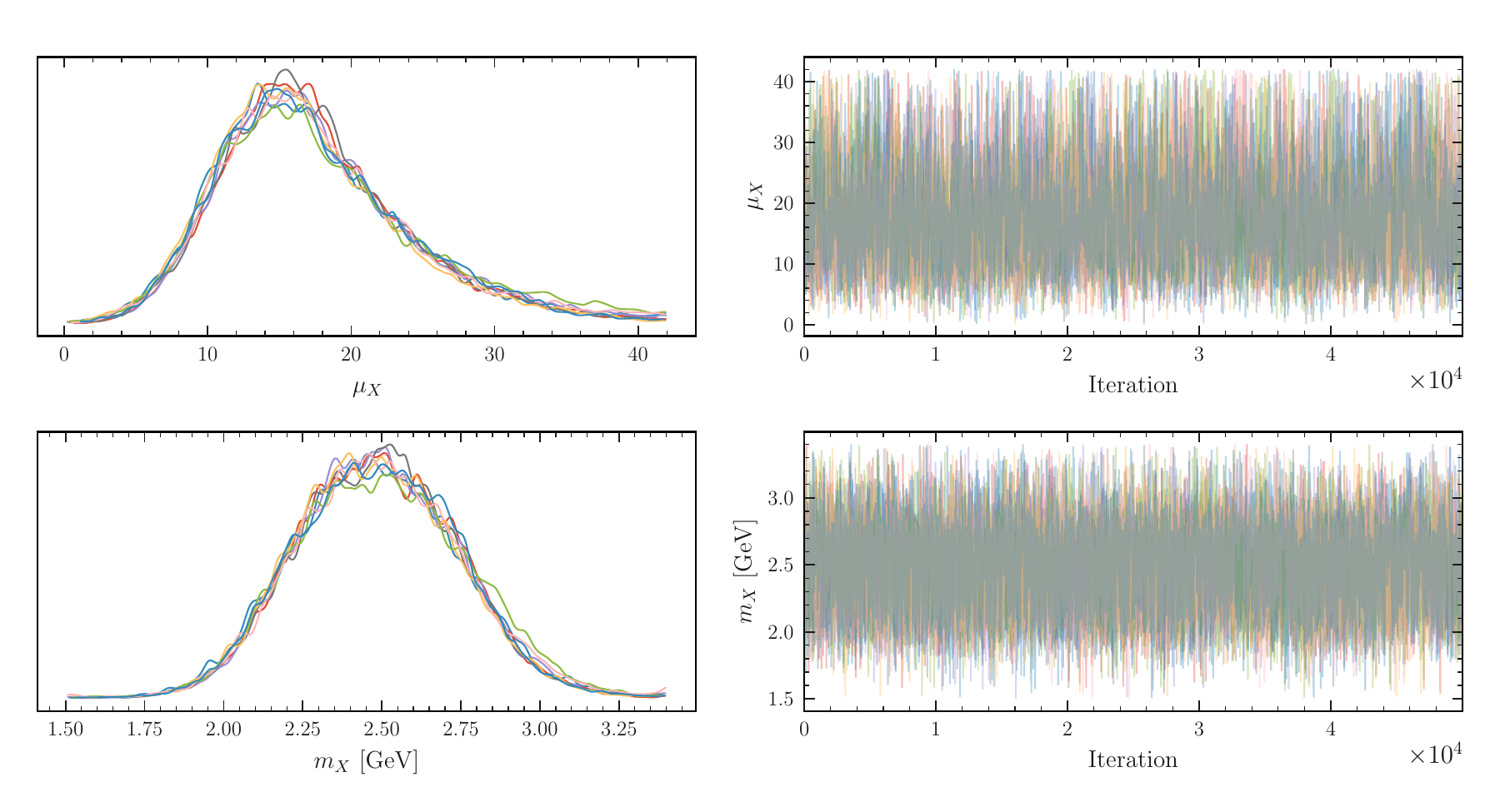}
    \caption{Chain-wise marginal posterior distributions (\textit{left}) and trace plots (\textit{right}) for the \BKX signal strength $\mu_X$ and mass $m_X$ in \cref{eq:bkx-width}, from the $\Gamma_X = 1.0~\text{GeV}$ posterior discussed in \cref{sec:bkx-results} (\cref{fig:bkx-posterior}). Each colour represents a different \ac{MCMC} chain. No trends or chain separation are visible.}
    \label{fig:trace-plots-bkx-10} 
\end{figure}

\begin{figure}
    \centering
    \includegraphics[width=\textwidth]{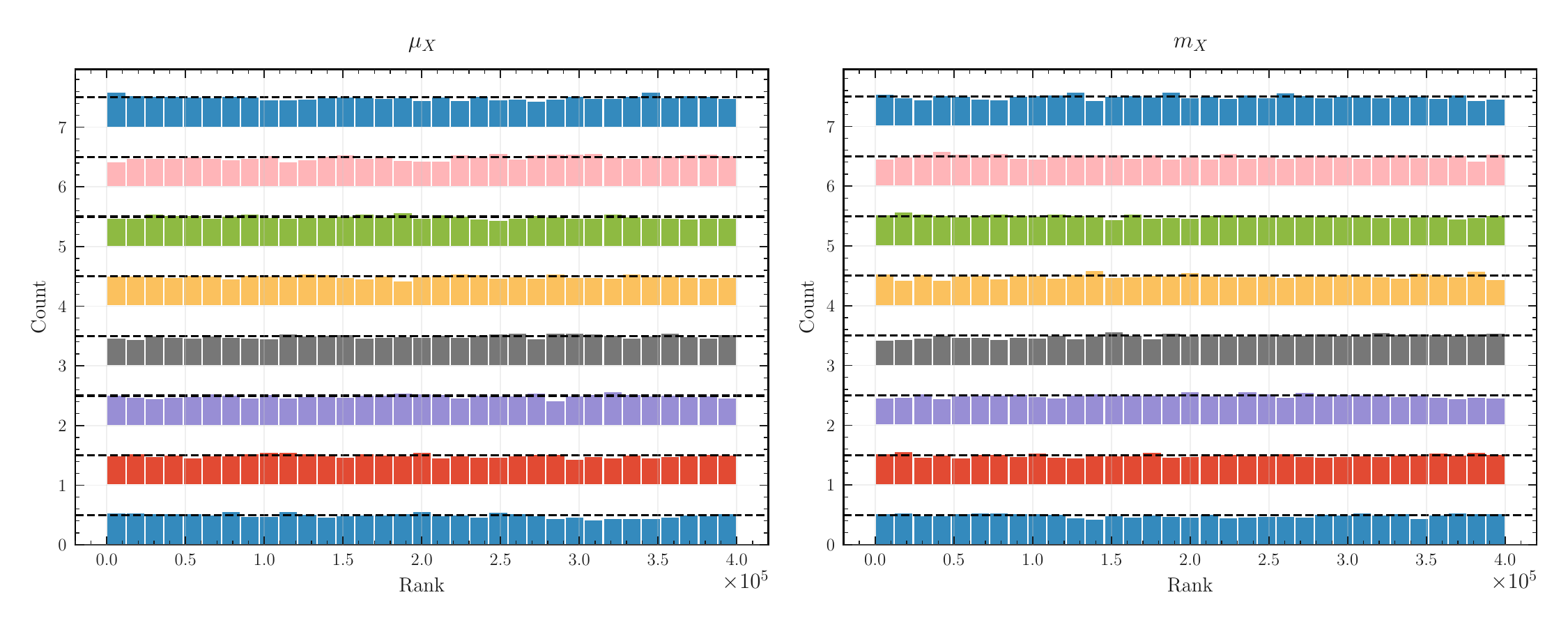}
    \includegraphics[width=\textwidth]{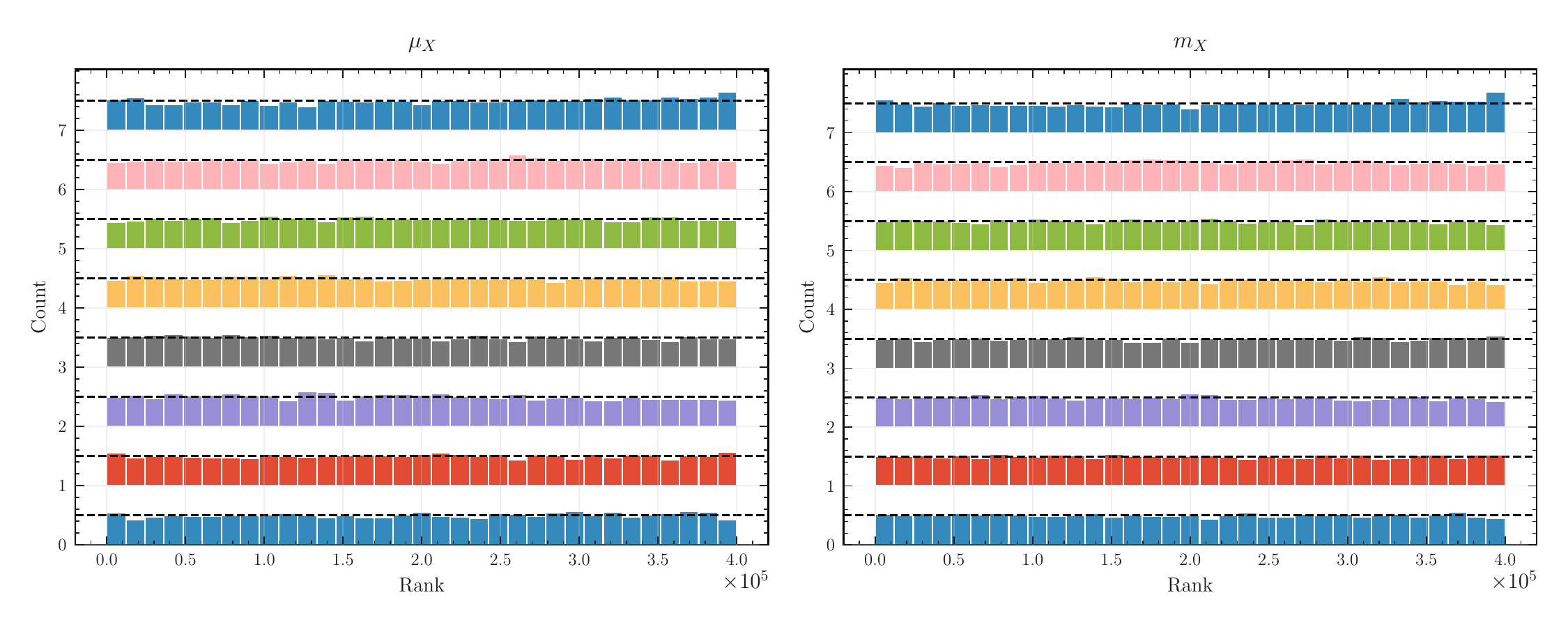}
    \includegraphics[width=\textwidth]{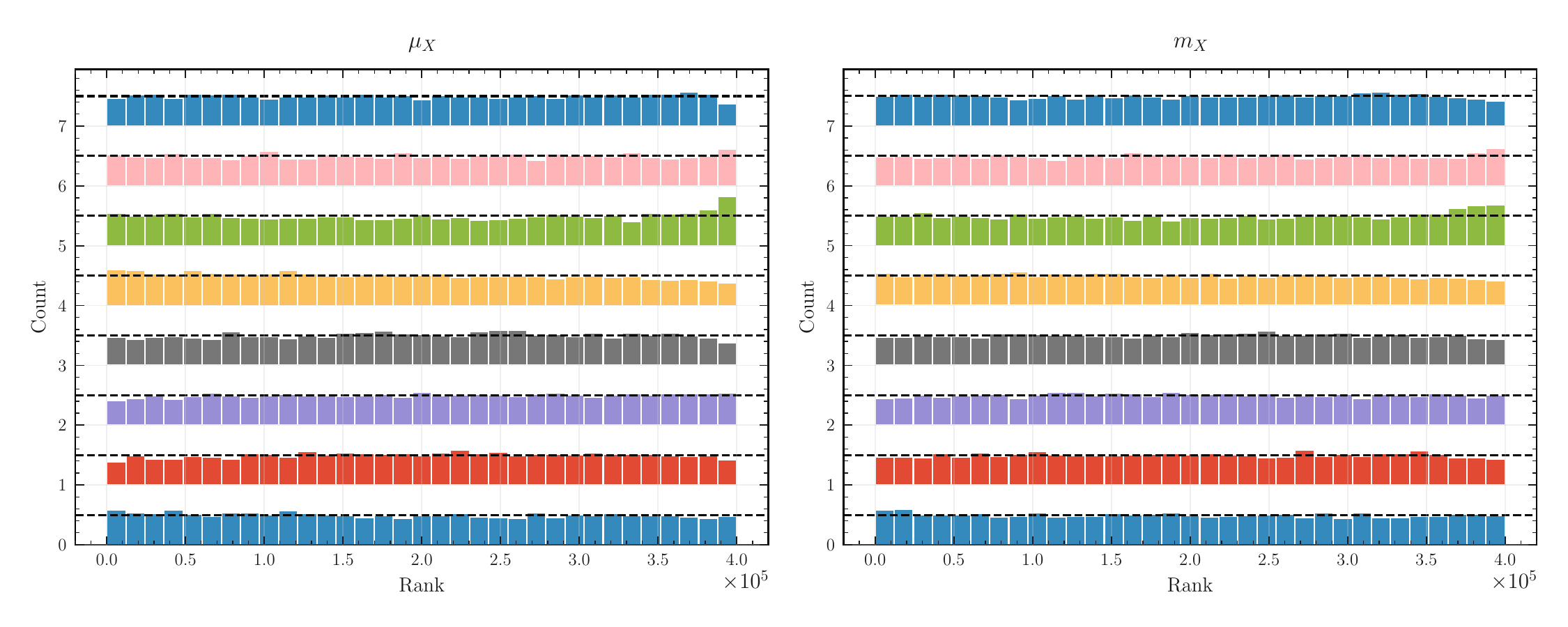}
    \caption{Rank plots for the \BKX signal strength $\mu_X$ and mass $m_X$ in \cref{eq:bkx-width}, from the posteriors discussed in \cref{sec:bkx-results} (\cref{fig:bkx-posterior}). The \textit{top}/\textit{middle}/\textit{bottom} two panels represent the $\Gamma_X = 0.1/0.5/1.0~\text{GeV}$ cases, respectively. Each colour represents a different \ac{MCMC} chain. The uniform distributions indicate good mixing.}
    \label{fig:rank-plots-bkx}
\end{figure}

Trace plots (see \cref{fig:trace-plots-bkx-01,fig:trace-plots-bkx-05,fig:trace-plots-bkx-10}) show ideal ``hairy caterpillar'' behaviour with rapid mixing and no visible trends. Rank plots (see \cref{fig:rank-plots-bkx}) display uniform distributions across all chains, confirming good mixing. 

These results provide confidence that the posterior constraints presented in \cref{tab:bkx results} are statistically robust and suitable for physics interpretation. The diagnostic validation ensures that reported credible intervals accurately reflect posterior uncertainty and that no additional sampling is required.

\section{Signal injection study}
\label{sec:bkx-injection}
\begin{figure}
    \centering
    \includegraphics[width=0.4\linewidth]{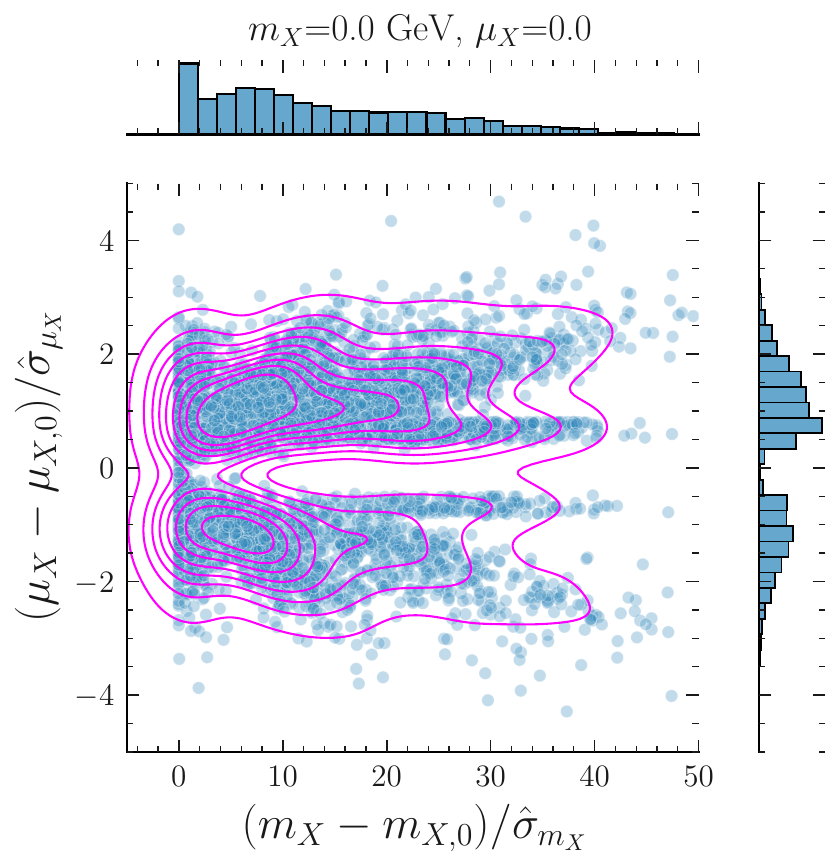}
    \includegraphics[width=0.4\linewidth]{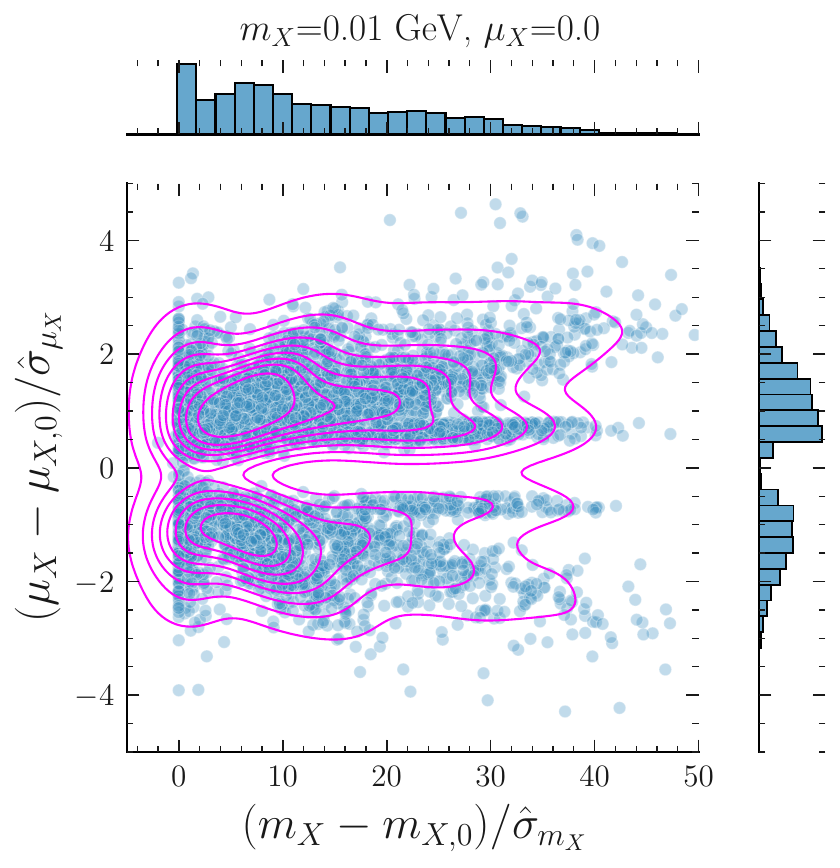}
    \includegraphics[width=0.4\linewidth]{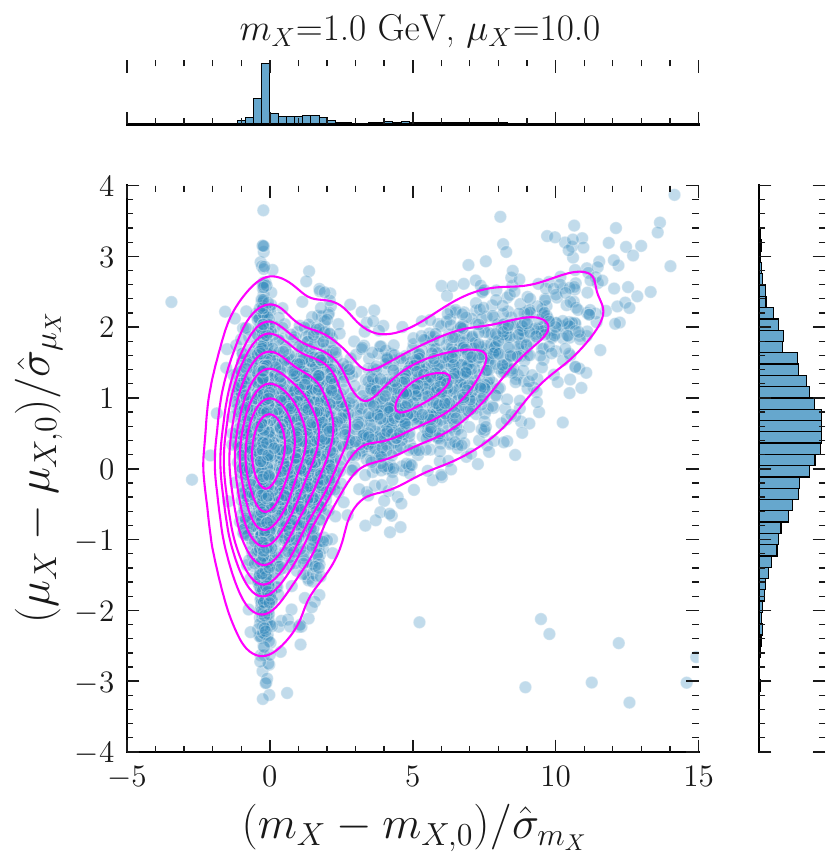}
    \includegraphics[width=0.4\linewidth]{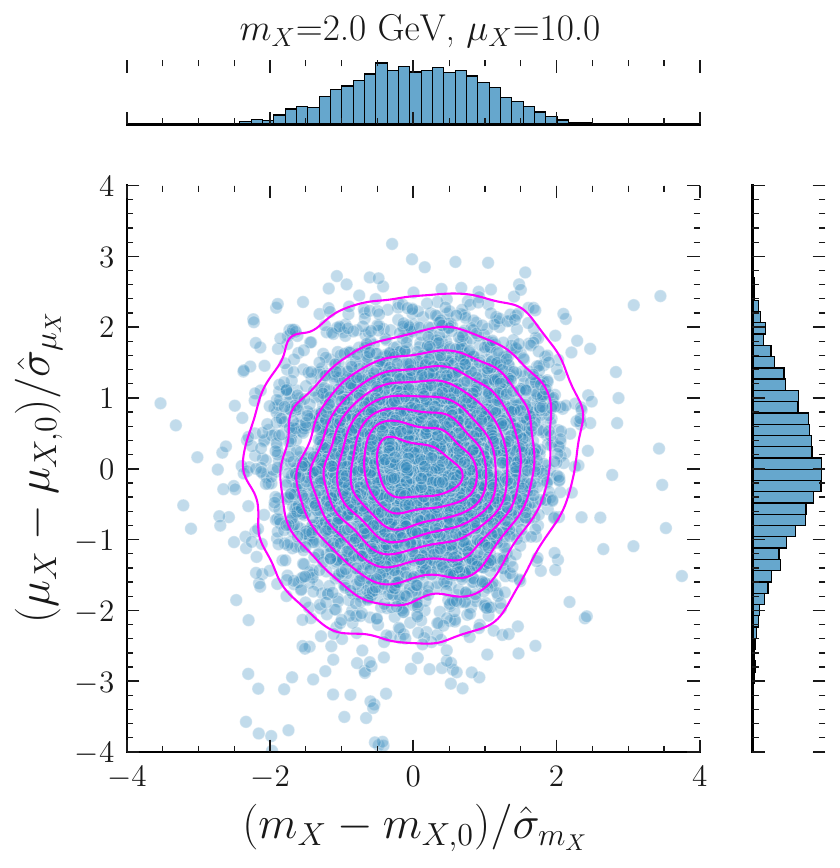}
    \includegraphics[width=0.4\linewidth]{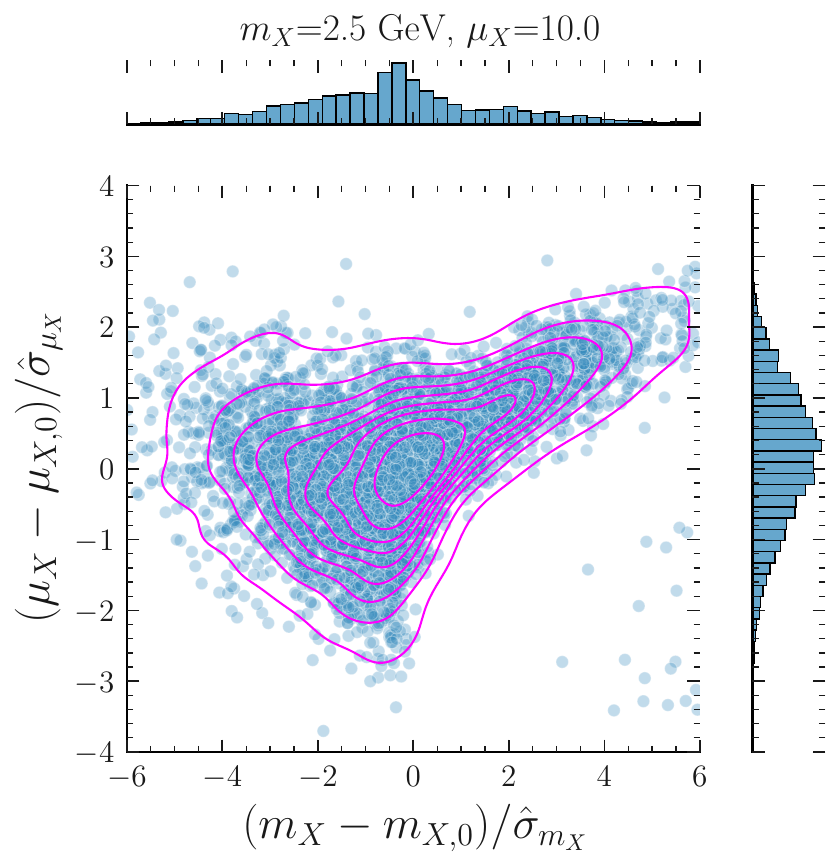}
    \includegraphics[width=0.4\linewidth]{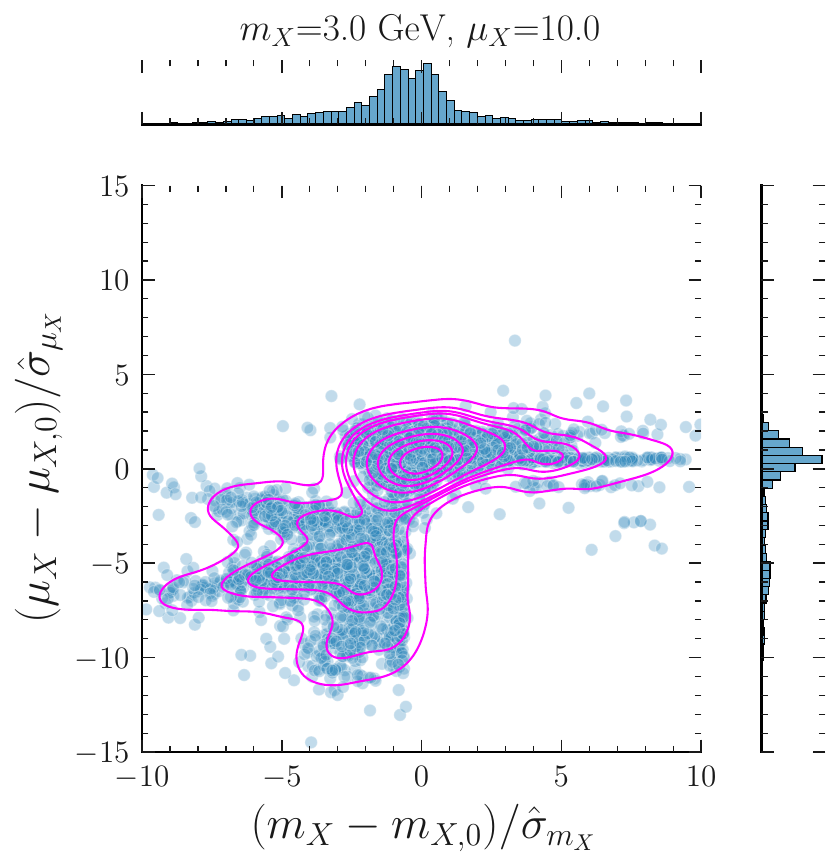}
    \caption{The pulls in $m_X$ and $\mu_X$ for each injection study, based on 5000 toy fits. The injected parameter point is displayed at the top of each plot.}
    \label{fig:injection}
\end{figure}

To quantify the ability to correctly identify a specific parameter point in the data, I perform an injection study in which the model is fitted to toy data drawn from the \ac{PDF} at a given point in $m_X$ and $\mu_X$. The model considered in this study corresponds to the case $\Gamma_X = 0.1~\text{GeV}$. In \cref{fig:injection}, I show the pulls
\begin{equation}
    \frac{\hat{\eta} - \eta_{0}}{\hat{\sigma}_{\eta}},
\end{equation}
in the 2-dimensional parameter space of $m_X$ and $\mu_X$.

Ideally, the pulls are expected to follow an uncorrelated multivariate normal distribution, as observed for the point $m_X = 2~\text{GeV}$ and $\mu_X = 10$. However, due to the complex nature of the model and the low efficiency for $m_X > 3~\text{GeV}$, distinct features arise depending on the injected point. For instance, if $\mu_X = 0$ exactly, $m_X$ becomes completely unconstrained, leading to the observed gap in the pulls for $\mu_X$ at the corresponding injected points. This strong parameter correlation also contributes to deviations from normality in the pull distributions. Additionally, the mass is bounded as $0 \leq m_X \leq 4.8~\text{GeV}$, resulting in a sharp cutoff of the pull distribution at the lower bound. Finally, it should be noted that $q^2_{\rm rec}$ is binned into only three intervals, $[-1, 4, 8, 25]~\text{GeV}^2$, which further limits the precision with which the mass can be determined.

%% file: chapters/appendix-knunu-update.tex
\section{Individual contributions}
\label{sec:update-contributions}
The work presented in \cref{sec:reinterpretation knunu mall,sec:reinterpretation knunu wet} is a joint effort by the Belle~II collaboration. The main proponents of this work are
Valerio Bertacchi (V. B.),
Yulan Fan,
Lorenz G\"artner (L. G.),
Eldar Ganiev,
Alexander Glazov (A. G.),
Yubo Han,
Danylo Kulakov,
Arsenii Kucher,
Meihong Liu,
Yuriy Onishchuk,
Sebastiano Raiz,
Niharika Rout,
Caspar Schmitt and
Slavomira Stefkova.

The theory predictions and decorrelated variations due to the form factor uncertainties presented in \cref{sec:update theory} were computed by L. G.

The statistical inference presented in \cref{sec:signal_extraction} was performed by L. G. in collaboration with V. B. and with guidance from A. G. In detail, the construction of the individual \pyhf models for all decay channels was refined by L. G. from the previous analysis~\cite{Belle-II:2023esi}. The combination of the \pyhf models and the initial fitting strategy were designed by L. G. The profiled likelihood scans for the uncorrelated, isospin average and 4-channel correlated models were also performed by L. G. The post-fit correlations were computed by L. G. and plotted by V. B. The impact of the systematic uncertainties on the \acp{POI} in \cref{tab:systImportance} were computed by V. B.

%% file: chapters/acronym.tex
\begingroup
\newcommand{\ThesisAcronym}[3]{\acro{#1}[#2]{#3}}
\newcommand{\ThesisAcronymPlural}[2]{\acrodefplural{#1}{#2}}
\begin{acronym}[LONGEST]
    \input{chapters/acronym-entries}
\end{acronym}
\endgroup


%% file: chapters/ai-disclaimer.tex
In the preparation of this thesis, I made limited use of language tools and large language models (LLMs), for non-substantive editorial purposes. These tools were used to correct typographical and grammatical errors, improve sentence clarity and fluency, and perform light paraphrasing to enhance readability. At no point were these tools employed to generate original content, conduct research, develop arguments, or contribute to the intellectual substance of this work. All ideas, analyses, and conclusions are entirely my own.

%% file: chapters/acknowledgements.tex
My deep gratitude goes to my advisors, collaborators, friends, family, and everyone else who has made the past years such a wonderful experience. The work presented in this thesis would not have been possible without the close guidance, idea-sparking discussions, and heart-warming support you gave me.

First and foremost I want to thank Prof. Dr. Thomas Kuhr for being a great supporter of my work, but also beyond. Our discussions always helped me to think clearly and find solutions. You encouraged me to pursue new ideas and created many opportunities for me, which made the last years extremely diverse and incredibly interesting.

Dr. Nikolai Krug, I want to thank for the many great discussions we had over the years. Watching you solve problems and deep-diving into ideas, while always remaining curious and playful, was truly inspiring. I appreciate the love we share for statistics and tough programming puzzles, and everything you taught me.

Prof. Dr. Slavomira Stefkova, you have been a guiding light since the early days of my PhD. I am amazed by your ability to always see things in a positive light, and find amazing compromises. I will never forget the fun we had in all our projects and trips together.

Prof. Dr. Danny van Dyk, your deep expertise in so many topics has always amazed me. It was always a pleasure to work with you, and of course I will never forget our summers in La Clusaz together.

Dr. M\'{e}ril Reboud, thank you for your guidance on so many occasions. You were always happy to discuss with me and support me in all our projects together. I loved our time in La Clusaz together and was super happy to have a fellow hiking companion.

Prof. Dr. Lukas Heinrich, thank you for being an inspiration in so many ways. I am grateful for our projects together and what I could learn from you, navigating and bringing together so many fields and topics.

I also want to thank Dr. Matthew Feikert and Dr. Giordon Stark for being inspiring colleagues over the years. You have taught me a lot about scientific software.

I want to thank Prof. Dr. Thomas Kuhr, Prof. Dr. Otmar Biebel, Prof. Dr. Danny van Dyk, Prof. Dr. Joseph Mohr and Dr. Sascha Mehlhase for agreeing to be on my defence committee. 

I want to thank everyone who has read and reviewed parts of this thesis, and has given me advice along the way. Special thanks go to Dr. Thomas Lueck, Dr. Nikolai Krug, Prof. Dr. Slavomira Stefkova, Dr. Markus Prim, and Caspar Schmitt.

I want to thank all my close colleagues who have made the last years an unforgettable journey. You have guided me through my initial days, until the very end of my thesis work. I felt the deepest support along every step of the way. 

Lastly, I want to thank my family and friends, without whom I would definitely not be where I am today. I owe you so much, and will always be thankful for the strength you provide me with. Your support means the universe to me.